\documentstyle[12pt]{article} \textheight=22cm
\newcommand{\beq}{\begin{equation}} \newcommand{\eeq}{\end{equation}}

\newcommand{\beqa}{\begin{eqnarray}}
\newcommand{\eeqa}{\end{eqnarray}} \newcommand{\ot}{\otimes}
\newcommand{\longto}{\longrightarrow} \newcommand{\nibun}{\frac{1}{2}}
\newcommand{\lmd}{\lambda} \newcommand{\Lmd}{\Lambda}
\newcommand{\nipi}{\frac{1}{2\pi i}} \newcommand{\ad}{{\rm ad}}
\newcommand{\Lie}{{\rm Lie}} \newcommand{\h}{{\rm Lie}(H)}
\newcommand{\liet}{{\rm Lie}(T)} \newcommand{\lietg}{{\rm Lie}(T_{G})}
\newcommand{\lieth}{{\rm Lie}(T_{H})} \newcommand{\g}{{\rm Lie}(G)}
\newcommand{\lieh}{h} \newcommand{\lieg}{{\rm g}}
\newcommand{\spl}{sl} \newcommand{\gl}{gl} \newcommand{\n}{n}
\newcommand{\HC}{H_{\!\C}} \newcommand{\Hc}{H_{\!\bf c}}
\newcommand{\semidir}{\tilde{\times}} \newcommand{\met}{\mbox{\sl g}}
\newcommand{\smet}{\mbox{\scriptsize {\sl g}}}
\newcommand{\bartial}{\bar \partial} \newcommand{\rA}{\mbox{\tiny A}}
 
\newcommand{\PP}{{\rm P}} \newcommand{\PPp}{{\rm P}_{\!+}}
\newcommand{\PPpk}{{\rm P}_{\!+}^{(k)}}
\newcommand{\PPptilk}{\PP_{\!+}^{(\tilk)}}
\newcommand{\Pv}{\PP^{\vee}} \newcommand{\tilP}{\tilde{\PP}}
\newcommand{\QQ}{{\rm Q}} \newcommand{\Qv}{\QQ^{\vee}}
\newcommand{\C}{{\bf C}} \newcommand{\R}{{\bf R}}  \newcommand{\Z}{{\bf Z}} \newcommand{\N}{\bf N}
\newcommand{\CP}{{\bf P}^1} \newcommand{\Ch}{{\rm C}}
 \newcommand{\alcv}{\widehat{{\rm C}}}
 \newcommand{\V}{{\rm V}}
\newcommand{\Waff}{W_{\!{\rm aff}}} \newcommand{\Waffh}{W'_{\!{\rm
aff}}} \newcommand{\Gmalcv}{\Gamma_{\alcv}} \newcommand{\tr}{{\rm tr}}
\newcommand{\ttr}{\,{}^{\tr}\!}  \newcommand{\trV}{\tr_{\!\!\:{}_V}}
\newcommand{\trP}{\tr_{\!\!\:{}_P}}
\newcommand{\trG}{\tr_{\!\!\:{}_G}}
\newcommand{\trH}{\tr_{\!\!\:{}_H}} \newcommand{\hp}{{\rm H}}
\newcommand{\Daff}{\Delta_{{\rm aff}}} \newcommand{\Vaff}{{\rm
V}_{\!{\rm aff}}} \newcommand{\PC}{P_{\!\C}} \newcommand{\A}{{\cal A}}
\newcommand{\AP}{{\cal A}_P} \newcommand{\Ac}{\A^{\circ}}
\newcommand{\G}{{\cal G}} \newcommand{\GP}{\G_P}
\newcommand{\GPC}{\G_{P_{\!\bf c}}} \newcommand{\Ph}{{\cal P}}
\newcommand{\NN}{{\cal N}} \newcommand{\NNc}{\NN^{\circ}}
\newcommand{\VV}{{\cal V}} \newcommand{\TT}{{\cal T}}
\newcommand{\dN}{d_{\NN}} \newcommand{\dSf}{d_{S_{\!f}}}
\newcommand{\dNf}{d_{\NN_{\!f}}} \newcommand{\bnu}{\bar \nu}
\newcommand{\lnu}{\mbox{\large $\nu$}} \newcommand{\blnu}{\mbox{\large
$\bar \nu$}} \newcommand{\un}{\mbox{\small {\sl U}}}
\newcommand{\sun}{\mbox{\scriptsize {\sl U}}}
\newcommand{\mm}{\mbox{\small {\sl M}}} \newcommand{\tilH}{\tilde{H}}
\newcommand{\kh}{k^{\!\bf c}} \newcommand{\HcovH}{\Hc\!/\!H}
\newcommand{\tilh}{{\tilde{h\,\,}\!\!}}  \newcommand{\tila}{\tilde{a}}
\newcommand{\tilb}{\tilde{b}} \newcommand{\LWZ}{{\cal L}_{\!{}_{\rm
W\!Z}}} \newcommand{\Pic}{{\rm Pic}} \newcommand{\Jac}{{\rm Jac}}
\newcommand{\Sgmtau}{\Sigma_{\tau}} \newcommand{\Ker}{{\rm Ker}}
\newcommand{\Vol}{{\rm vol}} \newcommand{\HH}{{\rm H}}
\newcommand{\tot}{{\rm tot}} \newcommand{\Aut}{{\rm Aut}}
 \newcommand{\vb}{\underline v}
\newcommand{\lbackslash}{\mbox{\large $\backslash$}}
\newcommand{\lslash}{\mbox{\large /}} \newcommand{\hgmm}{h_{\gamma}}
\newcommand{\dzz}{\left(\frac{dz}{z}\right)}
\newcommand{\DiffS}{Di\!f\!fS^1} \newcommand{\DiffoS}{Di\!f\!f_+S^1}
\newcommand{\DiffoSC}{Di\!f\!f_+\tilde{}S^1}

\newcommand{\bwhere}{\mbox{\Large ;}} \newcommand{\W}{{\rm W}}
\newcommand{\sgmV}{\sigma_{\!{}_V}}
\newcommand{\sleft}{{\scriptstyle(}}
\newcommand{\sright}{{\scriptstyle)}} \newcommand{\Dinf}{D_{\!\infty}}
 \newcommand{\Sgminf}{\Sigma_{\infty}}
\newcommand{\metau}{\met_{\tau}} \newcommand{\bz}{\bar \zeta}
\newcommand{\z}{\zeta} \newcommand{\pitau}{\frac{\pi}{\tau_2}}
\newcommand{\pinitau}{\frac{\pi}{2\tau_2}}
\newcommand{\sgmad}{\sigma_{\!\ad}} \newcommand{\OO}{{\cal O}}
\newcommand{\ooint}{\mbox{\footnotesize $\OO$}\!\!\!\!\!\!\!\:\int}
\newcommand{\Gree}{G^{z'}_{\bullet}\!}
\newcommand{\Green}{G^{z''}_{\bullet}\!\!}
\newcommand{\Chi}{\mbox{\large $\chi$}} \newcommand{\tilk}{\tilde{k}}
\newcommand{\coxh}{\lieh^{\vee}} \newcommand{\coxg}{\lieg^{\vee}}
\newcommand{\Rho}{\mbox{\large $\rho$}} \newcommand{\up}{\upsilon}
\newcommand{\dd}{d^2\!\!\:} \newcommand{\sq}{{\sl q}}
\newcommand{\sDelta}{{\sl \Delta}} \newcommand{\tinI}{\mbox{\tiny I}}
\newcommand{\tinJ}{\mbox{\tiny J}} 
\newcommand{\NP}{Nucl. Phys.\ } \newcommand{\PL}{Phys. Lett.\ }
\newcommand{\CMP}{Commun. Math. Phys.\ }

\newtheorem{th}{Theorem} \newtheorem{pn}[th]{Proposition}

\begin{document}
\baselineskip=0.63cm \normalsize

\renewcommand{\thefootnote}{\fnsymbol{footnote}}

\def\Komabanumber#1#2{\hfill \begin{minipage}{4cm} UT-Komaba #1
              \par\noindent #2 \end{minipage}}

\Komabanumber{94-3}{}

\vspace{2cm}
\begin{center}
{\huge {\sc On Global Aspects} }

\vspace{0.72cm} {\huge {\sc Of}}

\vspace{0.69cm} {\huge {\sc Gauged Wess-Zumino-Witten Model}}

\vspace{2.5cm} {\LARGE {\sc Kentaro Hori}}\footnote[2]{e-mail
address:\quad hori@tkyvax.phys.s.u-tokyo.ac.jp}

\vspace{1.8cm} {\Large {\sl Institute of Physics, University of Tokyo,
Komaba}}

\vspace{0.5cm} {\Large {\sl Meguro-ku, Tokyo 153, Japan}}

\vspace{5cm} {\Large January, 1994}
\end{center}

\newpage \renewcommand{\thefootnote}{\arabic{footnote}}

\renewcommand{\theequation}{0.\arabic{equation}}\setcounter{equation}{0}
{\large INTRODUCTION}

\vspace{1cm} \hspace{1.5cm} Soon after Albert Einstein has presented
the general theory of relativity, Hermann Weyl made an attempt to
derive the electromagnetic force as well as the gravitational force
from one common geometric structure of the space-time.\cite{Weyl} The
starting point of his theory is to insist that there is no preferred
scale structure in the universe, so that we have to choose a gauge at
each space-time point to measure the length of material bodies. Once a
gauge is chosen, scales at neighboring two points are compared with
the use of a one form $\varphi_{\mu}dx^{\mu}$. Namely, when a unit at
a point with coordinate $x^{\mu}$ is transported to a point with
coordinate $x^{\mu}+\Delta x^{\mu}$, it is observed as
$(1-\varphi_{\mu}(x)\Delta x^{\mu})$-times the unit there. If the
gauge is changed, the one form $\varphi_{\mu}dx^{\mu}$ and the metric
$g_{\mu \nu}dx^{\mu}dx^{\nu}$ are transformed as $$ \varphi_{\mu}\to
\varphi_{\mu}+\tau^{-1}\partial_{\mu}\tau\,,\qquad g_{\mu\nu}\to
\tau^{-1}g_{\mu \nu}\,,
$$ where $\tau$ is a function valued in positive real numbers. He
identified such one form $\varphi_{\mu}dx^{\mu}$ as the
electromagnetic-potential and developped a theory with the guiding
principle of gauge invariance.

His theory itself was not accepted by physical reasons but the
principle of gauge invariance survived over the years and became one
of the most important guiding principles in constructing candidates
for the fundamental theory of particles and fields.

After the birth of quantum mechanics, Weyl himself found that this
principle is applicable to the Dirac's theory of electrons provided
the `length' in the original argument is replaced by the `phase' which
has nothing to do with the metric $g_{\mu \nu}$. The theory of
electron field $\psi$ interacting with the electromagnetic fields is
invariant under the transformation $$ A_{\mu}\to
A_{\mu}+e^{-i\theta}\partial_{\mu}e^{i\theta}\,,\qquad \psi\to
e^{-i\theta}\psi\,, $$ where $-iA_{\mu}dx^{\mu}$ is the
electromagnetic potential and $e^{i\theta}$ is a function of the
space-time $M$ with values in the group $U(1)$ of phases. More
importantly, nothing like the phase of $\psi$ or the value of the
potential $-iA_{\mu}$ can be observed by physical measurements.

Proceeding analogously to the original argument, we may consider the
value $\psi(x)$ of the electron field at a point $x$ as the measured
value with respect to a chosen `$U(1)$-gauge' at $x$ denoted by a
symbol $s(x)$. The transformation $\psi(x)\to e^{-i\theta(x)}\psi(x)$
can then be considered as the effect of the change of gauge which is
denoted by $s(x)\to s(x)e^{i\theta(x)}$, so that the composite of
gauge and coefficient looks invariant $s(x)\cdot \psi(x)\equiv
s(x)e^{i\theta(x)}\cdot e^{-i\theta(x)}\psi(x)$. We are now tempted to
introduce the set $P$ of all $U(1)$-gauges over the space-time
points. There is a projection $\pi:P\to M$ such that the inverse image
$P_x=\pi^{-1}(x)$ of a point $x$ is the set of $U(1)$-gauges at
$x$. There is an action on $P$ of the group $U(1)$ coming from the
changing of gauges. A choice of gauge over $M$, or equivalently a
choice of gauge at each point of $M$ corresponds to the map $s:M\to P$
such that $s(x)$ is a gauge at $x$. Such a map is called a {\it
section } of $P$ or simply a gauge. If we can find a section $s$ with
respect to which the electron field behaves smoothly, we can identify
$P$ with the smooth manifold $M\times U(1)$ by $s(x)
e^{i\theta}\leftrightarrow (x,e^{i\theta})$.

If we define the set $P\!\times_{U(1)}\!V$ for $V=\C^4$ as the set of
equivalence classes in $P\times V$ under the equivalence relation
$(u,v)\sim (ue^{i\theta},e^{-i\theta}v)$, then, the electron field
$\Psi$ is a function on $M$ valued in $P\!\times_{U(1)}\!V$ where the
value $\Psi(x)$ at $x$ is the class represented by $(s(x),\psi(x))$
and denoted by $s(x)\cdot \psi(x)$. There is also a projection
$\pi_V:P\!\times_{U(1)}\!V\to M$ such that the inverse image
$V_x=\pi_V^{-1}(x)$ of a point is identified with the vector space
$V$.

If $x$ and $x'$ are two points in the space-time $M$, $\Psi(x)$ and
$\Psi(x')$ belong to different vector spaces $V_x$ and $V_{x'}$. With
the use of the potential $-iA_{\mu}dx^{\mu}$, the comparison of two
vectors becomes possible for infinitesimal separation between $x$ and
$x'$. Namely, if $x$ and $x'$ are `near' to each other with the
separation $\Delta x^{\mu}$, the vector $s(x)\cdot \psi(x)\in V_x$ is
transported to $s(x')\cdot(1-A_{\mu}(x)\Delta x^{\mu})\psi(x)\in
V_{x'}$ and compared with the vector $s(x')\cdot \psi(x')\in
V_{x'}$. It is easy to see that this transport is invariant under the
change of the defining gauge $s\to se^{i\theta}$ together with the
transformation of $A_{\mu}$ given above.  It is called the parallel
transport with respect to $A_{\mu}dx^{\mu}$. Taking the limit $\Delta
x^{\mu}\to 0$ of the difference, we obtain the covariant derivative
$d_{\!A}\Psi(x)=dx^{\mu}s(x)\cdot
(\partial_{\mu}+A_{\mu})\psi(x)$. The parallel transports assign to
each point $u$ of $P$ a subspace $H_uP$ (called the {\it horizontal
subspace}) of the tangent space $T_uP$ of $P$ consisting of vectors
tangent to the curves $s(x^{\mu}+t\Delta x^{\mu})e^{-tA_{\mu}(x)\Delta
x^{\mu}}\tilde{u}$ where $u=s(x)\tilde{u}$. This assignment is
$U(1)$-covariant in the sense that $H_uP$ is mapped by $e^{i\theta}\in
U(1)$ to $H_{ue^{i\theta}}P$. Hence the electromagnetic potential
$-iA_{\mu}dx^{\mu}$ gives rise to a geometric structure of the space
$P$ of gauges. In this sense, $A_{\mu}dx^{\mu}$ can be referred to as
the gauge field.

This observation has lead to the theory of connections of principal
bundles. It generalizes the above theory of gauges and gauge fields to
the situation in which the group $U(1)$ may be any Lie group $H$ and
in which there may not be a globally defined section $s:M\to
P$. Roughly speaking, a principal $H$-bundle $P$ is a set of all
$H$-gauges at points of $M$. A connection $A$ is to assign a
horizontal subspace to each tangent space of $P$ in $H$-covariant
way. If we choose an $H$-gauge $s:U\to P$ over an open set $U$ of $M$,
the connection $A$ defines a one form $A^s=A^s_{\mu}dx^{\mu}$ on $U$
valued in $\h$ called the gauge field, in the similar way as in the
case of electromagnetism. Under the change $s\to s g$ of gauges, $A^s$
is transformed as $$ A^s\longto A^{sg}=g^{-1}A^s g+g^{-1}dg\,.  $$ For
a representation $\rho :H\to GL(V)$ on a vector space $V$, we can
associate to a principal bundle $P$ over $M$ a {\it vector bundle}
$P\!\times_{H}\!V$ over $M$ defined by the equivalence relation
$(u,v)\sim (uh,\rho(h)^{-1}v)$ in $P\times V$. A connection $A$
determines a covariant derivative $d_{\!A}$ on sections of
$P\!\times_H\!V$ just as in the case of electron fields.

The above is a purely mathematical generalization. However, with the
aid of experimental supports, Yang-Mills fields which are
generalizations of the electromagnetic fields to the case of $H=SU(2)$
or $SU(3)$ have come to be believed to mediate some of the fundamental
interactions of nature.

According to R.P.Feynman, in such a quantum gauge theory, we have to
sum over all configurations that the gauge field can take. A question
arises : Should we sum over connections also of non-trivial principal
bundles? (A principal bundle $P$ over $M$ is said to be trivial when
we can find a globally defined continuous section $s:M\to P$.) An
affirmative answer was found in the1970s. The large mass of $\eta'$
meson may be explained by taking into account the gauge fields of all
possible kinds of $SU(3)$-bundles over $S^4$. Hence, it seems
important to investigate quantum gauge theory for topologically
non-trivial configurations and to ``sum over topologies''.

It seems that things become easier if we consider quantum gauge
theories in lower dimensions. Especially in two dimensions, it has
been observed\cite{BPZ} that exact treatment becomes possible for some
class of conformally invariant theories by a neat handling of the
infinite dimensional symmetry and its representation theory. In this
paper, we shall study exact relationship between the amplitudes for
topologically distinct configurations in gauged Wess-Zumino-Witten
(WZW) models --- a class of soluble conformaly invariant theories.

This work is inspired by the following
observation\cite{Gep,Moore-Seiberg,LVW}. The infinite conformal
symmetries of the gauged WZW model with compact target group $G$ and
compact gauge group $H$ are identified with the Virasoro generators by
the so called $G/H$ coset construction \cite{GKO}. This $G/H$ coset
constructed Virasoro algebra acts on the space of highest weight
vectors for the $H$-current algebra in the representation spaces of
the $G$-current algebra. A certain class of external automorphisms of
$G$ and $H$-current algebra, called {\it spectral flows}, keep
invariant the coset constructed Virasoro generators and the highest
weight vector conditions for the $H$-current algebra. Then, a spectral
flow induces a certain identification of states with respect to the
coset Virasoro algebra. At this stage, a natural question arises :
``Does this algebraic identification of states $\gamma|O\rangle\equiv
|O\rangle$ lead to the identification of the corresponding quantum
fields $\gamma O\equiv O$ ?''  To this problem, there has been no
answer with satisfactory explanation. The present paper gives the
following answer : ``Yes, provided that the gauge group $H$ is chosen
appropriately within the local isomorphism class and that the
contributions from all the topological types of $H$-bundles are taken
into account.'' A more refined statement is that a correlator for an
$H$-bundle $P$ with $\gamma O$-insertion at a point coinsides with a
correlator for another bundle $P\gamma$ with $O$-insertion at the same
point.

The above answer has the following significance. First of all, it
shows that a large number of soluble conformal field theories are
precisely described as lagrange field thories. Any of the unitary
minimal models (bosonic or supersymmetric), parafermionic models,
$N=2$ coset models or the twisted topological models and so on is
equivalent to a gauged WZW model (or WZW model and free fermionic
systems coupled to gauge field) with suitable choice of the gauge
group. For example, $(k+2,k+3)$-minimal model is equivalent to the
level $(k,1)$ WZW model with target $SU(2)\times SU(2)$ and gauge
group $SO(3)=SU(2)/{\Z}_2$ which acts on the target group by $h :
(g_1,g_2)\mapsto (hg_1h^{-1},hg_2 h^{-1})$. Second, it may help to
calculate the correlators. The calculation of a correlator for
$H$-bundle $P$ with $O$-insertion may be easier than the calculation
of the correlator for $H$-bundle $P\gamma^{-1}$ with insertion $\gamma
O$ or vise versa.

The answer is based on the obvious identification \beq \pi_0(LH)\cong
\pi_1(H)\,,
\label{pi0pi1}
\eeq between the set of connected components of the loop group $LH$
and the fundamental group of $H$. A spectral flow mentioned above can
be thought of as a representative of an element of $\pi_0(LH)$ and
$\pi_1(H)$ is identified with the set of topological types of
$H$-bundles. So it seems that we have only to interpret the spectral
flow as acting on the set of $H$-bundles through the above
identification (\ref{pi0pi1}), as implied in
\cite{Moore-Seiberg}. However, the story is more involved.

In the actual treatment of quantum gauge theory, we first perform the
integration along each orbit of the chiral gauge transformations
$A_{\bar z}\to h^{-1}A_{\bar z}h+h^{-1}\partial_{\bar z}h$
\cite{GawKup} which is analogous to Weyl's original gauge
transformations, and then sum up over the orbits (=over the
isomorphism classes of holomorphic $\HC$-bundles). Then, we arrive at
the following three systems coupled to the common background gauge
field representing an isomorphism class of holomorphic $\HC$-bundles :
$G$-WZW model with positive integral level, $\HC/H$-WZW model with
negative level and a free fermionic system called the ghost
system. Hence we must find a spectral flow that can be interpreted as
acting on the set of isomorphism classes of holomorphic
$\HC$-bundles. It turns out that this is possible only if $H$ is
abelian for which the moduli space of holomorphic $\HC$-bundles of one
topological type is isomorphic to that of another.

In general, the moduli spaces for different topological types are not
isomorphic. On the sphere, since there is only one holomorphic
$\HC$-bundle for each topological type that gives a non-negligible
contribution to the correlator, the spectral flow may be interpreted
as mapping one `moduli space' to another. However, the transformation
by a spectral flow on the set of gauge invariant fields is intricate
and unclear perhaps because the transformation on the representing
gauge fields is not canonically determined. This intricacy of the
spectral flow has been observed in ref.\cite{Nakatsu-Sugawara} which
deals with the twisted $N=2$ coset models on the sphere.

Hence, for a general compact group $H$, we need further
reformulation. The solution is to refer to the flag structure at the
insertion point of the field under consideration. We define a field
$O(f)$ called the {\it flag partner} of $O$ for each gauge invariant
field $O$ and for each flag $f$ at the insertion point. It is shown
that the integration of $O(f)$ over flags reproduces $O$. The spectral
flow is defined with respect to a flag and is shown to act on the set
of flag partners $O(f)\to \gamma O(f)$. The same spectral flow can be
interpreted to act on the set of isomorphism classes of holomorphic
$\HC$-bundles with flag structure at the insertion point. It is
checked in several examples that a component of the moduli space
relevant in the path-integral is transformed by such a spectral flow
to another relevant component of the moduli space. In this way, we
arrive at the above answer.

\vspace{0.3cm} The rest of this paper is organized as follows.

 In Chapter 1, we quantize the free fermionic systems with background
gauge fields. The representations of infinite dimensional groups on
the space of states are discussed and the spectral flow is
geometrically realized.

In Chapter 2, we deal with the WZW model of compact simply connected
target group $G$ coupled to background gauge field with gauge group
$G/Z_G$. WZW actions for topologically non-trivial $G/Z_G$-bundles are
constructed. Gauge covariant operator formalism is developped and the
spectral flow is again realized with geometric meaning.

In Chapter 3, we give a method to perform the integration over the
gauge fields. The structure of orbits for the chiral gauge
transformation group is reviewed. Neglecting irrelevant orbits, a
correlator is expressed as an integral over the moduli space of
holomorphic $\HC$-bundles. The latter half of the chapter is devoted
to the analysis of of $\HC/H$-WZW model which describes an integral
over each chiral gauge orbit.

In Chapter 4, we show the relations between correlators leading to the
field identifications. We first introduce the flag partners of gauge
invariant fields and find new integral expressions for correlators ---
integrals over the moduli space of holomorphic $\HC$-bundles with flag
structure at one point. In several examples, we give explicit
description of the moduli space of holomorphic $\HC$-bundles with flag
structure that gives non-negligible contribution to the
integral. Observing the above mentioned identification of the moduli
spaces, we obtain the desired relations.

Chapter 5 is the application of the theory of field
identification. The existence of a gauge invariant field fixed by a
spectral flow is shown to imply the non-vanishing of the partition
function for the corresponding $H$-bundle on the torus. The
calculation is attempted for the $SU(2)\times SU(2)$ WZW model coupled
to the gauge fields of the non-trivial $SO(3)$-bundle.

\vspace{0.4cm} The main part is Chapter 4. Among several new results,
the following are the principal products of the paper which play
essential roles in the derivation of the relation (\ref{FI}) that
leads to field identification:

\vspace{0.1cm} (i) the equation (\ref{intexpr}) which expresses a
dressed gauge invariant field as an integral over the flag manifold,

(ii) the new integral expression (\ref{newintexpr}) for correlation
functions.

\newpage {\large TABLE OF CONTENTS}

\vspace{1cm}
\noindent{\bf Chapter 1. Free Fermions\hfill \pageref{ch.1}}

\vspace{0.1cm} 1.1 Path Integral Quantization\hfill \pageref{1.1}

1.2 The Operator Formalism\hfill \pageref{1.2}

1.3 Field-State Correspondence and Spectral Flow \hfill \pageref{1.3}

1.4 Theories of Higher Ranks \hfill \pageref{1.4}

\vspace{0.3cm}
\noindent{\bf Chapter 2. Wess-Zumino-Witten Model \hfill
\pageref{ch.2}}

\vspace{0.1cm} 2.1 Path Integral Quantization\hfill \pageref{2.1}

2.2 The WZW Action for General Principal $H$-bundle\hfill
\pageref{2.2}

2.3 Covariant Operator Formalism\hfill \pageref{2.3}

2.4 The Spectral Flow\hfill \pageref{2.4}

\vspace{0.3cm}
\noindent{\bf Chapter 3. Integration over Gauge Fields\hfill
\pageref{ch.3}}

\vspace{0.1cm} 3.1 The Space of Gauge Fields\hfill \pageref{3.1}

3.2 The Path Integration\hfill \pageref{3.2}

3.3 WZW Model with Target Space $\HC/H$\hfill \pageref{3.3}

\vspace{0.3cm}
\noindent{\bf Chapter 4. Field Identification\hfill \pageref{ch.4}}

\vspace{0.1cm} 4.1 The Flag Partner\hfill \pageref{4.1}

4.2 A New Integral Expression\hfill \pageref{4.2}

4.3 The Moduli Space of Holomorphic Principal Bundles with Flag
Structure

\hspace{1cm} --- Examples\hfill \pageref{4.3}

4.4 The Hecke Correspondence\hfill \pageref{4.4}

4.5 Field Identification\hfill \pageref{4.5}

\vspace{0.3cm}
\noindent{\bf Chapter 5. Sample Calculations\hfill \pageref{ch.5}}

\vspace{0.1cm} 5.1 Differential Equations for Partition and
Correlation Functions

\hspace{0.7cm}of WZW Model on Torus\hfill \pageref{5.1}

5.2 Torus Partition Function for the Trivial Principal Bundle\hfill
\pageref{5.2}

5.3 Torus Partition Functions for Non-Trivial Principal Bundles\hfill
\pageref{5.3}

\vspace{0.3cm}
\noindent{\bf Concluding Remarks\hfill \pageref{conclude}}

\vspace{0.2cm}
\noindent{\bf Acknowledgement\hfill \pageref{acknowledge}}

\vspace{0.2cm}
\noindent{\bf Appendix 1. Lifting the Adjoint Action\hfill
\pageref{a.1}}

\vspace{0.2cm}
\noindent{\bf Appendix 2. Root Systems and Weyl Groups\hfill
\pageref{a.2}}

\vspace{0.2cm}
\noindent{\bf Appendix 3. Orthogonality of Characters\hfill
\pageref{a.3}}

\vspace{0.2cm}
\noindent{\bf References\hfill \pageref{reference}}

\vspace{0.2cm}
\noindent{\bf Index of Notation\hfill \pageref{indexofnotation}}

\newpage
\renewcommand{\theequation}{1.0.\arabic{equation}}\setcounter{equation}{0}

{\large CHAPTER 1.  FREE FERMIONS}\label{ch.1}

\vspace {1cm} \hspace{1.5cm} In the first chapter, we review the
theory of free fermionic systems on a Riemann surface. In this paper,
free fermions appear in two different ways : As a constituent of the
matter system and as a ghost system which is indispensable in the
consideration of gauge theories. In the present chapter, we first
consider the simplest case of one component fermions of spin
$(\lmd,1-\lmd)$ where $\lmd\in \nibun\Z$, and next consider the
general multi-component ones. We proceed from the functional integral
quantization to the operator formalism. After establishing the
field-state correspondence, we look at the effect of screening local
fields by external gauge fields of certain configurations. This leads
to the geometric definition of the spectral flow.

Now we start with an action integral \beq
I_{\Sigma,L}=\frac{i}{2\pi}\int_{\Sigma}
\psi_{-}\bar\partial_{\!A}\,\psi_{+}+\bar\psi_{-}\partial_{\!A}\,\bar\psi_{+}.
\label{fermiaction}
\eeq $\Sigma$ is a compact oriented two-dimensional Riemannian
manifold with metric $\met$ without boundary. The metric $\met$ makes
$\Sigma$ a Riemann surface with canonical bundle $K=T^*\Sigma_{\bf
C}^{(1,0)}$. Roughly speaking, $K$ is a holomorphic line bundle with
local holomorphic section $dz$ where $z$ is a local coordinate of the
Riemann surface $\Sigma$. We choose also a hermitian line bundle $L$
with $U(1)$-connection $d_{\!A}$. $\psi_{+}$ is an anti-commuting
field taking values in $K^{1-\lambda}\ot L$ where $1-\lambda$ is the
spin of $\psi_+$.  $\psi_{-}$ takes values in $K^{\lambda}\ot L^{-1}$
, $\bar \psi_{-}$ takes values in $\bar K^{1-\lambda}\ot L^{-1}$ and
$\bar \psi_{+}$ takes values in $\bar K^{\lambda}\ot L$. Note that
$K^{\lmd}$ is a holomorphic line bundle with local holomorphic section
$(dz)^{\lmd}$. If $\lmd$ is a half-integer, we must choose a sign,
i. e. we choose a spin structure within the $2^{2g}$-possibilities
where $g$ is the genus of $\Sigma$. $\bar\partial_{\!A}$ is the
$(0,1)$ part of the covariant exterior derivative on $K^{1-\lambda}\ot
L$: If we choose a local frame $(dz)^{1-\lambda}\! \ot \sigma$ and
express $\psi_{+}$ as $(dz)^{1-\lambda}\! \ot \! \sigma
\psi_{+}^{\sigma}$ , we have
$\bar\partial_{\!A}\,\psi_{+}=(dz)^{1-\lambda}\!\ot \!\sigma d\bar z
(\partial_{\bar z}+A^{\sigma}_{\bar z})\psi_{+}^{\sigma}$ where
$d_{\!A}\,\sigma =\sigma\cdot A^{\sigma}$. The operator $\bar
\partial_{\!A}$ determines a holomorphic structure on $K^{1-\lmd}\ot
L$ by the following statement : A local section $\psi$ is holomorphic
when $\bar \partial_{\!A}\psi =0$.

\renewcommand{\theequation}{1.1.\arabic{equation}}\setcounter{equation}{0}
\vspace{0.2cm}
\begin{center}
{\sc 1.1 Path Integral Quantization}\label{1.1}
\end{center}
\hspace{1.5cm} The metric $\met$ and the hermitian structure of $L$
introduce hermitian inner products on the spaces of sections such as
$\Omega^{0}(\Sigma,K^{1-\lambda}\ot L)$ or as
$\Omega^{0,1}(\Sigma,K^{1-\lmd}\ot L)$. These inner products in turn
define Laplace operators $\bar\partial_{A}^{\dag}\bar\partial_{A}$ on
$\Omega^{0}(\Sigma,K^{1-\lambda}\ot L)$ and
$\Omega^{0}(\Sigma,K^{\lambda}\ot L^{-1})$ and
$\partial_{A}^{\dag}\partial_{A}$ on $\Omega^{0}(\Sigma,\bar
K^{1-\lambda}\ot L^{-1})$ and $\Omega^{0}(\Sigma,\bar K^{\lambda}\ot
L)$. Then, using the spectral decomposition we can define a measure
for functional integration if we regularize in a suitable way the
infinite product of eigenvalues of the Laplacian (see
e.g. \cite{Ray-Singer}, \cite{Quillen}). That is, we can consider the
following path integral: \beq Z_{\Sigma,L}(\,\met,A \,; {\cal O}\,
)=\int {\cal D}_{\!\!\smet,A}[\psi_{+}{\bar \psi_{+}}\psi_{-}{\bar
\psi}_{-}] \,\, e^{-I_{\Sigma,L}}\, {\cal O},
\label{defcorr}
\eeq where ${\cal O}={\cal O}(\psi_{+},\bar \psi_{+}, \psi_{-}, \bar
\psi_{-})$ is any functional of finitely many $\psi_{\pm}$'s and
${\bar \psi}_{\pm}$'s. For the integral to be non vanishing, $\cal O$
must include fermionic integration parameters for the zero modes of
$\bar \partial_{\!A}$ and $\partial_{\!A}$. By the Riemann-Roch
theorem, the necessary condition for the non vanishing is $\sharp
\,{\cal O}=c_{1}(L) +(2\lambda -1)(1-g)$ , where we count the fermion
number by $\sharp \psi_{+} =1$ and $\sharp \psi_{-}=-1$. \footnote{
Here, $c_{1}(L)=\int_{\Sigma}c_{1}(L,d_{A})$\label{chern} is an
integer, not a cohomology class. We often use this abbreviated form.}

As follows from the translational invariance of the measure ${\cal
D}\psi\bar \psi$, inserted in the correlator, the field $\psi_+$
behaves as a holomorphic section of $K^{1-\lmd}\!\ot L$ over the
region without any other operator-insertion. In the same sense,
$\psi_-$ behaves holomorphically and $\bar \psi_{\pm}$ behave
anti-holomorphically.  Another important property is the behavior of
the correlation functions including $\psi_{+}(z)\psi_{-}(w)$ as $z$
approaches $w$: \beq Z_{\Sigma,L}(\, \met,A\,; \,
\psi_{+}(z)\psi_{-}(w)\ {\cal O}\ )\sim
G_{\smet,A}(z,w)\,Z_{\Sigma,L}(\, \met,A\,; {\cal O}\, ).  \eeq
$G_{\smet,A}(z,w)$ is the green function for the $\bar \partial_{A}$
operator. Using local holomorphic frames $dz$ and $\sigma$ of $K$ and
$L$, we can write the asymptotic behavior as \beq
\psi_{-}(z)\psi_{+}(w)=(dz)_{z}^{\lambda}\otimes
\!\sigma(z)^{-1}\left\{\, \frac{1}{z-w}\, +\,
:\!\psi_{-}^{\sigma}(z)\psi_{+}^{\sigma}(w)\!\! : \,
\right\}(dz)_{w}^{1-\lambda}\! \otimes \! \sigma(w).
\label{eqn:ord}
\eeq The expression $:\psi_{-}^{\sigma}(z)\psi_{+}^{\sigma}(w):$ is
regular as $z\to w$ and we call this the {\it regularized product} of
the coefficients $\psi_{-}^{\sigma}(z)$ and $\psi_+^{\sigma}(w)$.

\vspace{0.3cm} \underline{Anomaly and Ward Identities}

Next we look at the response of path integrals to variations of $\met$
and $A$. We define the {\it fermion number current} $J$ and the {\it
energy-momentum tensor} $T$ by \beq \delta Z_{\Sigma,L}(\,\met,A\,;
{\cal O}\, )=Z_{\Sigma,L}\Bigl(\,\met,A \,\bwhere \left[\:
\frac{1}{4\pi}\int_{\Sigma}\hbox{\small $\sqrt{\met}$} d^{2}x \,\delta
\met^{ab}T_{ab} +\frac{1}{2\pi i}\int_{\Sigma}J\delta A \: \right]
{\cal O}\, \Bigr).
\label{defTJ}
\eeq

We first consider the variation of $(\met, A)$ which preserves the
holomorphic structure of $(\Sigma,L)$. Let ${\phi}$ be a real valued
function on $\Sigma$ and $h:L\longto L$ be an automorphism of complex
line bundle. We consider the conformal transformation $\met \to \met
e^{\phi}$ and chiral gauge transformation $A\to A^{h}$ where the
latter is given by $\bar \partial_{\!A^{h}}=h^{-1}\bar
\partial_{\!A}h$. Note that the classical action is invariant under
these transformations together with the replacement $\psi_{\pm}\to
h^{\mp}\psi_{\pm}$ and $\bar \psi_{\pm}\to h^{\ast \pm}\bar
\psi_{\pm}$. In quantum theory, however, due to the regularization of
the infinite determinant, this invariance is only preserved up to a
prefactor \cite{Fujikawa}: \beq Z_{\Sigma,L}(\, \met e^{\phi}, A^{h}
\,; {\cal O}\, )=e^{I(\smet,A;\phi, hh^{\ast})}Z_{\Sigma,L}(\, \met,
A\,; h{\cal O}\,), \label{eqn:anom} \eeq where the exponent of the
prefactor is given by \beqa I( \, \met,A\, ; \phi,
hh^{\ast})&=&\frac{c_{\lambda}i}{48\pi}\int_{\Sigma}\left(\, \partial
\phi \bar \partial \phi +2R_{\Theta}\phi \, \right
)-\frac{i}{2\pi}\left(\lambda-\frac{1}{2}\right)\int_{\Sigma}\left(
\,F_{\!A}\phi-\partial\phi \bar \partial \varphi \right)\nonumber \\
&&\mbox{}-\frac{i}{4\pi}\int_{\Sigma}\left[\,\partial \varphi \bar
\partial \varphi
-2\left(F_{\!A}+\left(\lambda-\frac{1}{2}\right)R_{\Theta}\right)\varphi
\, \right].
\label{page:clmd}
\eeqa $c_{\lambda}=1-12\left(\lambda-\frac{1}{2}\right)^{2}$ is a
number called the central charge of the system. $\varphi$ is a real
valued function defined by $hh^{\ast}=e^{\varphi}$. $h{\cal O}$ is
given by $h{\cal O}(\psi_{\pm}, \bar \psi_{\pm})={\cal
O}(h^{-1}\psi_{+}, \psi_{-}h, h^{\ast}\bar \psi_{+}, \bar
\psi_{-}h^{\ast -1})$. $R_{\Theta}$ (resp. $\!F_{\!A}$) is the
curvature of $K^{-1}$ (resp. $\!L$) determined by $\met$
(resp. $\!A$). This phenomenon is called conformal and chiral
anomaly. Note also that the prefactor $I(\met,A;\phi,hh^{\ast})$ is
zero for the unitary automorphism $h:L\longto L$ where
$hh^{\ast}=1$. This fact is important when we consider gauge theory.

Taking the differentials at the identity element of the group of
conformal and chiral gauge transformations, we can obtain some usefull
identities out of the formula (\ref{eqn:anom}) of anomaly. One of them
is the current Ward identity: \beqa
\lefteqn{Z_{\Sigma,L}\Bigl(\,\met,A \,\bwhere \,\frac{1}{2\pi
i}\int_{\Sigma}\left(\, \epsilon \bar \partial J
-\epsilon^{\ast}\partial \bar J \, \right) {\cal O} \,
\Bigr)}\label{eqn:Ward J}\\ &&
=\frac{i}{2\pi}\int_{\Sigma}(\epsilon+\epsilon^{\ast})\!\! \left(
F_{\!A} +\left(\lambda-\frac{1}{2}\right)R_{\Theta} \right)
Z_{\Sigma,L}(\,\met,A\,; {\cal O}\, )+Z_{\Sigma,L}\Bigl(\,\met,A
\,\bwhere \, J(\epsilon){\cal O}+{\bar J}(\epsilon){\cal
O}\,\Bigr),\nonumber \eeqa where $\epsilon$ is an infinitesimal chiral
gauge transformation, that is, a section of ${\rm End}L$ (which is
trivial in this abelian case). $J(\epsilon)$ and ${\bar J}(\epsilon)$
are operators defined by $J(\epsilon){\cal O}=(\frac{d}{dt})_{0}{\cal
O}(e^{\mp t \epsilon}\psi_{\pm}, \bar \psi_{\pm})$ and ${\bar
J}(\epsilon){\cal O}=(\frac{d}{dt})_{0}{\cal O}(\psi_{\pm}, e^{\pm t
\epsilon^{\ast}}\bar \psi_{\pm})$. From this, we can read that the
current $J$ is a holomorphic one form and $\bar J$ is anti-holomorphic
one form on an open set without another field-insertion nor with
curvature. Then by the same equation, we see that if $O$ is a local
field inserted at $x\in \Sigma$ and $\epsilon$ is a holomorphic
function on a neighborhood $U$ of $x$ over which $\met$ and $A$ are
chosen to be flat , the following relations hold: \beq \frac{1}{2\pi
i}\oint_{x}\,\epsilon J\, O =J(\epsilon)O, \qquad \frac{-1}{2\pi
i}\oint_{x}\epsilon^{\ast}\bar J\, O ={\bar J}(\epsilon)O\, ,
\label{eqn:Jepsi} \eeq where the contour in $U$ encircles $x$
once. The equations (\ref{eqn:Jepsi}) generalize the definition of
$J(\epsilon)$ and ${\bar J}(\epsilon^{\ast})$ to the case where
$\epsilon$ is a meromorphic function with possible singularity at
$x$. Such $J(\epsilon)$ corresponds to the variation of $A$ which
amounts to some deformation of the holomorphic structure of $L$.

The formula (\ref{eqn:anom}) shows also that the $(1,1)$ part of the
energy momentum tensor is given by \beq T_{z\bar
z}=-\frac{c_{\lambda}}{12}R_{\Theta z\bar z}+\left( \, \lambda
-\frac{1}{2}\, \right)F_{A z\bar z}.
\label{eqn:T1.1}
\eeq We see that we have $T_{z\bar z}=0$ on a region where $\met$ and
$A$ are flat.

Next, we consider the directions of variation of $(\met,A)$ associated
to a diffeomorphism $f$ of $\Sigma$. The metric $\met$ is naturally
transformed to $f^{\ast} \!\met$ and consequently $K$ is transformed
to $f^{\ast}\! K$. But there is no natural way to transform $A$
preserving $L$ since there is no natural way to lift the action of $f$
to $L$. For a one parameter group of diffeomorphisms $f_{t}$ generated
by some vector field $v$, however, we can lift the action to $L$ : the
horizontal lift ${\tilde f}_{t}:L\longto L$ along the
paths.\label{page1:ftilde} This induces an automorphism
$f_{t}^{\sharp}$ of $\Omega^{\ast}( \Sigma, L)$ by
$f_{t}^{\sharp}\psi={\tilde f}_{t}^{-1}\circ \psi \circ f_{t}$ for
$\psi\in \Omega^*(\Sigma,L)$. The variation $A\to A_{t}$ is defined by
$d_{\!A_{t}}=f_{t}^{\sharp}d_{\!A}f_{t}^{\sharp -1}$ and we have
$\delta d_{\!A}=i(v)F_{\!A}$. Let $f_{t}^{\flat}: \Omega^{\ast}
(\Sigma, K^{\nu}\ot L)\to \Omega^{\ast}(\Sigma, f_{t}^{\ast}\!
K^{\nu}\ot L)$ be the transformation acting as $f_{t}^{\ast}$ on
$K$-part and $f_{t}^{\sharp}$ on $L$-part. We can see that the
variation $(\met,A)\to (f_{t}^{\ast}\! \met, A_{t})$ is isospectral in
the sense that $f_{t}^{\flat}$ preserves the spectrum of the
Laplacians. Therefore, we have an identity: \beq Z_{\Sigma,L}(\,
f_{t}^{\ast}\! \met, A_{t}\,; {\cal O}\, )=Z_{\Sigma,L}(\, \met, A\,;
f_{t}{\cal O}\, )\, , \eeq where $f_{t}{\cal O}$ is given by
$f_{t}{\cal O}={\cal O}( f_{t}^{\flat}\psi_{\pm}, f_{t}^{\flat}\bar
\psi_{\pm})$. Taking the derivative at $t=0$, we have \beqa
\lefteqn{Z_{\Sigma,L}\Bigl(\, \met, A \,\bwhere \, \frac{1}{2\pi
i}\int_{\Sigma}dzd{\bar z}\Bigl[ \, \partial_{\bar
z}v^{z}T_{zz}+\left(\nabla_{z}v^{z}+\nabla_{\bar z}v^{\bar
z}\right)T_{z\bar z}+\partial_{z}v^{\bar z}T_{\bar z\bar z} \,
\Bigr]{\cal O}\, \Bigr)}\label{eqn:Ward T}\\ &&+Z_{\Sigma,L}\Bigl(
\,\met,A \,\bwhere \,\frac{1}{2\pi i}\int_{\Sigma}dzd{\bar z}\Bigl(\,
J_{z}v^{z}F_{Az\bar z}+J_{\bar z}v^{\bar z}F_{Az\bar z}\, \Bigr) {\cal
O}\, \Bigr)=Z_{\Sigma,L}(\, \met, A\,; L_{v}{\cal O}\, )\, ,\nonumber
\label{eqn:T22}
\eeqa where $L_{v}{\cal O}$ is defined by $L_{v}{\cal
O}=(\frac{d}{dt})_{0}f_{t}{\cal O}$. As in the case of the current
$J$, $T_{zz}(dz)^{2}$ is holomorphic and $T_{\bar z\bar z}(d\bar
z)^{2}$ is antiholomorphic on an open set without curvature or without
field-insertion. When there is a local field insertion in a flat
neighborhood, the Ward identity (\ref{eqn:T22}) shows the following
relations: \beq \frac{1}{2\pi i}\oint_{x}dz v^{z}(z)T_{zz}O =L_{v}O \,
, \quad \frac{-1}{2\pi i}\oint_{x}d{\bar z}v^{\bar z}(\bar z)T_{\bar
z\bar z}O =L_{\bar v}O \, ,
\label{eqn:Lv}
\eeq where $O$ is an local field inserted at $x$ , the contours
encircle $x$ and $v=v^{z}(z)\partial/\partial z$ is a holomorphic
vector field around $x$. We can use these equations to define $L_{v}$
and $L_{\bar v}$ for $v$ a meromorphic vector field with possible
singularity at $x$. Such a $v$ corresponds to the deformation of
complex structure of $\Sigma$. We exemplify the action of $L_{v}$ on
$\psi_{\pm}$ for holomorphic $v$:
$$L_{v}\psi_{+}(x)=(1-\lambda)\frac{\partial v^{z}}{\partial
z}(x)\psi_{+}(x)+v^{z}(x)\partial_{\! A z}\psi_{+}(x)\, .$$
$$L_{v}\psi_{-}(x)=\lambda\frac{\partial v^{z}}{\partial
z}(x)\psi_{-}(x)+v^{z}(x)\partial_{\! A z}\psi_{-}(x)\, .$$

\vspace{0.3cm} \underline{The Currents in terms of Fundamental Fields}

Now let us try to find expressions for the current $J$ and energy
momentum tensor $T$ in terms of the fundamental fields $\psi_{+}$ and
$\psi_{-}$. By looking at the classical action $I$, we guess that
$J=\psi_{-}\psi_{+}$ and $T=-\lambda\psi_{-}\partial_{\!
A}\psi_{+}+(1-\lambda)\partial_{\! A}\psi_{-}\psi_{+}$. But since the
product of $\psi_{-}(z)$ and $\psi_{+}(w)$ have singularity as $z\to
w$ , we must make sense out of the products $\partial^{n}
\psi_{-}\partial^{m}\psi_{+}$. Natural candidates are the following :
\beq J=\, :\!\psi_{-}\psi_{+}\!:\, ,\qquad\quad T=-\lambda :\!
\psi_{-}\partial_{\! A}\,\psi_{+}\!:+(1-\lambda):\!\partial_{\!
A}\psi_{-} \psi_{+}\!: \, , \eeq where the {\it regularized products}
of the fields are defined by \beq :\!\psi_{-}\psi_{+}\!:\!(z)\,
=\lim_{\epsilon\to 0}\left\{ \,
\psi_{-}(\zeta_{\epsilon})\tau_{\zeta_{\epsilon},\,
z}^{(\lmd,-)}\tau_{z,\,
w_{\epsilon}}^{(1-\lmd,+)}\psi_{+}(w_{\epsilon})-\frac{dz}{\zeta_{\epsilon}-w_{\epsilon}}\,
\right\} \, ,
\label{defnorm}
\eeq
$$:\!\psi_{\!-}\partial_{\! A}\psi_{+}\!\!:\!(z)=\lim_{\epsilon\to
0}\left\{ \, \psi_{-}(\zeta_{\epsilon})\tau_{\zeta_{\epsilon}, \,
z}^{(\lmd,-)}\tau_{z,\, w_{\epsilon}}^{(2-\lmd,+)}\partial_{\!
A}\psi_{+}(\!w_{\epsilon}\!)
-\frac{(dz)^{2}}{(\zeta_{\epsilon}-w_{\epsilon})^{2}}+\frac{(dz)^{2}}{12}\left(
\, \partial_{z}\omega_{z}-2(\omega_{z}\!)^{2} \right)\, \right\}\, ,$$
$$:\!\partial_{\! A}\psi_{\!-}\psi_{+}\!\!:\!(z)=\lim_{\epsilon\to
0}\left\{ \, \partial_{\! A}\psi_{\!
-}(\zeta_{\epsilon})\tau_{\zeta_{\epsilon},\,z}^{(1+\lmd,-)}\tau_{z,\,w_{\epsilon}}^{(1-\lmd,+)}
\psi_{\! +}(\!w_{\epsilon}\!) +
\frac{(dz)^{2}}{(\zeta_{\epsilon}-w_{\epsilon})^{2}}-\frac{(dz)^{2}}{12}\!
\left( \partial_{z}\omega_{z}-2(\omega_{z}\!)^{2} \right) \, \right\}
.$$ In the above expressions, $\zeta_{\epsilon}$ , $z$ and
$w_{\epsilon}$ are on a geodesic $\tau$ centered by $z$ with equal
distance $\epsilon$. $\tau_{\zeta,\,z}^{(\mu,-)}$ is the parallel
transport of $K^{\mu}\! \ot \! L^{-1}$ along $\tau$ from $\zeta$ to
$z$, and $\tau^{(\nu,+)}_{z,\,w}$ is the parallel transport of
$K^{\nu}\! \ot \! L$ along $\tau$ from $w$ to
$z$. $\omega_{z}=\met^{z\bar z}\partial_{z}\met_{z\bar
z}$\label{page:ch1.omegaz} is the connection form of $K^{-1}$. We have
added the term $(dz)^{2}(\partial_{z}\omega_{z}\! -\!
2\omega_{z}^{2})/12$ to give
$-(dz)^{2}/(\zeta_{\epsilon}-w_{\epsilon})^{2}$ a covariant meaning in
the limit $\epsilon \to 0$. We have used the data given only by $g$
and $A$ except for the direction of the geodesic $\tau$. Using the
assymptotic behavior of the product $\psi_{-}(\zeta)\psi_{+}(w)$, we
find that the limit $:\psi_{\!-}\psi_{+}\!:$ is well defined, but
$:\psi_{\! -}\partial_{\! A}\psi_{+}\! :$ as well as $:\partial_{\!
A}\psi_{\! -}\psi_{+}\! :$ contains an ambiguous term which depends on
this choice of the direction and which is present if the curvature
$F_{\! A}$ and $R_{\Theta}$ are non zero at $z$.  We have decided to
discard this term.

In terms of the regularized product (\ref{eqn:ord}) of the
coefficients $\psi_{\pm}^{\sigma}$, they are expressed as \beqa
J_{z}\, \, &=&\, \,
\,:\psi_{-}^{\sigma}\psi_{+}^{\sigma}:-A^{\sigma}_{z}-\left(\lambda-\frac{1}{2}\right)\omega_{z}\,
, \label{eqn:Jz}\\
T_{zz}&=&-\lambda:\psi_{-}^{\sigma}\partial_{z}\psi_{+}^{\sigma}: +\,
(1-\lambda):\partial_{z}\psi_{-}^{\sigma}\psi_{+}^{\sigma}:\label{eqn:Tzz}\\
&&-A^{\sigma}_{z}:\psi_{-}^{\sigma}\psi_{+}^{\sigma}:+\left(\lambda-\frac{1}{2}\right)\partial_{z}A^{\sigma}_{z}+\frac{1}{2}A_{z}^{\sigma}A_{z}^{\sigma}
-\frac{c_{\lambda}}{12}\left(
\partial_{z}\omega_{z}-\frac{1}{2}\omega_{z}\omega_{z}\right)\,
.\nonumber \eeqa We can explicitly check that these are independent of
the choice of of local holomorphic frames $dz$ and $\sigma$. It is
easy to see that the above candidates of current and energy momentum
tensor satisfy the Ward identities (\ref{eqn:Ward J}),
(\ref{eqn:T1.1}), (\ref{eqn:Ward T}). Therefore, we conclude that
these are the right expressions for $J$ and $T$ in terms of the
fundamental fields.

\renewcommand{\theequation}{1.2.\arabic{equation}}\setcounter{equation}{0}
\vspace{0.3cm}
\begin{center}
{\sc 1.2 The Operator Formalism}\label{1.2}
\end{center}
\hspace{1.5cm} We proceed to the construction of the operator
formalism. We first calculate the commutation relations of the fields
$\psi_+$, $\psi_-$ and currents $J_z$, $T_{zz}$ over a coordinatized
annular open subset of $\Sigma$. After the argument on the path
integral over fields on a Riemann surface with circle boundary, the
space of states is defined. The set of commutation relation of
currents is then interpreted as an operator algebra of the
infinitesimal generators of infinite dimensional groups acting on the
space of states.

Let us now choose an annular neighborhood $U$ of a parametrized circle
$c:S^1 \to \Sigma$ with a complex coordinate system $z$ such that
$z(c(\theta))=e^{i\theta}$. We assume that $S=c(S^1)$\label{page1:S}
devides $\Sigma$ into two such parts $\Sigma_{\infty}$ and
$\Sigma_{0}$ that $\Sigma =\Sigma_{\infty}\cup \Sigma_{0}$ ,
$\Sigma_{\infty}\cap \Sigma_{0}=S$ and that $|z(U\cap
\Sigma_{\infty})|\geq 1$. We also assume that the metric is chosen to
be $\met^{z \bar z}=8\pi^2 |z|^2$ on $U$ and the connection $A$ is
chosen to be flat on $U$ with holonomy $e^{-2\pi i a}$ along $S$. It
is easy to show that we can choose a unitary frame $s$ of $L$ on $U$
such that $d_{\!A}s=s\cdot(-i a \, d\theta)$ where $z=re^{i
\theta}$. Then, $\sigma(z)=s(z)|z|^{-a}$ is a holomorphic frame of $L$
on $U$ : $d_{\! A}\sigma=\sigma\cdot(- a) dz/z$.

\vspace{0.3cm} \underline{Laurent Expansion of Fields and Currents}

Since $\psi_{+}$ and $\psi_{-}$ are holomorphic sections, we can take
the Laurent expansions on $U$ with the following coefficients: \beqa
\psi_{n}^{(+)}&=&\frac{1}{2\pi i}\oint \dzz^{\!\lambda}\!\! \ot
\sigma(z)^{-1}z^n\, \psi_{+}(z)\, ,\\ \noalign{\vskip0.1cm}
\psi_{n}^{(-)}&=&\frac{1}{2\pi i}\oint
\psi_{-}(z)\dzz^{\!1-\lambda}\!\!\!\! \ot \sigma(z)\, z^n.
\label{eqn:psi-}
\eeqa In the above expressions, the contour may be any circle
homotopic to $S$ provided the homotopy does not pass by any other
field insertion. From now on, we take the prescription of radial
ordering on $U$. That is, if $\Phi_{1}(z)$ and $\Phi_{2}(w)$ are local
operators inserted in $U$ with $|z|>|w|$, we always put $\Phi_{1}(z)$
to the left of $\Phi_{2}(w)$ in any correlation function. Then, from
the singular property of the product of $\psi_{+}$ and $\psi_{-}$ , we
see that the following relations hold : \beq \{ \psi_{n}^{(+)} ,
\psi_{m}^{(-)} \}=\delta _{n+m , 0} \, , \qquad\{ \psi_{n}^{(\pm)} ,
\psi_{m}^{(\pm)} \} =0 \, \, .  \eeq Therefore $\psi_{n}^{(+)}$'s and
$\psi_{n}^{(-)}$'s generate an infinite Clifford algebra. We can also
take the Laurent expansion of the regularized product of the
coefficients $\psi_{\pm}^{\sigma}$ introduced in (\ref{eqn:ord}): $$
:\!  \psi_{-}^{\sigma}(z)\psi_{+}^{\sigma}(w)\!\!:\,=\sum_{m, n}:\!
\psi_{m}^{(-)}\psi_{n}^{(+)}\!\!
:z^{-n-m-1}\left(\frac{w}{z}\right)^{\!-n-1+\lambda} \qquad
\mbox{for}\quad |z|>|w| , $$ \beq \mbox{where}\quad\qquad :\!
\psi_{m}^{(-)}\psi_{n}^{(+)}\!\!:\, \, = \left\{ \begin{array}{rl}
\psi_{m}^{(-)}\psi_{n}^{(+)} &\qquad \mbox {if $\quad n\geq
\lambda$}\\ \noalign{\vskip0.3cm} -\psi_{n}^{(+)}\psi_{m}^{(-)}
&\qquad \mbox {if $\quad n< \lambda$ \,\, .}
\end{array}\right.
\label{ch1:normalord}
\eeq Using this formula as well as the expressions (\ref{eqn:Jz}),
(\ref{eqn:Tzz}), and also $\omega_{z}=0$ and $A^{\sigma}=-a\, dz/z$ we
have the expansions $J_{z}(z)=\sum_{n}J_{n}z^{-n-1}$ and
$T_{zz}(z)=\sum_{n}L_{n}z^{-n-2}$ , where the coefficients are given
by \beqa J_{n}&=&\frac{1}{2\pi i}\oint dz\, z^{n}J_{z}=\sum_{m}:\!
\psi_{-m}^{(-)}\psi_{n+m}^{(+)}\! : +\delta_{n,0}\,a
+\delta_{n,0}\left(\lmd-\frac{1}{2}\right)\, \, ,\label{J}\\
L_{n}&=&\frac{1}{2\pi i}\oint dz\, z^{n+1}T_{zz} \label{L}\\
&=&\sum_{m}\left \{ (\lambda n+m ):\!
\psi_{-m}^{(-)}\psi_{n+m}^{(+)}\! : + a :\!
\psi_{-m}^{(-)}\psi_{n+m}^{(+)}\! :\right
\}+\delta_{n,0}\left(\!\left(\lambda -\frac{1}{2}\right)\!a
+\frac{1}{2}a^{2} -\frac{c_{\lmd}}{24}\right) \, \, .\nonumber \eeqa
In the radial ordering prescription, they satisfy the following
commutation relations : \beq [J_{n},\psi_{m}^{(\pm)}]=\mp
\psi_{m+n}^{(\pm)}\, ,\qquad [L_{n},
\psi_{m}^{(\pm)}]=-\left(\frac{1}{2}n+m\pm\left( \lambda
-\frac{1}{2}\right)n\mp a \right)\psi_{m+n}^{(\pm)} \,
,\label{eqn:JLpsi} \eeq \beq [J_{n},J_{m}] = n\delta_{n+m,0}\,
,\label{Heisenberg} \eeq \beq [L_{n}, J_{m}] = -mJ_{m+n}+\left(
\lambda -\frac{1}{2}\right) n^2 \delta_{m+n,0}\, , \eeq \beq [L_{n},
L_{m}] = (n-m)L_{m+n}+\frac{c_{\lambda}}{12}n^3 \delta_{m+n,0} \, .
\label{Virasoro}
\eeq Note that $J_{n}$'s generate the Heisenberg algebra, the Lie
algebra of the basic central extension of the group $LU(1)$ of smooth
loops in $U(1)$, and that $L_{n}$'s generate the Virasoro algebra of
central charge $c_{\lambda}$, the Lie algebra of a central extension
of the group $\DiffS$\label{page:DiffS} of diffeomorphisms of
$S^1$. In the following, we interpret this mixed
Clifford-Heisenberg-Virasoro algebra as an {\it operator algebra} on a
space of states.

\vspace{0.3cm} \underline{Path Integral for a Surface with Boundary
Circle}

We go over to the holomorphic representations of the fermionic path
integrals. Let $|\Sigma_{0},A_{0};{\cal O}_{0}\rangle$ be the result
of path integration over the fields $\psi_{\pm}$, $\bar \psi_{\pm}$ on
the interior of $\Sigma_{0}$, fixing $\psi_{+}|_{S}$ and $\bar
\psi_{-}|_S$ to be $\psi=\sum_{n}\psi_{n}z^{-n}(dz/z)^{1-\lambda}\!
\ot \!\sigma(z)$\label{page:psi(z)} and $\bar \psi=\sum_n \bar \psi_n
\bar z^{-n}(d\bar z/\bar z)^{1-\lmd}\!\ot \!\bar \sigma(z)$
respectively, where $\psi_{n}$ and $\bar \psi_{n}$ are grassmann odd
indeterminates : \beq |\Sigma_{0}, A_{0};{\cal O}_{0}\, \rangle =\int
_{\stackrel{\scriptstyle \psi=\psi_{\!+}|_S}{\bar \psi=\bar
\psi_{\!-}|_S}}\!\!{\cal D}_{\Sigma_{0}}\psi_{\! \pm}\bar
\psi_{\!\pm}\, \, e^{-I_{\Sigma_0}}{\cal O}_{0}(\psi_{\!\pm}\bar
\psi_{\!\pm}) \, .
\label{state0}
\eeq Similarly, we denote by $\langle \Sigma_{\infty},A_{\infty};{\cal
O}_{\infty}|$ the integral over fields on the interior of
$\Sigma_{\infty}$ with the same boundary conditions on $\psi_{+}|_S$
and $\bar \psi_{-}|_S$ : \beq \langle \,
\Sigma_{\infty},A_{\infty};{\cal O}_{\infty}| =
\int_{\stackrel{\scriptstyle \psi_{+}|_S =\psi}{\bar \psi_{\!-}|_S
=\bar \psi}}\! {\cal D}_{\Sigma_{\infty}}\!\psi_{\pm}\bar \psi_{\pm}\,
e^{-I_{\Sigma_{\infty}}}{\cal O}_{\infty}(\psi_{\pm}\bar \psi_{\pm}).
\label{stateinfty}
\eeq Then the total path integral is given by $\langle
 \Sigma_{\infty}, A_{\infty};{\cal O}_{\infty}|\Sigma_{0}, A_{0};
 {\cal O}_{0} \rangle$, where the pairing is the result of path
 integration $\int \prod_{n} d\psi_{n}d\bar \psi_n$ over the fields on
 the circle $S$.

We proceed for a while without rigor until we define the spaces to
which these states (\ref{state0}), (\ref{stateinfty}) belong and give
the definition of the pairing. Let us look at the effect of the
presence of the contour integrals $\psi_n^{(\pm)}$, $\bar
\psi_n^{(\pm)}$. Obviously, \beq |\Sigma_0,A_0;\psi_n^{(+)} {\cal
O}_0\,\rangle =\psi_n\wedge |\Sigma_0,A_0;{\cal O}_0 \,\rangle\,.
\eeq With the aid of the equation $\bar \partial_{\!A}\,\psi_{-}=0$
which holds on the interior of $\Sigma_0$, we see that \beq
|\Sigma_0,A_0;\hat{\epsilon}\psi_{n}^{(-)}{\cal O}_0 \,\rangle
=\left(\frac{d}{dt}\right)_{\!\!0}\!|\Sigma_0,A_0;{\cal O}_0\, \rangle
(\psi+t\hat{\epsilon}\upsilon_n)\,, \eeq where $\hat{\epsilon}$ is a
grassmann odd parameter and
$\upsilon_n=z^n(dz/z)^{1-\lmd}\!\ot\!\sigma$. In a word, the presence
of $\psi_n^{(-)}$ has the effect $i(\upsilon_n)$ of getting rid of the
indeterminate $\psi_{-n}$. Similar roles are played by $\bar
\psi_n^{(-)}$ and $\bar \psi_n^{(+)}$.

Now, let us consider the simple case with ${\cal O}_0=1$. Let
$\HH^{(\lmd,-)}$\label{page:Hlmd-} be the space $\Omega^0(S^1,{\bf
C})$ of smooth functions on $S^1$ which is identified, using the frame
$\dzz^{\!\lmd}\!\ot \sigma^{-1}$, with the space $\Omega^0(S,
K^{\lmd}\!\ot\! L^{-1})$ of sections of $K^{\lmd}\!\ot\!
L^{-1}|_S$. We consider the subspace $\W_{\Sigma_0}^{(\lmd,-)}\subset
\HH^{(\lmd,-)}$ identified with the space
$H^0(\Sigma_0,K^{\lmd}\!\ot\! L^{-1})|_S$ of sections that extend
holomorphically over $\Sigma_0$.  Since $\psi_{\pm}$ behaves
holomorphically on the interior of $\Sigma_0$, we have \beqa \oint
w\psi |\Sigma_0,A_0\,\rangle&=&0\qquad \mbox{for}\quad w\in
\W^{(\lmd,-)}_{\Sigma_0}\,,\label{cond.+}\\ \mbox{and }\hspace{3.3cm}
i(\upsilon)|\Sigma_0, A_0 \,\rangle &=&0\qquad \mbox{for}\quad
\upsilon \in \W^{(1-\lmd,+)}_{\Sigma_0}\,.\hspace{3.2cm}\label{cond.-}
\eeqa These conditions together with the anti-holomorphic counterparts
determine the state $|\Sigma_0,A_0\rangle$ up to constant. To see
this, we need to know more about the subspaces
$\W^{(\cdot\cdot,\cdot\cdot)}_{\Sigma_0}$. Let $\HH_+$
(resp.$\HH_-$)\label{page:H+} be the subspace of $\HH=\HH^{(\lmd,-)}$
consisting of one particle states $z^{n}(dz/z)^{\lmd}\ot \sigma^{-1}$
of positive frequency $n\geq 0$ (resp.negative frequency $n<0$).  As
the following argument shows, the subspace $\W_{\Sigma_0}^{(\lmd,-)}$
is, in a sense, near to $z^{-d}\HH_+$ for some integer $d$. Let us cap
the boundary $S$ of $\Sigma_{0}$ by the disc $D_{\infty}=\{z^{-1} \in
{\bf C}; |z|\geq 1\}$ and denote the resulting Riemann surface by
$\widehat{\Sigma}_0$. Extend the line bundle $L|_{\Sigma_0}$ to
$\widehat{\Sigma}_0$ by declairing that the frame $\sigma$ extend
holomorphically over $D_{\infty}$ and denote the result by
$\hat{L}_0$. Then, both \beqa \W_{\Sigma_0}^{(\lmd,-)}\cap
z^{1-\lmd}\HH_{-}&\cong& H^0(\widehat{\Sigma}_0,\hat{K}^{\lmd}_0\ot
\hat{L}_0^{-1})\\
\mbox{and}\hspace{3cm}\HH/(\W_{\Sigma_0}^{(\lmd,-)}+z^{1-\lmd}\HH_-)&\cong
&H^1(\widehat{\Sigma}_0,\hat{K}^{\lmd}_0\ot
\hat{L}_0^{-1})\hspace{3cm} \eeqa are finite dimensional and the
difference of the dimensions is given by the index
$-c_1(\hat{L}_0)+(2\lmd-1)(g_0-1)=:d+1-\lmd$, where $g_0$ is the genus
of $\widehat{\Sigma}_0$. This shows that $\W_{\Sigma_0}^{(\lmd,-)}$
has a dense subspace spanned by a sequence $\{w_n\}_{n=-d}^{\infty}$
of elements having finite ordre of poles at $z=\infty$: \beq
w_n=\dzz^{\lmd}\ot \sigma^{-1}\left( z^{s_n}+\sum_{m<s_n}z^mT_{m,n}
\right)\,, \eeq where $s_{n}=n$ for sufficiently large $n$. In this
sense we say that $\W_{\Sigma_0}^{(\lmd,-)}$ has virtual dimensions
$d$ relative to $\HH_+$. Similar argument shows also that
$\W_{\Sigma_0}^{(1-\lmd,+)}\subset \HH^{(1-\lmd,+)}$ have virtual
dimensions $-d$ relative to $z\HH_+^{(1-\lmd,+)}$. With respect to the
natural pairing of $\HH^{(1-\lmd,+)}$ and $\HH^{(\lmd,-)}$,
$\W_{\Sigma_0}^{(\lmd,-)}$ is contained in the ortho-complement
$\left( \W_{\Sigma_0}^{(1-\lmd,+)}\right)^{\perp}$ of
$\W_{\Sigma_0}^{(1-\lmd,+)}$. But the virtual dimension of the latter
relative to $\HH_+$ is $d$ and coinsides with that of the former. So
the pairing \beq \W_{\Sigma_0}^{(1-\lmd,+)}\times
\HH/\W_{\Sigma_0}^{(\lmd,-)}\longto {\bf C} \eeq is
non-degenerate. This fact together with the conditions (\ref{cond.+}),
(\ref{cond.-}) shows that the state $|\Sigma_0,A_0\rangle$ is
proportional to the ``product'' of infinite elements: \beq
|\Sigma_0,A_0\rangle \propto \prod_{n=-d}^{\infty}\oint w_n \psi
\prod_{n=-d}^{\infty}\oint \bar w_n\bar \psi\,,
\label{semi-inf1}
\eeq where $\{w_n\}_{n=-d}^{\infty}$ (resp.$\{\bar
w_n\}_{n=-d}^{\infty}$) is a base of the space of holomorphic sections
of $\hat{K}_0^{\lmd}\ot \hat{L}_0^{-1}$ (resp.anti-holomorphic
sections of $\hat{\bar K}_0^{\lmd}\ot \hat{\bar L}_0^{-1}$) having
pole at $z=\infty$ of finite ordre.

\vspace{0.3cm} \underline{ Definition of the Space of States and
Representations of Infinite Dimensional Groups}

Motivated by this ``semi-infinite product of grassmann odd
indeterminates'', we introduce infinite dimensional grassmann manifold
and determinant line bundle which provides a natural description of
the representation of loop groups on the space of states. The theory
is fully developped in \cite{S-W}, \cite{P-S} and \cite{S}. We choose
an inner product on the space $\HH$ of one particle states such that
$\{z^n(dz/z)^{\lmd}\ot \sigma^{-1}\}_{n\in {\bf Z}}$ forms an
orthonormal base and take the Hilbert space completion\footnote{In the
following, for any subspace we encounter, we always take the
completion. }. The grassmann manifold $Gr^{}_{\HH_+}$\label{page:Gr}
relative to the subspace $\HH_+$ is the set of subspaces {\it
comparable} with $\HH_+$ where we say that a subspace $\W\subset \HH$
is comparable with $\HH_+$ when the projection $\W\to \HH_+$ is a
Fredholm operator ( kernel and cokernel are both finite dimensional )
and the projection $\W\to \HH_-$ is a Hilbert-Schmidt operator ( for
some orthonormal base $\{ e_n\}$, $\sum_n ||pr_-(e_n)||^2<\infty$
). For example, $\W_{\Sigma_0}^{(\lmd,-)}$ is comparable with $\HH_+$
since the projection to $z^{1-\lmd}\HH_+$ ( and hence to $\HH_+$ ) has
finite dimensional kernel and cokernel as we have seen, and the
projection to $\HH_-$ can be seen to be Hilbert-Schmidt by the same
argument as in \cite{S-W}. The manifold $Gr^{}_{\HH_+}$ consists of
infinite number of connected components classified by the virtual
dimensions relative to $\HH_+$. The group $GL_{res}$ defined by
$GL_{res}=\{ f\in GL(\HH)\,;\, pr_{\mp}\!:\! f\HH_{\pm}\to \HH_{\mp}
\mbox{are both Hilbert-Schmidt}\}$ acts transitively on
$Gr^{}_{\HH_+}$. It contains as subgroups the group $\DiffoS$ of
orientation preserving diffeomorphisms of $S^1$ and the group $L{\bf
C}^*$ of smooth loops in $\C^*$ which act on the space $\HH$
by\footnote{It should be noted that the action of $\DiffoS$ is
well-defined only if $a$ is an integer. Otherwise,
(\ref{actiondiffoslc}) must be understood as an infinitesimal
representation. In the latter case, the operator on $\HH$ is generally
unbounded, see \cite{P-S}. } \beq f\in \DiffoS: w \to f^{\flat}w\quad
\mbox{and}\qquad h\in L{\bf C}^*:w\to w.h^{-1}.
\label{actiondiffoslc}
\eeq In the above expression, $w.h^{-1}\in \HH$ is defined by the
point-wise multiplication.

As in the finite dimensional case, we can define the determinant line
bundle $Det^{}_{\HH_+}\stackrel{\bf
C}{\longto}Gr^{}_{\HH_+}$.\label{page:Det} Since the determinant
cannot be defined for general operator on an infinite dimensional
space, $GL_{res}$ does not act on $Det^{}_{\HH_+}$ but its central
extension $G\tilde{L\:\:}_{\!\!\!\!res}$ does.\footnote{The extension
$1\to{\bf C}^*\to G\tilde{L\:\:}_{\!\!\!\!res}\to GL_{res}\to 1$
corresponds to the extension of the Lie algebra whose cocycle is given
by $$c\left(\pmatrix{ a_1 & b_1 \cr c_1 & d_1 \cr },\pmatrix{ a_2 &
b_2 \cr c_2 & d_2 \cr }\right)=\tr_{\!\HH_+}(b_2c_1-b_1c_2)$$ where
$\pmatrix{ a & b \cr c & d \cr }$ is a representation matrix with
respect to the decomposition $\HH=\HH_+\oplus\HH_-$.} The dual bundle
$Det_{\HH_+}^*$ has plenty of holomorphic sections. Hence we have a
representation of $G\tilde{L\:\:}_{\!\!\!\!res}$ on the space
$\Gamma=H^0(Gr^{}_{\HH_+},Det_{\HH_+}^*)$ and a representation on its
dual ${\cal F}_{\HH_+}=\Gamma^*$.\label{def:spaceofstates} An element
$w_{\bullet}$ in the determinant line $Det^{}_{\HH_+}\!(\W)$ over
$\W\in Gr^{}_{\HH_+}$ of virtual dimension $d$ is represented by a
base $\{ w_n\}_{n=-d}^{\infty}$ of $\W$ with the following property :
If we denote the integral $\nipi\oint w_n \psi$ by $\psi(w_n)$, the
semi-infinite product \beq
\psi(w_{\bullet})=\psi(w_{-d})\psi(w_{-d+1})\psi(w_{-d+2})\cdots
\cdots \eeq can be expanded $\psi(w_{\bullet})=\sum_{{\rm S}\in{\cal
S}_d}\pi_{\rm S}(w_{\bullet})\psi_{\rm S}$ by the ``monomials'' \beq
\psi_{\rm S}=\psi_{s_{-d}}\psi_{s_{-d+1}}\psi_{s_{-d+2}}\cdots \qquad
\mbox{for}\quad {\rm S}=\{ s_{-d},s_{-d+1}, s_{-d+2},\cdots \}
\label{ch1:monom}
\eeq where ${\cal S}_{d}$ is the set of subsets ${\rm S}=\{
s_{-d},s_{-d+1},\cdot\cdot \}\subset {\bf Z}$ such that $s_n<s_{n+1}$
for every $n$ and that $s_n=n$ for sufficiently large $n$. It can be
shown that the coefficients $\pi_{\rm S}(w_{\bullet})$ are square
summable: $\sum_{\rm S}|\pi_{\rm S}(w_{\bullet})|^2<\infty$. By the
evaluation map $\Gamma\times Det^{}_{\HH_+}\to \C$, we may consider
$\psi(w_{\bullet})$ as an element of ${\cal F}_{\HH_+}$ and the vector
space spanned by monomials $\{ \psi_{\rm S}\,;\,{\rm S}\in {\cal
S}=\cup_{d}{\cal S}_d\}$ can be shown to be a dense subspace of ${\cal
F}_{\HH_+}$.  Together with the observation (\ref{semi-inf1}), this
makes us to decide to take ${\cal F}_{\HH_+}$ as ( the holomorphic
part of ) the space of states $|\Sigma_0,A_0;{\cal O}_0\rangle$
appearing on the {\it outgoing} circle.

We can now identify the relations
(\ref{eqn:JLpsi})$\sim$(\ref{Virasoro}) as the commutation relations
of operators on ${\cal F}_{\HH_+}$. As was noted, it is natural to
interpret $\psi_+(w)$ as $\psi(w)\wedge$ for $w\in \HH=\HH^{(\lmd,-)}$
and $\psi_{-}(\upsilon)$ as $i(\upsilon)$ for $\upsilon \in
\HH^{(1-\lmd,+)}$. If $\tilde{g}\in G\tilde{L\:\:}_{\!\!\!\!res}$
covers the element $g\in GL_{res}$, then we have \beq
\tilde{g}\psi_+(w)\tilde{g}^{-1}=\psi_+(gw)\,,\qquad
\tilde{g}\psi_-(\upsilon)\tilde{g}^{-1}=\psi_-(g\upsilon)\,.
\label{JLpsiglobal}
\eeq If we define the groups $\DiffoSC$\label{page:DiffoSC} and
$\tilde{L{\bf C}}{}^*$ as the pull backs of
$G\tilde{L\:\:}_{\!\!\!\!res}$ by the embedding map of $\DiffoS$ and
$L{\bf C}^*$ to $GL_{res}$ respectively, they have representations on
the space ${\cal F}_{\HH_+}$ of states. Then, the commutation
relations (\ref{eqn:JLpsi}) are nothing but the infinitesimal version
of the relations (\ref{JLpsiglobal}), if $L_n$ and $J_n$ are
interpreted as the infinitesimal generators of that representations.

As ( the holomorphic part of ) the space of states $\langle
\Sigma_{\infty},A_{\infty};{\cal O}_{\infty}|$ appearing on the {\it
incoming} circle, we take ${\cal F}_{\HH_-}$ the dual of the space of
holomorphic sections of the dual determinant bundle $Det_{\HH_-}^*$
over the manifold $Gr^{}_{\HH_-}$. This is motivated by the fact that
the space $\W_{\Sigma_{\infty}}^{(\lmd,-)}$ of holomorphic sections
over $\Sigma_{\infty}$ is comparable with $\HH_-$. We arrange the
element of ${\cal F}_{\HH_-}$ as a sum of monomials $\{ \psi_{\rm
R}\,;\, {\bf Z}-{\rm R}\in {\cal S}\}$ of the form \beq \psi_{\rm
R}=\cdots\cdots \psi_{r_{-d-2}}\psi_{r_{-d-1}}\psi_{r_{-d}}\qquad
\mbox{for} \quad {\rm R}=\{ \cdots,r_{-d-2}, r_{-d-1},r_{-d}\}.  \eeq

Now we can define the pairing $\int\prod_nd\psi_n$ of the two spaces
${\cal F}_{\HH_-}$ and ${\cal F}_{\HH_+}$ in the obvious way: for
$\eta=\sum_{\rm R}\eta_{\rm R}\psi_{\rm R}\in {\cal F}_{\HH_-}$ and
$\xi=\sum_{\rm S}\xi_{\rm S}\psi_{\rm S}\in {\cal F}_{\HH_+}$, we put
\beq \langle \eta, \xi \rangle =\sum_{{\rm S}\in {\cal S}}(-1)^{\{{\bf
Z}-{\rm S}:{\rm S}\}}\eta_{{\bf Z}-{\rm S}}\xi_{\rm S}\,, \eeq where
$(-1)^{\{{\bf Z}-{\rm S}:{\rm S}\}}$ is the ratio of $\psi_{{\bf
Z}-{\rm S}}\psi_{\rm S}$ and $\psi_{{\bf Z}_{<0}}\psi_{{\bf Z}_{\geq
0}}$.

\newpage
\renewcommand{\theequation}{1.3.\arabic{equation}}\setcounter{equation}{0}
\vspace{0.3cm}
\begin{center}
{\sc 1.3 Field-State Correspondence and Spectral Flow}\label{1.3}
\end{center}
\hspace{1.5cm} In this section, we introduce the notion of field-state
correspondence. We also define the spectral flow as the transformation
of the space of states corresponding to the screening of local fields
by an external gauge field of certain configuration. It is identified
with the action of a certain element of the disconnected component of
the loop group $\tilde{L{\bf C}}^*$. To start with, we record such
configuration of gauge field. Since it will be frequently used in this
paper to define the spectral flow in various systems, we record it in
full generality.

\vspace{0.5cm} {\large The Basic Configuration}\label{basic}

\vspace{0.3cm} We consider a trivial principal bundle $P_0$ with a
compact gauge group $H$ over the unit disc $D_0=\{z\in {\bf
C}\,;\,|z|\leq 1\}$ provided with a trivialisation $s_0:D_0\to
P_0$. We choose a maximal torus $T$ of $H$ and denote its Lie algebra
by $\liet$. We also choose and fix a cut off function $\varrho :
D_0\to [0,1]$ such that $\varrho =0$ on a neighborhood of $z=0$ and
$\varrho =1$ on a neighborhood of the boundary $S=\partial D_0$. We
assume that $\varrho$ is rotationally invariant :
$\varrho=\varrho(|z|^2)$. The {\it basic gauge field} $A_{a}$ for
$ia\in \liet$ is given by the following form of covariant derivative
$d_{A_a}s_0=s_0\cdot A_a^{s_0}$ :\footnote{We define the angle
$\theta$ and the radius $r$ by $z=re^{i\theta}$. $s_0$ here is
represented as a frame of some vector bundle associated to $P_0$.}
\beq A^{s_0}_a=\varrho\, \frac{a}{2}\left(\frac{d\bar z}{\bar
z}-\frac{dz}{z}\right)=-\varrho\,ia\,d\theta\,.  \eeq

We can always find a holomorphic trivialization of the complexified
bundle $P_{0}^{\bf C}$ with group $H_{\bf C}$, that is, we can find
such a section $\sigma_0$ of $P_{0}^{\bf C}$ that $\bar
\partial_{\!A_a}\sigma_0=0$ for vector bundles $P_{0}^{\bf
C}\!\times_{H_{\!\bf c}}\! V$ coming from holomorphic representations
$H_{\!\bf C}\to GL(V)$. If we find a map $h_0:D_0\to H_{\bf C}$ which
satisfies $h_0\bar \partial h_0^{-1}=\varrho (a/2)d\bar z/\bar z$,
then $\sigma_0=s_0h_0$ is the desired section. Especially, we can find
the solution of the type $h_0=e^{f_0(|z|^2)}$ with $f_0(0)=0$. Such a
solution behaves as $h_0=c_{\varrho}^{-a} |z|^{-a}$ on a neighborhood
of the boundary $S$ where
$log\,c_{\varrho}=\nibun\int_0^1\varrho(x)dx/x$.

If $a\in i\liet$ lies in the lattice $\Pv=\frac{1}{2\pi
i}\exp^{-1}(1)\subset i\liet$, then we can find a horizontal section
(or a horizontal frame) $\sigma=s_0 e^{ia\theta}$ on a neighborhood of
$S$. This horizontal frame $\sigma$ defined around $S$ and the
holomorphic frame $\sigma_0$ defined on $D_0$ are related to each
other by a holomorphic transition function\label{page:ha(z)}
$h_a(z)=c_{\varrho}^{-a}z^{-a}$ : \beq \sigma_0(z)=\sigma (z)h_a(z)\,,
\eeq on a neighborhood of $S$.

\vspace{0.3cm} \underline{Field-State Correspondence}\label{ch1.FS}

Now let us introduce a one to one correspondence between local fields
and states. In a word, the state corresponding to a field is the
result of propagation of the field through the {\it standard disc}.
We take as the standard disc $D_0$, the disc with a complex coordinate
$z:D_0\stackrel{\cong}{\to}\{ z\in {\bf C}; |z|\leq 1 \}$ which gives
a parametrisation of the boundary $S=\partial D_0$ and with a metric
$\met_0=\met_{0z\bar z}dzd\bar z+c.c.$ which is canonically flat
$\met_0^{z\bar z}=2$ on a neighborhood of $z=0$ and behaves like
cylinder $\met_0^{z\bar z}=8\pi^2|z|^2$ near the boundary $S$. It has
a unit curvature $\frac{i}{2\pi}\int_{D_0}R(\met_0)=1$. Also we take
as the standard line bundle $L_{D_0}$, the trivial bundle with a
canonically flat connection $A_0$ and a horizontal frame $s_0$ which
provides a horizontal frame $\sigma$ near $S$.

 We insert a local field $O$ at the center $z=0$ of $D_0$. We denote
by $|O\rangle$ the state $|D_0,A_0\,;O(0)\rangle\in {\cal
F}_{H_+}\ot{\cal F}_{\bar H_+}$ corresponding to $O$ with respect to
the frame $\sigma$.  For $O=1$, the state $|1\rangle$ is determined by
the conditions (\ref{cond.+}), (\ref{cond.-}) up to proportionality
which depends on the metric $\met_0$. Because $\W_{D_0}^{(\lmd,-)}=\{
z^n\dzz^{\!\lmd}\!\!\ot \sigma^{-1};\,n\geq \lmd \}$ and
$\W_{D_0}^{(1-\lmd,+)}=\{ z^n\dzz^{\!1-\lmd}\!\!\ot \sigma\,;\,n\geq
1-\lmd\}$, the result is \beq
|1\rangle=|0\rangle\ot\overline{|0\rangle}\,, \eeq where the vacuum
$|0\rangle\in {\cal F}_{\HH_+}$ and $\overline{ |0\rangle}\in {\cal
F}_{\bar \HH_+}$ are given by \beq
|0\rangle=\psi_{\lmd}\psi_{\lmd+1}\psi_{\lmd+2}\cdots \,, \qquad
\overline{|0\rangle}=\bar \psi_{\lmd}\bar \psi_{\lmd+1}\bar
\psi_{\lmd+2}\cdots \,.  \eeq For a non trivial field $O\ne 1$, we can
determine $|O\rangle$ easily. For example, since the field
$\psi^{s_0}_+(0)$ can be obtained by contracting the contour integral
$\nipi \oint z^{-1}(dz)^{\lmd}\ot \sigma^{-1}\psi_+
=\psi^{(+)}_{\lmd-1}$, we have $|\psi_+^{s_0}\rangle
=\psi^{(+)}_{\lmd-1}|1\rangle=\psi_{\lmd-1}\psi_{\lmd}\psi_{\lmd+1}\cdots\ot
\overline{|0\rangle}$. In the similar way, we have $|\partial_z^n
\psi_+^{s_0}\rangle =n!\psi_{\lmd-n-1}\psi_{\lmd}\psi_{\lmd+1}\cdots
\ot \overline{|0\rangle}$ and
$|\psi_-^{s_0}\rangle=\psi_{-\lmd}^{(-)}|1\rangle
=\psi_{\lmd+1}\psi_{\lmd+2}\psi_{\lmd+3}\cdots\ot
\overline{|0\rangle}$ etc.

We now comment on the action of the generators $J_n$ and $L_n$. Since
the metric $\met_0$ is not flat, neither $J_z$ nor $T_{zz}$ can be
holomorphic.Taking into account the Levi-Civita coonection terms in
the expressions (\ref{eqn:Jz}) and (\ref{eqn:Tzz}), we have \beq
J_n|O\rangle =\left\vert \Bigl(J(z^n)+\Bigl(\lmd
-\nibun\,\Bigr)\delta_{n,0}\Bigr)O\right\rangle \,,\qquad L_n|O\rangle
=\left|\Bigl( L_{z^{n+1}\frac{\partial}{\partial
z}}-\frac{c_{\lmd}}{24}\delta_{n,0}\Bigr)O\right\rangle \,, \eeq where
the fields in the right hand sides are defined in (\ref{eqn:Jepsi}),
(\ref{eqn:Lv}). A field $O$ with eigen value $(\sDelta,\bar
\sDelta)$\label{page1:confdim} for $(L_{z\partial/\partial z},L_{\bar
z\partial/\partial \bar z})$ and eigen value $(\sq,\bar \sq)$ for
$(J(1),\bar J(1))$ is said to have conformal weight $(\sDelta,\bar
\sDelta)$ and charge $(\sq,\bar \sq)$. Then, for such a field $O$, we
have \beq J_0|O\rangle =\left(\sq+\lmd-\nibun \,\right)|O\rangle
\,,\qquad L_0|O\rangle =\left(\sDelta
-\frac{c_{\lmd}}{24}\,\right)|O\rangle \,, \eeq and similar equations
for $\bar J_0$ and $\bar L_0$.

\vspace{0.3cm} \underline{The Spectral Flow}

Now let us consider, instead of the canonically flat connection, the
basic configuration $A_a$ of gauge field for $H=U(1)$ such that $a\in
{\bf R}$ lies in the lattice $\frac{1}{2\pi i}exp^{-1}(1)={\bf Z}$. We
consider the line bundle $L_{D_0}$ to be associated to $P_0$ by the
representation $(u,c)\in U(1)\times {\bf C}\mapsto uc\in {\bf C}$. We
get a new version of the field-state correspondence $O\leftrightarrow
|O\rangle_a$ where the latter is the state $|D_0,A_a\,;O\rangle\in
{\cal F}_{\HH_+}\ot{\cal F}_{\bar \HH_+}$ with respect to the
horizontal frame $\sigma$ along $S$. We call the transformation
$\tilde{h_a}:|O\rangle \mapsto |O\rangle_a$ of the space ${\cal
F}_{\HH_+}\ot{\cal F}_{\bar \HH_+}$ of states the {\it spectral
flow}.\label{ch1.tilh} As the following argument shows, this operator
$\tilde{h_a}$ can be identified as the action of an element of the
group $\tilde{L{\bf C}}{}^*$ which is not in the identity component
but in the component of {\it winding number} $a$.

Since the frame $\sigma_0=\sigma h_a$ extend holomorphically over
$D_0$ and behaves as a unitary horizontal frame in a small
neighborhood of $z=0$, for $w\in \HH^{(\lmd,-)}$ and $\upsilon \in
\HH^{(1-\lmd,+)}$ we have \beq \psi_+(w)\tilde{h_a}|O\rangle
=\tilde{h_a}\psi_+(w.h_a)|O\rangle \,,\qquad\quad
\psi_-(\upsilon)\tilde{h_a}|O\rangle
=\tilde{h_a}\psi_-(h_a^{-1}.\upsilon )|O\rangle \,, \eeq and similar
equations for the anti-holomorphic part of the fields. In particular,
the state $|1\rangle_a$ satisfies the conditions
$\psi_n^{(+)}|1\rangle_a=0$ for $n\geq \lmd+a$ and
$\psi_n^{(-)}|1\rangle_a=0$ for $n\geq 1-\lmd -a$. So we have
$|1\rangle_a =|a\rangle \ot \overline{|a\rangle }$ where $|a\rangle
\in {\cal F}_{H_+}$ and $\overline{|a\rangle}\in{\cal F}_{\bar H_+}$
are given by \beq |a\rangle =\psi_{\lmd
+a}\psi_{\lmd+a+1}\psi_{\lmd+a+2}\cdots \,,\qquad \quad
\overline{|a\rangle}=\bar \psi_{\lmd+a}\bar \psi_{\lmd+a+1}\bar
\psi_{\lmd+a+2}\cdots \,.  \eeq

The intertwining relation of this spectral flow $\tilde{h_a}$ and the
currents $J$ and $T$ can be read in the formulae (\ref{J}) and
(\ref{L}) : \beqa J_n\,\tilde{h_a}|O\rangle &=&\tilde{h_a}
(J_n+a\delta_{0,n})|O\rangle\,,\\ L_n\tilde{h_a}|O\rangle
&=&\tilde{h_a}\!\left\{ L_n+a J_n
+\frac{1}{2}a^2\,\delta_{0,n}\right\}\!|O\rangle\,.  \eeqa

We may consider the state $|O\rangle_a$ to correspond in the standard
way to a new field $O_a$. If a field $O$ has conformal weight
$(\sDelta,\bar \sDelta)$ and charge $(\sq,\bar \sq)$, the new field
$O_a$ has conformal weight $({}^a\!\sDelta,{}^a\!\bar \sDelta)$ and
charge $({}^a\!\sq,{}^a\!\bar \sq)$ related to the old ones by the
following \beq {}^a\!\sDelta=\sDelta+\sq
a+\Bigl(\lmd-\nibun\Bigr)a+\nibun a^2\,,\qquad \quad
{}^a\!\sq=\sq+a\,,
\label{zxcv}
\eeq and similar equations for ${}^a\!\bar \sDelta$ and ${}^a\!\bar
\sq$.

 To summarize, {\it if a field $O$ is screened by the basic gauge
field $A_a$, it becomes another field $O_a$ where the change of
conformal weight and charge is given by (\ref{zxcv}). The
transformation $O\mapsto O_a$ corresponds to the spectral flow
$\tilde{h_a}:{\cal F}_{\HH_+}\to {\cal F}_{\HH_+}$ which comes from
the action of an element of the group $\tilde{L{\bf C}^*}$ of winding
number $a$.}

\renewcommand{\theequation}{1.4.\arabic{equation}}\setcounter{equation}{0}
\vspace{0.4cm}
\begin{center}
{\sc 1.4 Theories of Higher Ranks}\label{1.4}
\end{center}
\hspace{1.5cm} Let us next consider a system of multi-component
fermions. The classical action $I$ is the same as (\ref{fermiaction})
provided that the fields take values in $vector$ $bundles$. We
consider the vector bundles with strucures finer in general than
hermitian structure. So, let $H$ be a compact Lie group, and $V$ be a
finite dimensional complex vector space with a unitary representation
of $H$. Let $P$ be a principal $H$ bundle\label{page1:P} with
conncetion $A$ and $E$\label{vctbdle} be the associated $V$ bundle
$E=P\!\times_{H}\!V$. We consider four kinds of anti-commuting fields:
$$ \psi_{+}\in\Omega^{0}(\Sigma, K^{1-\lmd}\!\ot \!E)\, \, ,\qquad
\bar \psi_{+}\in \Omega^{0}(\Sigma, \bar K^{\lmd}\!\ot \!E) \, \, ,$$
\vspace{-1cm}
$$\psi_{-}\in \Omega^{0}(\Sigma, K^{\lmd}\ot E^{\ast}) \, \, , \qquad
\bar \psi_{-}\in \Omega^{0}(\Sigma, \bar K^{1-\lmd}\ot E^{\ast})\, \,
,$$ where $E^{\ast}$ is the dual bundle $P\!\times_{H}\!V^{\ast}$ of
$E$. In the process of quantisation, things are essentially the same
as in the case of one-component fermions but some modifications are
required.

Energy momentum tensor $T$ and $H$-current $J$ are defined by
(\ref{defTJ}) where $\delta A$ is a one form valued in the adjoint
bundle $\ad P=P\! \times_{H}\! \h$.\label{page:adP1} Therefore, the
current $J$ takes values in the coadjoint bundle $\ad P^{\ast}$. The
conformal and chiral anomaly (see equation (\ref{eqn:anom})) for the
transformation $\met \to e^{\phi}\met$, $\bar \partial_{\!A}\to
h^{-1}\bar \partial_{\!A}h$ where $h$ is an automorphism of $P_{\bf
C}$ is given by \beqa I( \, \met,A\, ; \phi,
hh^{\ast})\!\!&=&\!\!\frac{c_{\lmd,V}i}{48\pi}\! \int_{\Sigma}\!\left(
\partial \phi \bar \partial \phi +2R_{\Theta}\phi \right
)\!-\frac{i}{2\pi}\!\left(\lambda-\frac{1}{2}\right)\!\int_{\Sigma}\!\tr_{\!{}_E}\!\left(
F_{\!A}\phi-\partial\phi (hh^{\ast}\!)^{-1}\bar \partial (hh^{\ast}\!)
\right)\nonumber \\ \noalign{\vskip0.2cm}
&&+I_{E}\Bigl(A\!+\!\Bigl(\lmd-\nibun\,\Bigr)\Theta , hh^{\ast}\Bigr)
\, \, , \label{anommulti} \eeqa where
$c_{\lmd,V}=c_{\lmd}\mbox{dim}V$.\label{page:clmdv} $I_{E}$ in the
above equation is the Wess-Zumino-Witten action for the bundle $E$
which is determined by the following conditions: \beq
\left(\frac{d}{dt}\right)_{\!\!0}\!\!I_{\!E}\Bigl(A\!+\!\Bigl(\lmd
-\nibun\,\Bigr)\Theta, e^{t\epsilon}\, \Bigr)=\nipi
\int_{\Sigma}\tr_{\!{}_E}\!\left\{ \epsilon\, \Bigl(F_{\!A}+\Bigl(\lmd
-\nibun\,\Bigr)R_{\Theta}\Bigr)\right\}\, \, ,
\label{infanom}
\eeq \beq I_{\!E}\Bigl(A+\Bigl(\lmd-\nibun\,\Bigr)\Theta, \,
h_{1}h_{2}h_{2}^{\ast}h_{1}^{\ast}\,
\Bigr)=I_{\!E}\Bigl(A^{h_{1}}\!+\Bigl(\lmd -\nibun\,\Bigr)\Theta, \,
h_{2}h_{2}^{\ast}\,
\Bigr)+I_{\!E}\Bigl(A+\Bigl(\lmd-\nibun\Bigr)\Theta
,\,h_{1}h_{1}^{\ast}\, \Bigr) \, \, .
\label{assoc}
\eeq A description of the Wess-Zumino-Witten action is given in the
next chapter. The Ward identities satisfied by $J$ and $T$ have
exactly the same form as (\ref{eqn:Ward J}), (\ref{eqn:T1.1}) and
(\ref{eqn:Ward T}) though, this time, we need to put $\tr_{\!{}_E}$ in
suitable places. The expressions for $J$ and $T$ in terms of the
fundamental fields $\psi_{\pm}$, $\bar \psi_{\pm}$ are given by \beq
J\epsilon\, = \, :\!\psi_{\!-}\epsilon \, \psi_{\!+}\!\!: \, , \qquad
T=-\lmd :\!\psi_{\!-}\partial_{\!A}\psi_{\!+}\!\!:+\, (1-\lmd):\!
\partial_{\!A}\psi_{\!-}\psi_{\!+}\!\!: \, ,
\eeq where the definitions of the regularized products are the obvious
generalisations of (\ref{defnorm}). If we choose a local section
$\sigma$ of $P$ or $P_{\! \bf C}$, then it defines local frames :
$\sgmV$\label{page:sigmav} of $E$, $\sgmV^{-1}=\sigma_{\!{}_{V^*}}$ of
$E^{\ast}$, $\sigma_{\!\ad}$ of $\ad P$ or of $\ad P_{\!\bf C}$,
etc. With respect to these frames, we have the local expressions
$\psi_+=(dz)^{1-\lmd}\ot \sgmV\psi_+^{\sigma}$ and
$\psi_-=(dz)^{\lmd}\ot \sgmV^{-1}\psi_-^{\sigma}$ in terms of the
vector valued coefficients $\psi_+^{\sigma}(z)\in V$ and
$\psi_-^{\sigma}(z)\in V^*$. Using the obvious generalization to this
case of the definition (\ref{eqn:ord}) of the normal ordered product
of the coefficients, $J$ and $T$ are expressed as \beqa
J_{z}(\sigma_{\!\ad} X)\!& =&\,
:\psi_{-}^{\sigma}X\psi_{+}^{\sigma}\!:-\trV\left( A^{\sigma}_z X
\right)-\Bigl( \lmd-\nibun\,\Bigr)\trV(X)\omega_{z} \, \, ,\\
T_{zz}\,&=&-\lmd
:\psi_{-}^{\sigma}\partial_{z}\psi_{+}^{\sigma}\!:+(1-\lmd):\partial_{z}\psi_{-}^{\sigma}\psi_{+}^{\sigma
}\!: -\frac{c_{\lmd,V}}{12}\left(\partial_{z}\omega_{z}-\nibun
\omega_{z}\omega_{z}\right) \nonumber \\ && -:\psi_{-}^{\sigma
}A^{\sigma}_{ z}\psi_{+}^{\sigma }\!:+\trV\!\left(\Bigl( \lmd
-\nibun\,\Bigr)\partial_{z}A^{\sigma}_{ z} +\nibun A^{\sigma
}_{z}A^{\sigma }_{z} \right) \, .
\label{emmulti}
\eeqa

\vspace{0.3cm} \underline{Operator Formalism}

As in the case of one-component fermion, we choose a circle
$c:S^{1}\to S\subset \Sigma$ which devides $\Sigma$ into two parts
$\Sigma_{\infty}$ and $\Sigma_{0}$ with a complex coordinate $z$ on an
annular neighborhood $U$ of $S$. We choose a metric $\met$ flat on $U$
with the coefficient $\met^{z\bar z}=8\pi^2|z|^2$ and a connection $A$
of $P$ which is also flat on $U$ admitting a holomorphic section
$\sigma:U\to P_{\!\bf C}$ of the form $A^{\sigma}=-a\, dz/z$ where
$ia\in \h$. We consider the Laurent expansion of $\psi_{\pm}$. If $\{
e_{\tinI}\}_{\tinI\in I(V)}$\label{page:eI} is a base of $V$ dual to a
base $\{e^{\tinI}\}_{\tinI \in I(V)}$ of $V^*$ where $I(V)$ is an
indexing set, the coefficients
$\psi_{n}^{(+)}=e_{\tinI}\psi_{n}^{\tinI(+)}\in V$ and
$\psi_{n}^{(-)}=e^{\tinI}\psi_{n, \tinI}^{(-)}\in V^*$ are given by
\beqa \psi_{n}^{\tinI(+)}&=&\nipi \oint \dzz^{\!\lmd}\!\!\ot
\sgmV(z)^{-1}e^{\tinI}z^n\,\psi_{+}(z)\, \, ,\\
\psi_{n,\tinI}^{(-)}&=&\nipi \oint
\psi_{-}(z)\dzz^{\!\!1-\lmd}\!\!\!\!\ot \sgmV(z) e_{\tinI}z^n\, \, .
\label{eqn1.4:psipm}
\eeqa They satisfy the following anti-commutation relations in the
radial ordering prescription: \beq \{ \psi_{n,\tinJ}^{(-)},
\psi_{m}^{\tinI(+)}\}=\delta_{\tinJ}^{\tinI}\delta_{n+m,0} \, \,
,\qquad \{ \psi_{n}^{(\pm)}, \psi_{m}^{(\pm)}\}=0\, \, .  \eeq If we
define the normal ordered product of these coefficients by \beq
:\psi_{m,\tinI}^{(-)}\psi_{n}^{\tinJ (+)}\!:\, \,
=\left\{\begin{array}{rl}
\psi_{m,\tinI}^{(-)}\psi_{n}^{\tinJ(+)}&\qquad\mbox{if $n\geq \lmd$}
\\ \noalign{\vskip0.2cm}
-\psi_{n}^{\tinJ(+)}\psi_{m,\tinI}^{(-)}&\qquad \mbox{if $n\leq \lmd
-1$}\, \, ,
\end{array}\right.
\eeq the Laurent coefficients of the expansions of $J_{z}$ and
$T_{zz}$ can be written as \beqa J_{n}(X)\!\!&=&\!\!\nipi \oint dz\,
z^{n}J_{z}(\sigma_{\!\ad}X)
=\sum_{m}:\psi_{-m}^{(-)}X\psi_{n+m}^{(+)}\!:
+\delta_{n,0}\trV\!\!\left(aX\!+\!\Bigl(\lmd-\nibun\,\Bigr)X\right)
,\label{HJn}\\ L_{n}&=&\!\! \nipi \oint dz \, z^{n+1}T_{zz}
\label{HLn}\\ =\,&&\hspace{-1cm}\sum_{m}\left\{(\lmd n+m):\!
\psi_{-m}^{(-)}\psi_{n+m}^{(+)}\! :+:\!\psi_{-m}^{(-)}\, a\,
\psi_{n+m}^{(+)}\!:\right\}+\delta_{n,0}\trV\!\!\left( \!
\Bigl(\lmd-\nibun\,\Bigr)\!a\! +\nibun
a^{2}-\frac{c_{\lmd}}{24}\right)\, ,\nonumber \eeqa where $X\in
\h_{\bf C}$. The commutation relations satisfied by them are \beq
[J_{n}(X), \psi_{m}^{(\pm)}]=\mp X\psi_{n+m}^{(\pm)}\, ,\qquad
[L_{n},\psi_{m}^{(\pm)}]=-\left( \nibun n+m\pm\left(\lmd -\nibun
\right)n\mp a \right)\psi_{n+m}^{(\pm)}\, ,
\label{HpsiJ}
\eeq \beq [J_{n}(X),J_{m}(Y)] = J_{n+m}([X,Y])
+n\delta_{n+m,0}\trV(XY)\, \, ,\hspace{1.8cm}\label{HJJ} \eeq \beq
[L_{n},J_{m}(X)] = -J_{n+m}(mX-[a,X])+\left( \lmd -\nibun
\right)n^2\delta_{n,0}\trV(X)\, \, , \label{HLJ} \eeq \beq
[L_{n},L_{m}] = (n-m)L_{n+m}+\frac{c_{\lmd,V}}{12}n^3 \delta_{n+m,0}\,
\, .\hspace{3.7cm}\label{HLL} \eeq

We can consider these commutation relations to have come from a
projective representation of the group $LH_{\bf C}$ of loops in
$H_{\bf C}$ and the group $\DiffS$ of diffeomorphisms of $S^1$ on a
space of states. The choice of the space of states prodeeds just as in
the case of one component fermions. That is, the space of states is
chosen to be the dual ${\cal F}_{\HH_+}^V$ of the space of holomorphic
sections of the bundle $Det_{\HH_+}^*$ over the grassmann manifold
$Gr^{V}_{\HH_+}$. This grassmannian $Gr^V_{\HH_+}$ consists of
subspaces of the space $\HH^{(\lmd,V^*)}\cong \Omega^0(S,K^{\lmd}\ot
E^*)$\label{page:HlmdV} of one particle states comparable with the
subspace $\HH_+^{(\lmd,V^*)}$ of particles of positive frequency. We
should note that we have to choose a base $\{e^{\tinI}\}_{\tinI\in
I(V)}$ of $V^*$ and a total ordering $<$ of the indexing set $I(V)$ to
define the determinant line bundle $Det^{}_{\HH_+}$ (or its dual). The
space ${\cal F}_{\HH_+}^V$ is spanned by the semi-infinite products of
grassmann-odd indeterminates of the following form \beq
\psi_{n_1}^{\tinI_1}\psi_{n_2}^{\tinI_2}\psi_{n_3}^{\tinI_3}\cdots
\psi_{n_s}^{\tinI_s}\psi_{N}^V\psi_{N+1}^V\psi_{N+2}^V\cdots\,, \eeq
where $N$ is some integer and $\psi_n^V$ for $n\in {\bf Z}$ in the
above expression is given by \beq
\psi_n^V=\psi_n^{\tinJ_1}\psi_n^{\tinJ_2}\psi_n^{\tinJ_3}\cdots
\psi_n^{\tinJ_{{\rm dim}V}}\,\,\qquad ;\quad
\tinJ_1<\tinJ_2<\tinJ_3<\cdots <\tinJ_{{\rm dim}V}.  \eeq The contour
integrals $\psi_n^{\tinI(+)}$ and $\psi_{n,\tinI}^{(-)}$ acts on
${\cal F}_{\HH_+}^V$ by $\psi_n^{\tinI(+)}=\psi_n^{\tinI}\times$ and
$\psi_{n,\tinI}^{(-)}=i(\upsilon_{n,\tinI})$ where
$\upsilon_{n,\tinI}$ is the element $\dzz^{\!1-\lmd}\!\ot
\sigma_{\!{}_{\!V}}e_{\tinI}\in \HH^{(1-\lmd,V)}$ and $i(\upsilon)$ is
the inner derivation
$i(\upsilon)F(\psi)=(d/d\hat{\epsilon})F(\psi+\hat{\epsilon}\upsilon)$.

\newpage \vspace{0.3cm} \underline{Field -State Correspondence and
Spectral Flow}

Let $(D_0,\met_0,z)$ be the standard disc with metric and coordinate
introduced previously, and let $E_0^V=P_0\!\times_H\!V$ be a vector
bundle with fibre $V$ associated to the trivial $H$-bundle $P_0$ over
the disc $D_0$. We choose a canonically flat connection $A_0$ and a
frame $s_0$ horizontal with respect to $A_0$. Then we have a one to
one correspondence $O\leftrightarrow |O\rangle
=|\met_0,A_0;O(0)\rangle$ between fields and states. For example, we
have the state corresponding to the identity : $|1\rangle
=|0\rangle\ot \overline{|0\rangle }$ where $|0\rangle \in {\cal
F}_{\HH_+}^V$ and $\overline{|0\rangle }\in {\cal F}_{\bar
\HH_+}^{V^*}$ are given by \beq |0\rangle
=\psi_{\lmd}^V\psi_{\lmd+1}^V\psi_{\lmd+2}^V\cdots \,, \qquad
\overline{|0\rangle }=\bar \psi_{\lmd}^{V^*}\bar
\psi_{\lmd+1}^{V^*}\bar \psi_{\lmd+2}^{V^*}\cdots\,.  \eeq We now
choose a maximal torus $T$ of $H$. A field $O$ is said to have
conformal weight $\sDelta$ and charge $\sq\in \liet_{\bf
C}^*$\label{page2:sq} if it satisfies $L_{z(\partial/\partial
z)}O=\sDelta O$ and $J(x)O=\sq(x)O$ for any $x\in \liet_{\bf
C}$. Then, the state $|O\rangle$ corresponding to such a field $O$
satisfies \beq J_0(x)|O\rangle =\left( \sq(x)-\Bigl(\lmd
-\nibun\,\Bigr)\trV(x)\right)\!|O\rangle\,,\qquad L_0|O\rangle =\left(
\sDelta -\frac{c_{\lmd,V}}{24}\right)\!|O\rangle\,, \eeq for $x\in
\liet_{\bf C}$.

Next, we consider the {\it basic configuration} $A_a$ of gauge field
where $a\in i\liet$ is in the lattice $T^{\vee}=\nipi
\exp^{-1}(1)$. We get a new version of the correspondence
$O\leftrightarrow |O\rangle_a$ where the latter is the state
$|\met_0,A_a;O(0)\rangle \in {\cal F}_{\HH_+}^V\ot{\cal F}_{\bar
\HH_+}^{V^*}$ with respect to the horizontal frame $\sgmV$ along
$S$. The transformation $\tilde{h_a}:|O\rangle \to |O\rangle_a$ of
${\cal F}_{\HH_+}^V\ot{\cal F}_{\bar \HH_+}^{V^*}$ is called the {\it
spectral flow}. It is given by the action of an element of the group
$\tilde{LH}_{\bf C}$. ( The element may or maynot be in a non-trivial
connected component of the group. It depends. )

Since $\sigma_0=\sigma h_a$ extends holomorphically over $D_0$, the
intertwining relation of this spectral flow $\tilde{h_a}$ and the
currents $\psi_{\pm}$, $J_n$ and $L_n$ are given by the following :
\beq \psi_+(w)\tilde{h_a}|O\rangle =\tilde{h_a}\psi_+(w.h_a)|O\rangle
\,,\qquad \quad \psi_-(\upsilon)\tilde{h_a}|O\rangle
=\tilde{h_a}\psi_-(h_a.\upsilon)|O\rangle\,,
\label{spctrlflw1}
\eeq \vspace{-1cm} \beqa J_n(x)\tilde{h_a}|O\rangle
&=&\tilde{h_a}\!\left(J_n(x)+\delta_{n,0}\trV(ax)\right)|O\rangle\,,\\
\noalign{\vskip0.2cm} J_n(e_{\alpha})\tilde{h_a}|O\rangle
&=&c_{\varrho}^{\alpha(a)}\tilde{h_a}J_{n+\alpha(a)}(e_{\alpha})|O\rangle
\,,\\ L_n\tilde{h_a}|O\rangle &=&\tilde{h_a}\!\left\{
L_n+J_n(a)+\nibun \delta_{n,0}\tr_{\!{}_{\!V}}(a^2)\right\}\!|O\rangle
\,, \eeqa where $w\in \HH^{(\lmd,V^*)}$, $\upsilon\in
\HH^{(1-\lmd,V)}$, $x\in \liet_{\bf C}$ and $e_{\alpha}\in \h$ is a
root vector for a root $\alpha\in\Delta$\label{page1:roots} of $H$. We
take the base $\{e_{\tinI}\}_{\tinI \in I(V)}$ of $V$ consisting of
vectors of definite weight : $xe_{\tinI}=\lmd_{\tinI} (x)e_{\tinI}$
for $x\in \liet_{\bf C}$. This enables us to write down the state
$|1\rangle_a$ in a rather explicit way. With the aid of
(\ref{spctrlflw1}), we have
$|1\rangle_a=|a\rangle\ot\overline{|a\rangle}$ up to proportionality
where $|a\rangle\in{\cal F}_{\HH_+}^V$ and
$\overline{|a\rangle}\in{\cal F}_{\bar \HH_+}^{V^*}$ are given by \beq
|a\rangle=\prod_{\tinI \in
I(V)}\prod_{n=1}^{|\lmd_{\tinI}(a)|}\psi_{n,\tinI}^{(a)}|0\rangle
\,,\qquad \overline{|a\rangle}=\prod_{\tinI\in
I(V)}\prod_{n=1}^{|\lmd_{\tinI}(a)|}\bar
\psi_{n,\tinI}^{(a)}\overline{|0\rangle}\,, \eeq where
$\psi_{n,\tinI}^{(a)}$ in the right hand side of the first formula is
given by \beq \psi_{n,\tinI}^{(a)}=\left\{ \begin{array}{rl}
\psi_{\lmd -n}^{\tinI (+)} \quad&\quad \mbox{if $\lmd_{\tinI}(a)< 0$}
\nonumber \\ \noalign{\vskip0.2cm} \psi_{1-\lmd -n \,
,\tinI}^{(-)}&\quad \mbox{if $\lmd_{\tinI}(a)>0$}\, \, .
\end{array} \right.
\eeq $\bar \psi_{n,\tinI}^{(a)}$ is given in the similar way.

\noindent We may consider the state $|O\rangle_a$ to correspond in the
standard way to a new field $O_a$. If a field $O$ has conformal weight
$(\sDelta,\bar \sDelta)$ and charge $(\sq,\bar \sq)$, the new field
$O_a$ has conformal weight $({}^a\!\sDelta,{}^a\!\bar \sDelta)$ and
charge $({}^a\!\sq,{}^a\!\bar \sq)$ related to the old ones by \beq
{}^a\!\sDelta=\sDelta+\sq(a)+\trV\!\left( \Bigl(\lmd-\nibun \,\Bigr)a
+\nibun a^2\right)\,, \qquad {}^a\!\sq(x)=\sq(x)+\trV(ax)\,,
\label{xcvb}
\eeq and similar equations for ${}^a\!\bar \sDelta$ and ${}^a\!\bar
\sq$.

 To summarize, {\it if a field $O$ is screened by the basic gauge
field $A_a$, it becomes another field $O_a$ where the change of
conformal weight and charge is given by (\ref{xcvb}). The
transformation $O\mapsto O_a$ corresponds to the spectral flow
$\tilde{h_a}:{\cal F}_{\HH_+}^V\to {\cal F}_{\HH_+}^V$ which comes
from the action of an element of the group $\tilde{LH}_{\bf C}$.}

\newpage
\renewcommand{\theequation}{2.0.\arabic{equation}}\setcounter{equation}{0}

{\large CHAPTER 2. WESS-ZUMINO-WITTEN MODEL}\label{ch.2}

\vspace{1cm} \hspace{1.5cm} In this chapter, we study the WZW models
for compact groups coupled to external gauge fields of various
topological types. Introducing a line bundle over the loop group, we
construct the WZW action for topologically non-trivial
configurations. We develop a gauge covariant operator formalism and
establish the field-state correspondence. Finally, we describe the
spectral flow as the screening of fields by non-flat gauge field
without holonomy.

\vspace{0.5cm} Let $G$ be a compact, connected and simply connected
Lie group with center $Z_{G}$ and let $H$ be the quotient group
$G/Z_{G}$. $H$ acts on $G$ as an automorphism group by $(h,g)\to
hgh^{-1}$.\footnote{We denote the element $[h]$ of $H$ by the same
letter as a representative $h$ in $G$.} Let $P$ be a principal $H$
bundle\label{page2:P} over a compact Riemann surface $\Sigma$. Note
that the set of topological types of $H$-bundle over $\Sigma$ is
naturally identified with $\pi_1(H)\cong Z_G$. Let $\ad_{G}P$ be the
associated $G$ fibering $\ad_{G}P=P\!\times_{H}\!G$ and $\ad_{G_{\bf
C}}P$ be the $G_{\bf C}$ fibering. The WZW action
$I=I_{\Sigma,P}(A,g)$ is a functional of the triple of a Riemann
surface $\Sigma$, a connection $A$ of $P$ and a section $g$ of
$\ad_{G_{\bf C}}P$ which is characterized by the following relations:
\beq \left(\frac{d}{dt}\right)_{\!0}\!I_{\Sigma,P}(A\, ,e^{t
\epsilon}\, ) = \nipi\int_{\Sigma}\trP(\epsilon\, F_{\!A}) \, \, ,
\label{infchanom}
\eeq \beq
I_{\Sigma,P}(A^{h},g)=I_{\Sigma,P}(A,hgh^{\ast})-I_{\Sigma,P}(A,
hh^{\ast})\, \, .
\label{PW}
\eeq In the above expressions, $\epsilon$ is a section of $\ad
P=P\!\times_{H}\!\h$ and $h:P_{\bf C}\to P_{\bf C}$ is an automorphism
which gives a transformation
$\bartial_{\!A}\to\bartial_{\!A^{h}}=h^{-1}\bartial_{\!A}h$. $\trP$ is
the trace of $\mbox{End}(\ad P)$ normalized by $\tr_{\ad
P}=2\lieg^{\vee}\trP$ where $\lieg^{\vee}$ is the dual Coxeter number
of $G$.\footnote{If $G$ is not simple, $P$ decomposes as
$P=P_{1}\times_{\Sigma}\cdots \times_{\Sigma}P_{M}$ according to the
decomposition $G=\prod_{i=1}^{M}G_{i}$ into simple factors. Then,
$\tr_{\ad P}=2\lieg^{\vee}\trP$ should be understood as $\tr_{\ad
P_{i}}=2\lieg_{i}^{\vee}\tr_{P_{i}}$ for each $i$.} The latter
(\ref{PW}) is named as the Polyakov-Wiegmann (P-W) identity. These
properties are just what we would expect for the chiral anomaly in the
model of multi-component spin-half fermions (see (\ref{infanom}),
(\ref{assoc})). Indeed, WZW model was first introduced as the
non-abelian bosonization of spin-half fermions \cite{Witten}. If $P$
is topologically trivial, choosing a trivialization, we can write the
action in a rather explicit way: \beqa
I_{\Sigma}(A,g)&=&\frac{i}{4\pi}\int_{\Sigma}\tr\!\left(\partial
g^{-1}\!\bar \partial
g\right)-\frac{i}{12\pi}\int_{B_{\Sigma}}\tr\!\left(\,
\tilde{g}^{-1}\!d\tilde{g}\, \right)^{\!3}\nonumber \\
&&+\frac{i}{2\pi}\int_{\Sigma}\tr\!\left(\, g\partial
g^{-1}\!A''+A'g^{-1}\!\bar \partial g+A'g^{-1}\!A''g-A'A''\, \right)
\,\, ,
\label{trivwzw}
\eeqa where $B_{\Sigma}$ is a three manifold with boudary $\partial
B_{\Sigma}=\Sigma$, $\tilde{g}$ is an extension of $g :\Sigma \to G$
to $\tilde{g}:B_{\Sigma}\to G$. $A'$ and $A''$ are $\g_{\bf C}$ valued
$(1,0)$ and $(0,1)$ forms representing the connection $A$ with respect
to the chosen trivialization. $\tr$ is defined by
$2\lieg^{\vee}\tr(AB\cdots )=\tr_{\g}(\ad A\ad B \cdots)$ and induces
an invariant bilinear form in the space spanned by roots such that $||
\alpha ||^{2}=2$ for a long root $\alpha$. We can see that the above
action integral is independent of the choice of $B_{\Sigma}$ and
$\tilde{g}$ up to the addition of $2\pi i {\bf Z}$. If the $H$ bundle
$P$ is topologically non trivial, the expression for the action is
more involved. We will exhibit shortly how to construct it. In any
case, $e^{nI_{\Sigma,P}(A,g)}$ is well defined if $n\in {\bf Z}$.

\renewcommand{\theequation}{2.1.\arabic{equation}}\setcounter{equation}{0}
\vspace{0.35cm}
\begin{center}
{\sc 2.1 Path Integral Quantization}\label{2.1}
\end{center}
\hspace{1.5cm} Now, let us quantize the system with the classical
action $kI_{\Sigma,P}(A,g)$ for $k\in {\bf N}$\footnote{If $G$ is a
direct product $\prod_{i=1}^MG_i$ of simple groups, the level $k$ is a
sequence of natural numbers $k=(k_1,\cdots, k_M)$ and $kI_{\Sigma,P}$
should be understood as the sum $\sum_{i=1}^Mk_iI_{\Sigma,P_i}$ where
$I_{\Sigma,P_i}$ is the WZW action for the principal
$G_i/Z_{G_i}$-bundle $P_i$.} considering $A$ as an external gauge
field. A Riemannian metric $\met$ on $\Sigma$ defines an adjoint
invariant inner product on the vector space $\Omega^{0}(\Sigma, \ad
P)$\footnote{Remark that the pointwise composition makes the space
$\G_{P,G}=\Gamma(\Sigma,\ad_{G}P)$ an infinite dimensional Lie group
with Lie algebra $\Omega^{0}(\Sigma,\ad P)$.}  and in turn, this inner
product defines a left-right invariant Riemannian metric $h_{\smet}$
on the space $\G_{P,G}=\Gamma(\Sigma,\ad_{G}P)$ of smooth sections of
$\ad_GP$. We assume without proof that this metric $h_{\smet}$ defines
a left-right invariant measure ${\cal D}_{\!\smet}g$ on $\G_{P,G}$ :
\beq \int_{\G_{P,G}}\!\!{\cal D}_{\!\smet}g \,
f(agb)=\int_{\G_{P,G}}\!\!{\cal D}_{\!\smet}g \, f(g)\, \, ,
\label{leftright}
\eeq where $a$ and $b$ are sections of $\ad_{G_{\bf c}}P$ and $f$ is a
function on $\Gamma(\Sigma, \ad_{G_{\bf c}}P)$ which is well behaved
in some sense. The center of our interest is the behavior of the
function \beq Z_{\Sigma,P}(\, \met,A \,; {\cal O}\,
)=\int_{\G_{P,G}}\!\!\!{\cal D}_{\smet}g\,\, e^{-kI_{\Sigma,P}(A,g)}\,
{\cal O}(g)\, \, ,
\label{Feynmanwzw}
\eeq under the variation of $(\met , A)$. As the field insertion
${\cal O}$,we consider \beq {\cal
O}(g)=\prod_{l=1}^{s}\rho_{\Lambda_{l}}(g(x_{l}))\, \, ,
\label{wzwdef:fieldinsertion}
\eeq where $x_{1}, x_{2}, \cdots , x_{s}$ are points in $\Sigma$,
 $\Lambda_{l}$ parametrizes a unitary irreducible representation of
 $G$ on a vector space $V_{\Lambda_{l}}$ and $x\mapsto
 \rho_{\Lambda}(g(x))$ is the section of $\ad_{GL(V_{\Lambda})}P$
 determined by $g$. Note that $H$ acts on $GL(V_{\Lambda})$ by
 $(h,M)\to hMh^{-1}$ by the Schur's lemma.

\vspace{0.3cm} \underline{Ward Identities}

 The left-right invariance of the measure (\ref{leftright}) and the
 P-W identity (\ref{PW}) lead to the following response of $Z({\cal
 O})$ to the chiral gauge transformation $A\to A^{h}$ by
 $h:P_{\!\C}\to P_{\!\C}$ : \beq Z_{\Sigma,P}(\, \met , A^{h} ; {\cal
 O}\, )=e^{kI_{\Sigma,P}(A, hh^{\ast})}Z_{\Sigma,P}(\, \met , A\,;
 h{\cal O}\, )\, \, , \label{intPW} \eeq where $h{\cal O}$ is defined
 by $h{\cal O}(g)={\cal O}(h^{-1}\!gh^{\ast -1})$. This formula leads
 to the following Ward identities for the insertion of the current $J$
 (see (\ref{defTJ}) for the definition of the current) : \beqa
 \lefteqn{Z_{\Sigma,P}\Bigl(\, \met,A \,\bwhere\, \nipi
 \int_{\Sigma}(J\bartial_{\!A}\epsilon -\bar J
 \partial_{\!A}\epsilon^{\ast})\, {\cal O}\, \Bigr)} \label{onecurr}\\
 &&=\frac{ik}{2\pi}\int_{\Sigma}\trP\!\Bigl( (\epsilon
 +\epsilon^{\ast})F_{\!A}\, \Bigr) \, Z_{\Sigma,P}(\, \met , A\,;
 {\cal O}\, )+Z_{\Sigma,P}\Bigl(\, \met , A \,\bwhere\,
 J(\epsilon){\cal O}+\bar J(\epsilon){\cal O}\, \Bigr)\, \, ,\nonumber
 \eeqa \vspace{-0.9cm} \beqa \lefteqn{Z_{\Sigma,P}\Bigl(\, \met, A
 \,\bwhere
 \,\nipi\int_{\Sigma}\left(J\bartial_{\!A}\epsilon_{1}\right)
 J\epsilon_{2}\sleft x \sright \, {\cal O}\, \Bigr)}\label{twocurr}\\
 &&=k\,
 \trP\!\left(\partial_{\!A}\epsilon_{1}\epsilon_{2}\right)\!\sleft
 x\sright Z_{\Sigma,P}(\, \met, A\,; {\cal O}\,
 )+\frac{ik}{2\pi}\int_{\Sigma}\trP\!\left(
 \epsilon_{1}F_{\!A}\right)Z_{\Sigma,P}(\, \met,A\,;
 J\epsilon_{2}\sleft x\sright \, {\cal O}\, )\nonumber \\ &&\quad
 +Z_{\Sigma,P}\Bigl(\, \met,A \,\bwhere\,
 J[\epsilon_{1},\epsilon_{2}]\sleft x\sright \, {\cal
 O}+J\epsilon_{2}\sleft x\sright \, J(\epsilon_{1}){\cal O}\, \Bigr)\,
 \, ,\nonumber \eeqa where $J(\epsilon)$ and $\bar J(\epsilon)$ are
 defined by $J(\epsilon){\cal
 O}(g)=\frac{d}{dt}\Bigl|_{\!{}_0}\Bigr.{\cal O}(e^{-t\epsilon}\!g)$
 and $\bar J(\epsilon){\cal
 O}(g)=\frac{d}{dt}\Bigl|_{\!{}_0}\Bigr.{\cal
 O}(ge^{-t\epsilon^{\ast}})$. To derive (\ref{twocurr}), we have used
 the relation \beq \delta_{\!A}
 I_{\Sigma,P}(A,hh^{\ast})=\frac{i}{2\pi}\int_{\Sigma}\trP\!\left( \,
 hh^{\ast}\partial_{\!A}(hh^{\ast})^{-1}\delta A''+\delta A'
 (hh^{\ast})^{-1}\bartial_{\!A}(hh^{\ast}) \, \right)\, \, , \eeq
 which is easily proved for the action given in (\ref{trivwzw}) and
 which also holds for the action we shall construct for the
 topologically non-trivial $P$.

We assume that $Z({\cal O})$ behaves under the conformal
transformation $\met \to \met e^{\phi}$ as\footnote{$S_L(\met,\phi)$
is the Liouville functional
$\frac{i}{48\pi}\int_{\Sigma}(\partial\phi\bartial\phi
+2R_{\Theta}\phi)$ which has already appeared in Chapter 1.}  \beq
Z_{\Sigma,P}(\, \met e^{\phi},A\,; {\cal O}\,
)=e^{cS_{L}(\smet,\phi)-\sum_{l}\sDelta_{l}\phi(x_{l})}Z_{\Sigma,P}(\,
\met, A\,; {\cal O}\, )\, \, ,
\label{confWZW}
\eeq where $\sDelta_{l}$'s and $c=c_{G,k}$ are numbers to be
determined. The appearance of the factor $-\sDelta_{l}\phi(x_{l})$
could be understood to come from the renormalization of the field
$\rho_{\Lambda_{l}}(g(x_{l}))$ with respect to the scale determined by
the metric $\met$. This leads to the following expression for the
$(1,1)$-part of the energy momentum tensor : \beq T_{z\bar
z}=-\frac{c_{G,k}}{12}R_{\Theta z\bar z}-2\pi
i\sum_{l=1}^{s}\sDelta_{l}\delta_{x_{l}z\bar z}\, \, .
\label{T11wzw}
\eeq

Next we derive Ward identities for insertion of the energy momentum
tensor. Let $f_{t}:\Sigma \to \Sigma$ be a one parameter group of
diffeomorphisms generated by a vector field $v$ and let
$\tilde{f}_{t}:P\to P$\label{page2:ftilde} be its horizontal lift with
respect to the connection $A$. We consider the variation $(\met,A)\to
(f_{t}^{\ast}\met,\tilde{f}_{t}^{\ast}A)$ for which we think of $A$ as
the connection form on $P$. Then, since the automorphism
$f_{t}^{\sharp}$ of the Lie group $\G_{P,G}=\Gamma(\ad_{G}P)$ defined
by $(f_{t}^{\sharp}g)(x)=\tilde{f}_{t}^{-1}g(f_{t}(x))$ is an isometry
of the Riemannian manifolds $f_{t}^{\sharp}:(\, \G_{P,G},h_{\smet}\,
)\to (\G_{P,G},h_{f_{t}^{\ast}\smet}\, )$ and since we can prove the
invariance of the classical action
$I_{f_{t}^{\ast}\Sigma,P}(\tilde{f}_{t}^{\ast}A,f_{t}^{\sharp}g)=I_{\Sigma,P}(A,g)$,
we have the following identity : \beq Z_{\Sigma,P}(\,
f_{t}^{\ast}\met, \tilde{f}_{t}^{\ast}A\,; {\cal O}\,
)=Z_{\Sigma,P}(\, \met,A\,; f_{t}{\cal O}\, )\, \, ,
\label{diffwzw}
\eeq where $f_{t}{\cal O}$ is given by $f_{t}{\cal O}(g)={\cal
O}(f_{t}^{\sharp}g)$. With the aid of the expression (\ref{T11wzw})
and $\delta(\, \met\, ,A\, )=(\, L_{v}\met\, , i(v)F_{\!A}\, )$, the
infinitesimal version of the identity (\ref{diffwzw}) can be written
as \beqa \lefteqn{ Z_{\Sigma,P}\Bigl(\, \met, A \,\bwhere\, \nipi
\!\int_{\Sigma}dzd\bar z\!\left[ \, \partial_{\bar
z}v^{z}T_{zz}-\frac{c_{G,k}}{12}(\nabla_{\!z}v^{z}\!+\!\nabla_{\!\bar
z}v^{\bar z})R_{\Theta z\bar z}+\partial_{z}v^{\bar z}T_{\bar z \bar
z} \, \right] {\cal O}\, \Bigr)}\label{WardTwzw}\\
&&+Z_{\Sigma,P}\Bigl(\, \met,A \,\bwhere\, \nipi
\!\int_{\Sigma}\![(v^{z}\!J_{z}+v^{\bar z}\!\bar J_{\bar z})F_{\!A}]\,
{\cal O}\, \Bigr)=Z_{\Sigma,P}\Bigl(\, \met, A \,\bwhere \, L_{v}{\cal
O}+\sum_{l=1}^{s}\sDelta_{l}(\nabla_{\!z}v^{z}\!+\!\nabla_{\!\bar
z}v^{\bar z})_{x_{l}}{\cal O}\, \Bigr)\, \, ,\nonumber \eeqa where
$L_{v}$ is defined by $L_{v}{\cal O}=(\frac{d}{dt})_{0}f_{t}{\cal O}$.

\vspace{0.3cm} \underline{Local Expression of the Current and the
Sugawara Construction}

Let us take a local holomorphic section $\sigma$ of $P_{\! \bf C}$
over an open subset $U_{\sigma}\subset\Sigma$. With respect to this
section, the connection is represented by a $(1,0)$ form
$A^{\sigma}=A^{\sigma}_{z}dz$ and the curvature is given by
$F_{A}=\sigma_{\!\ad}\bartial A^{\sigma}$. From the first Ward
identity (\ref{onecurr}), we see that we can take the local
holomorphic part $J^{\sigma}$ of the current $J$ : \beq
J\sigma_{\!\ad} X=J^{\sigma}\!(X)-k\, \tr( A^{\sigma}X)\, \, ,
\label{holJ}
\eeq where $X\in \g_{\bf C}$. Looking at the second Ward identity
(\ref{twocurr}), we see the following singular behavior of the product
of two currents: \beq J_{z}^{\sigma}\!(X)(z)\,
J_{z}^{\sigma}\!(Y)(w)=\frac{k\,\tr(XY)}{(z-w)^{2}}+\frac{1}{z-w}J_{z}^{\sigma}\!([X,Y])(w)+:\!J_{z}^{\sigma}\!(X)(z)\,J_{z}^{\sigma}\!(Y)(w)\!:\,
\, ,
\label{JJsing}
\eeq where the last term in the right hand side is regular as $z\to
w$.

Classically the current and the energy momentum tensor are given by
$J\!\epsilon=k\trP(\!g\partial_{\!\!A}g^{-1}\!\epsilon)$ and
$T_{zz}=\frac{k}{2}\trP(g\partial_{\!A z}g^{-1}\!)^{2}$, and we see
that $T_{zz}$ has a quadratic expression in terms of currents :
$T_{zz}=\frac{1}{2k}(J_{z},J_{z})_{P}$ where $($ , $)_{P}$ is the
bilinear form on $(\ad P_{\!\bf C})^{\ast}$ induced by $\trP$. We
shall assume that in the quantum theory the energy momentum tensor
still has an expression quadratic in the currents. However, the
product $J_{z}(z)J_{z}(z)$ is seen to be ill-defined in view of the
equation (\ref{JJsing}), so we consider the following regularization :
\beq :\!(J,J\,)\!:\, =\lim_{\epsilon\to 0}\left\{
\left(\tau^{\zeta_{\epsilon}}_{z}J(\zeta_{\epsilon}),\tau^{w_{\epsilon}}_{z}J(w_{\epsilon})\right)_{\!\!{}_P}\!
-k\, \mbox{dim}G \left(
\frac{(dz)^{2}}{(\zeta_{\epsilon}-w_{\epsilon})^{2}}-\frac{(dz)^{2}}{12}(\partial_{z}\omega_{z}\!-\!2\omega_{z}^{2})
\right) \right\}\! .
\label{defnormWZW}
\eeq In the above expression, $\zeta_{\epsilon}$, $z$, and
$w_{\epsilon}$ are on a geodesic $\tau$ with equal distance
$d_{\met}(\zeta_{\epsilon},z)=d_{\met}(z,w_{\epsilon})=\epsilon$ and
$\tau^{w}_{z}$ is the parallel transform of $K\ot (\ad P_{\!\bf
C})^{\ast}$ along $\tau$ from $w$ to $z$. We add the term
$-(dz)^{2}(\partial_{z}\omega_{z}-2\omega_{z}^{2})/12$ to give
$(dz)^{2}/(\zeta_{\epsilon}-w_{\epsilon})^{2}$ a covariant meaning in
the limit $\epsilon\to 0$. Taking the assymptotic behavior
(\ref{JJsing}) into account, we see that the limit $\epsilon \to 0$ is
well defined up to a term which depends on the direction of the
geodesic $\tau$ and which we discard. This normal ordered product
$:\!(J,J)\!:$ has a local expression in terms of the holomorphic part
$J^{\sigma}$ (\ref{holJ}) of the current and their regularized product
given in (\ref{JJsing}): \beqa :\!(J_{z},J_{z})\!: &=&\eta^{\rm
ab}:\!J_{z}^{\sigma}(e_{\rm a})J_{z}^{\sigma}(e_{\rm b})\!:
\label{normJJwzw}\\ &&+2(k+\lieg^{\vee})\left\{ \,
-J_{z}^{\sigma}(A^{\sigma}_{z})+\frac{k}{2}\, \tr(A_z^{\sigma
2})-\frac{k\, \mbox{dim}G}{12(k+\lieg^{\vee})}\left(
\partial_{z}\omega_{z}-\frac{1}{2}\omega_{z}^{2}\right) \right\}\, \,
.\nonumber \eeqa where $\{e_{\rm a}\}$ is a base of $\g_{\bf C}$ and
$\eta^{\rm ab}\tr(e_{\rm b}e_{\rm c})=\delta_{\rm c}^{\rm a}$.

Now let us assume that $T_{zz}$ is proportional to
$:\!\!(J_{z},J_{z})\!\!:$. Inserting the expression (\ref{normJJwzw})
into the Ward identity (\ref{WardTwzw}), we see that we must have \beq
T_{zz}=\frac{1}{2(k+\lieg^{\vee})}:\!( J_{z},J_{z}\, )\!:\, \, ,
\label{defTzzwzw}
\eeq \beq \mbox{and}\hspace{1.2cm} c_{G,k}=\frac{k\,
\mbox{dim}G}{k+\lieg^{\vee}}\, \, ,\qquad \sDelta_{l}=\frac{C_{2}(
\Lambda_{l})}{2(k+\lieg^{\vee})}\, \, ,
\label{cwzw}
\eeq where $C_{2}(\Lambda)$ is a constant defined by
$C_{2}(\Lambda)=\eta^{\rm ab}\rho_{\Lambda}(e_{\rm
a})\rho_{\Lambda}(e_{\rm b})$. This assumption is non-trivial and
together with the Ward identities (\ref{onecurr}), (\ref{twocurr})
lead to the functional differential equations for $Z_{\Sigma,P}(\,
\met, A\,; \!{\cal O}\, )$ which, in the topologically trivial case,
gave the non-perturbative definitions of the partition and correlation
functions \cite{KZ}, \cite{E-O}, \cite{B}.

\renewcommand{\theequation}{2.2.\arabic{equation}}\setcounter{equation}{0}
\vspace{0.3cm}
\begin{center}
{\sc 2.2 The WZW Action for General Principal $H$-Bundle}\label{2.2}
\end{center}
\hspace{1.5cm} We choose a parametrized circle $c:S^{1}\to S\subset
\Sigma$\label{page2:S} which devides $\Sigma$ into two parts
$\Sigma_{\infty}$, $\Sigma_{0}$ with boundary $S=\partial
\Sigma_{0}=-\partial \Sigma_{\infty}$ and a complex coordinate $z$
defined on a neighborhood $U$ of $S$ just as in Chapter 1. In the next
section, we consider the $states$ $Z_{\Sigma_{0}}(A_{0};{\cal O}_{0})$
and $Z_{ \Sigma_{\!\infty}}\!( A_{\infty};{\cal O}_{\!\infty})$
formally given by the integral \beqa Z_{\Sigma_{0}}(A_{0};{\cal
O}_{0})(\gamma)&=&\int_{\gamma=g|_S}\!\!{\cal D}_{\Sigma_{0}}g \,
e^{-kI_{\Sigma_{0}}(A_{0},g)}\, {\cal O}_{0}(g)\, \, ,\label{wf}\\
Z_{\Sigma_{\!\infty}}\!(A_{\!\infty},{\cal
O}_{\!\infty})(\gamma)&=&\int_{g|_S=\gamma}\!\!\!{\cal
D}_{\Sigma_{\infty}}g \,
e^{-kI_{\Sigma_{\!\infty}}\!(A_{\!\infty},g)}\, {\cal
O}_{\!\infty}(g)\, \, ,\label{wf*} \eeqa and develop an operator
formalism which is {\it covariant} in a sense that becomes clear. To
do that, we first construct the weights $e^{-kI}$ for Riemann surfaces
with a circle boundary, following the way found by Felder, Gawedzki
and Kupiainen \cite{F-G-K}. This leads to a construction of the WZW
action $I_{\Sigma,P}(A,g)$ for general principal $H$-bundle $P$ .

\vspace{0.35cm} \underline{The Line Bundle $\LWZ$ over the Loop Group
$LG_{\!\C}$}\label{page:LGC}

\vspace{0.1cm} We start with defining the weight
$e^{-kI_{\Sigma_0}(A,g)}$ for a $\h$ valued one form $A$ and a
$G_{\!\bf C}$ valued function $g$ on $\Sigma_0$ with the boundary loop
$\gamma=g|_S\in LG_{\!\bf C}$. We ``cap'' the surface $\Sigma_0$ by
the disc $\Dinf=\{z^{-1}\in {\bf C};|z|^{-1}\leq 1\}$ and denote the
capped surface by $\hat{\Sigma}_0$. If we choose a map
$g_{\infty}:\Dinf \to G_{\!\bf C}$ with $g_{\infty}|_{-\partial
\Dinf}= \gamma$, we can define the action $I_{\hat{\Sigma}_0}(0\star
A,g_{\infty}\star g)$ by (\ref{trivwzw}) where $(0\!\star\!
A,g_{\infty}\!\star\! g)$ coincides with $(0,g_{\infty})$ on $\Dinf$
and with $(A,g)$ on $\Sigma_0$. Since we do not have a canonical way
to choose $g_{\infty}$, we consider the set $\Dinf G_{\!{\bf
C}}|_{\gamma}=\{ g_{\infty}\!:\!\Dinf \to G_{\!\bf
C}\,;\,g_{\infty}|_S=\gamma\}$ of all extensions and put a suitable
equivalence relation\footnote{The following is the explicit
description of the equivalence relation in $D_{\infty}G_{\!\bf
c}\times \C$ where $D_{\infty}G_{\!\bf c}$ is the group of smooth maps
of $D_{\infty}$ to $G_{\!\bf c}$ : $$(gh,c)\sim
(g,c\exp\{kI_{{\CP}}(h\star 1)+k\Gamma_{D_{\infty}}(g,h)\})$$ where
$g$, $h\in D_{\infty}G_{\!\bf c}$ such that $h|_{\partial
D_{\infty}}=1$ and $\Gamma_{D_{\infty}}(g,h)$ is given above.}  in
$\Dinf G_{\!\bf C}|_{\gamma}\times {\bf C}$ which gives a complex line
$\left(\LWZ^k\right)_{\!\gamma}$ so that the class \beq
e^{-kI_{\Sigma_0}(A,g)}=\{(g_{\infty},e^{-kI_{\hat{\Sigma}_0}(0\star
A,g_{\infty}\star g)})\} \in \left(\LWZ^k\right)_{\!\gamma}
\label{weightWZWD0}
\eeq is independent of the choice $g_{\infty}$ of extension. Gathering
all the loops $\gamma\in LG_{\!\bf C}$, we obtain a {\it line bundle}
$\LWZ^k=\bigcup_{\gamma}( \LWZ^k)_{\!\gamma}$. Therefore, the wave
function $\gamma \mapsto Z_{\Sigma_0}(A_0;{\cal O}_0)(\gamma)$ is a
{\it section} of $\LWZ^k|_{LG}\stackrel{\bf C}{\longto} LG$.

The group structure of $LG_{\!\bf C}$ lifts by
$\{(g_1,c_1)\}\{(g_2,c_2)\}=\{(g_1g_2,e^{-k\Gamma_{\Dinf}(g_1,g_2)}c_1c_2)\}$
to the semigroup structure of $\LWZ^k$ where
$\Gamma_{D_{\infty}}(g,h)$ is given by
$\frac{i}{2\pi}\int_{D_{\infty}}\tr(h\partial h^{-1}g^{-1}\bartial
g)$.\label{page:GammaSigma} The set $\bigl(\LWZ\bigr)^{\times}$ of
non-zero elements in $\LWZ$ is the basic central extension
$\tilde{LG}_{\!\bf C}$ of $LG_{\!\bf C}$. Most strikingly the weight
$e^{-kI_{\Sigma_0}}$ satisfies the following version of the
Polyakov-Wiegmann identity : \beqa
e^{-kI_{\Sigma_{0}}(A,hgh^{*})}&=&e^{-kI_{\Sigma_{0}}(A,h)}e^{-kI_{\Sigma_{0}}(A^{h},g)}e^{-kI_{\Sigma_{0}}(A,h^{*})}e^{-k\Gamma_{\Sigma_{0}}(A,h,h^{*})}\,
\, ,\label{locPW}\\
&;&\Gamma_{\Sigma_{0}}(A,h,h^*)=\frac{i}{2\pi}\int_{\Sigma_{0}}\tr\!\left(
h^*\partial_{\!A}h^{*-1}h^{-1}\bartial_{\!A}h\right)\,, \eeqa for
$G_{\!\bf C}$ valued functions $g$, $h$ on $\Sigma_0$. This identity
holds true if $h$ is $H_{\!\bf C}$ valued with contractible boundary
loop $h|_S=[\gamma]\in L_{0}H_{\!\bf C}\cong LG_{\!\bf C}/Z_{G}$,
though we needs some care to define $e^{-kI_{\Sigma_{0}}(A,h)}\!\ot\!
e^{-kI_{\Sigma_{0}}(A,h^{*})}\in (\LWZ^k)_{\gamma}\!\ot\!
(\LWZ^k)_{\gamma^{*}}$. If $h$ is $H$-valued, (\ref{locPW}) can be
written as

\beq \gamma^{-1}e^{-kI_{\Sigma_{0}}(A,g)}\gamma
=e^{-kI_{\Sigma_{0}}(A^{h},h^{-1}gh)} \, \, ,
\label{cov}
\eeq where $\gamma\in LG$ is a representative of the boundary value
$h|_{\partial \Sigma_{0}}\in L_{0}H$. In this sense we can say that
the weight $e^{-kI_{\Sigma_{0}}(A, g)}$ defined above is {\it gauge
covariant}.

{\it Remark}. The identity (\ref{cov}) holds true when $[\gamma]\in
LH$ is a non-contractible boundary value of an $H$ valued function $h$
defined on a neighborhood of $\partial \Sigma_0$ and when $g$ is
different from $1$ only on the same neighborhood. Note that the
pointwise adjoint action of $LH$ on $LG_{\!\bf C}$ lifts to a unique
automorphic action on the semigroup $\LWZ^k$ (see Appendix 1).

Quite in the similar way, we define a line bundle $\LWZ^{*k}$ so that
the class \beq
e^{-kI_{\Sgminf}(A,g)}=\{(g_0,e^{-kI_{\hat{\Sigma}_{\!\infty}}(A\star
0,g\star g_0)})\}\in \left(\LWZ^{*k}\right)_{g|_S}
\label{weightWZWDinfty}
\eeq is independent of the extension $g_0:D_0\to G_{\!\bf C}$ of a
$G_{\!\bf C}$ valued function $g$ on $\Sgminf$ where $A$ is a $\h$
valued one form on $\Sgminf$. The weight $e^{-kI_{\Sgminf}(A,g)}$
satisfies the P-W identity \beq
e^{-kI_{\Sigma_{\!\infty}}\!(A,hgh^{*})}=e^{-kI_{\Sigma_{\!\infty}}\!(A,h)}e^{-kI_{\Sigma_{\!\infty}}\!(A^{h},g)}e^{-kI_{\Sigma_{\!\infty}}\!(A,h^{*})}e^{-k\Gamma_{\Sigma_{\!\infty}}\!(A,h,h^{*})}
\, \, ,
\label{loc*PW}
\eeq with respect to the composition law in $\LWZ^{*k}$ analogous to
the one in $\LWZ^k$. Note that we can identify $\LWZ^{*k}$ as the dual
$\LWZ^{-k}$ of $\LWZ^k$ by the pairing \beq
\{(g_0,c_0)\}.\{(g_{\infty},c_{\infty})\}=c_0c_{\infty}e^{kI_{{\bf
P}^1}(g_{\infty}\star g_0)}\in {\bf C}.  \eeq

\vspace{0.3cm} \underline{Construction of the Action}

We come back to the closed Riemann surface
$\Sigma=\Sigma_{\!\infty}\cup \Sigma_{0}$. If
$g_{\!\infty}:\Sigma_{\infty}\to G$ and $g_{0}:\Sigma_{0}\to G$ are
restrictions of a smooth map $g:\Sigma \to G$, and if $\g$ valued one
forms $A_{\!\infty}\in \Omega^{1}(\Sigma_{\!\infty},\g)$, $A_{0}\in
\Omega^{1}(\Sigma_{0},\g)$ are restrictions of $A\in
\Omega^{1}(\Sigma,\g)$, then, the pairing of
$e^{-kI_{\Sigma_{\!\infty}}(A_{\!\infty},g_{\!\infty})}\in \LWZ^{*k}$
and $e^{-kI_{\Sigma_{0}}(A_{0},g_{0})}\in \LWZ^k$ reproduces the
weight \beq
e^{-kI_{\Sigma}(A,g)}=e^{-kI_{\Sigma_{\!\infty}}(A_{\!\infty},g_{\infty})}.e^{-kI_{\Sigma_{0}}(A_{0},g_{0})}\in
{\bf C}\, \, , \eeq for the trivial $H$ bundle over the closed surface
$\Sigma$ which is given in (\ref{trivwzw}).

To construct the action integral for the topologically non trivial $H$
bundle, let us choose an open neighborhood $U_0$ (resp.$U_{\infty}$)
of $\Sigma_0$ (resp.$\Sgminf$). We consider smooth maps
$g_{\infty}:U_{\infty}\to G$, $g_{0}:U_{0}\to G$ and one forms
$A_{\!\infty}\in \Omega^{1}(U_{\!\infty},\g)$, $A_{0}\in
\Omega^{1}(U_{0},\g)$ related through the transition function
$h_{\infty 0}: U_{\!\infty}\cap U_{0}\to H$ by \beqa A_{0}&=&h_{\infty
0}^{-1}A_{\!\infty}h_{\infty 0}+h_{\infty 0}^{-1}dh_{\infty 0}\, \,
,\label{relA}\\ g_{0}&=&h_{\infty 0}^{-1}\, g_{\infty}h_{\infty
0}\qquad \qquad\mbox{on}\quad U_{\!\infty}\cap U_{0} \, .\label{relg}
\eeqa The data $( U_{\!\infty}, U_{0}, h_{\infty 0})$ determines an
$H$-bundle $P$ over $\Sigma$. A connection $A$ of $P$ is provided by
the pair $\{ A_{\infty},A_0\}$ of one forms and $\{g_{\infty},g_0\}$
gives a section $g$ of $\ad_GP$. We cannot take a pairing of
$e^{-kI_{\Sigma_{\!\infty}}(A_{\!\infty},g_{\infty})}$ and
$e^{-kI_{\Sigma_{0}}(A_{0},g_{0})}$ because the base elements
$\gamma_{\infty}=g_{\infty}|_{S}$, $\gamma_{0}=g_{0}|_{S}$ do not
coincide but related by $\gamma_{0}=\gamma_{\infty
0}^{-1}\gamma_{\infty }\gamma_{\infty 0}$ where $\gamma_{\infty
0}=h_{\infty 0}|_S\in LH$. Fortunately, since the group $LH$ acts on
the semigroup $\LWZ^{k}$ (Appendix 1), we can take \beq
e^{-kI_{\Sigma,P}(A,g)}=e^{-kI_{\Sigma_{\!\infty}}(A_{\!\infty},g_{\infty})}.\ad\gamma_{\infty
0}\!\left(\, e^{-kI_{\Sigma_{0}}(A_{0},g_{0})}\, \right) \in {\bf C}\,
,
\label{nontrivwzw}
\eeq as a candidate for the WZW action $I_{\Sigma,P}(A,g)$ for general
$H$ bundle $P$.

We can check that this satifies the conditions (\ref{infchanom}) and
(\ref{PW}). (\ref{infchanom}) is obvious. The global P-W identity
(\ref{PW}) for the above action (\ref{nontrivwzw}) follows from
(\ref{locPW}) and (\ref{loc*PW}) since $\ad\gamma_{\!\infty 0}$
preserves the structure of semigroup of $\LWZ^k$ and the paring of
$\LWZ^{*k}$ and $\LWZ^{k}$ satisfies
$(\tilde{\gamma}'_{1}\tilde{\gamma}'_{2}).(\tilde{\gamma}_{1}\tilde{\gamma}_{2})=\tilde{\gamma}'_{1}.\tilde{\gamma}_{1}\cdot
\tilde{\gamma}'_{2}.\tilde{\gamma}_{2}$ for $\tilde{\gamma}'_{i}\in(
\LWZ^{*k})_{\!\gamma_{i}}$ and $\tilde{\gamma}_{i}\in (
\LWZ^{k})_{\!\gamma_{i}}$ where $\gamma_{1}$, $\gamma_{2}\in LG_{\!
\bf C}$. Therefore, we conclude that (\ref{nontrivwzw}) is the proper
definition of the weight for the WZW model.

\renewcommand{\theequation}{2.3.\arabic{equation}}\setcounter{equation}{0}
\vspace{0.4cm}
\begin{center}
{\sc 2.3 Covariant Operator Formalism}\label{2.3}
\end{center}
\hspace{1.5cm} Now we go on to the description of the space of states
and of the action of loop groups. As was noticed, the wave functions
(\ref{wf}) and (\ref{wf*}) are sections of $\LWZ^k$ and $\LWZ^{*k}$
respectively over the loop space $LG$. We rewrite the formal
definitions as \beqa Z_{\Sigma_{0}}(A_{0},{\cal
O}_{0})(\gamma)&=&\int_{1=g|_S}\!\!{\cal D}_{\Sigma_{0}}g\,
e^{-kI_{\Sigma_{0}}(A_{0},g_{0}g)}{\cal O}_{0}(g_{0}g) \, ,\\
Z_{\Sigma_{\!\infty}}\!(A_{\!\infty},{\cal
O}_{\!\infty})(\gamma)&=&\int_{g|_S=1}\!\!\!{\cal
D}_{\Sigma_{\infty}}g \,
e^{-kI_{\Sigma_{\!\infty}}\!(A_{\!\infty},g_{\infty}g)}\, {\cal
O}_{\!\infty}(g_{\infty}g)\, \, , \eeqa where $g_0:\Sigma_0\to G$ and
$g_{\infty}:\Sgminf\to G$ are any maps satisfying $g_{0}|_{\partial
\Sigma_{0}}=g_{\infty}|_{-\partial \Sigma_{\!\infty}}=\gamma$. Then,
we see that these sections can be extended to holomorphic sections
over the complexified loop space $LG_{\!\bf C}$ by admitting $g_0$ and
$g_{\infty}$ to be $G_{\!\bf C}$ valued.\footnote{Let ${\cal Y}\subset
LG_{\!\bf C}$ be an open subset and $g_{\infty}:{\cal Y}\to
D_{\!\infty}G_{\!\bf C}$ be a holomorphic map such that
$g_{\infty}(\gamma)|_{-\partial D_{\infty}}=\gamma$. Then the section
$\gamma\mapsto \{(g_{\infty}(\gamma),1)\}$ of $\LWZ^k$ over ${\cal Y}$
is by definition holomorphic.}

 {\it Remark}. We may as well change the $H$-frame on $S$. If a new
frame $s_1$ is related to the original one $s_0$ by
$s_1=s_0\gamma_{01}$ where $\gamma_{01}\in LH$, the wave function
$\Phi^{(s_1)}\in H^0(LG_{\!\bf C},\LWZ^k)$ with respect to $s_1$ is
related to the original one $\Phi^{(s_0)}\in H^0(LG_{\!\bf C},\LWZ^k)$
by \beq
\Phi^{(s_1)}(\gamma)=\gamma_{01}^{-1}\Phi^{(s_0)}(\gamma_{01}\gamma\gamma_{01}^{-1})\gamma_{01}
\eeq which is due to the covariance (\ref{cov}) of the weight.

\vspace{0.3cm} \underline{Representation of Loop Group on the Space of
States}

Since the line bundle $\LWZ^{k}$ is acted on by the group
$\tilde{LG}_{\!\bf C}\cong\bigl(\LWZ\bigr)^{\times}$ from the left and
from the right through the covering map $\bigl(\LWZ\bigr)^{\times}\to
\bigl(\LWZ^k\bigr)^{\times}$, there are representations of
$\tilde{LG}_{\!\bf C}$ on the space $H^{0}(LG_{\!\bf C},\LWZ^{k})$ of
sections---the left and the right representations : \beq
J(\tilde{\gamma}_1)\bar
J(\tilde{\gamma}_2)\Phi(\gamma)=\tilde{\gamma}_1\Phi(\gamma_1^{-1}
\gamma\gamma_2^{*-1})\tilde{\gamma}_2^* \, ,
\label{wzwdef:reprJ}
\eeq where the anti-holomorphic anti-automorphism
$\tilde{\gamma}\mapsto \tilde{\gamma}^*$\label{page:gamma*} of $\LWZ$
is defined by $\{(g,c)\}^*=\{(g^*,c^*)\}$ where $g\in
D_{\!\infty}G_{\!\C}$ and $c\in \C$. The P-W identity (\ref{locPW})
shows that, for any smooth map $h:\Sigma_{0}\longto H_{\!\bf C}$ with
contractible boundary value $[\gamma] \in L_{0}H_{\!\bf C}\cong
LG_{\!\bf C}/Z_{G}$, the element
$\tilde{\gamma}=e^{-I_{\Sigma_{0}}(A_{0}^{h},h^{-1})-\frac{1}{2}\Gamma_{\Sigma_{0}}(A_{0},h,h^{*})}\in
(\tilde{LG}_{\!\bf C})_{\gamma}$ gives a transformation of
$H^0(LG_{\!\bf C}, \LWZ^{k})$ such that \beq J(\tilde{\gamma}){\bar
J}(\tilde{\gamma})Z_{\Sigma_{0}}(A_{0};{\cal
O})=Z_{\Sigma_{0}}(A_{0}^{h};h^{-1}{\cal O})\, \, .
\label{bbbb}
\eeq By taking the differentials of this equation, we can identify the
currents with the infinitesimal generators of the representations $J$,
$\bar J$. To see this explicitly, we take the connection $A_{0}$ to be
flat on a neighborhood $V_0$ of $S$ with holonomy $e^{-2\pi i a}\in H$
along $S$ where $a\in \h$. We can take a holomorphic frame $\sigma$ of
the $H_{\!\bf C}$ bundle $P_{\!\bf C}$ over $V_0$ with respect to
which the connection $A_{0}$ is represented as $A_0^{\sigma}=-a\,
dz/z$ and which gives an $H$-frame on $S$. Let $\varrho:\Sigma_{0}\to
[0,1]$ be a cut off function such that $\varrho=1$ on $S$ and
$supp(\varrho)\subset V_0$. For $n\in {\bf Z}$ and $v\in \g$, we
consider the vector $\tilde{vz}^{\!n}\in\mbox{Lie}(\tilde{LG}_{\!\bf
C})$ tangent to the curve $e^{-I_{\Sigma_{0}}(h_t)}$ in
$\tilde{LG}_{\!\bf C}$ where $h_t=e^{t\varrho vz^n}$. Then the
equation (\ref{bbbb}) leads to \beq
J(\tilde{vz}^{\!n})Z_{\Sigma_{0}}^{(\sigma)}(A_{0}; {\cal
O}\,)=Z_{\Sigma_{0}}^{(\sigma)}(A_{0};J\!\sigma_{\!n}(v){\cal O}\, )\,
,
\label{lll}
\eeq
\beq \mbox{where} \qquad\qquad J\!\sigma_{\!n}(v)=\nipi \oint_S J
\sigma_{\ad} z^n v =J_{n}^{\sigma}(v)+k\, \tr(av)\delta_{n,0}\, ,
\label{page:Jsgmn}
\eeq
and similar equations for $\bar J$. $J^{\sigma}_n(v)$ is a Laurent
coefficient of the holomorphic part $J^{\sigma}$ of the current given
in (\ref{holJ}). These $J\!\sigma_{\!n}(v)$'s generate of course the
affine Kac-Moody algebra with center $k$ which could be seen by using
the operator product expansions (\ref{JJsing}). Note also that the
Laurent coefficient of the energy momentum tensor is expressed as \beq
L_{n}=\frac{1}{2(k+\lieg^{\vee})}\sum_{m\in {\bf Z}}\eta^{\rm
ab}:\!J_{-m}^{\sigma}(e_{\rm a})J_{n+m}^{\sigma}(e_{\rm
b})\!:+J_{n}^{\sigma}(a)+\frac{k}{2}\tr(a^2)\delta_{n,0}-\frac{c_{G,k}}{24}\delta_{n,0}\,
\, ,
\label{virgenwzw}
\eeq \beq \mbox{where}\qquad
:\!J_{n}^{\sigma}(X)J_{m}^{\sigma}(Y)\!:\, \, \, =\left\{
\begin{array}{rl} J_{n}^{\sigma}(X)J_{m}^{\sigma}(Y)&\quad \mbox{if
$n\leq m$}\\ \noalign{\vskip 0.2cm} J_{m}^{\sigma}(Y)J_{n}^{\sigma}(X)
&\quad \mbox{if $m< n$}
\end{array}\right.\, \, .
\eeq These $L_{n}$'s generate the Virasoro algebra of central charge
$c_{G,k}$ given in (\ref{cwzw}).

Finally we comment on the choice of the space of states. Due to the
relation (\ref{bbbb}), the state
$Z_{\Sigma_0}=Z_{\Sigma_0}(A_0=0;{\cal O}=1)$ is invariant under
$J(e^{-I_{\Sigma_0}(h)})$ and $\bar J(e^{-I_{\Sigma_0}(h)})$ for any
holomorphic map $h:\Sigma_0\to G_{\!\bf C}$. In other words,
$Z_{\Sigma_0}$ is annihilated by
$J(\tilde{\W}_{\Sigma_0}^{(0,\lieg)})$ and $\bar
J(\tilde{\W}_{\Sigma_0}^{(0,\lieg)})$ where
$\W_{\Sigma_0}^{(0,\lieg)}\subset \HH^{(0,\lieg)}={\rm Lie}(LG_{\!\bf
C})$ is the space of $\g_{\!\bf C}$-valued functions on S that extends
holomorphically over $\Sigma_0$. This space
$\W_{\Sigma_0}^{(0,\lieg)}$ is comparable with the subspace
$\HH_+^{(0,\lieg)}$ in the same sense as in Chapter 1. Motivated by
this observation, we take as the space of states, the space ${\cal
H}^{(G,k)}\subset H^0(LG_{\!\bf C},\LWZ^k)$ spanned by a family of
vectors each of which is annihilated by $J(\tilde{\W})$ and $\bar
J(\tilde{\W})$ for some subspace $\W\subset \HH^{(0,\lieg)}$
comparable with $\HH_+^{(0,\lieg)}$.\label{page:spstWZW} This subspace
${\cal H}^{(G,k)}$ is preserved by $J(\tilde{LG}_{\!\bf C})\times \bar
J(\tilde{LG}_{\!\bf C})$ since $LG_{\!\bf C}$ is contained in
$GL_{res}(\HH^{(0,\lieg)})$. Though it has not yet been proved, it is
likely that the Virasoro generators (\ref{virgenwzw}) can act on this
space ${\cal H}^{(G,k)}$.

\vspace{0.3cm} \underline{Field-State Correspondence}

Now we consider the state at the boundary $S$ of the standard disc
$D_0=(D_0,\met_0)$ (see Chapter 1) with a local field insertion. We
denote by $|O\rangle$ the state $Z^{(s_0)}_{D_0}(A_0;O(0))$
corresponding to the local field $O$ where $s_0$ is a frame on $D_0$
which is horizontal with respect to the canonically flat connection
$A_0$. We consider the field $O_{\Lambda}(g)=\rho_{\Lambda}(g(0))$ for
a unitary irreducible representation $\rho_{\!\Lmd}:G\to
GL(V_{\Lmd})$. Since $A_0$ is invariant under the chiral gauge
transformation by holomorphic maps from $D_0$ to $G_{\!\bf C}$, the
relation (\ref{bbbb}) shows that\footnote{The P-W identity
(\ref{locPW}) and hence the relation (\ref{bbbb}) can be seen to hold
when we take $h$ and $h^*$ to be independent.}  \beq
J(e^{-I_{D_{0}}(g_1)}){\bar
J}(e^{-I_{D_{0}}(g_2)})|O_{\!\Lambda}\rangle
=|\rho_{\!\Lambda}(g_1(0)^{-1})O_{\!\Lambda}\,
\rho_{\!\Lambda}(g_2(0)^{*-1})\rangle \, \, ,
\label{aaaa}
\eeq for $G_{\!\bf C}$ valued holomorphic functions $g_1$ and $g_2$ on
$D_{0}$. We recall that $e^{-I_{D_0}(g)}$ in the above expressions is
an element $e^{-I_{D_0}(0,g)}$ of $\tilde{LG}_{\!\C}$ over the
boundary loop $g|_S$ of $g:D_0\to G_{\!\C}$. (See
eq. (\ref{weightWZWD0}) for the definition)

Let us now choose a maximal torus $T_{G}$ of $G$ and $T=T_{G}/Z_{G}$
of $H$. We also choose a chambre $\Ch$ in $\V=i\liet$ (see Appendix 2
for the notations and basics on the root system and the Weyl
group). We denote by $B^{+}$\label{page:B+} the Borel subgroup of
$LG_{\!\C}$ consisting of boundary loops $g|_S$ of holomorphic maps
$g:D_{0}\to G_{\!\C}$ such that $g(0)$ are in the Borel subgroup
$B_0^+$\label{page:B0} of $G_{\!\bf C}$ determined by $\Ch$. We also
denote by $N^+$ the subgroup of $B^+$ consisting of boundary loops
$g|_S$ of holomorphic maps $g$ with $g(0)\in N_{0}^+$, where $N_0^+$
is the maximal unipotent subgroup of $B_0^+$. $B^+$ is embedded into
$\tilde{LG}_{\!\C}$ by $g|_{S}\in B^+ \mapsto e^{-I_{D_{0}}(g)}\in
\widetilde{B}^+$.  We take as the index $\Lambda$ the highest weight
of the representation $\rho_{\Lambda}$. If we denote by
$\Lmd^*=-w_{0}\Lmd$\label{page:Lmd*} the highest weight of the dual
representation ${}^t \!\rho_{\Lmd}^{-1}:G_{\!\C}\to
GL(V_{\Lmd}^*)$\footnote{$w_{0}\in W$ is the element of maximum length
with respect to the simple reflections determined by $\Ch$.}, the
above equation (\ref{aaaa}) shows that the state
$\Phi_{\Lmd}=|(O_{\Lambda^*})^{\!-\Lambda}_{\,\,-\Lambda}\rangle$ is a
highest weight state\footnote{ We say that a state $\Phi$ is a
$highest$ $weight$ $state$ $with$ $weight$ $(\Lmd_{L}, \Lmd_{R})$ of
the left-right representation when it is annihilated by the
infinitesimal generators of $\widetilde{N}^{+}\subset
\tilde{L}G_{\!\C}$ $$\hspace{2cm}J_{n}(v)\Phi=\bar
J_{n}(v)\Phi=J_{0}(e_{\alpha})\Phi=\bar J_{0}(e_{\alpha})\Phi=0\, ,
\qquad \mbox{ for $n>0$, $v\in \g_{\!\C}$, $\alpha \in
\Delta_{+}$}\label{page1:Delta+} \, ,$$ $$\mbox{and}\qquad \qquad
\quad J_{0}(t)\Phi=\Lmd_{L}(t)\Phi\, \, ,\quad \bar
J_{0}(t)\Phi=\Lmd_{R}(t^*)\Phi \, , \qquad \mbox{for}\quad t\in
\liet_{\!\C} .\hspace{3cm}$$ where $J_n(v)=J(\tilde{vz}^{\!n})$ and
$\bar J_n(v)=\bar J(\tilde{vz}^{\!n})$ for $n\in{\bf Z}$ and $v\in
\g_{\!\bf C}$.  }  with weight $(\Lmd,\Lmd)$ of the left-right
representations $J\times \bar J$. Hence we have \beqa
\Phi_{\!\Lmd}(g_1g_2^*|_S)
&=&e^{\!-\Lmd}(g_1(0))\left\{J(e^{-I_{D_{0}}(g_1)})\bar
J(e^{-I_{D_{0}}(g_2)})\Phi_{\!\Lmd}\right\}(g_1g_2^*|_S)\label{hws} \\
&=&e^{\!-\Lmd}(g_1(0))\,e^{-kI_{D_{0}}(g_1)}\Phi_{\!\Lmd}(1)e^{-kI_{D_{0}}(g_2^*)}=
c_{\Lmd} e^{\!-\Lmd}(g_1(0))\,e^{-kI_{D_{0}}(g_1g_2^*)}\,,\nonumber
\eeqa for $g_1|_S\in B^+$, $g_2|_S\in N^+$ and
$c_{\Lmd}=\Phi_{\!\Lmd}(1)$ is a complex number. Since the subset
$B^+(N^+\!)^*$ is known to be open dense in $LG_{\!\C}$ \cite{P-S},
there is at most one such highest weight state. In \cite{F-G-K}, it is
shown that this section $g_1g_2^*|_S\mapsto
e^{\!-\Lmd}(g_1(0))e^{-kI_{D_{0}}(g_1g_2^*)}$ of $\LWZ^k$ over
$B^+(N^+\!)^*$ extends to a section over the whole space $LG_{\!\C}$
if and only if $0\leq (\Lambda,\alpha )\leq k$ for any $\alpha\in
\Delta_{+}$. Such a weight $\Lmd$ is said to be integrable at level
$k$ and the set of integrable weight at level $k$ is denoted by
$\PPpk$.\label{page:Pk+} As is shown in \cite{F-G-K}, the highest
weight state of $J\times \bar J$ must have the equal left-right
highest weight $\Lmd_{L}=\Lmd_{R}$.

The state $\Phi_{\!\Lmd}$ for $\Lmd\in \PPpk$ generates an irreducible
$\tilde{LG}_{\!\C}\times \tilde{LG}_{\!\C}$ module ${\cal
H}^{(k)}_{\Lmd}\subset {\cal H}^{(G,k)}$ which is isomorphic to
$L_{(\Lmd,k)}\ot \overline{L_{(\Lmd,k)}}$ where $L_{(\Lmd,k)}$
($resp.$ $\overline{L_{(\Lmd,k)}}$) is the holomorphic ($resp.$
anti-holomorphic) irreducible representation of the group
$\tilde{LG}_{\!\C}$ at level $k$ with highest weight $\Lmd\in
\PPpk$.\label{page:LLmdk} We see that this subspace ${\cal
H}_{\Lmd}^{(k)}$ corresponds to the current descendant fields in view
of the relations \beq J_{n}(v)|O\rangle=|J_{n}^{s_{0}}(v)O\rangle \,,
\qquad \bar J_{n}(v)|O\rangle =|\bar J_{n}^{s_{0}}(v^*)O\rangle\,,
\eeq which follow from (\ref{lll}) and analogous equation for $\bar
J$. We do not know whether the direct sum $\bigoplus_{\Lmd\in
\PPpk}{\cal H}_{\Lmd}^{(k)}$ is dense in ${\cal H}^{(G,k)}$ or not,
but if it {\it is} dense, we have a one to one correspondence
$O\leftrightarrow |O\rangle$ of scaling fields and states of definite
$(L_0,\bar L_0)$ values.

\vspace{0.3cm} \underline{Gluing}

Let us comment on the pairing of the state
$Z_{\Sgminf}(A_{\infty};{\cal O}_{\infty})\in \check{\cal H}^{(G,k)}$
at the {\it incoming} circle and the state $Z_{\Sigma_0}(A_0;{\cal
O}_0)\in {\cal H}^{(G,k)}$ at the {\it outgoing}
circle\label{page2:spstWZW}, where we say that a boundary circle $S$
of an oriented two manifold $\Sigma$ is incoming (resp. outgoing) when
$S=-\partial \Sigma$ (resp. $S=\partial \Sigma$). The space
$\check{\cal H}^{(G,k)}$ is a certain subspace of $H^0(LG_{\!\C},
\LWZ^{*k})$ which shall be made precise shortly.

 Na\"{\i}vely, the pairing is given by the path-integration over the
loops in $G$ : $\langle \Psi,\Phi\rangle=\int_{LG}{\cal
D}\gamma\Psi(\gamma).\Phi(\gamma)$. If we could define it, due to the
left-right invariance of the measure, it would satisfy the relation
\beq \langle J(\tilde{\gamma}'_1)\bar
J(\tilde{\gamma}'_2)\Psi,J(\tilde{\gamma}_1)\bar
J(\tilde{\gamma}_2)\Phi \rangle
=(\tilde{\gamma}'_1.\tilde{\gamma}_1)\langle \Psi,\Phi\rangle
(\tilde{\gamma}_2'^*.\tilde{\gamma}_2^*) \,,
\label{Haar}
\eeq for $\tilde{\gamma}'_i\in \bigl(\LWZ^*\bigr)^{\times}$ and
$\tilde{\gamma}_i\in \bigl(\LWZ\bigr)^{\times}$ ($i=1,2$) where the
representation $J\times \bar J$ of $\bigl(\LWZ^*\bigr)^{\times}\times
\bigl(\LWZ^*\bigr)^{\times}$ on the space $H^0(LG_{\!\bf
C},\LWZ^{*k})$ is defined in a similar way as in the case of $\LWZ^k$.

We now sketch the construction of such a pairing defined on some
subspaces of $H^0(LG_{\!\bf C},\LWZ^{*k})$ and $H^0(LG_{\!\bf
C},\LWZ^k)$. The basic ingredients are the hermitian inner product
$(\,,\,)$ defined on a dense subspace of ${\cal H}_{\Lmd}^{(k)}$ for
each $\Lmd\in \PPpk$ \cite{P-S} such that \beq
(J(\tilde{\gamma}_1)\bar
J(\tilde{\gamma}_2)\Phi_1,\Phi_2)=(\Phi_1,J(\tilde{\gamma}_1^*)\bar
J(\tilde{\gamma}_2^*)\Phi_2) \eeq and the anti-holomorphic
homomorphism $\natural:\LWZ^{*k}\to \LWZ^k$ which is anti-linear on
each fibre and covers the involution $\gamma\mapsto \gamma^{*-1}$ of
$LG_{\!\bf C}$. Then, the pairing would be given by $\langle
\Psi,\Phi\rangle = (\overline{\Psi},\Phi)$ where the anti-linear map
$\Psi \mapsto \overline{\Psi}$ from $H^0(LG_{\!\bf C},\LWZ^{*k})$ to
$H^0(LG_{\!\bf C},\LWZ^k)$ is defined by
$\overline{\Psi}(\gamma)=\natural\left(\Psi(\gamma^{*-1})\right)$. If
the map $\natural$ is given by \beq
\natural\{(g_0,c)\}=\{(g_{\infty}^{*-1},c^*e^{-kI_{{\bf
P}^1}(g_{\infty}^{*-1}\star g_0^{*-1})+2kK_{D_0}(g_0^{*-1})})\}\,,
\eeq where $K_D(g)=\frac{i}{4\pi}\int_D\tr(\partial g^{-1}\bartial
g)$, then this pairing would satisfy the condition (\ref{Haar}).

As the space $\check{\cal H}^{(G,k)}$ of states at the incoming
circle, we take the space spanned by vectors each of which is
annihilated by $J(\tilde{\W}_-)$ and $\bar J(\tilde{\W}_-)$ for some
subspace $\W_-\subset \HH^{(0,\lieg)}$ comparable with the space
$\HH_-^{(0,\lieg)}$. This space contains the sum $\bigoplus_{\Lmd\in
\PPpk}\check{\cal H}_{\Lmd}^{(k)}$ of lowest weight representations
where $\check{\cal H}_{\Lmd}^{(k)}$ is generated by the lowest weight
vector $\Phi\check{}_{\Lmd}$ which is the wave function corresponding
to the insertion of $(O_{\Lmd})^{\!\Lmd}_{\,\Lmd}$ in $D_{\infty}$ at
$z=\infty$. We can explicitly check that
$\overline{\Phi\check{}_{\Lmd}}\propto \Phi_{\Lmd}$ which proves that
$Z_{{\bf P}^1,{\rm triv}}(
(O_{\Lmd})^{\!\Lmd}_{\,\Lmd}(\infty)(O_{\Lmd^*})^{\!-\Lmd}_{\,-\Lmd}(0))\ne
0$ and the pairing \beq \bigoplus_{\Lmd\in \PPpk}\check{\cal
H}_{\Lmd}^{(k)}\times \bigoplus_{\Lmd\in \PPpk}{\cal
H}_{\Lmd}^{(k)}\longto {\bf C} \eeq is densely defined and satisfies
the condition (\ref{Haar}).

\newpage
\renewcommand{\theequation}{2.4.\arabic{equation}}\setcounter{equation}{0}
\vspace{0.4cm}
\begin{center}
{\sc 2.4 The Spectral Flow}\label{2.4}
\end{center}
\hspace{1.5cm} We consider again the state at the boundary $S$ of the
standard disc $D_0$ with field insertion $O(0)$. This time we suppose
that the gauge field is in the {\it basic configuration} $A_a$ such
that $a\in i\liet$ lies in the lattice $\Pv=\frac{1}{2\pi
i}\exp^{-1}(1)$. To observe the state, we stand on the horizontal
frame $\sigma$ along $S$ which is related to the original frame $s_0$
by \beq s_0=\sigma\gamma_{10}\,, \hspace{2cm};\qquad
\gamma_{10}(e^{i\theta})=n_{\!w_{\sigma}}^{-1}e^{-ia\theta}\,.  \eeq
where $n_{\!w_{\sigma}}\in G$ is in the normalizer
$N_{T_G}$\label{page:NTG} of $T_G$ that represents an element
$w_{\sigma}\in W$.\footnote{Although we have now introduced the twist
$n\in N_{T_G}$ only to relate the spectral flow to the affine Weyl
group, twists of certain kind play essential roles in the theory of
field identification.} The state $Z_{D_0}^{(\sigma)}(A_a;O(0))$ of
interest is related to the state $Z_{D_0}^{(s_0)}(A_a;O(0))$ with
respect to the original frame by
$Z_{D_0}^{(\sigma)}(A_a,O(0))(\gamma_1)=\ad\gamma_{10}Z_{D_0}^{(s_0)}(A_a;O(0))(\ad\gamma_{10}^{-1}\gamma_1)$.  Since
$A_0''=h_0\bartial h_0^{-1}$, the original state is given by
$J(\tilde{\gamma}_0)\bar J(\tilde{\gamma}_0) \Phi_O$ for
$\tilde{\gamma}_0=e^{-I_{D_0}(h_0)-\frac{1}{2}\Gamma_{D_0}(h_0,h_0^*)}$
(see equation (\ref{bbbb})) where $\Phi_O=|O\rangle$ corresponds to
$O$ in the standard way. Hence we have \beq
Z_{D_0}^{(\sigma)}(A_a;O(0))(\gamma_1)=\tilde{\gamma}_{\sigma}\Phi_O(\gamma_{\sigma}^{-1}\gamma_1
\gamma_{\sigma}^{*-1})\tilde{\gamma}_{\sigma}^*\,, \eeq where
$\gamma_{\sigma}=\gamma_{10}\gamma_0\in LH_{\!\bf C}$ is the boundary
loop $h_{\sigma}|_S$ of the holomorphic transition function
$\sigma_0=\sigma h_{\sigma}$ given by
$h_{\sigma}(z)=n_{\!w_{\sigma}}^{-1}c_{\varrho}^{-a} z^{-a}$. We call
this transformation $\Phi \mapsto\tilh_{\sigma}\Phi$ ;
$\bigl(\tilh_{\sigma}\Phi\bigr)(\gamma_1)=\tilde{\gamma}_{\sigma}\Phi(\gamma_{\sigma}^{-1}\gamma_1
\gamma_{\sigma}^{*-1})\tilde{\gamma}_{\sigma}^*$ of the space ${\cal
H}^{(G,k)}$ of states the {\it spectral flow}\label{ch2.tilh}
corresponding to the transition function $h_{\sigma}$.

\vspace{0.3cm} \underline{Calculation of $\tilh_{\sigma}\Phi$}

We calculate how the highest weight state $\Phi_{\Lmd}$ given in
(\ref{hws}) is transformed by $\tilh_{\sigma}$. Note that the loop
$\gamma_{\sigma}\in LH_{\!\bf C}$ represents an element of the affine
Weyl group $\Waffh$ of $LH$. As explained in Appendix 2
(\ref{Waff'decompo}), we can write it as a product
$\gamma_{\sigma}=[\gamma_{\diamond}]\gamma$ where
$\gamma_{\diamond}\in LG_{\!\bf C}$ represents an element of the
affine Weyl group $\Waff$ of $LG$ and $\gamma\in LH$ represents an
element of the group $\Gmalcv$\label{page1:Gmalcv} which preserves the
decomposition of the set $\Daff$ of affine roots into ${\Daff}_+$ and
${\Daff}_-$. This leads to
$\tilh_{\sigma}\Phi_{\Lmd}=J(\tilde{\gamma}_{\diamond})\bar
J(\tilde{\gamma}_{\diamond})\tilh\Phi_{\Lmd}$ where $h$ is a $H_{\!\bf
C}$-valued holomorphic function with $h|_S=\gamma$ which is expressed
as $h(z)=z^{-\mu}n_{\!w}$ using $\mu \in \Pv$ and $n_{\!w}\in
N_{T_H}$.

We see how the state $\tilh \Phi_{\Lmd}\in {\cal H}^{(G,k)}$ looks
over the open dense set $B^+\left(N^+\right)^*\subset LG_{\!\bf
C}$. So we take holomorphic maps $g_1$ and $g_2$ from $D_0$ to
$G_{\!\bf C}$ such that $g_1(0)\in B^+_0$ and $g_2(0)\in N^+_0$. Since
$\ad\gamma$ preserves the subgroups $B^+$ and $N^+$, we see that
$\gamma^{-1}g_1|_{S} \gamma$ and $\gamma^{-1}g_2|_{S}\gamma$ extend as
holomorphic functions $h^{-1}g_1h$ and $h^{-1}g_2h$ on $D_0$ such that
$(h^{-1}g_1h)(0)\in B^+_0$ and $(h^{-1}g_2h)(0)\in N^+_0$. Hence we
have \beq \left(
\tilh\Phi_{\Lmd}\right)(g_1g_2^*|_S)=e^{\!-\Lmd}\left((h^{-1}g_1h)(0)\right)\ad
\gamma\Bigl(
e^{-kI_{D_0}\left((h^{-1}g_1h)(h^{-1}g_2h)^*\right)}\Bigr)\,.  \eeq

If $g_1(0)\equiv e^{t_0}\in T_G$ mod.$\!N^+_0$ , we find that
$(h^{-1}g_1h)(0)\equiv e^{w^{-1}t_0}$ mod.$\!N^+_0$, which gives \beq
e^{\!-\Lmd}\left((h^{-1}g_1h)(0)\right)=e^{-w\Lmd(t_0)}\,.  \eeq

Applying the transformation rule of the adjoint action of $LH$ on
$\LWZ^k$ given in Appendix 1, we have \beq \gamma
e^{-kI_{D_0}(h^{-1}g_1h)}\gamma^{-1}=\{(\check{g_1},e^{-kI_{{\bf
P}^1}\left( (\check{h}^{-1}\check{g_1}\check{h})\star(h^{-1}g_1
h)\right)+kc(\check{h},\check{g_1})})\}\,,
\label{page1:cocycle}
\eeq where $\check{h}$ is an extension of the loop $\gamma$ to a map
$\check{h}:D_{\!\infty}-\{\infty\}\to H$ given by
$\check{h}(re^{i\theta})=\gamma(e^{i\theta})$ and $\check{g_1}$ is an
extension $\check{g_1}:D_{\!\infty}\to G_{\!\bf C}$ of $g_1$ defined
in the following way: If $g_1$ is expressed as $g_1(z)=e^{v(z)}$ where
$v:D_0\to \g_{\!\bf C}$ is a holomorphic map, we take
$\check{g_1}(re^{i\theta})=e^{\varrho_{\infty}(r)v(e^{i\theta})}$
using a cut-off function $\varrho_{\infty}:D_{\infty}\to [0,1]$ such
that $\varrho_{\infty}|_S=1$ and $\varrho_{\infty}(\infty)=0$.  For
this choice of $\check{g_1}$, we see that $I_{{\bf
P}^1}((\check{h}^{-1}\check{g_1}\check{h})\star
(h^{-1}g_1h)=K_{D_{\!\infty}}(\check{h}^{-1}\check{g_1}\check{h})$ and
$K_{D_{\!\infty}}(\check{g_1})=I_{{\rm P}^1}(\check{g_1}\star
g_1)$. Note also that $\check{h} d\check{h}^{-1}=i\mu d\theta$ and
that
$d\check{g_1}\check{g_1}^{-1}|_r=\check{g_1}^{-1}d\check{g_1}|_r=d\varrho
v(e^{i\theta})$. From these, it follows that \beqa -\lefteqn{I_{{\bf
P}^1}\!\left( (\check{h}^{-1}\check{g_1}\check{h})\star(h^{-1}
g_1h)\right) +c(\check{h},\check{g_1})}\nonumber \\
&=&-K_{\!D_{\!\infty}}(\check{g_1})-\frac{i}{2\pi}\int_{[1,\infty)\times
S^1}\!\!\!\!\!\tr\!\left(d\varrho\, t_0 \,i\mu d\theta\right)=
-I_{{\bf P}^1}( \check{g_1}\star g_1) -\tr(t_0\mu)\,.  \eeqa Hence, we
have \beq \gamma e^{-kI_{D_0}(h^{-1}g_1h)}\gamma^{-1}=e^{-k\,\tr(\mu
t_0)}e^{-kI_{D_0}(g_1)}\,.  \eeq

Doing the similar calculation for $\ad\gamma
e^{-kI_{D_0}\left((h^{-1}g_2h)^*\right)}$ and combining all the
results, we obtain the expression of the section $\tilh \Phi_{\Lmd}$
over $B^+(N^+)^*\subset LG_{\!\bf C}$ : \beq
\Bigl(\tilh\Phi_{\Lmd}\Bigr) (g_1g_2^*|_S)=e^{-w\Lmd(t_0)-k\,\tr(\mu
t_0)}e^{-kI_{D_0}(g_1g_2^*)}\,.  \eeq

Since the adjoint action of $LH$ on $\LWZ^k$ is continuous, this state
must extend to the whole space $LG_{\! \C}$ if the state
$\Phi_{\!\Lmd}$ does. Therefore, if $\Lmd\in \PPpk$, the weight
$w\Lmd+k\ttr \mu$ is necessarily in $\PPpk$ and the state
$\tilh\Phi_{\!\Lmd}$ is the highest weight state $\Phi_{\gamma\Lmd}$
with that highest weight \beq \gamma\Lmd =w\Lmd +k\ttr
\mu\,\,,\label{newweight1} \eeq where the weight $\ttr \mu\in \PP$ is
given by $\ttr\mu(v)=\tr(\mu v)$.

\newpage \vspace{0.2cm} \underline{The Spectral Flow as an Algebra
Automorphism}

The above result could be seen by looking at the response of the state
to the action of $J(\tilde{LG}_{\!\C})\times \bar
J(\tilde{LG}_{\!\C})$. Instead of doing the direct calculation, we
exploit the relation (\ref{lll}) of the current $J$ and the
infinitesimal generators of $J(\tilde{LG}_{\!\C})$.\footnote{ We only
look at the left representation $J$ since everything is the same for
the right representation $\bar J$.} In the situation of ours we have
\beqa
J_n(v)Z_{D_0}^{(\sigma)}(A_a;O)&=&Z_{D_0}^{(\sigma)}(A_a;J_n^{\sigma}(v)O)\,,\\
;\quad J_n^{\sigma}(v)=\nipi&&\hspace{-1cm} \oint
z^ndz\left\{J_z^{\sigma_0}(h_{\sigma}^{-1}vh_{\sigma})+\frac{1}{z}k\,\tr(ah_{\sigma}^{-1}vh_{\sigma})\right\}\,,
\eeqa where $J_z^{\sigma_0}dz$ is the local holomorphic part
(\ref{holJ}) of the current with respect to the frame $\sigma_0$
defined on $D_0$. Taking account of this and the expression
(\ref{virgenwzw}) for the energy momentum tensor, we have the
following intertwining relations between the spectral flow
$\tilh_{\sigma}$ and the generators $L_n$, $J_n(v)$ : \beqa
L_n\tilh_{\sigma}|O\rangle &=&\tilh_{\sigma}\!\Bigl\{
L_n+J_n(a)+\frac{k}{2}\tr(a^2)\delta_{n,0}\Bigr\}\!|O\rangle
\,,\label{spctrlwzwLn}\\ J_n(t)\tilh_{\sigma}|O\rangle
&=&\tilh_{\sigma}\!\Bigl\{
J_n(w_{\sigma}t)+k\delta_{n,0}\tr(aw_{\sigma}t)\Bigr\}\!|O\rangle\,,\label{spctrlwzwJnt}\\
J_n(e_{\alpha})\tilh_{\sigma} |O\rangle&=&\tilh_{\sigma}
J_{n+w_{\sigma}\alpha(a)}(n_{\!w_{\sigma}}e_{\alpha}n_{\!w_{\sigma}}^{-1})|O\rangle
c_{\varrho}^{w_{\sigma}\alpha(a)}\,,\label{spctrlwzwJnr} \eeqa where
$n\in {\bf Z}$, $t\in \liet_{\!\C}$ and $\alpha\in \Delta$. Note that
(\ref{spctrlwzwLn}) and (\ref{spctrlwzwJnt}) for $n=0$ derive the
transformation rule of $[\gamma_{\sigma}]=e^{ia\theta}w_{\sigma}$ on
the Lie algebra $\hat{\V}$ of the torus $U(1)\times \tilde{T}_G$ given
in the formula (\ref{affact}) in Appendix 2. These show that a state
$\Phi$ of weight $({\cal E},\lmd,k)$\label{page:weightlmd} is
transformed to another state $\tilh_{\sigma} \Phi$ of weight
$(\gamma_{\sigma}{\cal E},\gamma_{\sigma}\lmd,k)$ where \beq
\gamma_{\sigma}{\cal E}={\cal E}+\lmd(a)+\frac{k}{2}\tr(a^2)\,,\qquad
\gamma_{\sigma}\lmd=w_{\sigma}^{-1}(\lmd+k\ttr a)\,.\label{newweight2}
\eeq

If the loop $\gamma_{\sigma}$ represents an element of $\Gmalcv$ from
the start, (\ref{spctrlwzwJnr}) shows that a highest weight state
$\Phi_{\Lmd}$ is transformed to another highest weight state
$\tilh_{\sigma}\Phi_{\Lmd}$. Under the identification
$\mu=w_{\sigma}^{-1}a$ and $w=w_{\sigma}^{-1}$, we see that the two
results on the new weight, (\ref{newweight1}) and (\ref{newweight2})
coincide. Moreover, a calculation shows that the $L_0$ value
$\sDelta_{\gamma\Lmd}-\frac{c_{G,k}}{24}$ is given by $\gamma{\cal E}$
of (\ref{newweight2}) with ${\cal
E}=\sDelta_{\Lmd}-\frac{c_{G,k}}{24}$ and $\lmd=\Lmd$ where
$\sDelta_{\Lmd}$ is given by (\ref{cwzw}) or \beq
\sDelta_{\Lmd}=\frac{(\Lmd,\Lmd+2\rho)}{2(k+\lieg^{\vee})}\,,
\label{page2:confdim}
\eeq where $\rho\in \PP$ is half the sum of positive root of $G$.

We can see in a more transparent manner the transformation
$\Lmd\mapsto \gamma\Lmd$ of highest weight for $\gamma\in \Gmalcv$
where we identify the loop $\gamma$ with the corresponding element of
the Weyl group $\Waffh$. As explained in Appendix 2, the group
$\Gamma_{\alcv}$ is a permutation group of the set ${\rm
B}(\alcv)=\{\, \hat{\alpha}_{0},\cdots , \hat{\alpha}_{l}\,
\}$\label{page1:BCaff} of simple affine roots and it acts on
$\{0,1,\cdots , l\}$ by $\gamma\hat{\alpha}_{i}=\hat{\alpha}_{\gamma
i}$ for $i=0,1,\cdots ,l$. We introduce the {\it fundamental affine
weights} $\hat{\Lmd}_{0},\hat{\Lmd}_{1},\cdots ,\hat{\Lmd}_{l}\in
\hat{\V}$ by \beq \hat{\Lmd}_{0}=(0,0,1)\,\,,\qquad
\hat{\Lmd}_{i}=(0,\Lmd_{i}, m_{i})\,,\quad i=1,\cdots ,l\,, \eeq where
$\Lmd_{1},\cdots ,\Lmd_{l}\in \PP$ are the fundamental weights defined
by $\Lmd_{i}(\alpha_{j}^{\vee})=\delta_{i,j}$\footnote{ For each root
$\alpha\in \Delta$, we associate the coroot $\alpha^{\vee}\in
\Qv$\label{page1:Qv} defined by $\ttr\alpha^{\vee}=2
\alpha/||\alpha||^2$. } and $m_1,\cdots, m_l\in \N$ are the
coefficients of the coroot
$\tilde{\alpha}^{\vee}=\sum_{i=1}^lm_i\alpha_i^{\vee}$ for the highest
root $\tilde{\alpha}$\label{page1:highestroot}. Then, in view of the
relations
$2(\hat{\Lmd}_{i},\hat{\alpha}_{j})/||\hat{\alpha}_{j}||^2=\delta_{i,j}$
and $\gamma\hat{\alpha}_{i}=\hat{\alpha}_{\gamma i}$ we see that
$\gamma\hat{\Lmd}_{i}\equiv \hat{\Lmd}_{\gamma i}$ modulo ${\R}\times
\{0\}\times \{0\}$. The affine weight
$\hat{\Lmd}=(\sDelta_{\Lmd},\Lmd,k)$ for $\Lmd\in \PPpk$ is expressed
as $\hat{\Lmd}=\sum_{i=0}^{l}n_i\hat{\Lmd}_i-\sDelta_{\Lmd}\delta$
where $n_i$'s are non-negative integers and $\delta=(-1,0,0)$. Then,
the new weight is given by \beq
\gamma\hat{\Lmd}=\sum_{i=0}^{l}n_i\hat{\Lmd}_{\gamma
i}-\sDelta_{\gamma\Lmd}\delta\,.  \eeq

{\it Remark}. We may consider the state $\tilh_{\sigma}|O\rangle$ to
correspond in the standard way to a new field $\gamma_{\sigma}O$. If
the field $O$ has conformal weight $(\sDelta_O,\bar \sDelta_O)$ and
charge $(\Lmd_O,\bar \Lmd_O)$, the new field $\gamma_{\sigma}O$ has
conformal weight $(\gamma_{\sigma}\sDelta_O,\gamma_{\sigma}\bar
\sDelta_O)$ and charge $(\gamma_{\sigma}\Lmd_O,\gamma_{\sigma}\bar
\Lmd_O)$. A primary field $O$ is transformed to another primary field
$\gamma_{\sigma}O$ if $\gamma_{\sigma}$ represents an element of
$\Gmalcv\subset \Waffh$.

\newpage
\renewcommand{\theequation}{3.0.\arabic{equation}}\setcounter{equation}{0}

{\large CHAPTER 3. INTEGRATION OVER GAUGE FIELDS}\label{ch.3}

\vspace{1cm} \hspace{1cm} Let $H$ be a compact Lie group and let $M$
denote the {\it matter} field theory which we discussed in the
previous chapters : the system of free fermions taking values in a
vector bundle $E$ with structure group $H$, the WZW model for a group
$G$ such that $H$ is a closed subgroup of $G/Z_G$, or some combination
of these systems. We consider in this chapter the quantization problem
of gauge theory. That is, we give a method to perform the integration
\beq Z_{\Sigma,P}(\met\,;{\cal O})=\int_{{\cal A}_P/{\cal
G}_P}\!\!{\cal D}_{\!\scriptsize
\met}A\,Z_{\Sigma,P}^{M}(\met,\!A\,;{\cal O})\,,
\label{intgauge}
\eeq over gauge equivalence classes ${\cal A}_P/{\cal G}_P$ of
connections on a principal $H$-bundle $P$.  The integrand is the
correlation function of the matter theory $M$ with $\GP$-invariant
insertion ${\cal O}$ of local fields. We first perform the integration
over each orbit of the group $\GPC$ of chiral gauge transformations
and then, sum up over the orbits. The first step naturally leads to
the WZW model with negative level, targetting the symmetric space
$H_{\!\C}/H$. The latter half of this chapter is devoted to the
analysis of this theory.

\renewcommand{\theequation}{3.1.\arabic{equation}}\setcounter{equation}{0}
\vspace{0.3cm}
\begin{center}
{\sc 3.1 The Space of Gauge Fields}\label{3.1}
\end{center}
\hspace{1.5cm} We give a description of the structure of $\GPC$-orbits
in the space ${\cal A}_P$ of gauge fields and argue that we can
neglect some orbits in the integration (\ref{intgauge}). To start
with, we note on the relation of holomorphic structures of the
$H_{\!\C}$-bundle $\PC$ and connections of $P$. A connection $A$ of
$P$ determines a holomorphic structure $J_{\!A}$ of $\PC$ : A local
section $\sigma$ of $\PC$ is holomorphic with respect to $J_A$ when
$\bartial_{\!A}\,\sigma=0$ where $\sigma$ is represented as a local
frame of the vector bundle associated to $\PC$ through a faithful
holomorphic representation of $H_{\!\C}$. Conversely, any holomorphic
structure $J$ of $\PC$ determines a ``hermitian connection'' $A(J)\in
\AP$ such that $J_{\!A(J)}=J$. So we can identify the set of
holomorphic structures of $\PC$ and the set $\AP$ of connections of
$P$. Two holomorphic structures $J_1$ and $J_2$ are {\it isomorphic}
when there is an automorphism $h\in\GPC$ which gives a bi-holomorphic
map $h:(\PC,J_1)\to (\PC,J_2)$. In other words, $A_1$ and $A_2$
determine isomorphic holomorphic structures in $\PC$ when
$h^{-1}\bartial_{\!A_2} h=\bartial_{\!A_1}$. So, we can identify the
set of isomorphism classes of holomorphic structures of $\PC$ and the
set $\AP/\GPC$ of orbits of the chiral gauge transformation group.

It should be noticed that the space $\A_P$ is given a structure of
complex manifold in such a way that the action of the group $\GPC$ is
holomorphic : The complex structure at each tangent space
$\Omega^1(\Sigma,\ad P)$ is given by the operator $*$ which determines
an involution $*d\bar z=id\bar z$, $*dz=-idz$ on the space of
differentials.

\vspace{0.3cm} \underline{On the Sphere}

We begin with the case in which $\Sigma$ is the complex projective
line $\CP$. It is covered by two complex planes --- $z$-plane and
$w$-plane where $z$ and $w$ are related by $zw=1$. We denote by
$D_0\subset \CP$ the unit disc in the $z$-plane.

\vspace{0.2cm}
\noindent( For $H=U(1)$ )

First, we consider the simplest case $H=U(1)$. We start with the
topological classification. A principal $U(1)$ bundle $P$ admits local
sections $s_0$ and $s_{\infty}$ defined on $U_0^{\epsilon}$ and
$U_{\!\infty}^{\epsilon}$ for some $\epsilon>0$ respectively where
$U_0^{\epsilon}$ (resp. $\!U_{\infty}^{\epsilon}$) is an open
neighborhood of $D_0$ (resp. $\!\CP-D_0$) consisting of $z$ with
$|z|<1+\epsilon$ (resp. $\!w$ with $|w|<1+\epsilon$). They are related
by the transition rule $s_0=s_{\infty}h_{\infty 0}$ where $h_{\infty
0}$ is a function on $U_0^{\epsilon}\cap U_{\!\infty}^{\epsilon}$
valued in $U(1)=\{ e^{i\phi}\}$. A topological invariant of $P$ is
given by the winding number of $h_{\infty 0}$ \beq
a=\frac{i}{2\pi}\oint_Sh_{\infty 0}^{-1}dh_{\infty 0}\,, \eeq which
must be an integer. Then, $h_{\infty 0}$ can be written as $h_{\infty
0}=e^{-ia\theta +i\phi_{\infty 0}}$ where $\phi_{\infty 0}$ is a real
valued function on $U_0^{\epsilon}\cap
U_{\infty}^{\epsilon}$. \footnote{As before, the radius $r$ and the
angle $\theta$ are related to the complex coordinate $z$ by
$z=re^{i\theta}$.} Using the cut-off function $\varrho:S^2\to [0,1]$
which is identically $0$ on $S^2-U_{\infty}^{\epsilon}$ (resp. $\!1$
on $S^2-U_0^{\epsilon}$), we have the following transition rule \beq
s_0e^{-i\varrho \phi_{\infty 0}}=s_{\infty}e^{i(1-\varrho)
\phi_{\infty 0}}e^{-ia\theta}\,.  \eeq Thus, the topological type of a
principal $U(1)$-bundle is determined by the winding number $a\in
{\Z}$. Such a classifacation holds for any surface $\Sigma$.

We proceed to the classifiction of holomorphic
$H_{\!\C}=\C^*$-bundles. A holomorphic $\C^*$-bundle $\Ph$ admits
local holomorphic sections $\sigma_0$ and $\sigma_{\infty}$ defined on
$U_0^{\epsilon}$ and $U_{\infty}^{\epsilon}$ for some $\epsilon>0$
respectively which are related by the transition rule
$\sigma_0=\sigma_{\infty}f_{\infty 0}$ where $f_{\infty 0}$ is a
holomorphic function on $U_0^{\epsilon}\cap U_{\infty}^{\epsilon}$
valued in $\C^*$. If $a=\displaystyle{\frac{i}{2\pi}\oint}f_{\infty
0}^{-1}df_{\infty 0}$, $f_{\infty 0}$ can be written as $f_{\infty
0}(z)=z^{-a}e^{x_{\infty 0}(z)}$ in terms of a holomorphic function
$x_{\infty 0}$ on $U_0^{\epsilon}\cap U_{\infty}^{\epsilon}$. Then,
taking the Laurent expansion $x_{\infty
0}(z)=\displaystyle{\sum_{-\infty}^{\infty}}x_nz^n$, we have the
following transition rule : \beq
\sigma_0(z)e^{-\sum_{n=0}^{\infty}x_nz^n}=\sigma_{\infty}(z)e^{\sum_{n=1}^{\infty}x_{-n}w^n}z^{-a}\,.  \eeq
This shows that for each $\C^*$-bundle $\PC$, there is only one
holomorphic structure up to isomorphism. Thus, the space $\A_P$ of
gauge fields is itself a single $\GPC$-orbit.

\newpage \vspace{0.2cm}
\noindent ( For simple $H$ )

Next we consider the concrete example in which $H=SU(n)/{\Z}_n$ where
${\Z}_n$ is the center of $SU(n)$ consisting of identity matrices
multiplied by $n$-th roots of unity. For each $j\in{\cal
J}_0=\{0,1,\cdots,n-1\}$\label{page1:calJ0}, we take an $H$-bundle
$P^{(j)}$ which admits a section $s_0$ on the $z$-plane $U_0$ and a
section $s_{\infty}$ on the $w$-plane $U_{\infty}$ related by
$s_0=s_{\infty}h_{\infty 0}$ on $U_0\cap U_{\infty}$ where the
transition function $h_{\infty 0}:U_0\cap U_{\infty}\to H$ is
represented by a multi-valued map $\tilde{h}_{\infty 0}$ to $SU(n)$
such that $\tilde{h}_{\infty 0}(re^{i\theta+2\pi i})=e^{2\pi
i\frac{j}{n}}\tilde{h}_{\infty 0}(re^{i\theta})$. Then,
$\{P^{(j)}\,;\,j\in {\cal J}_0\}$ is the set of topologically distinct
$H$-bundles over $S^2$. Such a classification holds for any surface
$\Sigma$.

A holomorphic $H_{\!\C}=SL(n,{\C})/ {\Z}_n$ bundle over $\CP$ is
described by the transition rule $\sigma_0=\sigma_{\infty}f_{\infty
0}$ where $\sigma_0$ (resp. $\!\sigma_{\infty}$) is a holomorphic
section over $U_0$ (resp. $\!U_{\infty}$) and the holomorphic map
$f_{\infty 0}:U_0\cap U_{\infty}\to H_{\!\C}$ is also represened by a
multi-valued map to $SL(n,\C)$. By the Birkhoff factorization theorem
\cite{Grothendieck}, \cite{P-S}, we may assume that $f_{\infty 0}$ is
given by $f_{\infty0}(z)=z^{-a}$ where $a$ belongs to $\Pv$, that is,
$a$ is a traceless diagonal matrix \beq a=\left(\begin{array}{cccc}
a_1& & \\ &\ddots& \\ & & a_n
\end{array}\right)    \qquad\mbox{with}\quad
a_i+\frac{j}{n}\in{\Z}\quad(i=1,\cdots,n)\,,
\label{adiag}
\eeq for some $j\in {\cal J}_0$. This holomorphic bundle denoted by
${\cal P}_{[a]}$ is isomorphic to $(P_{\!\C}^{(j)},J_{\!A(a)})$ for
some $A(a)\in {\cal A}_{P^{(j)}}$. The theorem of Birkhoff shows that
${\cal P}_{[a]}$ is isomorphic to ${\cal P}_{[a']}$ if and only if
$a'\in Wa$ where $W$ is the permutation group of $\{ a_1, \cdots ,
a_n\}$. Thus, the discrete set $\Pv/W$ indexes the set of isomorphism
classes of holomorphic $H_{\!\C}$-bundles over $\CP$ and it follows
that the set ${\cal A}_{P^{(j)}}/{\cal G}_{P^{(j)}_{\!\bf c}}$ of
$\G_{P^{(j)}_{\!\bf c}}$-orbits is indexed by $\Pv_j/W$ where
$\Pv_j$\label{page:Pvj} is the set of matrices $a\in \Pv$ whose
entries differ from $\displaystyle{-\frac{j}{n}}$ by integers.

 We calculate the dimension of the group ${\rm Aut}{\cal P}_{[a]}$ of
holomorphic automorphisms of ${\cal P}_{[a]}$. An element $f\in {\rm
Aut}{\cal P}_{[a]}$ is given by a pair of holomorphic maps $f_0:U_0\to
H_{\!\C}$ and $f_{\infty}:U_{\infty}\to H_{\!\C}$ that are related by
$f_0(z)=z^af_{\infty}(z)z^{-a}$. We see that the matrix element
$(f_0)^{\!i}_j$ is a span of $1, z, \cdots, z^{a_i-a_j}$ if $a_i\geq
a_j$ and $(f_0)^{\!i}_j=0$ if $a_i<a_j$ which shows that ${\rm
dim}{\rm Aut}{\cal
P}_{[a]}=n-1+\sum_{i<j}\Bigl(\,\delta_{a_i,a_j}\!+1+|a_i-a_j|\,\Bigr)$. It
is minimized by $a=\mu_0,\mu_1,\cdots,\mu_{n-1}$ where $\mu_j\in
\Pv_j$ is given by \beq (\mu_j)_i=1-\frac{j}{n}\quad\mbox{for}\quad
i=1,\cdots,j,\quad \mbox{and}\quad (\mu_j)_i=-\frac{j}{n}\quad
\mbox{for}\quad i=j+1,\cdots,n\,.
\label{ch3.muj}
\eeq Since ${\rm Aut}{\cal P}_{[a]}$ for $a\in \Pv_j$ is naturally
isomorphic to the isotropy subgroup of ${\cal G}_{P^{(j)}_{\!\C}}$ at
$A(a)\in {\cal A}_{P^{(j)}}$, ${\cal A}_{P^{(j)}}$ contains a single
orbit $\A_{\mu_j}=A(\mu_j)\cdot{\cal G}_{P^{(j)}_{\!\bf c}}$ of
maximal dimension. The codimension $d_a$ of the orbit $\A_a$ in
$\A_{P^{(j)}}$ is hence given by \beq
d_a=\sum_{i<j}\Bigl(\,\delta_{a_i,a_j}-1+|a_i-a_j|\,\Bigr)=\sum_{a_i>a_j}(a_i-a_j-1)\,.
\label{page:da}
\eeq Note that the orbit $\A_{\mu_j}$ has a complement in ${\cal
A}_{P^{(j)}}$ of codimension $1$ for $j=0$ and $n$ for $j=1,\cdots ,
n-1$. Hence, in the integration (\ref{intgauge}) for $P=P^{(j)}$, we
have only to take into account the contribution of the orbit
$\A_{\mu_j}$, if $Z^M_{\CP,P^{(j)}}$ is smoothly defined over the
whole space ${\cal A}_{P^{(j)}}$.

If $H$ is the quotient group $\tilde{H}/Z_{\tilde{H}}$ for a compact
simple simply connected group $\tilde{H}$, the story is almost the
same. We follow the notation in Appendix 2. The set of distinct
topological $H$-bundle is represented by $\{P^{(j)}\}_{j\in {\cal
J}_0}$\label{page2:calJ0} where $P^{(j)}$ is an $H$-bundle given by a
transition rule $s_0=s_{\infty}e^{-i\mu_j \theta}$. The set of
isomorphism classes of holomorphic $H_{\!\C}$-bundle over $\CP$ is
given by $\{{\cal P}_{[a]}\,;\,a\in \Pv\!\cap \overline{\Ch}\}$ where
${\cal P}_{[a]}$ admits local holomorphic sections $\sigma^{(a)}_0$
and $\sigma^{(a)}_{\infty}$ related by
$\sigma^{(a)}_0=\sigma^{(a)}_{\infty}z^{-a}$. The dimension of the
group of holomorphic automorphisms of ${\cal P}_{[a]}$ is given by
$l+\sum_{\alpha\in \Delta_+}\bigl(\,
\delta_{\alpha(a),0}+1+|\alpha(a)|\,\bigr)$ which is minimized by
$\{\mu_j\}_{j\in {\cal J}_0}$\label{page2:muj} (see proposition 2,
Appendix 2). This shows that each ${\cal A}_{P^{(j)}}$ containes a
single ${\cal G}_{P_{\!\bf c}^{(j)}}$-orbit $\A_{\mu_j}$ of maximal
dimension. The codimension of the orbit $\A_a$ corresponding to ${\cal
P}_{[a]}$ is given by \beq d_a=\sum_{\alpha(a)>0}(\alpha(a)-1)\,.
\eeq Hence we have only to take into account the contribution of the
orbit $\A_{\mu_j}$ in the integration (\ref{intgauge}) for
$P=P^{(j)}$.

\vspace{0.3cm} \underline{On a Surface of Genus $\geq 1$}

If the genus $g$ of the surface $\Sigma$ is larger than zero, the set
of $\GPC$-orbits in the space of connections of an $H$-bundle $P$ is
not in general a discrete set. But there is an efficient way developed
by Atiyah and Bott \cite{A-B} to classify the orbits by a discrete
set.

The basic notion in the classification is the semi-stability of vector
bundles. We say that a holomorphic vector bundle $E$ is ({\it
semi}-){\it stable} if for any proper holomorphic subbundle $F$ of
$E$, we have $\mu(F)<\mu(E)$ ($\mu(F)\leq \mu(E)$) where
$\mu(E)$\label{page:slope} is the {\it slope} of $E$ defined by
$\mu(E)=c_1(E)/{\rm rank}(E)$. A vector bundle is not in general
semi-stable but has a unique filtration $0=E_0\subset E_1 \subset
\cdots \subset E_r=E$ by subbundles, called the {\it canonical
filtration} of $E$, such that the quotients $D_i=E_i/E_{i-1}$ are
semi-stable and satisfy $\mu(D_i)>\mu(D_{i+1})$, see Harder and
Narasimhan \cite{HN}. A holomorphic $\HC$-bundle $\Ph$ admits its {\it
canonical parabolic reduction} $\Ph_Q$, that is, a reduction to a
parabolic subgroup $Q$ of $\HC$ such that $\ad\Ph_Q/\ad_{\n_Q}\Ph_Q$
is semi-stable and of slope zero and that the canonical filtration of
$\ad\Ph/\ad\Ph_Q$ (resp. $\! \ad_{\n_Q}\Ph_Q$) has quotients of
negative slopes (resp. positive slopes) where $n_Q$ is the Lie algebra
of the unipotent radical of $Q$.

We now define the {\it type} of a holomorphic principal bundle. We
choose a maximal torus $T$ of $H$ and a chambre $\Ch$ in
$i\liet$. Note that an element $\mu$ of $\overline{\Ch}$ determines a
parabolic subgroup $Q$ of $\HC$ by the statement ``a root $\alpha$ of
$\HC$ is a root of the unipotent radical $\n_Q$ of $Q$ if and only if
$\alpha(\mu)>0$''. For each one dimensional representation $\chi$ of a
parabolic subgroup $Q$, we denote by $\lmd_{\chi}$ the linear form on
$\V=i\liet$ such that $\chi(e^t)=e^{\lmd_{\chi}(t)}$ for $t\in
\V_{\!\C}$. A holomorphic $\HC$-bundle $\Ph$ is said to be of type
$\mu \in \overline{\Ch}$,\label{page:type} when the canonical
parabolic reduction of $\Ph$ is to the parabolic subgroup $Q$
determined by $\mu$ and for any representation $\chi:Q\to {\C}^*$, the
line bundle $\chi(\Ph_Q)=\Ph_Q\!\times_{Q}\!\C$ associated to $\Ph_Q$
through $\chi$ has the first Chern number $\lmd_{\chi}(\mu)$ : \beq
c_1(\chi(\Ph_Q))=\lmd_{\chi}(\mu)\,.  \eeq For example, the
holomorphic $\HC$-bundle $\Ph_{[a]}$ on the Riemann sphere is of type
$a\in \Pv$. There exists an $\HC$-bundle of type $\mu$ only if $\mu$
is an `integral point', that is, $\lmd_{\chi}(\mu)\in {\Z}$ for every
character $\chi$ of $Q$. Thus, the gauge fields are classified with
respect to the types and we have the disjoint union \beq
\AP=\bigcup_{\mu:{\rm type}}\A_{\mu}\,, \eeq where $\A_{\mu}$ consists
of $\GPC$-orbits corresponding to type $\mu$ holomorphic structures of
$P$. A simple index calculation shows that $\A_{\mu}$ is a submanifold
of $\AP$ of codimension \beq
d_{\mu}=\sum_{\alpha(\mu)>0}(\alpha(\mu)+g-1)\,.
\label{page:dmu}
\eeq Hence, for $g\geq 1$ we have only to take into account of the
submanifold $\A_0=\A_{ss}$\label{page:Ass} corresponding to the set of
semi-stable $\HC$-bundles. (A holomorphic $\HC$-bundle $\Ph$ is said
to be {\it semi-stable} when $\ad\Ph$ is semi-stable.)

\vspace{0.5cm} For $H=U(1)$, since every holomorphic
$H_{\!\C}=\C^*$-bundle is stable, we take the whole space $\A_P$ into
account. The set of isomorphism classes of holomorphic $\C^*$-bundles
is given by the sheaf cohomology group $H^1(\Sigma,{\cal
O}_{\Sigma}^{\times})$ called the Picard group of $\Sigma$ and denoted
by $\Pic\Sigma$ where ${\cal O}_{\Sigma}^{\times}$ is the sheaf of
germs of holomorphic functions valued in $\C^*$.\label{page:calOSigma}
The homomorphism ${\cal O}_{\Sigma}\to {\cal O}_{\Sigma}^{\times}$
with kernel ${\Z}$ given by $f\mapsto e^{2\pi i f}$ induces the long
exact sequence of sheaf cohomology groups : \beq 0\to
H^1(\Sigma,{\Z})\to H^1(\Sigma,{\cal O}_{\Sigma})\to H^1(\Sigma,{\cal
O}_{\Sigma}^{\times})\to H^2(\Sigma,{\Z})\to 0\,, \eeq where the last
projection counts the winding number. The Jacobian variety
$\Jac\Sigma=H^1(\Sigma, {\cal
O}_{\Sigma})/H^1(\Sigma,{\Z})$\label{page:JacSigma} is a complex torus
of dimension $g$. Thus, we have the description of the set of
holomorphic $\C^*$-bundles by the exact sequence \beq 0\to \Jac\Sigma
\to \Pic\Sigma \stackrel{c_1}{\to}{\Z}\to 0\,.
\label{descript.pic}
\eeq Let us choose a base point $x_0\in \Sigma$. An element ${\cal
O}(x_0)\in \Pic\Sigma$\label{page:calOx} of winding number one is
defined by the transition rule \beq
\sigma_0(z)=\sigma_{\infty}(z)z^{-1}\,, \eeq where $z$ is a local
coordinate with $z(x_0)=0$, $\sigma_0$ is a section over the
coordinate neighborhood and $\sigma_{\infty}$ is a section over
$\Sigma-\{x_0\}$. This bundle ${\cal O}(x_0)$ generates a subgroup of
$\Pic\Sigma$ which is mapped isomorphically onto ${\Z}$ by the
projection $c_1$ in (\ref{descript.pic}). In other words, `tensoring
by ${\cal O}(x_0)^a$' induces an isomorphism
$(\Pic\Sigma)_0=\Jac\Sigma\to(\Pic \Sigma)_a=c_1^{-1}(a)$ of complex
manifolds. This isomorphism plays an important role in the theory of
field identification which we discuss in the next chapter.

\vspace{0.5cm} If $H$ is simple, the semi-stable orbits are further
classified. We can show that for $g\geq 2$, $\A_{ss}$ contains a
submanifold $\A_s$\label{page:As} with a complement of codimension
$\geq 1$ where $\A_s$ consists of $\GPC$-orbits corresponding to
stable bundles.(See \cite{Ramanathan} for the definition of stability
of principal bundles). The theorem of Narasimhan and Seshadri
\cite{N-S}, \cite{Donaldson}, \cite{Ramanathan} states that the set
$\A_F^{irr}$ of irreducible flat connections is included in the space
$\A_s$ consisting of stable orbits and that the inclusion map induces
the bijection
$\A_F^{irr}/\GP\stackrel{\cong}{\to}\A_s/\GPC=\NN_P^{st}$ of the
quotients. It is also known that $\NN_P^{st}$ is a complex orbifold of
dimension ${\rm dim}H(g-1)$ and has a natural compactification
$\NN_P=\A_{ss}/\!\sim$ which can be identified with the set $\A_F/\GP$
of all flat connections.\footnote{ $A_1\sim A_2$ when $A_1$ and $A_2$
determine holomorphic bundles $\Ph_1$ and $\Ph_2$ respectively such
that $\ad\Ph_1$ and $\ad\Ph_2$ have filtrations whose quotients are
stable and coincide with each other up to permutation.}

If $g=1$, $\A_{ss}$ contains a submanifold $\A_{ss}^{\circ}$ with a
complement of codimension $\geq 1$ : A holomorphic bundle $\Ph$
corresponds to an orbit in $\A_{ss}^{\circ}$ when $\ad\Ph$ is
decomposed into a direct sum of distinct stable bundles. The quotient
space $\A_{ss}^{\circ}/\GPC$ is a complex orbifold and it has a
natural compactification $\NN_P=\A_{ss}/\!\sim$ which can be
identified with $\A_F/\GP$.

\vspace{0.3cm} \underline{Example --- Flat $SO(3)$-Connections over
the Torus}\label{ex.ch3}

We explicitly describe the spaces $\NN_P$ of flat connections of the
trivial and the non-trivial $H=SO(3)$-bundles on the torus
$\Sigma_{\tau}=\C/({\Z}+\tau{\Z})$ of period $1$ and $\tau$ where
$\tau_2={\rm Im}\tau >0$. We denote by $\z$ the coordinate of this
plane $\C$. We choose a homology base $A$, $B : [0,1]\to \Sgmtau$
defined by $\z(A_t)=t$ and $\z(B_t)=t\tau$, which also determines a
set of generators of the fundamental group $\pi_1\Sigma={\Z}^2$. A
flat connection of an $H$-bundle $P$ determines (up to conjugation) a
holonomy representation $\rho : \pi_1\Sigma \to H$. It is determined
by $a=\rho(A)$ and $b=\rho(B)$ that commute with each other.

If $P$ is trivial, $a$ and $b$ are represented by elements $\tila$ and
$\tilb$ of $\tilH=SU(2)$ that also commute with each other. By
conjugation if necessary, we may as well assume that $\tila$ and
$\tilb$ are diagonal matrices \beq \tila=\pmatrix{ e^{2\pi i \phi} & 0
\cr 0 & e^{-2\pi i\phi}\cr }\,,\qquad \tilb=\pmatrix{ e^{2\pi i\psi} &
0 \cr 0 & e^{-2\pi i\psi} \cr }\,.  \eeq Such holonomy is provided by
the gauge field of the following form : \beq
A_u=\Bigl(\frac{\pi}{\tau_2}u\,d\bar \zeta-\frac{\pi}{\tau_2}\bar
u\,d\zeta\Bigr)\pmatrix{ 1 & 0 \cr 0 & -1 \cr }\,,
\label{flatconnSO(3)triv}
\eeq where $u=\psi-\tau \phi$. $A_{u'}$ is $\GP$-equivalent to $A_u$
if it is a gauge transform of $A_u$ by element $g\in \GP$ of the
following form : \beq g(\zeta=x+\tau y)=n_w \pmatrix{ e^{\pi i(nx+my)}
& 0 \cr 0 & e^{-\pi i(nx+my)} \cr }\,, \eeq where $n_w\in SU(2)$
represents an element of the Weyl group $W=S_2$ and $n$, $m$ are
integers. (This $g$ represents a single valued function on $\Sgmtau$
valued in $SO(3)$.) Thus, $A_{u'}$ is equivalent to $A_u$ if and only
if $u'=\pm u -\frac{m}{2}+\tau \frac{n}{2}$ for some $n$, $m\in {\Z}$
which shows that \beq \NN_P=\C\lslash
\!\!\left\{\bigl(\mbox{$\nibun{\Z}+\frac{\tau}{2}{\Z}$}\bigr)\semidir
\{\pm1\}\right\}\,.  \eeq It is topologically a sphere and is a
complex orbifold with four singularities
$u=0,\frac{1}{4},\frac{\tau}{4},\frac{\tau+1}{4}$ of order $2$. The
orbifold $\NN_P-\{u=0\}$ corresponds to the quotient space
$\A_{ss}^{\circ}/\GPC$ in the above general argument.

If $P$ is non-trivial, $a$ and $b$ are represented by elements
$\tila$, $\tilb$ of $\tilH=SU(2)$ that do not commute but satisfy \beq
\tila \tilb\tila^{-1}\tilb^{-1}=\pmatrix{ -1 & 0 \cr 0 & -1 \cr }\,.
\eeq There is only one such pair $(\tila,\tilb)$ modulo conjugation :
\beq \tila=\pmatrix{ i & 0 \cr 0 & -i \cr }\,,\qquad \tilb=\pmatrix{
0& -1 \cr 1 & 0\cr }\,.
\label{holonomnontriv}
\eeq Hence, for the non-trivial $SO(3)$-bundle $P$ over the torus,
\beq \NN_P=\{\mbox{one point}\}\,.  \eeq

It should be noted that, unlike in the abelian case, $\NN_{{\rm
triv.}}$ is not isomorphic to $\NN_{{\rm non-triv.}}$ and that even
the dimensions are different on torus. This is the general
situation. For general simple group $H$ without center, $\NN_{{\rm
triv.}}$ on the torus $\Sgmtau$ is given by \beq \NN_{{\rm
triv.}}=\liet_{\!\C}/(\Pv+\tau\Pv)\semidir W\,, \eeq and hence of
dimension ${\rm rank}H$. But for each $j\in {\cal J}$ in the
terminology of Appendix 2, we have a non-trivial $H$-bundle $P^{(j)}$
and we can see that
$\dim\NN_{P^{(j)}}=\dim\Ker(w_jw_0-1)$\label{page1:wjw0} which is
strictly less than the rank of $H$.

\renewcommand{\theequation}{3.2.\arabic{equation}}\setcounter{equation}{0}
\vspace{0.3cm}
\begin{center}
{\sc 3.2 The Path Integration}\label{3.2}
\end{center}
\hspace{1.5cm} Now we perform the integration (\ref{intgauge}). To
define the measure for integration, we introduce metrics on the spaces
$\AP$ and $\GPC$. So, let us consider the tangent spaces \beq
\Bigl(T_{\!A}\AP\!\Bigr)_{\!\C}^{\!(1,0)}\cong\Omega^{0,1}(\Sigma,\ad
P_{\!\C})\,, \qquad
\Bigl(T_h\GPC\Bigr)_{\!\C}^{\!(1,0)}\cong\Omega^0(\Sigma,\ad
P_{\!\C})\,, \eeq at points $A\in \AP$ and $h\in \GPC$ where $a\in
\Omega^{0,1}(\Sigma,\ad P_{\!\C})$ is tangent to the curve
$\bartial_{\!A_t}=\bartial_{\!A}+ta$ at $t=0$ and $\epsilon\in
\Omega^0(\Sigma,\ad P_{\!\C})$ is tangent to the curve
$h_t=he^{t\epsilon+\bar t \epsilon^*}$ at $t=0$. We define inner
products on those spaces by
$(a_1,a_2)=\frac{i}{2\pi}\int_{\Sigma}\trP(a_1^*a_2)$ and by
$(\epsilon_1,\epsilon_2)=\frac{1}{2\pi}\int_{\Sigma}*\trP(\epsilon_1^*\epsilon_2)$
where $\trP$ is a suitably normalized $H$-invaiant fibre metric of
$\ad P$. Then, $\AP$ becomes a $\GP$ invariant K\"ahler manifold and
$\GPC$ becomes a Hermitian manifold invariant under the left
translations by elements of $\GPC$ and under the right translations by
elements of $\GP$.

\vspace{0.3cm} \underline{Local Parametrization of Gauge
Fields}\label{page:localparam}

As have been noticed in the previous section, we may neglect some
class of $\GPC$-orbits in the integration (\ref{intgauge}).  So, let
$\Ac_P\subset \AP$ be the submanifold that we take into account and
let $\NNc_P$ denote the quotient space $\NNc_P=\Ac_P/\GPC$ and let
$\dN$ be its dimension. If $H=U(1)$, $\A^{\circ}_P$ is the whole space
$\A_P$ and $\NN^{\circ}_P$ is the connected component of the Picard
group $\Pic\Sigma$ for the winding number
$c_1(P\!\times_{\!U(1)}\!\C)$. This gives the value $\dN=g$. If the
group $H$ is simple, we have

$(g=0)$\, $\Ac_P=\A_{\mu_j}$ for $P=P^{(j)}$, $\NNc_P=$ one point,

$(g=1)$\, $\Ac_P=\A_{ss}^{\circ}$\,, \,$\NNc_P=$ complex orbifold
$\subset \NN_P$\,, \, $0\leq \dN \leq {\rm rank}H$,

$(g\geq 2)$\, $\Ac_P=\A_s$\,,\, $\NNc_P=$ complex orbifold $\subset
\NN_P$\,, \, $\dN={\rm dim}H(g-1)$.

\noindent We may as well restrict our attention to the non-singular
points of $\NNc_P$ and we assume hence-forth that
$\NNc_P$\label{page:NNcP} is a complex manifold. For every point
$u_0\in \NNc_P$, we can take a neighborhood $U$ of $u_0$ in $\NNc_P$
with a holomorphic family $\{A_u\}_{u\in U}$\label{page:holfam} of
representatives, that is, a holomorphic map $U\to \Ac_P$, $u\mapsto
A_u$ such that $A_u\GPC=u$. We denote by $\A_U$ the inverse image of
$U$ by the projection $\Ac_P\to \NNc_P$ and define the surjective map
\beq f:U\times \GPC\longto \A_U \eeq by $f(u,h)=A_u^h$. This is not in
general injective due to the symmetries of $\A_u$. In fact, if
$S_u={\rm Aut}\bartial_{\!A_u}$ is the group of automorphisms of
$(P_{\!\C},J_{\!A_u})$, $f(u',h')=f(u,h)$ if and only if $u'=u$ and
$h'\in S_u h$. We denote by $d_S$ the dimension of $S_u$ which is
constant on $U$ since $d_S-\dN$ is the index $\dim H(1-g)$ of the
operator $\bartial_{\!A_u}$ acting on the sections of the adjoint
bundle $\ad \PC$.

Now, we coordinatize the spaces $\AP$ and $U\times \GPC$ in a way that
makes easy to pull back the measure on $\AP$ by the map $f$. Let
$(u^1,\cdots, u^{\dN})$ be a complex coordinate system on $U$. Then,
$\lnu_{\!1}(u),\cdots,\lnu_{\!\dN}(u)$ defined by
$\lnu_{\!a}(u)=(\partial/\partial u^a)A_u^{0,1}$\label{page:nuu}
determine a base of the tangent space \beq
T^{(1,0)}_uU=H^{0,1}_{\bartial_{\!A_u}}(\Sigma,\ad P_{\!\C}), \eeq for
each point $u\in U$. We choose a base $a^1(u),\cdots , a^{\dN}(u)$ of
the cotangent space \beq
\Bigl(T^{(1,0)}_uU\Bigr)^*=H_{\bartial_{\!A_u}}^{\,0}\!(\Sigma,\ad
P_{\!\C}\!\ot\! K)\,, \eeq and a base $\epsilon_1(u),\cdots\!,
\epsilon_{d_S}(u)$\label{page:epsilonu} of the space ${\rm
Lie}(S_u)=H_{\bartial_{\!A_u}}^0\!(\Sigma, \ad P_{\!\C})$ of
infinitesimal symmetries of $\bartial_{\!A_u}$.

At the point $f(u,h)=A_u^h$, we choose an orthonormal base $\{
a_n(u,h)\}_{n\in \N}$ of the tangent space ${\rm Im}\{
\bartial_{\!A_u^h}:\Omega^0 _{\ad P_{\!\bf c}}\to \Omega^{0,1}_{\ad
P_{\!\bf c}}\}$ of the $\GPC$-orbit through $A_u^h$ and an orthonormal
base $\{\epsilon_n(u,h)\}_{n\in \N}$ of the space $\Bigl( h^{-1}{\rm
Lie}S_u h\Bigr)^{\perp}\subset \Omega^0(\Sigma,\ad P_{\!\C})$ normal
to the symmetries of $\bartial_{\!A_u^h}$. If we put
$a_{b-\dN}(u,h)=h^*a^b(u)^*h^{*-1}$ and
$\epsilon_{i-d_S}(u,h)=h^{-1}\epsilon_i(u)h$, then we have a base of
$T^{\!(1,0)}_{\!A_u^h}\!\AP$ and a base of $T^{\!(1,0)}_h\GPC$ given
respectively by \beq \Bigl\{\,a_n(u,h)\,\Bigr\}_{n=1-\dN}^{\infty}
\quad \mbox{and}\quad
\Bigl\{\,\epsilon_n(u,h)\,\Bigr\}_{n=1-d_S}^{\infty}\,.  \eeq We
introduce complex coordinate systems :
$x=(x^{1-\dN},\cdot\cdot,x^0,x^1,\cdots)$ on a neighborhood of $A_u^h$
in $\A_U$ and $t=(t^{1-d_S},\cdot\cdot,t^0,t^1,\cdots)$ on a
neighborhood of $h$ in $\GPC$ such that \beqa
A(x)^{(0,1)}&=&\bigl(A_u^h\,\bigr)^{(0,1)}+\sum_{n=1-\dN}^{\infty}x^n
a_n(u,h)\,,\\ \mbox{and}\quad\qquad h(t)&=&{\rm
exp}\Biggl\{\sum_{i=1}^{d_S}t^{i-d_S}\epsilon_i(u)\Biggr\}h\,{\rm
exp}\Biggl\{\sum_{n=1}^{\infty}t^n\epsilon_n(u,h)\Biggr\}\,.  \eeqa
Then, the pull backs $\delta f^{(0,1)}=h^{-1}\delta A_u
h+\bartial_{\!A_u^h}(h^{-1}\delta h)$ of differentials $dx^n$ are
expressed as \beqa f^*dx^{a-\dN}&=&\sum_{b,c=1}^{\dN}M^{a\bar
b}(u,h)\!\left\langle a^b(u),\lnu_{\!c}(u)\right\rangle
du^c\,,\label{fdta}\\
f^*dx^n&=&\sum_{m=1}^{\infty}\left(a_n(u,h),\bartial_{\!A_u^h}\epsilon_m(u,h)\right)dt^m+\sum_{c=1}^{\dN}\left(a_n(u,h),h^{-1}\lnu_{\!c}(u)
h\right)du^c\,,
\label{fdtn}
\eeqa where $M^{a\bar b}(u,h)$ is the matrix element of the inverse of
$M_{\bar
ab}(u,h)=\Bigl(a_{a-\dN}\!(u,h)\!,a_{b-\dN}\!(u,h)\Bigr)$.
$\langle\,\,\,,\,\,\,\rangle$
in the equation (\ref{fdta}) is the natural pairing given by $\langle
a,\lnu\rangle=\nipi\int_{\Sigma}\tr(a\lnu)$.\label{ch3:pair}

\newpage \vspace{0.2cm} \underline{The Measure ${\cal D}A$}

We denote by $\widetilde{{\cal D}A}$ (resp. $\!\widetilde{{\cal D}h}$)
the volume element of the Hermitian space $\AP$ (resp. $\!\GPC$). At
the points $A_u^h\in \AP$ and $h\in \GPC$, we have the formal
expressions in terms of the coordinates $\{x^n\}_{n=1-\dN}^{\infty}$
and $\{t^n\}_{n=1-d_S}^{\infty}$ : \beqa \left(\widetilde{{\cal
D}A}\right)_{\!\!A_u^h}&=&\det\Bigl(M_{\bar a
b}(u,h)\Bigr)\!\!\prod_{n=1-\dN}^{\infty}\!d^2\!x^n\,,\\
\left(\widetilde{{\cal D}h}\right)_h&=&\det\left(S_{\bar i
j}(u,h)\right)\prod_{n=1-d_S}^{\infty}\!\dd t^n\,, \eeqa where we
denote by $d^2\!x^n$ the real two form $idx^n\wedge d\bar x^n$
(similarly for $d^2 t^n$ and we use such notation in the rest of the
paper) and the matrix $S_{\bar i j}(u,h)$ is given by
$\Bigl(\epsilon_{i-d_S}(u,h),\epsilon_{j-d_S}(u,h)\Bigr)$. By the
formulae (\ref{fdta}), (\ref{fdtn}), the pull back
$f^*\widetilde{{\cal D}A}$ at the point $(u,h)\in U\times \GPC$ is
expressed as, \beqa f^*\widetilde{{\cal
D}A}_{(u,h)}&=&\frac{\left\vert \,\det\!\left\langle
a^b(u),\lnu_{\!c}(u)\right\rangle\right\vert^2}{\det\Bigl(M_{\bar a
b}(u,h)\Bigr)}\left\vert
\,\det\!\!\left(a_n(u,h),\!\bartial_{\!\!A_u^h}\epsilon_m(u,h)\right)\!\right\vert^2\!\prod_{c=1}^{\dN}\dd
u^c \!\prod_{n=1}^{\infty}\dd t^n \nonumber\\ &=&\prod_{c=1}^{\dN}\dd
u^c \left\vert \,\det\!\left\langle a^b(u),\lnu_{\!c}(u)\right\rangle
\right\vert^2\frac{\det'\Bigl(\,\bartial_{\!A_u^h}^{\dag}\bartial_{\!A_u^h}\,\Bigr)}{\det
M(u,h) \det S(u,h)}\det S(u,h)\prod_{n=1}^{\infty}\dd t^n \nonumber \\
&=&\prod_{c=1}^{\dN}\dd u^c\left\vert \,\det\!\left\langle
a^b(u),\lnu_{\!c}(u)\right\rangle
\right\vert^2\frac{\det'\Bigl(\,\bartial_{\!A_u}^{\dag}\bartial_{\!A_u}\,\Bigr)}{\det\Bigl(a^a(u),a^b(u)\Bigr)\det\Bigl(\epsilon_i(u),\epsilon_j(u)\Bigr)}\nonumber\\
&&\times \,\,{\rm exp}\!\left\{I_{\ad P}(A_u,hh^*)\right\}\det S(u,h)
\prod_{n=1}^{\infty}\dd t^n \,,
\label{pullback}
\eeqa where we have used in the last step the formulae
(\ref{eqn:anom}), (\ref{anommulti}) for the chiral anomaly. In the
above expressions, ${\det}'(D^{\dag}\!D)$ denotes the regularized
determinant of $D^{\dag}\!D$ restricted to its positive eigen-space.

We now introduce a function $F_u:\GPC\to {\C}^{d_S}$ which satisfies
the following condition : On each $S_u$-orbit in $\GPC$, $F_u$ takes
the value zero at one and only one point and that at each zero point
$h$ the differential $F_{u,h}:{\rm Lie}S_u\to
{\C}^{d_S}$\label{page:Fu} defined by
$F_{u,h}(\epsilon)=\left(\frac{d}{dt}\right)_{\!0}F_u(e^{t\epsilon}h)$
is a linear isomorphism. If the point $(u,h)\in U\times \GPC$
satisfies $F_u(h)=0$, we have the equality : \beq
\prod_{i=1}^{d_S}\delta^{(2)}(t^{i-d_S})=\delta^{(2d_S)}\Bigl(F_u(h(t))\Bigr)\!\left\vert
\,\det\!\left(F^i_{u,h(t)}\bigl(\epsilon_j(u)\bigr)\right)\!\right\vert^2\,.
\eeq
Since the role of this factor is to fix the finite dimensional `gauge
degrees of freedom', we call it the residual gauge-fixing term and the
function $F_u$ is called the {\it residual gauge-fixing function}.
Making use of this function, the pull back of the measure
$\widetilde{{\cal D}A}$ by the isomorphism $f_{\star}:\cup_{u\in
U}S_u\backslash \GPC\longto \A_U$ is found to be \beqa
f_{\star}^*\widetilde{{\cal D}A}_{(u,S_uh)}&=&\prod_{a=1}^{\dN}\dd u^a
\int_{S_u}\!\widetilde{{\cal D}h}\,e^{I_{\ad P}(A_u,hh^*)}\,
\delta^{(2d_S)}\Bigl(F_u(h)\Bigr)\nonumber\\
&&\times\,{\det}'\Bigl(\,\bartial_{\!A_u}^{\dag}\bartial_{\!A_u}\,\Bigr)\frac{\left\vert
\,\det\!\left(F^i_{u,h}\bigl(\epsilon_j(u)\bigr)\right)\!\right\vert^2}{\det\Bigl(\epsilon_i(u),\epsilon_j(u)\Bigr)}\frac{\left\vert
\,\det\!\left\langle a^b(u),\lnu_{\!c}(u)\right\rangle
\right\vert^2}{\det\Bigl(a^a(u),a^b(u)\Bigr)}\,.
\label{measure22}
\eeqa

Let us now introduce a spin $(1,0)$ free fermionic system $(b,c)$
taking values in the adjoint bundle $\ad P_{\!\C}$ which we call the
{\it adjoint ghost system}. Then, we have another expression for the
measure given in (\ref{measure22}) : \beqa f_{\star}^*\widetilde{{\cal
D}A}_{(u,S_uh)}&=&\prod_{a=1}^{\dN}\dd u^a
\int_{S_u}\!\widetilde{{\cal D}h}\,e^{I_{\ad P}(A_u,hh^*)}\,
\delta^{(2d_S)}\Bigl(F_u(h)\Bigr)\nonumber\\
&&\times\,\,Z_{\Sigma,P}^{\,{\rm gh}}\biggl(\,\met\,, A_u \,;
\prod_{i=1}^{d_S}F^i_{u,h}(c)\bar F^i_{u,h}(\bar c)
\prod_{a=1}^{\dN}\Bigl\langle b,\lnu_{\!a}(u)\Bigr\rangle\!
\left\langle \bar b, \blnu_{\!a}(u)\right\rangle\,\biggr)\,.
\label{pullback2}
\eeqa As it should be, this is independent of the choice of the family
$A_u$ of gauge fields or of the residual gauge-fixing functions
$F_u$. We can also check that this is invariant by the right
translation by elements of $\GP$.

\vspace{0.2cm} \underline{Expression for the Integral
(\ref{intgauge})}

We shall express the integral (\ref{intgauge}) as an integral over the
moduli space $\NN^{\circ}$, the integrand of which is determined by a
composition of three kinds of quantum field theories---the matter
theory $M$, the adjoint ghost system and the $H_{\!\C}/H$ WZW model
which we shall study in the next section.

 Before doing that, we fix a definite model for the matter field
theory $M$. Let us take a semi-simple simply connected compact Lie
group $G$ such that the group $H$ is embedded into its adjoint group
$G/Z_G$ and we choose a half-integer $\lmd\in \nibun {\Z}$ and a
unitary representation $\rho :H\to U(V)$ on a finite dimensional
hermitian vector space $V$. We consider as the matter $M$ the combined
system of the WZW model at level $k\in {\N}$ with the target group $G$
and the system of free fermions taking values in vector bundles with
fibre $V$ and structure group $\rho(H)$.

Due to the PW identity (\ref{intPW}) and the chiral anomaly
(\ref{eqn:anom}), the integrand of (\ref{intgauge}) is given by \beqa
Z_{\Sigma,P}^M(\met,A_u^h;{\cal
O})&=&e^{I_{\Sigma}^M({\smet},A_u;hh^*)}Z_{\Sigma,P}^M(\met,A_u;h{\cal
O}) \,,\label{Manom}\\ &&\hspace{-2cm};\,\,
I_{\Sigma}^M(\met,A\,;hh^*)=kI_{\Sigma,P}^G(A,hh^*)+I_E\Bigl(A+\Bigl(\lmd-\nibun\Bigr)\Theta
\,;hh^*\Bigr)\,,\label{MHC} \eeqa where $I_{\Sigma,P}^G$ is the WZW
action for the group $G$ (we consider here that $hh^*$ is
$G_{\!\C}$-valued) and $I_E$ is the WZW action corresponding to the
vector bundle $E=P\times_H V$. The gauge invariance of the field
insertion ${\cal O}$ implies that $h{\cal O}$ is invariant under
$h\mapsto h\un$; $\un\in \GP$. This shows that it has the dependence
of the form $h{\cal O}={\cal O}(hh^*,\mm)$ where $\mm$ denotes the
fields in the matter theory $M$. Since $S_u$ preserves the connection
$A_u$, we may as well assume that we can expand the dressed insertion
$h{\cal O}$ as $h{\cal O}=\sum_m{\cal O}^m(hh^*){\cal O}_m(\mm)$ where
for each symmetry $h_s\in S_u$, ${\cal O}^m$ and ${\cal O}_m$
transform as \beq h_s{\cal O}^m={\cal O}^me^{I_{\Sigma,P}^M(
\smet,A_u;h_sh_s^*)}\quad\mbox{and}\quad h_s{\cal O}_m={\cal
O}_me^{-I_{\Sigma,P}^M({\smet},A_u;h_sh_s^*)}\,,
\label{symselect}
\eeq respectively. (If necessary, we select out the components of
$h{\cal O}$ that transform in the above way.)

After factoring out the volume element of the gauge transformation
group $\GP$, we integrate $f_{\star}^*\widetilde{{\cal D}A}\,
Z_{\Sigma,P}^M(\met,A_u^h\,;{\cal O})$ over each space $S_u\backslash
\GPC/\GP$. Then, we obtain the following measure on $U$ : \beqa
\Omega_{\Sigma,P}^M(\,\met\,;{\cal O})&=&\prod_{a=1}^{\dN}\dd u^a
\int_{\GPC/\GP}\!\!\!{\cal D}h \, e^{-I_{\Sigma,P}^{H_{\!\bf
c}\!/\!H}({\smet},A_u, hh^*)}\,\delta^{(2d_S)}\Bigl(F_u(h)\Bigr){\cal
O}^m(hh^*)\label{ch3.intmeasure}\\ &&\times\,\,Z_{\Sigma,P}^{M+{\rm
gh}}\biggl(\,\met\,,A_u\,; \prod_{i=1}^{d_S}F^i_{u,h}(c)\bar
F^i_{u,h}(\bar c) \prod_{a=1}^{\dN}\Bigl\langle
b,\lnu_{\!a}(u)\Bigr\rangle\!  \Bigl\langle \bar b,
\blnu_{\!a}(u)\Bigr\rangle {\cal O}_m\,\biggr)\,,\nonumber \eeqa where
the exponent $I_{\Sigma,P}^{H_{\!\bf c}\!/\!H}$ is given by
$-I_{\Sigma,P}^M-I_{\ad P}$ and ${\cal D}h$ is the volume element of
the homogeneous Riemannian manifold $\GPC/\GP$. Since this form
$\Omega_{\Sigma,P}^M$ is independent of the choice of families
$\{A_u\}_{u\in U}$ and $\{F_u\}_{u\in U}$, $\Omega_{\Sigma,P}^M$
extends to a well-defined measure of the moduli space
$\NN^{\circ}$. Thus, we have the following expression for the integral
(\ref{intgauge}) : \beq Z_{\Sigma,P}(\,\met\,;{\cal
O})=\int_{\NN^{\circ}_P}\Omega_{\Sigma,P}^M(\,\met\,;{\cal O})\,.
\label{oldintexpr}
\eeq {\it Remark}. Though it is not realistic, we have assumed in the
above argument that the function $\delta\Bigl(F_u(h)\Bigr)$ is right
$\GP$-invariant. Some care is needed if the zero point set of $F_u$ is
transversal to the $\GP$-orbits.

\renewcommand{\theequation}{3.3.\arabic{equation}}\setcounter{equation}{0}
\vspace{0.3cm}
\begin{center}
{\sc 3.3 WZW Model with Target Space $H_{\!\C}/H$}\label{3.3}
\end{center}
\hspace{1.5cm} The integration over each $\GPC$-orbit has thus lead to
a system of quantum field theory---the system of fundamental fields
$hh^*\in \GPC/\GP$ with the action $I_{\Sigma,P}^{\Hc\!/\!H}$. In the
final section, we study this system in the simplest cases where $H$ is
$U(1)$ or $H$ is simple. The general case is essentially a ``direct
sum'' of such abelian and simple models.

To start with, we write down the action $I_{\Sigma,P}^{\Hc/H}$ in such
simplest cases. For the abelian case $H=U(1)$, we introduce a new
variable $X$ defined by $hh^*=e^X$. We denote by $e^{ix}\mapsto
e^{ix\mu}$ and $e^{ix}\mapsto e^{ix\mu_{{}_V}}$ the embedding map
$\imath : H\to G/Z_G$ and the representation $\rho : H\to GL(V)$
respectively, where $\mu\in \Pv$ and $\mu_{{}_V}\in \gl(V)$. Then, we
have the following expression for the action : \beq
I_{\Sigma,P}^{\Hc/H}(\met,A\,;e^X)=\frac{i}{4\pi}\int_{\Sigma}\left\{\kh
(\partial X\bartial X-2f_A
X)-2c^1\Bigl(\lmd-\nibun\Bigr)R_{\Theta}X\right\}\,,
\label{Rwzwaction}
\eeq where $\kh=k\trG(\mu^2)+\trV(\mu_{{}_V}^2)$ and
$c^1=\trV(\mu_{\!{}_V})$.\footnote{ The symbol $\trG$ is the
normalised trace of the universal envelooping algebra of $\g$. See
Chapter 2.}  This is the free boson system with the background charge
provided by the curvatures $f_A$ and $R_{\Theta}$ of the gauge field
$A$ and the metric $\met$ respectively.

If $H$ is simple, since it coincides with its commutator subgroup
$[H,H]$, the term $\tr_{\!{}_E}\epsilon R_{\Theta}$ in the formula
(\ref{infanom}) disappears. Hence we see that the action is
proportional to the WZW action for the universal covering group
$\tilH$ : \beqa I_{\Sigma,P}^{\Hc/H}(\met,A\,;hh^*)&=&-\kh
I_{\Sigma,P}^{\tilH}(A,hh^*)\,,\\ ;\quad
\kh&=&kr^H_G+r^H_V+2\lieh^{\vee}\,,\label{neglev} \eeqa where $r_G^H$
and $r_V^H$ are ratios defined by \beq \trG(\imath X\imath
Y)=r_G^H\trH(XY)\qquad \mbox{and}\qquad
\trV(\rho(X)\rho(Y))=\trH(XY)\,,
\label{reltrace}
\eeq for $X$, $Y\in \h$. By this observation, we call this model the
{\it $\HC/H$ WZW model at level $-\kh$}.

\vspace{0.2cm} \underline{Fusion Rule as Integrability Condition}

Let $H$ be simple. Unlike in the ordinary quantum field theory, the
path-integration \beq Z_{\Sigma,P}^{\HcovH}(\,\met,A\,;{\cal
O})=\int_{\GPC/\GP}\!\!\!\!{\cal D}h\,\,e^{\kh
I_{\Sigma,P}^{\tilH}(A_u,hh^*)}\delta\Bigl(F_u(h)\Bigr)\left\vert \det
F_{u,h}^i(\epsilon_j)\right\vert^2{\cal O}(hh^*)\,, \eeq is
ill-defined for generic insertion ${\cal O}$ of local fields. This is
due to the fact that $\HC/H$ WZW model is a field theory {\it induced}
by the integration over gauge fields and that for the integral to be
well-defined, the integrand need to be finite everywhere on $\AP$ or
have at most mild singularities. For illustration, we consider a
concrete example---two and three point functions on the sphere of a
model with $G=SU(2)$, $V=\{0\}$ and $H=SO(3)$.

We look at the behavior of the integrand $Z_{\CP,{\rm
triv.}}^G(\,\met,A\,;{\cal O})$ as the gauge field $A$ lying in the
dense orbit $\A_0$ approaches a point in the orbit $\A_{\sigma_3}$ of
codimension one. Recall that if $A$ lies in the dense orbit ;
$A^{0,1}=h^{-1}\bartial h$, $h:\CP\to \HC$, then we have \beq
Z_{\CP,{\rm triv.}}^G(A\,;{\cal O})=e^{kI_{\CP}(hh^*)}{\cal
O}(hh^*)^{\Hc}Z_{\CP,{\rm triv.}}^G(0\,;{\cal O}(g)^{\Hc})\,, \eeq
where ${\cal O}(hh^*)^{\Hc}{\cal O}(g)^{\Hc}$ is the invariant
component of $h{\cal O}$ with respect to the global symmetry group
$\HC$ of the trivial connection $0$ (see (\ref{symselect})). Let us
consider a family $\{A_{(c)}\}_{c\in \C}$ of gauge fields given by
$A_{(c)}^{0,1}=(1-c)A_{(0)}^{0,1}$ where \beq
A_{(0)}^{0,1}=\frac{d\bar z}{(1+|z|^2)^2}\pmatrix{ -z & z^2 \cr 1 & z
\cr } \eeq lies in the orbit $\A_{\sigma_3}$. If $c\ne 0$, $A_{(c)}$
lies in the dense orbit $\A_0$ : $A_{(c)}^{0,1}=h_c^{-1}\bartial h_c$
where \beq h_c=\frac{1}{1+|z|^2}\pmatrix{ c^{\nibun}+c^{-\nibun}|z|^2
& z(c^{\nibun}-c^{-\nibun}) \cr \bar z(c^{\nibun}-c^{-\nibun}) &
c^{-\nibun}+c^{\nibun}|z|^2 \cr }\,.  \eeq A direct calculation shows
that $e^{kI(h_ch_c^*)}=|c|^{2k}e^{k(1-|c|^2)}$. Let ${\cal O}_{jj}$
and ${\cal O}_{j_1j_2j_3}$ be the field insertions
$\tr_{\!j}g(\infty)\tr_{\!j}g(0)$ and
$\tr_{\!j_1}g(\infty)\tr_{\!j_2}g(1)\tr_{\!j_3}g(0)$ respectively
where $\tr_{\!j}$ is the trace in the spin $j$ representation of
$SU(2)$. Then we have the following asymptotic behavior : \beqa
e^{kI_{\CP}(h_ch_c^*)}{\cal O}_{jj}(h_ch_c^*)^{\Hc}&\sim&
|c|^{2(k-2j)}e^{k(1-|c|^2)}\qquad \mbox{as $c\to 0$,}\\
e^{kI_{\CP}(h_ch_c^*)}{\cal
O}_{j_1j_2j_3}(h_ch_c^*)^{\Hc}&\sim&|c|^{2(k-j_1-j_2-j_3)}e^{k(1-|c|^2)}\quad\mbox{as
$c\to 0$}.  \eeqa These are finte in the limit $c\to 0$ if and only if
$j\leq \displaystyle{\frac{k}{2}}$ and $j_1+j_2+j_3\leq k$.

A detailed analysis by Gaw\c edzki \cite{GawQuad} shows that these are
precisely the convergence condition of the integrals \beq
\int_{\G_{\!\bf c}/\G}\!\!\!{\cal
D}h\,\,e^{(k+4)I_{\CP}(hh^*)}\delta(hh^*){\cal O}(hh^*)^{\Hc}\,, \eeq
for ${\cal O}={\cal O}_{jj}$ and for ${\cal O}={\cal O}_{j_1j_2j_3}$
respectively. Fortunately, if $j>\displaystyle{\frac{k}{2}}$ and
$j_1+j_2+j_3>k$, the correlators $Z_{\CP,{\rm triv.}}^G(0;{\cal
O}(g)^{\Hc})$ for ${\cal O}={\cal O}_{jj}$ and ${\cal O}={\cal
O}_{j_1j_2j_3}$ are identically zero due to the fusion rule found by
Gepner and Witten \cite{GW}. Hence, the integrand $Z_{\CP,{\rm
triv.}}^G(\,\met,A\,;{\cal O})$ or the path-integral of the total
system never diverges.

It has been conjectured by Gaw\c edzki that this phenomenon generally
occurs. That is, the fusion rule of WZW model for compact group
$\tilH$ provides a convergence condition for the $\HC/H$ WZW model.

\newpage \vspace{0.2cm} \underline{Direct Path Integration}

If $H=U(1)$, the symmetry group $S_u$ is always the group $\C^*$ of
constant multiplication. Hence, to fix the gauge degrees of freedom is
just to fix the constant mode of the field $X$. Thus we are left with
the integral \beq \int_{Map(\Sigma,\R)}\!\!\!\!{\cal
D}X\,\,e^{-I_{\Sigma}({\smet},A;X)}\delta(X(x_0)){\cal O}(e^X)\,, \eeq
where $x_0\in \Sigma$ is any base point and the insertion ${\cal O}$
satisfies ${\cal O}(e^{X+X_0})=e^{-I_{\Sigma}({\smet},A;X_0)}{\cal
O}(e^X)$. This is essentially a Gaussian integral which can be
performed exactly if we suitably renormalize the local fields
$e^{\alpha X(z)}$.

If $H$ is simple, the direct calculation gets more difficult. So far
it has been done only for topologically trivial $H$-bundles on the
sphere and on the torus by reducing the problem to iterative Gaussian
integrals (see Gaw\c edzki \cite{GawQuad} and Gaw\c edzki and
Kupiainen \cite{GawKup}). This is possible because we can take in that
case an Iwasawa-like decomposition of the group $\GPC$ which is
point-wisely preserved by holonomies of the background connection
$A_u$. For higher genus and for non-trivial $H$-bundle, a good
coordinatization of the homogeneous space $\GPC/\GP$ has not been
found.\footnote{Gaw\c edzki has recently announced \cite{Gawnew} that
he has found such good coordinatization (=good choice of the family of
background gauge fields $A_u$) for the $SU(2)$-bundle over each
Riemann surface of genus $\geq 2$.} However, using the field
identification which shall be proved in the next Chapter, the
calculation for general $H$-bundle is reduced to a problem for the
trivial bundle.


\vspace{0.2cm} \underline{Ward Identities and the Sugawara
Construction}

Instead of doing the direct path-integration, we can get an amount of
information out of the Sugawara construction together with the Ward
identities for current and energy-momentum tensor. We list below the
useful identities : \beqa &(i)& \mbox{ Ward identities (\ref{onecurr})
and (\ref{twocurr}) for the $H$-current},\hspace{4cm}\label{WardJHC}\\
&(ii)& \mbox{ Expression (\ref{T11wzw}) for $T_{z\bar z}$},\\ &(iii)&
\mbox{ Ward identity (\ref{WardTwzw}) for $T_{zz}$ and $T_{\bar z\bar
z}$}, \eeqa where $k$ and $G$ in the formulae in Chapter 2 should be
replaced by $-\kh$ and $\tilH$ respectively. These are essentially the
results of the left $\GPC$-invariance of the measure ${\cal D}h$. The
Ward identities (\ref{WardJHC}) lead to the local expression \beq
J\sigma_{\ad}v=J^{\sigma}(v)+\kh\,\tr(A^{\sigma}v)\qquad;\quad
\bartial J^{\sigma}(v)=0\,, \eeq for the current with respect to a
local holomorphic frame $\sigma$. We also make the assumption \beq
T_{zz}=-\frac{1}{\kh-\lieh^{\vee}}:(J_z,J_z):
\label{SugHC}
\eeq on the relation of the current and the energy-momentum tensor
where the product $:J_zJ_z:$ on the right hand side is defined as in
(\ref{normJJwzw}) with the above mentioned replacement.

\vspace{0.2cm} \underline{Space of States and Representation of Loop
Group}

 We give a description of the space of states of the model. As in the
case of WZW model for compact groups, we construct a natural line
bundle ${\cal L}$ over the space $L\HC/LH$ of configuration on a
circle $S^1$ and take as the space of states the space $\Omega^0(
L\HC/LH,{\cal L})$ of smooth sections. We also discuss on the
representation of a loop group on this space.

\vspace{0.2cm}
\noindent( For $H=U(1)$ )

We record here the description of the space of states which is
applicable to any $\kh>0$ and any $c^1\in \C$. The construction of the
line bundle ${\cal L}=\LWZ^{-1}=\Bigl(D_{\!\infty}\R\times
\C\Bigr)/\sim$\label{page:LWZ-} is essentially the same as in Chapter
2. Therefore, we only list below the basic results : \beqa \mbox{( The
definition of $\sim$ )}&&
(X_{\infty}+Y,1)\sim(X_{\infty},e^{I_{D_{\infty}}(Y)+\Gamma_{D_{\infty}}(X_{\infty},Y)})\,,\hspace{2.2cm}\\
;\hspace{1cm}&&\hspace{-1cm}X_{\infty},Y :D_{\infty}\to \R\,,\quad
Y|_{-\partial D_{\!\infty}}=0\,,\nonumber \\ &&\hspace{-1cm}
I_{D_{\infty}}(Y)=\frac{i}{4\pi}\int_{D_{\infty}}\!\!\!\partial
Y\bartial Y\,,\quad
\Gamma_{D_{\!\infty}}(X,Y)=\frac{i}{2\pi}\int_{D_{\!\infty}}\!\!\!\partial
Y \bartial X\,.  \eeqa \beqa \mbox{( Definition of the weight
)}&&e^{-I_{\Sigma_0}(X)}
=\{(X_{\!\infty},e^{-I_{\hat{\Sigma}_0}(X_{\!\infty}\star
X)})\}\,,\hspace{3cm}\\ ;\hspace{1cm}&&\hspace{-1cm} X:\Sigma_0\to
\R\,, \quad X|_{\partial\Sigma_0}=X_{\infty}|_{-\partial
D_{\!\infty}}\,.\nonumber \eeqa \beq \hspace{0.5cm}\mbox{( The
composition law )}\quad
\{(X_1,c_1)\}\{(X_2,c_2)\}\!=\!\{(X_1\!+X_2,c_1c_2e^{-\Gamma_{D_{\!\infty}}(X_1,X_2)})\}.  \eeq
\beq \mbox{( PW identity )}\qquad
e^{-I_{\Sigma_0}(X_1+X_2)}=e^{-I_{\Sigma_0}(X_1)}e^{-I_{\Sigma_0}(X_2)}e^{-\Gamma_{\Sigma_0}(X_1,X_2)}\,.\hspace{1.8cm}
\eeq The bundle $\LWZ^{*-1}$ is also defined in the similar way (see
Chapter 2.) and we have the \beq \mbox{(Pairing)}\qquad
\{(X_0,c_0)\}.\{(X_{\!\infty},c_{\infty})\}=c_0c_{\infty}e^{I_{\CP}(X_{\!\infty}\star
X_0)}\,,\hspace{3.2cm} \eeq of elements of $\LWZ^{*-1}$ and
$\LWZ^{-1}$ over the same loop $X_0|_{\partial
D_0}=X_{\!\infty}|_{-\partial D_{\!\infty}}$. We denote by
$\tilde{L\R}$ the group $\Bigl(\LWZ^{-1}\Bigr)^{\times}$ of invertible
elements. It has the following representation on the space
$\Omega^0(L\R,\LWZ^{-1})$ of states : \beqa \mbox{( Left and right
representations )}&& \hspace{-0.5cm}\mbox{For} \,\,\, \Phi\in
\Omega^0(L\R,\LWZ^{-1})\, \,\mbox{and}\,\,\,
\tilde{x}_L,\tilde{x}_R\in \tilde{L\R}\,, \nonumber \\
\Bigl(J(\tilde{x}_L)\bar J(\tilde{x}_R)\Phi
\Bigr)(x)&=&\tilde{x}_L\Phi(x-x_L-x_R)\tilde{x}_R^*\,.  \eeqa

We coordinatize the configuration space $L\R$ by \beq
x(\theta)=x_0+\sum_{n=1}^{\infty}\left( x_n e^{in\theta}+\bar
x_ne^{-in\theta}\right)\,, \eeq where $x_0$ is a real parameter and
$x_n$'s are complex parameters. A global frame
$\sigma_{D_0}:L\R\longto \LWZ^{-1}$ is defined by \beq
\sigma_{D_0}(x)=e^{-I_{D_0}(X_x)}\quad\mbox{where}\quad
X_x(z)=x_0+\sum_{n=1}^{\infty}\left(x_nz^n+\bar x_n\bar z^n\right)\,.
\label{globalframeRwzw}
\eeq Let $J_n$ and $\bar J_n$ be the generators of $J$ and $\bar J$
respectively corresponding to the vector tangent to the curve
$\tilde{x}_n(t)=e^{-I_{D_0}(tX_n)}$ where $X_n$ is a function on $D_0$
defined by $X_n(z)=\varrho(|z|^2)z^n$ ($\varrho$ is a cut off function
such that $\varrho(0)=0$ and $\varrho(1)=1$). If we denote by
$\Phi^{(0)}$ the coefficient of a section $\Phi$ of $\LWZ^{-1}$ with
respect to the frame $\sigma_{D_0}$ ; $\Phi=\sigma_{D_0}\Phi^{(0)}$,
then a calculation shows that \beqa
\left(J_n\Phi\right)^{(0)}&=&\left\{\begin{array}{rl}
\displaystyle{-\frac{\partial}{\partial
x_n}\Phi^{(0)}}\,,\hspace{2cm}&\mbox{if}\quad n\geq 0\\
\noalign{\vskip0.2cm} \displaystyle{\Bigl(-\frac{\partial}{\partial
\bar x_{|n|}}+|n|x_{|n|}\,\Bigr)\Phi^{(0)}}\,,&\quad\mbox{if}\quad
n\leq -1\,,
\end{array}\right.\\
\mbox{and}\qquad \left(\bar
J_n\Phi\right)^{(0)}&=&\left\{\begin{array}{rl}
\displaystyle{-\frac{\partial}{\partial \bar
x_n}\Phi^{(0)}}\,,\hspace{2cm}&\mbox{if}\quad n\geq 0\\
\noalign{\vskip0.2cm} \displaystyle{\Bigl(-\frac{\partial}{\partial
x_{|n|}}+|n|\bar x_{|n|}\,\Bigr)\Phi^{(0)}}\,,&\quad\mbox{if}\quad
n\leq -1\,.
\end{array}\right.\hspace{1.8cm}
\eeqa Note that the subspace spanned by the $J$-descendants of the
state $\sigma_{D_0}$ is isomorphic to the well-known {\it Boson Fock
space}. When completed with respect to a certain topology, it is
preserved by the representation $J$ of the loop group $\tilde{L\R}$.

\vspace{0.2cm}
\noindent ( For simple $H$ )

We take as the natural line bundle the restriction ${\cal
L}$\label{page:calL} of the bundle $\LWZ^{-\kh}\to L\tilH_{\!\C}$ to
the loops of the form $\gamma\gamma^*$ for $\gamma\in
L\HC$. Equivalently, ${\cal L}$ is the pull back of $\LWZ^{-\kh}$ by
the map $[\gamma]\in L\HC\mapsto \gamma\gamma^*\in L\tilH_{\!\C}$. We
take as the space of states the space $\Omega^0(L\HC/LH,{\cal L})$ of
sections. The group $\tilde{LH}_{\!\C}$ acts on this space by \beq
\left({\cal
J}(\tilde{\gamma})\Phi\right)(\gamma_1\gamma_1^*)=\tilde{\gamma}\Phi(\gamma^{-1}\gamma_1\gamma_1^*\gamma^{*-1})\tilde{\gamma}^*\,.
\label{HCwzwdef:reprJ}
\eeq Note that $\tilde{L\tilH}_{\!\C}\times \tilde{L\tilH}_{\!\C}$
does not act on this space. However, the infinitesimal action induces
the representation of the complexification
$\left(\tilde{L\lieh}_{\!\C}\right)_{\!\C}$ of the Lie algebra
$\tilde{L\lieh}_{\!\C}$. Decomposing
$\left(\tilde{L\lieh}_{\!\C}\right)_{\!\C}$ into $(1,0)$ and $(0,1)$
subspaces, we have the mutually commuting representations of negative
level $-\kh$ : \beq J={\cal
J}^{(1,0)}:\left(\tilde{L\lieh}_{\!\C}\right)_{\!\C}^{\!(1,0)}\longto
\gl\left(\Omega^0({\cal L})\right)\quad\mbox{and}\quad\bar J={\cal
J}^{(0,1)}:\left(\tilde{L\lieh}_{\!\C}\right)_{\!\C}^{\!(0,1)}\longto
\gl\left(\Omega^0({\cal L})\right)\,.  \eeq The action of these
generators can be identified with the action of the current in the
$\HC/H$ WZW theory.

\vspace{0.3cm} \underline{States Corresponding to Some Fields and
Spectral Flow}

Let $(D_0,\met_0,z)$ be the standard disc with coordinate and with
metric of unit curvature $\frac{i}{2\pi}\int_{D_0}R(\met_0)=1$ (see
Chapter 1, section 2). We insert a local field at $z=0$ and we find
explicit expression for the wave function at the boundary circle $S$
of $D_0$. Also, we describe the spectral flow as the screening of
local field by certain $H$-gauge field with integral curvature.

\vspace{0.4cm}
\noindent ( For $H=U(1)$ )

We come back to the definite model with the action given in
(\ref{Rwzwaction}). Rescaling the field $X$ by $Y=\sqrt{\kh}X$, the
action is expressed as \beq
I_{\Sigma}(\,\met,A\,;Y)=\frac{i}{4\pi}\int_{\Sigma} \biggl\{ \partial
Y\bartial Y-2\Bigl( \sqrt{\kh}
f_A+\frac{c^1}{\sqrt{\kh}}\Bigl(\lmd-\nibun\Bigr) R_{\Theta} \Bigr) Y
\biggr\} \,.  \eeq When a local field $O$ is inserted at $z=0$, the
effect of propagation through the standard disc $(D_0,\met_0)$ with
the basic gauge field $A_a$ can be seen by looking at the state
$Z_{D_0}(\,\met_0,A_a\,;O(0))\in \Omega^0(L\R,\LWZ^{-1})$ which is
formally defined by \beq
Z_{D_0}(\,\met_0,A_a\,;O(0))(y)=\int_{y=Y|_S}\!\!\!{\cal
D}Y\,\,e^{-I_{D_0}(\smet_0,A_a;Y)}O\Bigl(\frac{1}{\sqrt{\kh}}Y\Bigr)(0)\,,
\eeq where the weight is given by
$e^{-I_{D_0}(\smet_0,A_a;Y)}=\{(Y_{\!\infty},e^{-I_{\CP}(0\star\smet_0,0\star
A_a; Y_{\!\infty}\star Y)})\}\in \LWZ^{-1}$. For a certain kind of
field insertion $O$, we can detemine such states up to constant. We
first consider $O=1$. In this case, by changing the variable $Y$ by
$Y+Y_y$ (see the definition in (\ref{globalframeRwzw})) and using the
PW identity $e^{-I_{D_0}(Y+Y_y)}=e^{-I_{D_0}(Y_y)}e^{-I_{D_0}(Y)}$, we
see that the path-integral factors out and we get \beq
Z_{D_0}(\,\met_0,A_a\,;1)(y)=e^{-I_{D_0}(Y_y)}\exp\!\left\{
\frac{i}{2\pi}\int_{D_0}Y_y\Bigl( \sqrt{\kh}
f_{A_a}+\frac{c^1}{\sqrt{\kh}}\Bigl(\lmd-\nibun\Bigr) R(\met_0)
\Bigr)\right\}\,.  \eeq Since $A_a$ is the basic gauge field and the
metric $\met_0$ has the unit curvature, we may as well assume that we
can find a function $\varphi$ on $D_0$ such that
$\varphi(z)=\log|z|^2+$const. on a neighborhood of $S$ and that \beq
R(\met_0)=\partial\bartial \varphi\,,\qquad
f_{A_a}=a\partial\bartial\varphi\,.  \eeq Then, we see that
$\displaystyle{\frac{i}{2\pi}\int}_{\!D_0}Y_y\Bigl( \sqrt{\kh}
f_{A_a}+\frac{c^1}{\sqrt{\kh}} \Bigl(\lmd-\nibun\Bigr)R(\met_0)
\Bigr)=y_0\Bigl( \sqrt{\kh}
a+\frac{c^1}{\sqrt{\kh}}\Bigl(\lmd-\nibun\Bigr) \Bigr)$. With respect
to the natural coordinates $x_0=y_0/\sqrt{\kh}$, $x_n=y_n/\sqrt{\kh}$,
the state is expressed as \beq
Z_{D_0}(\,\met_0,A_a\,;1)(x)=\sigma_{D_0}(\sqrt{\kh}x)e^{(\kh a
+c^1(\lmd-\nibun))x_0}\,.  \eeq Quite in the similar way, we obtain
\beq Z_{D_0}(\,\met_0,A_a\,;e^{-\Lmd
X(0)})(x)=\sigma_{D_0}(\sqrt{\kh}x)e^{(-\Lmd+\kh a
+c^1(\lmd-\nibun))x_0}\,.
\label{stateafterspctrlflw}
\eeq In particular, the state corresponding in the standard way to the
field $e^{-\Lmd X}$ is given by \beq |\Lmd\rangle
=\sigma_{D_0}(\sqrt{\kh}x)e^{(-\Lmd +c^1(\lmd-\nibun))x_0}\,.  \eeq If
it is screened by the basic gauge field $A_a$, the state
$|\Lmd\rangle_a$ appearing at the boundary $S$ is given by
(\ref{stateafterspctrlflw}), that is, \beq |\Lmd\rangle_a=|\Lmd-\kh
a\rangle\,.  \eeq We call this map $|O\rangle\to |O\rangle_a$ the
spectral flow generated by the gauge field $A_a$.

\vspace{0.4cm}
\noindent( For simple $H$ )

In the theory of field identification we shall develop in the next
Chapter, two kinds of fields are of special importance. The first kind
are the matrix elements of finite dimensional representations of $H$
admissible with respect to the fusion rules : \beq
\rho_{\!\lmd}(hh^*)^{\!m}_{\,\bar m}\qquad;\quad \lmd\in
\PPp^{(\kh-2\lieh^{\vee})}\,.
\label{ch3dress}
\eeq These come from the dressing $h{\cal O}$ of the field insertion
${\cal O}$ of the matter field theory $M$.

The second kind are the fields of the following
form\footnote{$\rho$\label{page:halfroots} is the half the sum of
positive roots of $H$.}: \beq
\left|e^{\lmd+2\rho}(b(h))\right|^{2}\qquad; \quad \lmd\in
\PPp^{(\kh-2\lieh^{\vee})}\,,
\label{ch3partner}
\eeq where $b(h)$ is the `Borel part' of the Iwasawa decomposition
$h=b(h)\un(h)$ ; $b(h)\in B_0^+$ and $\un(h)\in H$. It is assumed that
a flag structure at the insertion point is specified. In the sense
that becomes clear in Chapter 4, the field (\ref{ch3partner})
constitutes a part of the {\it flag partner} of a dressed field
composed of fields of the form (\ref{ch3dress}).

Inserting those fields $\rho_{\!\lmd}(hh^*)^{\!-\lmd^*}_{\,-\lmd^*}$
and $\Bigl| e^{\lmd+2\rho}(b(h))\Bigr|^2$ at the center $z=0$ of the
standard disc $(D_0,\met_0)$ with the canonically flat connection
$A_0=0$, we odtain the states
$\left|(O_{\!\lmd})_{\,-\lmd^*}^{\!-\lmd^*}\right\rangle$ and
$|-\lmd-2\rho\rangle$ at the boundary $S=\partial D_0$. These generate
highest weight representations $\VV^{\lmd^*}$,
$\VV_{-\lmd-2\rho}\subset\Omega^0(L\HC/LH,{\cal L})$ of the Lie
algebra
$\left(\tilde{L\lieh}_{\!\C}\right)_{\!\C}^{\!(1,0)}\!\!\oplus\!\left(\tilde{L\lieh}_{\!\C}\right)_{\!\C}^{\!(0,1)}$
with highest weights $\lmd^*$ and $-\lmd-2\rho$ respectively. It
should be noted that these subspaces are not stable by the action of
the group $\tilde{L\tilH}_{\!\C}$. This can be seen in view of the
transformation rule (\ref{affact*}) of weights by the action of the
affine Weyl group $\Waff$ where the $L_0$-value is not bounded from
below due to the negativity of the level.

We now look at the effect of the presence of non-flat gauge field on
$D_0$. We take an element $e^{-i\mu\theta}w\in \Gmalcv$. Let the field
$\Bigl| e^{\lmd+2\rho}(b(h))\Bigr|^2$ be inserted into the disc
$(D_0,\met_0)$ with the basic gauge field $A_{w^{-1}\mu}$. We observe
the resulting state on $S$ standing on the horizontal frame $\sigma$
related to the original frame $s_0$ by $s_0=\sigma
e^{-i\mu\theta}n_w$. The observed state is the transform
$\tilh\Phi_{-\lmd-2\rho}$\label{ch3.tilh} of the state
$\Phi_{-\lmd-2\rho}=|-\lmd-2\rho\rangle$ by the spectral flow $\tilh$
corresponding to $h(z)=z^{-\mu}n_w$ which is defined exactly in the
same way as in Chapter 2. The state $\Phi_{-\lmd-2\rho}$ is given by
\beq \Phi_{-\lmd-2\rho}(\gamma\gamma^*)=e^{\kh
I_{D_0}(bb^*)}\left|e^{\lmd+2\rho}(b(0))\right|^2\,, \eeq where
$\gamma$ is the boundary loop of a holomorphic function $b:D_0\to \HC$
with $b(0)\in B_0^+$. A direct calculation shows that the new state
$\tilh\Phi_{-\lmd-2\rho}$ is given by
$|-w(\lmd+2\rho)-\kh\ttr\mu\rangle$. Since
$\hat{\rho}=(0,\rho,\lieh^{\vee})$ is $\Gmalcv$-invariant, we have
$w\rho+\lieh^{\vee}\ttr\mu=\rho$ which shows that the new state is the
same as $|-(w\lmd+(\kh-2\lieh^{\vee})\ttr\mu)-2\rho\rangle$. We note
that the space $\VV_{-\lmd-2\rho}$ is transformed by this spectral
flow to $\VV_{-(w\lmd+(\kh-2\lieh^{\vee})\ttr\mu)-2\rho}$ which
follows from the fact that the group $\Gmalcv$ preserve the positive
affine roots.

\newpage
\renewcommand{\theequation}{4.0.\arabic{equation}}\setcounter{equation}{0}

{\large CHAPTER 4. FIELD IDENTIFICATION}\label{ch.4}

\vspace{1cm} \hspace{1.5cm} The fundamental group $\pi_1(H)$ of a
compact group $H$ acts simply transitively on the set of isomorphism
classes of topological $H$-bundles over each surface $\Sigma$ ;
$\pi_1(H)\ni \gamma : P\mapsto P\gamma$. We show that the same group
$\pi_1(H)$ acts on the set of gauge invariant local fields of the
theory ; $\pi_1(H)\ni \gamma :O\mapsto \gamma O$, in such a way that
the following holds \beq Z_{\Sigma,P\gamma}(\,\met\,;O_1\cdots O_s
O)=Z_{\Sigma,P}(\,\met\,;O_1\cdots O_s\gamma O)\,.
\label{FI}
\eeq This may be referred to as the field identification phenomenon
since correlators with $O$-insertion and with $\gamma O$-insertion
coincides with each other if we sum up over topologies. We cannot see
such relation by a glance at the integral expressions
(\ref{oldintexpr}) for the correlators because the moduli spaces
$\NNc_{P}$ and $\NNc_{P\gamma}$ are not in general isomorphic and in
certain cases even the dimensions of them are different. In the first
half of this chapter, we shall find a new integral expression suited
to treat the problem. The basic notions in the reformulation are the
flag partner of each gauge invariant local field and the flag
strucures on a principal holomorphic $H_{\!\C}$-bundle. Making use of
the (conjectured) identification between moduli spaces of holomorphic
$\HC$-bundles of topological types $\PC$ and $(P\gamma )_{\!\C}$ with
flag structures at the insertion point, we show the relation
(\ref{FI}) which lead to the field identification.

\renewcommand{\theequation}{4.1.\arabic{equation}}\setcounter{equation}{0}
\vspace{0.2cm}
\begin{center}
{\sc 4.1 The Flag Partner}\label{4.1}
\end{center}
\hspace{1.5cm} As the first step to the required reformulation, we
express the dressed local field $hO$ as an integral over the flag
manifold of $H$ or of the fibre of the $H$-bundle over the insertion
point. The integrand may be referred to as the {\it flag partner} of
$hO$.

\vspace{0.3cm} \underline{Flag Manifold and the Borel-Weil-Bott
Theorem}

We first recall some basics on the theory of representation of compact
groups \`a la Borel, Weil and Bott which facilitates our description
of the flag partner.

\vspace{0.2cm} Let $Fl(H)$ be the ensemble of choices of maximal tori
and chambres : \beq Fl(H)=\left\{\, (T,\Ch)\,\mbox{\Large ;}
\begin{array}{ll}\mbox{$T$ is a maximal torus of $H$} \\
\noalign{\vskip0.1cm} \mbox{ and $\Ch$ is a chambre in
$i\liet$}\end{array}\,\right\}\,.  \eeq We choose and fix a pair
$(T,\Ch)\in Fl(H)$. Then, the set $Fl(H)$ is identified with the set
$H/T$ of right cosets of $T$ and becomes a manifold called the {\it
(generalized) flag manifold} of $H$. Furthermore, it can be given a
structure of homogeneous complex manifold since the embedding of $H$
into the complexification $\HC$ induces the isomorphism \beq
H/T\stackrel{\cong}{\longto}\HC/B \eeq of smooth manifolds where
$B$\label{page:B} is the Borel subgroup of $\HC$ determined by
$(T,\Ch)$.

Let us take a linear form $\lmd\in \V^*$ on $\V=i\liet$ that gives a
character $e^{\lmd}:T\to U(1)$ by $e^{2\pi i v}\mapsto e^{2\pi
i\lmd(v)}$. Then, it also gives a character $e^{\lmd}:B\to \C^*$ and
we can define a homogeneous holomorphic line bundle \beq
L_{-\lmd}=\HC\times_B \C \longto Fl(H)\,,
\label{def:L-lmd}
\eeq by the equivalence relation $(hb,c)\sim (h,e^{-\lmd}(b)c)$ where
$h\in \HC$, $b\in B$ and $c\in \C$. We denote by $h\cdot c\in
L_{-\lmd}$ the equivalence class represented by $(h,c)\in \HC\times
\C$. The theorem of Borel, Weil and Bott states that the space
$H^0(Fl(H),L_{-\lmd})$ of holomorphic sections is an irreducible
$\HC$-module $V_{\lmd^*}$ of highest weight $\lmd^*=-w_0\lmd$, which
is non-zero if and only if $\lmd$ takes non-negative values on
$\Ch$. See the ref. \cite{Bott} and also \cite{Kostant}.

Note that the line bundle $L_{-\lmd}$ is equipped with an
$H$-invariant hermitian metric $(\,\,,\,\,)_{-\lmd}$ such that an
element $h$ of $H$ determines a unitary frame $h\cdot 1$ ; $(h\cdot
c_1,h\cdot c_2)_{-\lmd}=\bar c_1c_2$. Note also that there exists an
$H$-invariant volume form $\Omega$\label{page:volFLH} on $Fl(H)=H/T$
since $T$ is compact. Then, we see that an $H$-invariant hermitian
inner product $(\,\,,\,\,)_{Fl(H)}$ on the space
$V_{\lmd^*}=H^0(Fl(H),L_{-\lmd})$ of sections is defined by \beq
(\psi_1,\psi_2)_{Fl(H)}=\frac{1}{\Vol
Fl(H)}\int_{Fl(H)}(\psi_1,\psi_2)_{-\lmd}\Omega\,.  \eeq Hence, the
representation $\rho_{\lmd^*}:H\to GL(V_{\lmd^*})$ becomes unitary.

 Let $\{ e_m\,;\,m\in \tilP_{\lmd}\}$ be an orthonormal base of the
hermitian space $V_{\lmd}$ consisting of weight vectors where
$\tilP_{\lmd}$ is an indexing set. We always take the weight $\lmd$
itself as the index for the highest weight vector. Denoting by
$(h)^{\!m_1}_{m_2}$ the matrix element $(e_{m_1},\rho_{\lmd}(h)
e_{m_2})$ of $\rho_{\lmd}(h)$, we put \beq
\psi^m(hB)=h\cdot(h)^{\!m}_{\lmd}\,,
\label{ch4:holsec}
\eeq for each $m\in \tilP_{\lmd}$. Then, $\{\, \psi^m\,;\,m\in
\tilP_{\lmd}\}$ forms a base of the space $H^0(Fl(H),L_{-\lmd})$ of
sections. It is also an orthogonal base with respect to the hermitian
product $(\,\,,\,\,)_{Fl(H)}$ : \beqa
(\psi^{m_1},\psi^{m_2})_{Fl(H)}&=&\frac{1}{\Vol
Fl(H)}\int_{Fl(H)}(\psi^{m_1},\psi^{m_2})_{-\lmd}\Omega \nonumber\\
&=&\frac{1}{\Vol
H}\int_H\left(\psi^{m_1}(hB),\psi^{m_2}(hB)\right)_{-\lmd}dh\nonumber
\\ &=&\frac{1}{\Vol H}\int_H
\overline{(h)^{\!m_1}_{\lmd}}(h)^{\!m_2}_{\lmd} dh =\frac{1}{\dim
V_{\lmd}}\delta^{m_1,m_2}\,,\label{Peter-Weyl} \eeqa where $dh$ is the
Haar measure of $H$ and we have used the theorem of Peter and Weyl in
the last step. This orthogonality formula (\ref{Peter-Weyl}) shall be
used in the following argument.

\vspace{0.3cm} \underline{Gauge Invariant Local Fields and the Dressed
Fields}\label{invfields}

We specify the set of gauge invariant fields in the matter theory
$M$. Recall that, in the theory $M$, there exists a one to one
correspondence between local fields and states. The space ${\cal H}^M$
of states is decomposed into a direct sum \beq {\cal
H}^M=\bigoplus_{\Lmd}L_{\Lmd}^M\ot\overline{L_{\Lmd}^M}\,, \eeq of
irreducible components of the left and right representations of the
symmetry group $G^M$ of the theory. For example, if $M$ is a free
fermionic theory, $G^M=G\tilde{L_{res}}(\HH^M)$ where $\HH^M$ is the
space of left moving one particle states and there is only one
irreducible component. If $M$ is the WZW model at level $k\in {\N}$
with simply-connected target group $G$, $G^M$ is the loop group
$\tilde{LG}_{\!\C}$ and the sum $\displaystyle{\bigoplus}_{\Lmd}$ is
over dominant weights $\Lmd\in \PPp^{(k)}$. For the matter theory $M$
we are considering, the Kac-Moody algebra $\tilde{L\lieh}_{\!\C}$ is
contained in the Lie algebra of the symmetry group $G^M$. We can
decompose the space $L_{\Lmd}^M$ into irreducible representations of
$\tilde{L\lieh}_{\!\C}$ \`a la Goddard, Kent and Olive \cite{GKO}:
\beq L_{\Lmd}^M=\bigoplus_{\lmd}L_{\Lmd,\lmd}\ot L_{\lmd}^H\,.
\label{branchingrule}
\eeq where the space $L_{\Lmd,\lmd}$ is the subspace of $L_{\Lmd}^M$
consisting of highest weight states of weight
$(\lmd,\kh-2\lieh^{\vee})$ with respect to
$\tilde{L\lieh}_{\!\C}$. Each non-zero element of $L_{\Lmd,\lmd}$
generates the integrable representation $L_{\lmd}^H$ whose lowest
energy subspace is identified with the unitary representation
$V_{\lmd}$ of a covering group $\tilH$ of $H$.

Let an element $\Phi_{\lmd}$ of $L_{\Lmd,\lmd}\ot
\overline{L_{\Lmd,\lmd}}$ correspond in the standard way to a field
denoted by $(O_{\lmd})^{\!\lmd}_{\lmd}$. The $J(N_0^-)\times \bar
J(N_0^-)$ descendants of $(O_{\lmd})^{\!\lmd}_{\lmd}$ form a $(\dim
V_{\lmd})^2$-dimensional representation $\{ (O_{\lmd})^{\!\bar
m}_m\,;\,m,\bar m \in \tilP_{\lmd}\}$ of $\tilH_{\!\C}\times
\tilH_{\!\C}$ : \beq J(h_L)\bar J(h_R)(O_{\lmd})^{\!\bar
m}_m=(h_R^*)^{\!\bar m}_{\bar m'}(O_{\lmd})^{\!\bar
m'}_{m'}(h_L)^{\!m'}_m\,.  \eeq Note that $\HC$ can act on this space
by $h\mapsto J(\tilde{h})\bar J(\tilde{h})$, where
$\tilde{h}\in\tilH_{\!\C}$ represents $h\in \HC$.

We now recall that the gauge invariance condition for the local field
$O$ is stated as (see the equations in (\ref{eqn:Jepsi})) \beqa \left(
J(v)+\bar J(v) \right)O&=&0\qquad \mbox{for $v\in \h$}\,,\\ J(z^n
v)O=\bar J(z^n v)O&=&0\qquad \mbox{for $v\in \h$ and $n\geq 1$}\,.
\eeqa By Schur's lemma, we can define a one to one correspondence
between the gauge invariant local fields and the states in the
subspace \beq
\bigoplus_{\Lmd,\lmd}L_{\Lmd,\lmd}\ot\overline{L_{\Lmd,\lmd}}\,, \eeq
of ${\cal H}^M$ : The state $\Phi_{\lmd}$ chosen above corresponds to
the gauge invariant field given by \beq \frac{1}{\dim
V_{\lmd}}\tr_{\lmd}O_{\lmd}=\frac{1}{\dim V_{\lmd}}\sum_{m\in
\tilP_{\lmd}}(O_{\lmd})^{\!m}_m\,, \eeq where $\tr_{\lmd}$ is the
trace over the vector space $V_{\lmd}$. The dressed field is then
given by \beq \frac{1}{\dim
V_{\lmd}}h\left(\tr_{\lmd}O_{\lmd}\right)=\frac{1}{\dim
V_{\lmd}}\tr_{\lmd}\left(O_{\lmd}hh^*\right)\,, \eeq where we recall
that $h{\cal O}$ is the field insertion defined by $h{\cal O}(g)={\cal
O}(h^{-1}gh^{*-1})$ (see the explanation of the eq. (\ref{intPW})).

For example, if the matter $M$ is the WZW model targetting $G$ and if
$H=G/Z_G$, then we have $L_{\Lmd,\lmd}=\delta_{\Lmd,\lmd}\C$. A
generator of $L_{\lmd,\lmd}\ot\overline{L_{\lmd,\lmd}}$ corresponds to
$\tr_{\lmd}\left(g^{-1}\right)$ and the dressed field is given by
$\tr_{\lmd}\left(g^{-1}hh^*\right)$.

\vspace{0.3cm} \underline{Integral Expression of Gauge Invariant
Fields}

We now express the dressed field $\frac{1}{\dim
V_{\lmd}}\tr_{\lmd}\left(O_{\lmd}hh^*\right)$ as an integral over the
flag manifold $Fl(H)$. We introduce a field valued differential form
$\Omega_{\lmd}(hh^*)$ on $Fl(H)$ of top degree defined by the
following statement : At the point $h_1B\in Fl(H)$ represented by
$h_1\in H$, it is related to the $H$-invariant volume form $\Omega$ by
\beq \left(
\Omega_{\lmd}(hh^*)\right)_{h_1B}=h_1(O_{\lmd})^{\!\lmd}_{\lmd}\Bigl|
e^{\lmd+2\rho}(b(h_1^{-1}h))\Bigr|^2\Omega_{h_1B}\,,
\label{flagmeasure1}
\eeq where $b(h_1^{-1}h)\in B$ is the Borel-part of the Iwasawa
decomposition $\HC=B\cdot H$ of $h_1^{-1}h\in \HC$. As shall be proved
shortly, the following relation (R) holds true \beq \mbox{ (R)}\qquad
\pounds_{h^{-1}}^*\Omega_{h^{-1}h_1B}=\Bigl|
e^{2\rho}(b(h_1^{-1}h))\Bigr|^2\Omega_{h_1B}\,,
\label{page:relR}
\eeq where $\pounds_{h^{-1}}$ denotes the left translation
$\pounds_{h^{-1}}:Fl(H)\to Fl(H)$ defined by
$\pounds_{h^{-1}}h_1B=h^{-1}h_1B$. Then, we have \beq
\left(\pounds_{h}^*\Omega_{\lmd}(hh^*)\right)_{h^{-1}h_1B}=h_1(O_{\lmd})^{\!\lmd}_{\lmd}\Bigl|e^{\lmd}(b(h_1^{-1}h))\Bigr|^2
\Omega_{h^{-1}h_1B}\,.  \eeq The Iwasawa decomposition
$h^{-1}h_1=\un\, b(h_1^{-1}h)^{-1}$ shows that we have a
representative $\un\in H$ of $h^{-1}h_1B$ such that
$h^{-1}h_1(O_{\lmd})^{\!\lmd}_{\lmd}=\un
(O_{\lmd})^{\!\lmd}_{\lmd}\Bigl| e^{-\lmd}(b(h_1^{-1}h))\Bigr|^2$. It
then follows that \beq \left(
\pounds_{h}^*\Omega_{\lmd}(hh^*)\right)_{\sun B}=h\un
(O_{\lmd})^{\!\lmd}_{\lmd}\,\Omega_{\sun B}=h(O_{\lmd})^{\!\bar m}_m
(\un^{-1})^{\!\lmd}_{\bar m}(\un)^m_{\lmd}\,\Omega_{\sun B}\,, \eeq
which amounts to the following identity of top differential forms :
\beq \pounds_h^*\Omega_{\lmd}(hh^*)=h(O_{\lmd})^{\!\bar m}_m
(\psi^{\bar m},\psi^m)_{-\lmd}\Omega\,, \eeq where $\psi^m$ is a
holomorphic section of the line bundle $L_{-\lmd}$ over the flag
manifold $Fl(H)$ with the expression (\ref{ch4:holsec}). Integrating
this over $Fl(H)$, we finally have the main result of this section :
\beq \frac{1}{\Vol Fl(H)}\int_{Fl(H)}\Omega_{\lmd}(hh^*)=\frac{1}{\dim
V_{\lmd}}\tr_{\lmd}\left(O_{\lmd}hh^*\right)\,,
\label{intexpr}
\eeq where we have used the orthogonality formula (\ref{Peter-Weyl}).

\vspace{0.2cm} {\it ( Proof of the relation {\rm (R)} )}
\hspace{0.45cm} Using the Iwasawa decomposition of $h_1^{-1}h$, we see
that to prove (R) is equivalent to prove the following relation for
$b\in B$ : \beq
\pounds_b^*\Omega_B=\Bigl|e^{-2\rho}(b)\Bigr|^2\Omega_B\,.  \eeq Since
the $(1,0)$-tangent space at the origin $B$ of $Fl(H)$ is isomorphic
to $\Lie(\HC)/\Lie B$, we have only to show that $e^{-2\rho}(b)$ is
the coefficient of $\displaystyle{\bigwedge_{-\alpha\in
\Delta_-}}e_{-\alpha}$ in the weight vector decomposition of
$\displaystyle{\bigwedge_{-\alpha\in \Delta_-}}\ad
b(e_{-\alpha})$.\footnote{We denote by $e_{\alpha}$ a root vector
corresponding to $\alpha\in \Delta$.} To show this, we introduce a
total ordering $<$ in the set $\Delta_-$ of negative roots such that
\beq \mbox{ if $\,\,-\alpha<-\beta$ , $\,\,-\alpha\notin -\beta
+\Delta_+$ .} \nonumber \eeq It is easy to see that there exists such
ordering. Then, the root vector decomposition of $\ad b(e_{-\beta})$
does not involve $e_{-\alpha}$ if $-\alpha<-\beta$. This completes the
proof.

\vspace{0.3cm} \underline{The Flag Partner}\label{4.1.fl}

Suppose that the dressed gauge invariant field $\frac{1}{\dim
V_{\lmd}}\tr_{\lmd}(O_{\lmd}hh^*)$ is inserted into a correlator
$Z_{\Sigma, P}^{\tot}(\met,A\,;\cdots)$ of the combined system of the
matter theory $M$, the $\HC/H$-WZW model and the adjoint ghost
system. We assume that the background gauge field $A$ is chosen to be
flat on a disc $D_0$ around the insertion point $x\in \Sigma$. If we
choose a horizontal frame $s_0$ on $D_0$, we can define the
field-state correspondence and the inserted field may be expressed as
an integral (\ref{intexpr}) over the flag manifold $Fl(H)$.

To get an intrinsic formula which does not refer to such choice of the
frame, we introduce the flag manifold $P_x/T\cong{\PC}_x/B$ of the
fibre $P_x$ over $x\in \Sigma$ which is denoted by
$Fl(P_x)$.\label{page:FLPx} Since the choice $s_0(x)\in P_x$
determines an isomorphism \beq Fl(H)\longto Fl(P_x)\,,
\label{mapflag}
\eeq by $hB\mapsto s_0(x)hB$, we find it more natural to think of
$\Omega_{\lmd}(hh^*)$ a measure on $Fl(P_x)$ rather than a measure on
$Fl(H)$. It is rewritten as \beq
\Omega_{\lmd}(hh^*)_f=(O_{\lmd})^{\!\lmd}_{\lmd}(f)\Bigl|
e^{\lmd+2\rho}(b_f(h))\Bigr|^2\Omega_f\,,
\label{flagmeasure}
\eeq where $f=sB\in Fl(P_x)$ is a flag over $x$ represented by an
$H$-frame $s\in P_x$. The field $(O_{\lmd})^{\!\lmd}_{\lmd}(f)$
correspond to the state $\Phi_{\lmd}\in
L_{\Lmd,\lmd}\ot\overline{L_{\Lmd,\lmd}}$ with respect to the
horizontal frame that coincides with $s\in P_x$ at $x$.  $b_f(h)$ is
the Borel part of the Iwasawa decomposition of the coefficient $h^s$
of $h(x)=sh^s$.  As it should be, we can check that these fields
$(O_{\lmd})^{\!\lmd}_{\lmd}(f)$ and $b_f(h)$ do not depend on the
choice of the representative $s\in P_x$ of the flag $f$. The volume
form $\Omega$ on $Fl(H)$ is sent by the map (\ref{mapflag}) to an
$\ad_HP_x$-invariant volume form on $Fl(P_x)$ denoted again by
$\Omega$.

Making use of the ghost fields, we may further rewrite the measure
$\Omega_{\lmd}(hh^*)$ in another form. Let us take an open subset
$U_F$ of $Fl(P_x)$ which is coordinatized by complex parameters
$f^1,\cdots , f^{|\Delta_+|}$. We assume that there is a family $\{
\sigma_f\}_{f\in U_F}$ of holomorphic sections of
$(\PC,\bartial_{\!A})$ over the neighborhood $D_0$ of $x$ such that
$\sigma_f(x)B=f$. Then, the symbol $(\partial \sigma_f/\partial
f^{\alpha})\sigma_f^{-1}$ determines a holomorphic section
$\nu_{\!\alpha}(f)$\label{page:nuf} of $(\ad\PC,\bartial_{\!A})$ over
$D_0$. Using the singular behavior (\ref{eqn:ord}) of the product
$b(z)c(w)$ of ghosts as $z\to w$, we obtain the following expression
\footnote{Here and henceforth, we denote the normalized contour
integral $\nipi\oint$ by ${\displaystyle \ooint}$.} for
$\Omega_{\lmd}(hh^*)$ on $U_F$ : \beq
\Omega_{\lmd}(hh^*)_f=\prod_{\beta=1}^{|\Delta_+|}d^2\!f^{\beta}
\ooint_xb\,\nu_{\!\beta}(f) \ooint_x\bar b \,\bnu_{\!\beta}(f)
O_{\lmd}(f)\,, \eeq \beq \mbox{
where}\hspace{2cm}O_{\lmd}(f)=(O_{\lmd})^{\!\lmd}_{\lmd}(f)\left|
e^{\lmd+2\rho}(b_f(h))\right|^{\!2}\prod_{-\alpha<0}c_s^{-\alpha}\bar
c_s^{-\alpha}\,.\hspace{2.8cm}
\label{flagmeasuregh}
\eeq In the expression (\ref{flagmeasuregh}), $c_s^{-\alpha}$ is the
coefficient of the ghost $c(x)=s\cdot e_a c_s^a$ with respect to an
$H$-frame $s$ such that $sB=f$. Note that the product
$\prod_{-\alpha<0}c_s^{-\alpha}\bar c_s^{-\alpha}$ does not depend on
the choice of the representative $s\in P_x$ of the flag $f$.

We call the field $O_{\lmd}(f)$ given in (\ref{flagmeasuregh}) the
{\it flag partner} of the dressed invariant field $\frac{1}{\dim
V_{\lmd}}\tr_{\lmd}(O_{\lmd}hh^*)$. It is a local field in the
combined system of the matter field theory $M$, the $\HC/H$-WZW model
and the adjoint ghost system. It should be noted that it is defined
only after a flag $f$ over the insertion point $x$ is specified.

\vspace{0.1cm} {\it Remark}. \hspace{0.1cm} Though it plays less
important role in this paper, we introduce a fermionic current \beq
Q=J^Mc+J^{\Hc/H} c +\nibun :\!J^{\rm gh}c\!:\,, \eeq where $J^M$,
$J^{\Hc/H}$ and $J^{\rm gh}$ are the $H$-currents of the matter field
theory $M$, the $\HC/H$-WZW model and the adjoint ghost system
respectively. The product $:\!J^{\rm gh}c\!:$ is defined by the point
splitting regularization as in the equations (\ref{defnorm}),
(\ref{defnormWZW}), though this time we do not need to subtract any
singular term. It is possible to see that the current $Q$ is
meromorphic as well as gauge invariant. The product of $Q(z)$ with the
dressed invariant field $\frac{1}{\dim
V_{\lmd}}\tr_{\lmd}(O_{\lmd}hh^*)(w)$ or with the flag partner
$O_{\lmd}(f)(w)$ is regular as $z\to w$.

In the literature \cite{BMP}, $\hat{t}$-relative $d=\nipi\oint
Q$-cohomology groups are calculated by choosing suitable
$\tilde{L\lieh}_{\!\C}$-modules for the $\HC/H$-part to construct the
cochain complex. They include as a non-trivial element, the state of
the form \beq |\lmd\rangle_M\ot|-\lmd-2\rho\rangle_{\Hc/H}\ot
\prod_{-\alpha<0}c_0^{-\alpha}|0\rangle_{{\rm gh}}\,, \eeq where
$|\lmd\rangle_M$ is a state in the matter theory $M$ which is highest
with weight $(\lmd,\kh-2\lieh^{\vee})$ with respect to
$\tilde{L\lieh}_{\!\C}$, $|-\lmd-2\rho\rangle_{\Hc/H}$ is a highset
weight state with weight $(-\lmd-2\rho,-\kh)$ in a suitably chosen
$\tilde{L\lieh}_{\!\C}$-module and $|0\rangle_{{\rm gh}}$ is the
natural vacuum of the ghost sector. This state seems to correspond to
the left moving part of our flag partner $O_{\lmd}(f)$ where the flag
$f$ is represented by a frame that determines the field-state
correspondence. It should be emphasized however that, to calculate
correlators using $O_{\lmd}(f)$, in general we must integrate over all
flags with suitable $b$-ghost insertions.

\renewcommand{\theequation}{4.2.\arabic{equation}}\setcounter{equation}{0}
\vspace{0.3cm}
\begin{center}
{\sc 4.2 A New Integral Expression}\label{4.2}
\end{center}
\hspace{1.5cm} We combine the result of Chapter 3 and the result of
the preceding section. The correlation function (\ref{intgauge}) is
expressed as an integral over a certain space which is generically a
flag bundle over the moduli space of holomorphic $\HC$-bundles.

\vspace{0.3cm} \underline{Combination of the Results}

Let $O$ be a gauge invariant field of the matter field theory $M$. We
consider the correlation function $Z_{\Sigma,P}(\met\,;{\cal O}O(x))$
of $O$ inserted at $x\in \Sigma$ and other gauge invariant fields
${\cal O}=O_1(x_1)\cdots O_s(x_s)$ inserted elsewhere. The result of
Chapter 3 states that the correlator is given by an integral of a
measure $\Omega_{\Sigma,P}^M(\met,{\cal O}O(x))$ over the moduli space
$\NNc_P$ of holomorphic $\HC$-bundles of topological type $\PC$. If we
choose a coordinatized open subset $U$ with a holomorphic family
$\{A_u\}_{u\in U}$ of representing gauge fields, the measure on $U$ is
given by \beqa
\lefteqn{\hspace{0.5cm}\Omega_{\Sigma,P}^M(\,\met\,;{\cal
O}\,O(x))_u}\label{page:Ztot}\\ &=&\prod_{a=1}^{\dN}\dd
u^a\,Z_{\Sigma,P}^{\tot}\!\biggl(\,\met,A_u\,\mbox{\Large ;}\,
\delta^{(2d_S)}\!\Bigl(F_u(h)\Bigr)
\!\prod_{i=1}^{d_S}F^i_{u,h}(c)\bar F^i_{u,h}(\bar
c)\!\prod_{a=1}^{\dN}\!\Bigl\langle b,\lnu_{\!a}(u)\Bigr\rangle\!
\Bigl\langle \bar b, \blnu_{\!a}(u)\Bigr\rangle \tilde{\cal O}\,
\tilde{O}(x)\,\biggr) ,\nonumber \eeqa where $\tilde{\cal O}$ and
$\tilde{O}$\label{page:Otilde} are the dressed gauge invariant
fields. We recall that $Z_{\Sigma,P}^{\rm tot}(\met,A\,;{\cal O})$ is
the correlation function for the {\it total system} --- the combined
system of the matter system $M$, the $\HC/H$-WZW model and the adjoint
ghost system.

If we use the result of the section 4.1 which expresses the dressed
field $\tilde{O}$ as an integral over the flag manifold $Fl(P_x)$, we
see that the following measure on $U\times Fl(P_x)$ reproduces the
above measure $\Omega_{\Sigma,P}^M$ after the integration along each
$Fl(P_x)$ : \beqa \tilde{\Omega}_{\Sigma,P,x}^M(\,\met\,;{\cal
O}\,O)_{(u,f)}\!\!\! &=&\!\!\!\prod_{a=1}^{\dN}\dd
u^a\prod_{\alpha=1}^{|\Delta_+|}d^2\!f^{\alpha}\,Z_{\Sigma,P}^{\tot}\!\biggl(\,\met,A_u\,\mbox{\Large
;} \,\delta\Bigl(F_u(h)\Bigr)\!\prod_{i=1}^{d_S}F^i_{u,h}(c)\bar
F^i_{u,h}(\bar c) \label{omegatilde}\\
&&\quad\quad\quad\times\prod_{a=1}^{\dN}\!\Bigl\langle
b,\lnu_{\!a}(u)\Bigr\rangle\! \Bigl\langle \bar b,
\blnu_{\!a}(u)\Bigr\rangle \prod_{\alpha=1}^{|\Delta_+|}\ooint_x
b\nu_{\!\alpha}(f)\ooint_x\bar b \bnu_{\!\alpha}(f)\,\tilde{\cal O}\,
O(f)\,\biggr) ,\nonumber \eeqa where $O(f)$ denotes the flag partner
of the dressed field $\tilde{O}(x)$.

At this stage, however, it is not obvious whether this form
$\tilde{\Omega}_{\Sigma,P,x}^M$ on $U\times Fl(P_x)$ extends to a well
defined form on some flag bundle over the moduli space $\NNc_P$.

\vspace{0.3cm} \underline{Transformation Properties of the Form
$\tilde{\Omega}_{\Sigma,P,x}^M$}

To find an answer to this question, let $\{A_{{}_1u}\}_{u\in U}$ and
$\{A_{{}_2u}\}_{u\in U}$ be two families of representatives that are
related by \beq A_{{}_1u}=A_{{}_2u}^{h_{{}_{21}u}}\,, \eeq through a
family $\{h_{{}_{21}u}\}_{u\in U}$ of chiral gauge
transformations. The groups $S_{{}_iu}=\Aut\bartial_{\!A_{{}_iu}}$ of
symmetries are then related by
$S_{{}_1u}=h_{{}_{21}u}^{-1}S_{{}_2u}h_{{}_{21}u}$. Hence, if
$\{F_{{}_1u}\}$ is a family of residual gauge-fixing functions for
$\{S_{{}_1u}\}$, $F_{{}_2u}(h)=F_{{}_1u}(h_{{}_{21}u}^{-1}h)$
determines a family $\{F_{{}_2u}\}$ of residual gauge-fixing functions
for the symmetries $\{S_{{}_2u}\}$. Correlators of the total system
with the backgrounds $A_{{}_1u}$ and $A_{{}_2u}$ are related by \beqa
\lefteqn{Z_{\Sigma,P}^{\tot}\!\biggl(\,A_{{}_1u}\,\mbox{\Large ;}\,
\delta\!\Bigl(F_{{}_1u}(h)\Bigr)
\!\prod_{i=1}^{d_S}F^i_{{}_1u,h}(c)\bar F^i_{{}_1u,h}(\bar
c)\tilde{\cal O}\prod\bigl( b; \bar b \bigr) O(f)\,\biggr)} \\
&=&\!\!Z_{\Sigma,P}^{\tot}\!\biggl(A_{{}_2u}\,\mbox{\Large ;}\,
\delta\!\Bigl(F_{{}_2u}(h)\Bigr)
\!\prod_{i=1}^{d_S}\!F^i_{{}_2u,h}(c)\bar F^i_{{}_2u,h}(\bar
c)\tilde{\cal O}\prod\!\bigl( h_{{}_{\!21}u}^{-1}bh_{{}_{\!21}u};
h_{{}_{\!21}u}^*\bar b h_{{}_{\!21}u}^{*-1}\bigr)
h_{{}_{\!21}u}\!O(f)\biggr),\nonumber \eeqa where $\prod(b;\bar b)$ is
any functional of the fields $b$ and $\bar b$. To derive this
relation, we have used the fact that the chiral anomaly is absent in
the total system : $kr_G^H+r_V^H-\kh+2\lieh^{\vee}=0$. (See the
equation (\ref{neglev}).)

Making use of the Iwasawa decomposition of $h_{{}_{\!21}u}(x)\in
\ad_{\Hc}\PC$ with respect to the flag $f\in Fl(P_x)$, it is possible
to see that \beq h_{{}_{\!21}u}O(f)=O(h_{{}_{\!21}u}f)\,, \eeq where
the action of $\GPC$ on the flag manifold $Fl(P_x)=(\PC)_x/B$ is
induced by the action on $(\PC)_x$. Now, it is enough to note the
relation \beq \delta A_{{}_1u}^{(0,1)}=h_{{}_{\!21}u}^{-1}\delta
A_{{}_2u}^{(0,1)}h_{{}_{\!21}u}+\bartial_{\!A_{{}_1u}}\!\!\left(h_{{}_{\!21}u}^{-1}\delta
h_{{}_{\!21}u}\right)\,, \eeq to see that the form
$\tilde{\Omega}_{\Sigma,P,x}^M$ on $\{1\}\!\times \!U\!\times Fl(P_x)$
with the representative $\{A_{{}_1u}\}_{u\in U}$ coincides with one on
$\{2\}\!\times \!U\!\times \!Fl(P_x)$ with the representative
$\{A_{{}_2u}\}_{u\in U}$, under the identification of the two spaces
given by \beq (1,u,f)\longleftrightarrow (2,u,h_{{}_{\!21}u}f)\,.
\eeq

\vspace{0.3cm} \underline{The Space $\NNc_{P,x}$}

Let $\{U_i\}$ be an open covering of the moduli space $\NNc_P$ such
that each $U_i$ is endowed with a holomorphic family
$\{A_{{}_iu}\}_{u\in U_i}$ of representing gauge fields. If $U_i$ and
$U_j$ intersect with each other, we can choose a family
$\{h_{{}_{\!ij}u}\}_{u\in U_i\cap U_j}$ of chiral gauge
transformations such that $A_{{}_ju}=A_{{}_iu}^{h_{{}_{\!ij}u}}$.

If the symmetry group $S_{{}_iu}=\Aut\bartial_{\!A_{{}_iu}}$ is
trivial everywhere, the families
$\displaystyle{\bigcup_{i,j}\{h_{{}_{\!ij}u}\}}$ necessarily satisfy
the triangle identities : \beq
\hspace{3cm}h_{{}_{\!ij}u}h_{{}_{\!jk}u}=h_{{}_{\!ij}u}\, ,\qquad
\mbox{for}\quad u\in U_i\cap U_j\cap U_k\,.  \eeq Hence we can define
the flag bundle
$\NNc_{P,x}\mbox{\large$\stackrel{Fl(P_x)}{\longto}$}\NNc_P$ by
identifying $(i,u,f)\in \{i\}\!\times \!U_i\!\times Fl(P_x)$ and
$(j,u,h_{{}_{\!ji}u}f)\in \{j\}\!\times\!U_j\!\times\!Fl(P_x)$ if
$u\in U_i\cap U_j$. Then, $\tilde{\Omega}_{\Sigma,P,x}^M$ extends to a
measure $\Omega_{\Sigma,P,x}^M$ on the flag bundle $\NNc_{P,x}$ and we
have \beq Z_{\Sigma,P}(\met\,;{\cal
O}\,O(x))=\int_{\NNc_{P,x}}\Omega_{\Sigma,P,x}^M(\,\met\,;{\cal
O}\,O)\,.  \eeq

If the symmetry groups $S_{{}_iu}$ are non-trivial and act
non-trivially on the flag manifold $Fl(P_x)$, the situation is
subtle. It may be possible that any choice of families
$\displaystyle{\cup_{i,j}\{h_{{}_{\!ij}u}\}}$ does not satisfy the
triangle identities and even if there exists a good choice, it is
highly non-canonical.

To avoid such subtlety, we shall perform the integration along each
$S_{u}$-orbit, expecting to obtain a well-defined measure on the space
of $S_{u}$-orbits. However, the group $S_{{}_iu}$ is generically
non-compact and each quotient $S_u\!\lbackslash Fl(P_x)$ is not even
Hausdorff. We now argue that we can select out certain orbits with
good quotients. An orbit through the flag $f$ is the homogeneous space
$S_u/S_{u,f}$ where $S_{u,f}$ is the group of symmetries of $A_u$ that
fix the flag $f$. Since the upper semi-continuity theorem for $\dim
S_{u,f}$ is expected to hold, it may be possible to find an open dense
subset $S_u^{\mbox{\tiny max}}\!F_x$ of $Fl(P_x)$ consisting of
$S_u$-orbits of maximum dimensions. In the integration along
$S_u$-orbits in $Fl(P_x)$, we have only to take into account the
contribution of the subset $S_u^{\mbox{\tiny max}}\!F_x$, unless the
form $\tilde{\Omega}_{\Sigma,P,x}^M$ has distributional supports in
the complement $Fl(P_x)-S_u^{\mbox{\tiny max}}\!F_x$. Expecting the
quotient $S_u\!\lbackslash S_u^{\mbox{\tiny max}}\!F_x$ or the family
$\displaystyle{\bigcup_{u\in U}}S_u\!\lbackslash S_u^{\mbox{\tiny
max}}\!F_x$ of quotients to be a good space (such as manifold or
orbifold), we introduce the space \beq
\NNc_{P,x}=\biggl(\bigcup_i\bigcup_{u\in
U_i}\{i\}\!\times\!U_i\!\times\!S_{{}_iu} \!\lbackslash
S_{{}_iu}^{\mbox{\tiny max}}\!F_x\biggr)\mbox{\LARGE /}\!\sim\,,
\label{defdomain}
\eeq where the equivalence relation is given by $(i,u,[f])\sim
(j,u,[h_{{}_{\!ji}u}f])$ for $u\in U_i\cap U_j$. Then, we expect that
the integration of $\tilde{\Omega}_{\Sigma,P,x}^M$ along each $S_u$
orbit becomes a well-defined measure $\Omega_{\Sigma,P,x}^M$ on the
space $\NNc_{P,x}$. This shall be established shortly.

Note that the space $\NNc_{P,x}$ can be considered as a subset of
$\A_P\!\times_{\GPC}\!Fl(P_x)$, where the group $\GPC$ acts
holomorphically on the product space $\A_P\times Fl(P_x)$ by
$((A,f),h)\mapsto (A^h,h^{-1}f)$. As we shall see in the next section,
this space can be identified with a moduli space of certain
holomorphic objects --- holomorphic $\HC$-bundles with flag structure
at one point $x$.

\vspace{0.3cm} \underline{Assumption of the Existence of Holomorphic
Family}

To pave the way to find a new expression for the integral
(\ref{intgauge}), we study local properties of the space $\NNc_{P,x}$.

First, we introduce notations. Recall that $\dN$ and $d_S$ denote the
dimensions of the moduli space $\NNc_P$ and the symmetry group $S_u$
for $u\in \NNc_P$ respectively. We denote by $\dSf$ the dimension of
the symmetry group $S_{u,f}$ where the pair $(A_u,f)$ represents an
element of $\NNc_{P,x}$. Then, the expected dimension $\dNf$ of the
space $\NNc_{P,x}$ is given by $\dNf=\dN+|\Delta_+|-d_S+\dSf$.

We assume without proof that the following holds : For a generic point
$v_0\in \NNc_{P,x}$, we can find a coordinatized neighborhood $\VV$ of
$v_0$ in $\NNc_{P,x}$ in such a way that there is a family
$\{\,(A_v,f_v)\!\in\! \A_P\!\times\! Fl(P_x)\,\}_{v\in \VV}$ of
representatives depending holomorphically on the coordinates
$v^1,\cdots, v^{\dNf}$.\label{page:vvv} We take a family of
holomorphic trivializations $\sigma_0(v)$ over a neighborhood $U_0$ of
$x$ such that $\sigma_0(v,x)B=f_v$. We also take a family
$\sigma_{\infty}(v)$ of holomorphic trivializations over the open
subset $U_{\infty}=\Sigma-\{x\}$. We take such families $\sigma_0(v)$
and $\sigma_{\infty}(v)$ of trivializations so that the transition
function $h_{\infty 0}(v)$ defined by
$\sigma_0(v)=\sigma_{\infty}(v)h_{\infty 0}(v)$ depends
holomorphically on the coordinates $v^{\rA}$.

We choose a coordinate system $v^1,\cdots, v^{\dNf}$ on $\VV$ in such
a way that the last $(\dNf-\dN)$-tuples of coordinates
$v^{\dN+\alpha}=v^{\dN+1},\cdots,v^{\dNf}$ deform only flags, that is,
$\partial \sigma_{\infty}(v)/\partial v^{\dN+\alpha}=0$. Then, we see
that the section $\nu_{\!\alpha}(v)=\Bigl(\partial
\sigma_0(v)/\partial v^{\dN+\alpha}\Bigr)\sigma_0(v)^{-1}$ of $\ad\PC$
defined on $U_0$ is expressed as \beq
\nu_{\!\alpha}(v)=\sigma_0(v)\cdot h_{\infty
0}(v)^{-1}\frac{\partial}{\partial v^{\dN+\alpha}}h_{\infty 0}(v)\,.
\eeq

Let us choose trivializations $s_0$ and $s_{\infty}$ of $P|_{U_0}$ and
$P|_{U_{\infty}}$ respectively that do not depend on $v$. We introduce
the functions $h_0(v)$ and $h_{\infty}(v)$ that relate the sections
$\sigma_0(v)$, $s_0$ of $\PC|_{U_0}$ by $\sigma_0(v)=s_0h_0(v)^{-1}$
and the sections $\sigma_{\infty}$, $s_{\infty}$ of
$\PC|_{U_{\infty}}$ by
$\sigma_{\infty}(v)=s_{\infty}h_{\infty}(v)^{-1}$ respectively. Since
the connection $A_v$ is represented by $\bartial_{\!A_v}s_0=s_0\cdot
h_0(v)^{-1}\bartial h_0(v)$ on $U_0$ and by
$\bartial_{\!A_v}s_{\infty}=s_{\infty}\cdot h_{\infty}(v)^{-1}\bartial
h_{\infty}(v)$ on $U_{\infty}$, the deformation of $A_v$ for the
variation of $v$ is expressed as \beqa \delta
A_v^{(0,1)}&=&\bartial_{\!A_v}\Bigl(s_0\cdot h_0(v)^{-1}\delta
h_0(v)\Bigr)\qquad\mbox{on $U_0$}\\
&=&\bartial_{\!A_v}\Bigl(s_{\infty}\cdot h_{\infty}(v)^{-1}\delta
h_{\infty}(v)\Bigr)\qquad\mbox{on $U_{\infty}$}\,.  \eeqa Hence, if
$b$ is a section of $K\ot \ad\PC$ which is holomorphic with respect to
$\bartial_{\!A_v}$ on the supports of $s_0\cdot h_0(v)^{-1}\delta
h_0(v)$ and $s_{\infty}\cdot h_{\infty}(v)^{-1}\delta h_{\infty}(v)$,
we have \beqa \nipi\int_{\Sigma} b\,\delta A_v^{(0,1)}&=&\nipi
\oint_xb\Bigl(s_0\cdot h_0(v)^{-1}\delta h_0(v)- s_{\infty}\cdot
h_{\infty}(v)^{-1}\delta h_{\infty}(v)\Bigr)\\ &=&\nipi
\oint_xb\,\sigma_0(v)\cdot h_{\infty 0}(v)^{-1}\delta
h_{\infty0}(v)\,, \eeqa where the contour encircles the point $x$.

\vspace{0.3cm} \underline{The New Integral Expression}

We shall obtain a measure $\Omega_{\Sigma,P,x}^M(\met\,;{\cal O}\,O)$
by integrating the form $\tilde{\Omega}_{\Sigma,P,x}^M$ along each
$S_u$-orbit in $Fl(P_x)$ that corresponds to a point in $\NNc_{P,x}$.

We start with the coordinatization. We take a holomorphic family
$\{\,(A_v,f_v)\,\}_{v\in \VV}$ on an open set $\VV$ in $\NNc_{P,x}$
which is coordinatized as in the above argument. We denote by $\vb$
the first $\dN$-tuples of $v$ : $\vb=(v^1,\cdots,v^{\dN})$. We may
think that the connection $A_v$ depends only on $\vb$ and hence we
write the symmetry groups and the residual gauge-fixing functions by
$S_{\vb}=\Aut\bartial_{\!A_v}$ and $F^i_{\vb}$. We may as well assume
that we can find a submanifold $\TT_{\vb}$ of $S_{\vb}$ that is
projected diffeomerphically onto $S_{\vb}/S_{\vb,f_v}$ for every $v\in
\VV$. (If it does not exist, we take some open covering $\{V_i\}$ of
$\cup_v S_{\vb}/S_{\vb,f_v}$ and argue in the same way on each $V_i$.)
We choose a local coordinate $t^1,\cdots, t^{d_S-\dSf}$ on an open set
in $\TT_{\vb}$.

The measure $\tilde{\Omega}_{\Sigma,P,x}^M$ is then expressed as \beqa
\tilde{\Omega}_{\Sigma,P,x}^M(\,\met\,;{\cal O}\,O)_{(v,h_t\!f_v)}
\!=\!\prod_{\rA=1}^{\dNf}\dd
v^{\rA}\hspace{-0.6cm}&&\hspace{-0.2cm}\prod_{i=1}^{d_S-\dSf}\!\!\!d^2
t^i\,Z_{\Sigma,P}^{\tot}\!\biggl(\,\met,A_v\,\mbox{\Large ;}
\delta\Bigl(F_{\vb}(h)\Bigr)\!\!\prod_{i=1}^{d_S}\!F^i_{\vb,h}(c)\bar
F^i_{\vb,h}(\bar c) \\
\times\hspace{0.2cm}&&\hspace{-0.7cm}\prod_{\rA=1}^{\dNf}\!\ooint_x
\!b\,\nu_{\!\rA}(v)\ooint_x \!\bar b\, \bnu_{\!\rA}(v)\!\!\!
\prod_{i=1}^{d_S-\dSf}\!\!\!\ooint_x \!b\,\nu_{\!i}(v,t)\ooint_x\!\bar
b \,\bnu_{\!i}(v,t)\,\tilde{\cal O}\, O(h_t\!f_v)\,\biggr) ,\nonumber
\eeqa where $\nu_{\!\rA}(v)$ and $\nu_{\!i}(v,t)$ are the holomorphic
sections of $(\ad\PC,\bartial_{\!A_v})$ on the subset $U_0\cap
U_{\infty}$ defined by \beqa \nu_{\!\rA}(v)&=&\sigma_0(v)\cdot
h_{\infty0}(v)^{-1}\!\!\frac{\partial}{\partial v^{\rA}}h_{\infty
0}(v)\,, \label{nuA}\\ \nu_i(v,t)&=&\Bigl(\frac{\partial}{\partial
t^i}h_t \Bigr)h_t^{-1}.  \eeqa The absence of chiral anomaly shows
that the above measure $\tilde{\Omega}_{\Sigma,P,x}^M(\,\met\,;{\cal
O}\,O)$ at $(A_v,h_t\!f_v)$ is expressed as \beqa
\prod_{\rA=1}^{\dNf}\dd
v^{\rA}\hspace{-0.3cm}\prod_{i=1}^{d_S-\dSf}\!\!\!d^2
t^i\,&&\hspace{-0.7cm}Z_{\Sigma,P}^{\tot}\!\biggl(\,\met,A_v\,\mbox{\Large
;}
\delta\Bigl(F_{\vb}(h_th)\Bigr)\!\!\prod_{i=1}^{d_S}\!\!F^i_{\vb,h_th}(h_tch_t^{-1})\bar
F^i_{\vb,h_th}(h_t^{*-1}\bar ch_t^*) \\
\times\hspace{0.2cm}&&\hspace{-0.7cm}\prod_{\rA=1}^{\dNf}\!\ooint_x
\!b\,\nu_{\!\rA}(v)\ooint_x \!\bar b\, \bnu_{\!\rA}(v)\!\!\!
\prod_{i=1}^{d_S-\dSf}\!\!\!\ooint_x
\!h_tbh_t^{-1}\nu_{\!i}(v,t)\ooint_x\!h_t^{*-1}\bar b h_t^*
\,\bnu_{\!i}(v,t)\,\tilde{\cal O}\, O(f_v)\,\biggr) .\nonumber \eeqa

At this stage, we change the order of the integrations : We consider
the field $h\in \GPC$ to be fixed and first integrate over
$\TT_{\vb}$. For each $h\in \GPC$, let $h_{\vb}(h)$ be the unique
element of $S_{\vb}$ such that $F_{\vb}(h_{\vb}(h)h)=0$. We denote by
$h_0(h)$ the unique element of $\TT_{\vb}$ which lies in
$h_{\vb}(h)S_{\vb,f_v}$. We choose a neighborhood $U_{\TT}$ of
$h_0(h)$ in $\TT_{\vb}$ and reparametrize the image ${\C}^{d_S}$ of
the function $F_{\vb}$ in such a way that the followings hold :

\vspace{0.1cm} (i) If $h'\in h_0(h)S_{\vb,f_v}$ satisfies
$F_{\vb}^1(h'h)=\cdots =F_{\vb}^{\dSf}(h'h)=0$, then $h'=h_{\vb}(h)$.

\vspace{0.05cm} (ii) For each $h''\in U_{\TT}$,
$F_{\vb}^{\dSf+1},\cdots,F_{\vb}^{d_S}$ are constant along
$h''S_{\vb,f_v}h$.

\vspace{0.1cm}
\noindent Roughly speaking, $F_{\vb}^i$ for $1\leq i \leq \dSf$ varies
in the direction of $S_{\vb,f_v}$-orbit and $F^j_{\vb}$ for $\dSf<
j\leq d_S$ varies in the direction of $\TT_{\vb}\cong S_{\vb}/S_{\vb,
f_v}$. The first condition (i) shows that the functions
$F^i_v(h)=F^i_{\vb}(h_0(h)h)$\label{page:Fv} for $1\leq i\leq \dSf$
play the role of residual gauge-fixing functions for the symmetry
groups $S_{\vb,f_v}$.

If we integrate over $U_{\TT}$, we see that the delta function serves
the factor \beq \left|\det\!\left(\frac{\partial}{\partial
t^j}F_{\vb}^{\dSf+i}\!(h_th)\right)\right|^{\!-2}\delta^{(2\dSf)}\!\Bigl(F_v(h)\Bigr)\,.\label{servedet}
\eeq At the same time, we deform the contours of the integrals
$\displaystyle{\ooint_xb\,h_t^{-1}\!\frac{\partial}{\partial
t^i}h_t}$, then, the field $b$ meets the $c$-insertions
$F^j_{\vb,h_th}(h_tch_t^{-1})$. The contour integrals around these
$c$-insertions serve a determinant factor that cancels with the
determinant in the formula (\ref{servedet}).

Thus we have reached to the following measure on $\VV$ : \beqa
\lefteqn{\Omega_{\Sigma,P,x}^M(\,\met\,;{\cal O}\,O)_v}
\label{newform}\\ &=&\prod_{\rA=1}^{\dNf}\dd
v^{\rA}\,Z_{\Sigma,P}^{\tot}\!\biggl(\,\met,A_v\,\mbox{\Large ;}
\,\delta\Bigl(F_v(h)\Bigr)\!\prod_{i=1}^{\dSf}\!F^i_{v,h}(c)\bar
F^i_{v,h}(\bar c) \prod_{\rA=1}^{\dNf}\ooint_x
\!b\,\nu_{\!\rA}(v)\ooint_x \!\bar b\, \bnu_{\!\rA}(v)\,\tilde{\cal
O}\, O(f_v)\,\biggr) ,\nonumber \eeqa We can check that this
expression is independent on the choice of the holomorphic family
$\{\,(A_v,f_v)\,\}_{v\in \VV}$. This shows that the form
$\Omega_{\Sigma,P,x}^M(\,\met\,;{\cal O}\,O)$ extends to a
well-defined measure on the space $\NNc_{P,x}$. We have thus obtained
the new integral expression for the correlation function : \beq
Z_{\Sigma,P}(\,\met\,;{\cal
O}\,O(x))=\int_{\NNc_{P,x}}\!\!\Omega_{\Sigma,P,x}^M(\,\met\,;{\cal
O}\,O)\,.
\label{newintexpr}
\eeq

\renewcommand{\theequation}{4.3.\arabic{equation}}\setcounter{equation}{0}
\vspace{0.6cm}
\begin{center}
{\sc 4.3 The Moduli Space Of Holomorphic Principal Bundles\\ With Flag
Structure --- Examples}\label{4.3}
\end{center}
\hspace{1.5cm} As has been noticed in the preceding section, the space
$\NNc_{P,x}$ introduced in the course of finding the expression
(\ref{newintexpr}) is a subset of the set
$\A_P\!\times_{\GPC}\!Fl(P_x)$ of $\GPC$-orbits in the product space
$\A_P\!\times \!Fl(P_x)$. This set $\A_P\!\times_{\GPC}\!Fl(P_x)$ can
naturally be identified with the set of isomorphism classes of certain
holomorphic objects --- holomorphic $\HC$-bundles with quasi-flag
structure at one point. Using this fact, we give an explicit
description of the space $\NNc_{P,x}$ for some simple cases.

\vspace{0.4cm} \underline{Holomorphic $\HC$-Bundles with Quasi-Flag
Structure}

We fix a maximal torus $T$ of $H$ and a chambre $\Ch$ and we denote by
$B$ the corresponding Borel subgroup of $\HC$. For a holomorphic
$\HC$-bundle $\Ph$ over $\Sigma$, a choice of flag $f\in \Ph_x/B$ at
$x\in \Sigma$ is called a {\it quasi-flag structure} of $\Ph$ at
$x$.\cite{Mehta-Seshadri} Two holomorphic $\HC$-bundles with
quasi-flag structure at $x $ say $(\Ph_1,f_1)$ and $(\Ph_2,f_2)$ are
said to be {\it isomorphic} when there is an isomorphism $\Ph_1\longto
\Ph_2$ of holomorphic $\HC$-bundles which sends the flag $f_1$ to
$f_2$. Notice that the set of isomorphism classes of quasi-flag
structures of a holomorphic $\HC$-bundle $\Ph$ at $x$ is given by the
set $\Aut\Ph\lbackslash \Ph_x /B$ where $\Aut\Ph$ is the group of
automorphisms of $\Ph$. In the rest of the paper, we shall abbreviate
the term `quasi-'.

As in the case without flags, for a principal $H$-bundle $P$, the set
$\A_P\!\times_{\GPC}\!Fl(P_x)$ of $\GPC$-orbits in the space
$\A_P\!\times \!Fl(P_x)$ can naturally be identified with the set of
isomorphism classes of holomorphic $\HC$-bundles of topological type
$\PC$ with quasi-flag structure at $x$.

A method is given to characterize a holomorphic $\HC$-bundle with flag
structure that represents a class identified with an element of
$\NNc_{P,x}$. For a holomorphic $\HC$-bundle $\Ph$ with flag structure
$f$ at $x$, we denote by $\Aut(\Ph,f)$ the group of automorphisms of
$\Ph$ that preserve the flag $f$. Then, $(\Ph,f)$ represents a class
that is identified with an element of $\NNc_{P,x}$ if and only if
$\Ph$ represents a class identified with an element of $\NNc_P$ and
$\dim\Aut(\Ph,f)\leq \dim\Aut(\Ph,f')$ for other choices $f'$ of
flags.

\vspace{0.32cm} \underline{On the Sphere}

We classify the holomorphic principal bundles over the complex
projective line $\CP$ with quasi-flag structure at one point. We
follow the notation of section 3.1.

We start with the example in which $H=SU(n)/\Z_n$ or
$\HC=PSL(n,\C)$. We choose the Borel subgroup $B=B_0^+$ that is
represented by the set of upper triangular matrices. Recall that the
holomorphic $\HC$-bundles on $\CP$ are classified using the Birkhoff
factorization theorem. A detailed statement of the same theorem also
classifies the holomorphic $\HC$-bundles with flag structure at, say,
$z=0$. Let $(\Ph,f)$ be a holomorphic $\HC$-bundle with a flag $f$ at
$z=0$. Let $\sigma_0$ be a frame on the $z$-plane $U_0$ such that
$\sigma_0(0)B=f$, and let $\sigma_{\infty}$ be a frame on the
$w$-plane $U_{\infty}$. These are related by the transition rule
$\sigma_0=\sigma_{\infty}h_{\infty 0}$ where $h_{\infty 0}$ is
represented by a multivalued holomorphic map from $U_0\cap U_{\infty}$
to $SL(n,\C)$. The theorem states \cite{P-S} that there is a unique
element $a\in \Pv$ such that \beq h_{\infty
0}(z)=b_-(z)z^{-a}b_+(z)^{-1}\,, \eeq where $b_-$ extends to a map
from $U_{\infty}$ and $b_+$ extends to a map from $U_0$ such that
$b_+(0)\in B$. Thus, the set of isomorphism classes of holomorphic
$\HC$-bundles with quasi-flag structure at $z=0$ is identified with
the discrete set $\Pv$. For each $a\in \Pv$, we denote by
$\Ph_a$\label{page:Pha} the pair $(\Ph_{[a]},f_a)$ of an $\HC$-bundle
$\Ph_{[a]}$ with the transition rule
$\sigma_0^{(a)}=\sigma_{\infty}^{(a)}z^{-a}$ and a flag
$f_a=\sigma_0^{(a)}(0)B$ at $z=0$.

Now let us calculate the dimension of the symmetry group
$\Aut\Ph_a\subset \Aut\Ph_{[a]}$. Recall that an element $h$ of $\Aut
\Ph_{[a]}$ is represented with respect to the frame $\sigma_0^{(a)}$
by an $SL(n,\C)$-valued function of $z$ whose $i$-$j$-th entry
$(h_0)^{\! i}_j(z)$ is a span of $1, z,\cdots,z^{a_i-a_j}$ if $a_i\geq
a_j$ and zero if $a_i<a_j$. This element $h$ belongs to $\Aut\Ph_a$ if
$h_0(0)\in B$, that is, if $(h_0)_j^{\!i}(0)=0$ for $i>j$. Thus, we
see that the dimension is given by \beq \dim \Aut \Ph_a=\dim \Aut
\Ph_{[a]}-\sum_{\stackrel{i>j}{a_i\geq
a_j}}1=n-1+\sum_{i<j}\left(\,|a_i-a_j|+\theta_{a_i,a_j}\right)\,, \eeq
where $\theta_{x,y}=0$ if $x<y$ and $\theta_{x,y}=1$ if $x\geq y$. An
element $a\in \Pv$ minimizes this value in its permutation class if
and only if the entries satisfy $a_1\leq a_2\leq\cdots\leq a_n$. That
is, $\dim\Aut\Ph_a\leq \dim\Aut\Ph_{wa}$ for any $w\in W=S_n$
$\Longleftrightarrow$ $a\in \Pv\cap (-\overline{\Ch})$.

Remember that there is a smooth $SU(n)/\Z_n$-bundle $P^{(j)}$ for each
$j\in {\cal J}_0$ such that $\NNc_{P^{(j)}}$ is one point $\{
\Ph_{[\mu_j]}\}$. By the above statement of the Birkhoff theorem, the
set of distinct flag structures on $\Ph_{[\mu_j]}$ is identified with
the Weyl orbit $W\mu_j$. Let $n_{w_jw_0}$ denote the matrix given by
\beq \pmatrix{ 0 & {\bf 1}_j \cr {\bf 1}_{n-j}\hspace{-0.3cm} & 0 \cr
}(-1)^{\frac{j(n-j)}{n}}\,, \eeq which represents the element $w_jw_0$
of $W$. In the above expression, ${\bf 1}_j$ denotes the $j\times j$
identity matrix. Since $a=\ad n_{w_jw_0}^{-1}\mu_j=(w_jw_0)^{-1}\mu_j$
satisfies $a_1\leq \cdots \leq a_n$, it is the unique element in the
orbit $W\mu_j$ that minimizes the dimension of the symmetry group
$\Aut\Ph_a$. Hence $\NNc_{P^{(j)},x}$\label{page:NNcPjx} consists of
one point which is represented by $\Ph_j=(\Ph_{[\mu_j]},f_j)$ where
$\Ph_{[\mu_j]}$ is an $\HC$-bundle with the transition rule \beq
\sigma_0(z)=\sigma_{\infty}(z)z^{-\mu_j}n_{w_jw_0}\,, \eeq and $f_j$
is the flag $\sigma_0(0)B$.

For general centerless simple group $H$, the story is essentially the
same. We choose a maximal torus $T$ and a chambre $\Ch$ (see Appendix
2 for notations). The corresponding Borel subgroup is denoted by
$B$. Isomorphism classes of holomorphic $\HC$-bundles with flag
structure at $z=0$ are indexed by the lattice $\Pv$ : Each $a\in \Pv$
indexes an isomorphism class represented by $\Ph_a=(\Ph_{[a]},f_a)$
where $\Ph_{[a]}$ is an $\HC$-bundle with the transition rule
$\sigma_0^{(a)}=\sigma_{\infty}^{(a)}z^{-a}$ and $f_a$ is the flag
$\sigma_0^{(a)}(0)B$ at $z=0$. The dimension of the group $\Aut\Ph_a$
of automorphisms is given by \beq
\dim\Aut\Ph_a=l+\sum_{\alpha>0}\left(\,|\alpha(a)|+\theta_{\alpha(a),0}\,\right)\,,
\eeq which is minimized, within each Weyl orbit, by a unique element
in $-\overline{\Ch}$. For each $j\in {\cal J}_0$, $\NNc_{P^{(j)},x}$
consists of one point which is represented by
$\Ph_j=(\Ph_{[\mu_j]},f_j)$ where $\Ph_{[\mu_j]}$ is an $\HC$-bundle
with the transition rule \beq
\sigma_0(z)=\sigma_{\infty}(z)z^{-\mu_j}n_{w_jw_0}\,, \eeq and $f_j$
is the flag $\sigma_0(0)B$. In the above, $n_{w_iw_0}$ is an element
of $N_T$ that represents the element $w_jw_0\in W$.

\vspace{0.3cm} \underline{On Torus with $H=SO(3)$}

We next consider the case in which $\Sigma$ is the torus and the gauge
group $H$ is $SO(3)$. This time, we realize the torus $\Sigma_{\tau}$
by $\C^*\lslash q^{\Z}$ where $q^{\Z}$ is the subgroup of $\C^*$
generated by $q=e^{2\pi i\tau}$.\label{page:agen.q} This is obtained
by the previous realization $\C/(\Z+\tau \Z)$ through the exponential
map $\zeta\mapsto z=e^{-2\pi i\zeta}$.

First, we describe several holomorphic $PSL(2,\C)$-bundles over the
torus $\Sigma_{\tau}$.

We recall that a flat $SO(3)$ connection on the trivial bundle is
represented by the holonomies $\tila=e^{2\pi i \phi \sigma_3}$ and
$\tilb=e^{2\pi i \psi \sigma_3}$ where $\sigma_3$ is one of the
Pauli's matrices. The corresponding holomorphic $PSL(2,\C)$-bundle is
obtained by identifying the points in $\C\times PSL(2,\C)$ in the
following way :
$(\zeta,g)\equiv(\zeta+1,\tila^{-1}g)\equiv(\zeta+\tau,\tilb^{-1}g)$. If
we introduce a frame $\sigma_u^{(0)}(z)=(\zeta, e^{-2\pi i\phi \zeta
\sigma_3})$ which is single valued along the closed loop $\zeta \to
\zeta+1$, this bundle denoted by $\Ph_u^{(0)}$ is described by the
transition rule : \beq \sigma_u^{(0)}(zq)=\sigma_u^{(0)}(z)e^{-2\pi i
u\sigma_3}\,, \qquad u=\psi-\tau \phi\,.
\label{ch4.transtrivu}
\eeq Recall that $\Ph_u^{(0)}$ and $\Ph_{u'}^{(0)}$ are isomorphic if
and only if $u'\equiv\pm u$ modulo $\nibun \Z+\frac{\tau}{2}\Z$.

In the similar way, the holomorphic $PSL(2,\C)$-bundle corresponding
to the unique flat $SO(3)$-bundle of non-trivial topology is described
by the transition rule \beq \sigma_F^{(1)}(zq)=\sigma_F^{(1)}(z)
\pmatrix{ 0 & q^{-\frac{1}{4}}z^{-\frac{1}{2}} \cr
-q^{\frac{1}{4}}z^{\nibun} & 0 \cr }\,,
\label{transflatnontriv}
\eeq and is denoted by $\Ph_F^{(1)}$.

There is a topologically trivial semi-stable $PSL(2,\C)$-bundle
$\Ph_{00}^{(0)}$ which does not come from a flat
$SO(3)$-connection. It is described by the transition rule \beq
\sigma_{00}^{(0)}(zq)=\sigma_{00}^{(0)}(z)\pmatrix{ 1 & 1 \cr 0 & 1
\cr }\,.  \eeq This is not isomorphic but is equivalent to
$\Ph_0^{(0)}$ ( $\Ph_{00}^{(0)}\sim \Ph_0^{(0)}$ ; see the footnote in
the section 3.1). In fact, $\{\Ph_u^{(0)}\}_u\cup \{\Ph_{00}^{(0)}\}$
is the set of all semi-stable $PSL(2,\C)$-bundles with trivial
topology.

Finally, there is a parametrized family $\{\Ph_u^{(1)}\}$ of
topologically non-trivial holomorphic $PSL(2,\C)$-bundles. The bundle
$\Ph_u^{(1)}$ is described by the transition rule \beq
\sigma_u^{(1)}(zq)=\sigma_u^{(1)}(z)e^{2\pi
i(u+\frac{1}{4})\sigma_3}z^{-\nibun \sigma_3}\,.
\label{transnonflnontrivu}
\eeq It is not even semi-stable. $\Ph_u^{(1)}$ and $\Ph_{u'}^{(1)}$
are isomorphic if and only if $u'\equiv u$ modulo
$\nibun\Z+\frac{\tau}{2}\Z$.  Of course there are many holomorphic
bundles of other {\it types}, though we do not list them up.

For every holomorphic $\HC$-bundle $\Ph$ described by the transition
rule $\sigma(zq)=\sigma(z)h_q(z)$, an automorphism of $\Ph$ is given
by a holomorphic map $h:\C^*\to \HC$ such that
$\sigma(zq)h(zq)=\sigma(z)h(z)h_q(z)$ or
$h(zq)=h_q(z)^{-1}h(z)h_q(z)$. We list below the group
$\Aut\Ph_{\star}^{(\epsilon)}$ of automorphisms of the bundle
$\Ph_{\star}^{(\epsilon)}$ given above. The typical elements of the
automorphism groups are represented with respect to the frames
$\sigma_{\star}^{(\epsilon)}$ : \beqa \Aut\Ph_u^{(0)}&\cong&\left\{
\begin{array}{ll} \C^*\qquad \pmatrix{ c & 0 \cr 0 & c^{-1} \cr
}\!,\,\,\,c\in \C^*,\quad &\mbox{if $u\equiv\!\!\!\!\!\!/\,\,\,\,
0,\frac{1}{4},\frac{\tau}{4},\frac{1+\tau}{4}$}\\
\noalign{\vskip0.2cm} PSL(2,\C)\qquad h\in PSL(2,\C),\quad &\mbox{if
$u\equiv 0$}\\ \noalign{\vskip0.2cm} \C^*\semidir \Z_2\quad \pmatrix{
c & 0 \cr 0 & c^{-1} \cr }\!,\,\,\pmatrix{ 0 & c \cr -c^{-1} & 0 \cr
}\!,\,\,c\in \C^*,\quad &\mbox{if $u\equiv \frac{1}{4}$}\\
\noalign{\vskip0.2cm} \C^*\semidir \Z_2\quad \pmatrix{ c & 0 \cr 0 &
c^{-1} \cr }\!,\,\,\pmatrix{ 0 & \!\!cz^{\nibun} \cr
-c^{-1}z^{-\nibun} & \!\!0 \cr }\!,\,\,\, c\in \C^*,\,\, &\mbox{if
$u\equiv \frac{\tau}{4},\,\frac{1+\tau}{4}$}
\end{array} \right.\\
\Aut\Ph_{00}^{(0)}&\cong&\C\qquad \pmatrix{ 1 & x \cr 0 & 1 \cr }\!,
\,\,\,\,x\in \C,\\
\Aut\Ph_F^{(1)}&\cong&\Z_2\!\times\Z_2=\left\{\pmatrix{ 1 & 0 \cr 0 &
1 \cr }\!,\,\,\pmatrix{ i & 0 \cr 0 & -i \cr }\!,\,\,\pmatrix{ 0 & i
\cr i & 0 \cr }\!,\,\,\pmatrix{ 0 & -1 \cr 1 & 0 \cr
}\right\},\label{symflatnontriv}\\ \Aut\Ph_u^{(1)}&\cong&B_0^-\qquad
\pmatrix{ c & \!\!0 \cr x\vartheta_{\!\tau,u}(z) & \!\!c^{-1} \cr
}\!,\,\,\, c\in \C^*,\,x\in \C\,, \eeqa where $\vartheta_{\!\tau,u}$
is the theta function given by
$\vartheta_{\!\tau,u}(z)=\vartheta(\tau,
\zeta+2u+\frac{1+\tau}{2})$.\footnote{$\vartheta(\tau,
\zeta)$\label{thetafcn} is the Riemann's theta function defined by
$\vartheta(\tau,\zeta)=\sum_{n\in \Z}q^{\nibun n^2}z^{-n}$ where
$q=e^{2\pi i\tau}$ and $ z=e^{-2\pi i \zeta}$.} Note that
$\vartheta_{\!\tau,u}(1)=0$ if and only if $u\equiv 0$.

We next consider the flag structures on these bundles
$\Ph_{\star}^{(\epsilon)}$ at the point $z=1$. A flag at $z=1$ is
identified with a ray in the vector space $\C^2$ by the following map
: \beq \left [\matrix{ x_1\cr x_2\cr }\right ]\in
\CP=\left(\C^2-\{0\}\right)\!\mbox{\Large /}\C^* \longmapsto
\sigma_{\star}^{(\epsilon)}(1)\!\pmatrix{ x_1 & x_3 \cr x_2 & x_4 \cr
}\!\!B\in
\left(\Ph_{\star}^{(\epsilon)}\right)_{\!\!z=1}\!\!\mbox{\Large /}B\,,
\eeq where the matrix in the right hand side is made unimodular by
choosing suitable numbers $x_3$ and $x_4$. It should be recalled that
$\Bigl(\Ph_{\star}^{(\epsilon)}, \left [\matrix{ x_1\cr x_2\cr }\right
]\Bigr)$ is isomorphic to
$\Bigl(\Ph_{\star}^{(\epsilon)},\left[h(1)\left (\matrix{ x_1\cr
x_2\cr }\right )\right]\Bigr)$ if $h(z)$ represents an automorphism
$\sigma_{\star}^{(\epsilon)}(z)\mapsto
\sigma_{\star}^{(\epsilon)}(z)h(z)$ of
$\Ph_{\star}^{(\epsilon)}$. Having these in mind we obtain the
following list of the flag structures at $z=1$ on the $PSL(2,\C)$
bundles $\Ph_{\star}^{(\epsilon)}$ :

\vspace{0.5cm}
\noindent On topologically trivial semi-stable bundles we have \beq
\begin{array}{ll}
\mbox{if
$u\equiv\!\!\!\!\!\!/\,\,\,\,0,\frac{1}{4},\frac{\tau}{4},\frac{\tau+1}{4}$,}\\
\noalign{\vskip0.2cm} \left\{\begin{array}{ll} \left(\Ph_u^{(0)},\left
[\matrix{ 1\cr 1\cr }\right ]\right)\!,\,\,\Aut=\{1\},\\
\noalign{\vskip0.2cm} \left(\Ph_u^{(0)},\left [\matrix{ 1\cr 0\cr
}\right ]\right)\!,\,\,\Aut=\C^*,\\ \noalign{\vskip0.2cm}
\left(\Ph_u^{(0)},\left [\matrix{ 0\cr 1\cr }\right
]\right)\!,\,\,\Aut=\C^*,
\end{array}\right.
\end{array}\,
\begin{array}{ll}
\noalign{\vskip-1.1cm} \mbox{if $u\equiv
\frac{1}{4},\frac{\tau}{4},\frac{1+\tau}{4}$,}\\ \noalign{\vskip0.2cm}
\left\{\begin{array}{ll} \left(\Ph_u^{(0)},\left [\matrix{ 1\cr 1\cr
}\right ]\right)\!,\,\,\Aut=\Z_2,\\ \noalign{\vskip0.2cm}
\left(\Ph_u^{(0)},\left [\matrix{ 1\cr 0\cr }\right
]\right)\!,\,\,\Aut=\C^*,
\end{array}\right.
\end{array}\,
\begin{array}{ll}
\noalign{\vskip0.7cm} \left\{\begin{array}{ll} \left(
\Ph_{00}^{(0)},\left [\matrix{ 0\cr 1\cr }\right
]\right)\!,\,\,\Aut=\{1\},\\ \noalign{\vskip0.2cm}
\left(\Ph_{00}^{(0)},\left [\matrix{ 1\cr 0\cr }\right
]\right)\!,\,\,\,\Aut=\C,
\end{array} \right. \\
\noalign{\vskip0.2cm} \,\,\,\,\,\,\,\left(\Ph_0^{(0)},\left [\matrix{
1\cr 0\cr }\right ]\right)\!,\,\, \Aut=B_0^+.
\end{array}
\label{listtriv}
\eeq On the topologically non-trivial semi-stable bundle coming from
the flat $SO(3)$-bundle, we have \beq \left(\Ph_F^{(1)},\left
[\matrix{ 1\cr y\cr }\right ]\right)\!,
\,\,\Aut=\left\{\begin{array}{ll} \{1\}\,\,&\mbox{if
$y\equiv\!\!\!\!\!\! /\,\,\,\, 0,1,i$,}\\ \noalign{\vskip0.2cm}
\Z_2\,\,&\mbox{if $y\equiv 0,1,i$,}
\end{array}\right.\,\,\,
\hspace{6cm}
\label{listflat}
\eeq where the flag structures $\left [\matrix{ 1\cr y\cr }\right ]$
and $\left [\matrix{ 1\cr y'\cr }\right ]$ are isomorphic to each
other if and only if $y'\equiv y$, that is, $y'$ coincides with one of
$y,-y,y^{-1}$ or $-y^{-1}$.

\noindent On the one parameter family of topologically non-trivial non
semi-stable bundles, we have \beq
\begin{array}{ll}
\noalign{\vskip-0.4cm} \mbox{if $u\equiv\!\!\!\!\!\! /\,\,\,\,0$,}\\
\noalign{\vskip0.3cm} \left\{\begin{array}{ll} \left(\Ph_u^{(1)},\left
[\matrix{ 1\cr 0\cr }\right ]\right)\!, \,\,\Aut=\C^*,\\
\noalign{\vskip0.2cm} \left(\Ph_u^{(1)},\left [\matrix{ 0\cr 1\cr
}\right ]\right)\!,\,\,\Aut=B_0^-,
\end{array}\right.
\end{array}\,
\mbox{and if $u\equiv 0$,}\,\,\left\{\begin{array}{ll}
\left(\Ph_0^{(1)},\left [\matrix{ 1\cr 1\cr }\right ]\right)\!,\,\,
\Aut=\C,\\ \noalign{\vskip0.2cm} \left(\Ph_0^{(1)},\left [\matrix{
1\cr 0\cr }\right ]\right)\!,\,\,\Aut=B_0^-,\\ \noalign{\vskip0.2cm}
\left(\Ph_0^{(1)},\left [\matrix{ 0\cr 1\cr }\right
]\right)\!,\,\,\Aut=B_0^-.
\end{array}\right.\hspace{5cm}
\label{listnontriv})
\eeq In the above expressions, each ``$\Aut$'' denotes the
automorphism group of the corresonding holomorphic bundle with flag
structure.

Recall that the moduli space $\NNc_{{\rm triv}}$ for the topologically
trivial bundle is represented by the family
$\{\Ph_u^{(0)}\}_{u\equiv\!\!\!\!/\,\,\,
0,\frac{1}{4},\frac{\tau}{4},\frac{1+\tau}{4}}$ of $PSL(2,\C)$-bundles
and the moduli space $\NNc_{{\rm non-triv}}$ for the topologically
non-trivial bundle is one point $\{\Ph_F^{(1)}\}$. By looking at the
dimensions of the symmetry group of the flag structures listed above,
we see that \beqa \NNc_{{\rm triv},x}&=&\left\{\left(\Ph_u^{(0)},\left
[\matrix{ 1\cr 1\cr }\right ]\right)\right\}_{u\equiv\!\!\!\!
/\,\,0,\frac{1}{4},\frac{\tau}{4},\frac{\tau+1}{4}}\!\!\!\!\cong\C\mbox{\LARGE
/}\!\Bigl(\mbox{$\nibun\Z+\frac{\tau}{2}\Z$}\Bigr)\semidir \Z_2-\{
\mbox{4-points} \},\\ \NNc_{{\rm non-triv},
x}&=&\left\{\left(\Ph_F^{(1)},\left [\matrix{ 1\cr y\cr }\right
]\right)\right\}_{ y\in \C}\cong \left(\Z_2\times \Z_2\right)
\!\mbox{\LARGE $\backslash$} \CP\,.
\label{trivnontriv}
\eeqa If $\NNc_{{\rm triv},x}$ is compactified by attaching the points
$\left(\Ph_u^{(0)},\left [\matrix{ 1\cr 1\cr }\right ]\right)$,
$u=\frac{1}{4},\frac{\tau}{4},\frac{\tau+1}{4}$ and
$\left(\Ph_{00}^{(0)},\left [\matrix{ 0\cr 1\cr }\right ]\right)$,
then we see that the compactified moduli space $\overline{\NNc_{{\rm
triv},x}}$ coinsides topologically with the moduli space $\NNc_{{\rm
non-triv},x} \cong S^2$. Moreover, it seems that the families of
automorphism groups coincide with each other : Generically there is no
non-trivial symmetry, but there are three points with $\Aut\cong
\Z_2$. In the next section, we shall see that this is not an accident
by constructing a natural bijection between $\overline{\NNc_{{\rm
triv},x}}$ and $\NNc_{{\rm non-triv},x}$. In fact, this is an
essential step to observe the field identifications.

\renewcommand{\theequation}{4.4.\arabic{equation}}\setcounter{equation}{0}
\vspace{0.4cm}
\begin{center}
{\sc 4.4 The Hecke Correspondence}\label{4.4}
\end{center}
\hspace{1.5cm} Let us remind ourselves of the action of the
fundamental group $\pi_1(H)$ on the set of isomorphism classes of
topological $H$-bundles over a surface $\Sigma$.

Let $P$ be a principal $H$-bundle over $\Sigma$. We take a
trivialization $s_0$ of $P$ on a disc $D_0$ in $\Sigma$ and denote by
$s$ the trivialization restricted to the boundary $S=\partial
D_0$. Let us take a closed loop $\gamma :S\to H$. Deleting from $P$
the restriction $P|_{D_0^{\circ}}$ over the interior $D_0^{\circ}$ of
the disc $D_0$, we attach the trivial bundle $D_0\times H$ to the rest
$P-P|_{D_0^{\circ}}$ by the identification \beq
s_0'(z)\longleftrightarrow s(z)\gamma(z) \qquad z\in S, \eeq where
$s_0'$ is a trivialization of $D_0\times H$. The topological type of
the resulting principal $H$-bundle $P\gamma$ depends only on the
topological type of $P$ and the homotopy type of $\gamma$. This
operation determines the action of $\pi_1(H)$, since the set of
homotopy types of maps $S\to H$ is identified with the group
$\pi_1(H)$ under an arbitrary orientation preserving parametrization
$[0,1]\to S$. In view of the fact that $P-P|_{D_0^{\circ}}$ is
trivializable, it is easy to see that this action is free and
transitive.

As a last step to derive the field identification, we show that this
action lifts to an action on the set of isomorphism classes of
holomorphic $\HC$-bundles with quasi-flag structure at one
point. Namely, for topologically distinct $H$-bundles $P$ and $P'$, we
find a way to identify the sets $\A_P\!\times_{\GPC}\!Fl(P_x)$ and
$\A_{P'}\!\times_{\G_{P'_{\!\bf c}}}\!Fl(P'_x)$. We also argue that,
under this identification, the moduli spaces $\NNc_{P,x}$ and
$\NNc_{P',x}$ have a chance to correspond essentially to each
other. This is presened as a conjecture and is verified in several
examples.

\vspace{0.3cm} \underline{The Lift in the Abelian Case}

We choose and fix a neighborhood $U_0$ of a point $x$ in $\Sigma$
provided with a complex coordinate $z$ such that $z(x)=0$.

First, we recall the situation in the abelian case $H=U(1)$,
$\pi_1H=\Z$. In this case, we can find a lift of the action of
$\pi_1(H)$ to an action on the set of isomorphism classes of
holomorphic $\C^*$-bundles without reference to any additional
structure such as {\it flag}. For each $a\in \Z$, `tensoring by ${\cal
O}(x)^a$' induces an isomorphism of $\Pic_{c_1}\Sigma$ onto
$\Pic_{c_1+a}\Sigma$, where $\Pic_{c_1}\Sigma$ is the moduli space of
holomorphic $\C^*$-bundles over $\Sigma$ of winding number $c_1\in
\Z$. It is given in the following way.

Let $\Ph$ be a holomorphic $\C^*$-bundle of winding number $c_1$. We
take a (holomorphic) trivialization $\sigma_0$ of $\Ph|_{U_0}$ and
denote by $\sigma$ the restriction of $\sigma_0$ to the open set
$U_0-\{x\}$. By a surgery of $\Ph$, we construct another $\C^*$-bundle
$\Ph'$ of winding nimber $c_1+a$. We delete from $\Ph$ the fibre
$\Ph_x$ over $x$ and attach the trivial bundle $U_0\times \C^*$ to the
rest $\Ph-\Ph_x$ by the following identification : \beq
\sigma_0'(z)\longleftrightarrow \sigma(z)z^{-a}\qquad z\in U_0-\{x\},
\eeq where $\sigma_0'$ is a trivialization of $U_0\times \C^*$. Then,
the resulting space $\Ph'$ can be given a structure of holomorphic
$\C^*$-bundle. The isomorphism class of the bundle $\Ph'$ does not
depend on the choice of the trivialization $\sigma_0$ used in the
construction nor on the choice of the coordinate $z:U_0\to
\C$. Moreover, two isomorphic bundles $\Ph_1\cong \Ph_2$ are mapped to
two isomorphic bundles $\Ph_1'\cong\Ph_2'$. Since any holomorphic
family $\{\Ph_t\}$ is obviously mapped to another holomorphic family
$\{\Ph'_t\}$, we have obtained an isomorphism $\Pic_{c_1}\Sigma \to
\Pic_{c_1+a}\Sigma$ giving rise to the action of $\pi_1U(1)\cong \Z$
on $\Pic\Sigma$.

\vspace{0.3cm} \underline{The Lift for Simple Groups}

Now we consider the case in which $H$ is simple and centerless. We use
the notations introduced in the Appendix 2.

We choose and fix a maximal torus $T$ of $H$ and a chambre
$\Ch$. Recall that the fundamental group $\pi_1(H)$ is isomorphic to
the center $\Pv/\Qv$ of the universal covering group $\tilH$. The
group $\Pv/\Qv$ is in turn isomorphic to the subgroup $\Gmalcv$ of the
affine Weyl group $\Waffh$ of $LH$ consisting of elements that
preserve the alc\^ove $\alcv$. If an element $\gamma$ of $\Gmalcv$ is
represented by a loop $e^{i\theta}\mapsto\gamma(e^{i\theta})$, the
holomorphic extention $\hgmm : U_0-\{x\}\to \HC$\label{page:hgmm}
satisfies the following : If $b:U_0\to \HC$ is a holomorphic map such
that $b(x)\in B=B_0^+$, the map $\hgmm b\hgmm^{-1}:U_0-\{x\}\to \HC$
extends to a holomorphic map from $U_0$ to $\HC$ whose value at $x$
belongs to $B$. It may be recalled that $\hgmm$ is expressed as
$\hgmm(z)=z^{-\mu_j}n_{w_jw_0}$ for some $j\in {\cal J}_0$ where
$n_{w_jw_0}\in N_T$ represents the element $w_jw_0\in W$.

For a holomorphic $\HC$-bundle $\Ph$ over $\Sigma$ with flag structure
$f$ at $x$, a trivialization $\sigma_0$ of $\Ph$ over a neighborhood
of $x$ is said to be {\it admissible with resect to $f$} or simply
{\it admissible} when the value $\sigma_0(x)$ at $x$ represents the
flag $f$, that is, when $f=\sigma_0(x)B$.

Let $\gamma$ be an element of $\Gmalcv$ and let $\hgmm(z)$ be the
holomorphic extention of a representative loop of $\gamma$. A choice
of admissible trivialization $\sigma_0$ of $(\Ph,f)$ over the
coordinate neighborhood $U_0$ determines the following surgery of
$\Ph$ which gives another holomorphic $\HC$-bundle over $\Sigma$ with
flag structure at $x$. We delete the fibre $\Ph_x$ over $x$ from $\Ph$
and attach the trivial bundle $U_0\times\HC$ to the rest $\Ph-\Ph_x$
by the identification \beq \sigma_0'(z)\longleftrightarrow
\sigma(z)\hgmm(z)\qquad z\in U_0-\{x\}\,, \eeq where $\sigma_0'$ is a
trivialization of $U_0\times \HC$ and $\sigma$ denotes the restriction
of $\sigma_0$ to the open set $U_0-\{x\}$. Then, the resulting space
can be given a structure of a holomorphic $\HC$-bundle denoted by
$\Ph'$. We can also give $\Ph'$ a flag structure $f'$ at $x$ by saying
that the trivialization $\sigma_0'$ is admissible. We call this
operation the {\it $\gamma$-surgery with respect to $\sigma_0$} or
simply {\it $\gamma$-surgery}. Another choice of admissible
trivialization $\tilde{\sigma}_0$ determines another $\gamma$-surgery
of $\Ph$ which gives a bundle $(\tilde{\Ph}',\tilde{f}')$ isomorphic
to $(\Ph',f')$: If $\tilde{\sigma}_0$ is related to $\sigma_0$ by
$\tilde{\sigma}_0=\sigma_0b$ through a holomorphic map $b:U_0\to \HC$
with $b(x)\in B$, an isomorphism $(\tilde{\Ph}',\tilde{f}')\to
(\Ph',f')$ is defined by $\tilde{\sigma}'_0\to
\sigma_0'\hgmm^{-1}b\hgmm$ over $U_0$ and $id: \Ph|_{\Sigma-\{x\}}\to
\Ph|_{\Sigma-\{x\}}$ over $\Sigma-\{x\}$. Moreover, $\gamma$-surgeries
of isomorphic bundles with isomorphic flag srtucture
$(\Ph_1,f_1)\cong(\Ph_2,f_2)$ lead also to isomorphic bundles with
isomorphic flag structure $(\Ph_1',f_1')\cong (\Ph_2,'f_2')$ : If an
isomorphism $h_{21}:(\Ph_1,f_1)\to (\Ph_2,f_2)$ is represented over
$U_0$ by $\sigma_{{}_{\!1}0}\to\sigma_{{}_{\!2}0}(h_{21})_0$ with
respect to admissible trivializations, $(h_{21})_0:U_0\to \HC$ is a
holomorphic map with $(h_{21})_0(x)\in B$. Therefore, an isomorphism
$h_{21}' : (\Ph'_1,f'_1)\to (\Ph'_2,f'_2)$ is defined by the map
$\sigma'_{{}_{\!1}0}\to \sigma'_{{}_{\!2}0}\hgmm^{-1}(h_{21})_0\hgmm$
over $U_0$ and by $h_{21}|_{\Sigma-\{x\}}$ over $\Sigma-\{x\}$. Thus,
we see that the $\gamma$-surgery induces the transformation
$\{(\Ph,f)\} \longto
\{(\Ph',f')\}=\{(\Ph,f)\}\gamma_x$\label{page:gammax} of the set of
isomorphism classes of holomorphic $\HC$-bundles with flag structure
at $x$. It is easy to see that this transformation is uniquely
determined by $\gamma$ and $x$.

Since these transformations $\{\gamma_x\,;\,\gamma\in \Gmalcv\}$
apparently preserve the composition law of the group $\Gmalcv$, they
determine an action of $\Gmalcv$ on the set of holomorphic
$\HC$-bundles with flag structure at $x$. As is obvious by the
construction, this is a lift of the action of $\pi_1(H)\cong \Gmalcv$
on the set of isomorphism classes of topological $H$-bundles. In other
words, $\gamma_x$ maps the set $\A_P\!\times_{\GPC}\!\!Fl(P_x)$
bijectively onto the set $\A_{P\gamma}\!\times_{\G_{P\gamma_{\!\bf
c}}}\!\!Fl((P\gamma)_x)$.

\vspace{0.3cm} \underline{The Conjecture}

One important property of the map $\gamma_x$ is that it preserves the
automorphism groups. Namely, if the bundle $(\Ph',f')$ is obtained by
the $\gamma$-surgery of the bundle $(\Ph,f)$, $\Aut(\Ph,f)$ is
isomorphic to $\Aut(\Ph',f')$. This can be seen by putting
$(\Ph_1,f_1)=(\Ph_2,f_2)$ in the preceding argument on the
isomorphisms.

Recall that an element of the moduli space $\NNc_{P,x}$ is identified
with an isomorphism class represented by a holomorphic $\HC$-bundle
$\Ph$ with flag structure $f$ at $x$ that satisfies the following
conditions : $\Ph$ represents a class identified with an element of
$\NNc_P$ and $\dim\Aut(\Ph,f)\leq \dim\Aut(\Ph,\tilde{f})$ for every
flag $\tilde{f}$ at $x$.

Having these in mind, we make the following conjecture : {\it There is
a method to compactify the moduli space $\NNc_{P,x}$ by attaching
suitable points of $\A_P\!\times_{\GPC}\!\!Fl(P_x)$ in such a way that
for each $\gamma\in \Gmalcv$, the compactified moduli space
$\overline{\NNc_{P,x}}$ is mapped isomorphically onto another space
$\overline{\NNc_{P\gamma,x}}$ by $\gamma_x$.} If this holds true, we
have the following double fibration \beq
\begin{array}{ccccc}
&&\!\!\!\!\!\!\!\overline{\NNc_{P,x}}\cong\overline{\NNc_{P\gamma,x}}\!\!\!\!\!\!\!\!\!&&\\
\noalign{\vskip0.2cm} &\mbox{\Large $\swarrow$}&&\mbox{\Large
$\searrow$}&\\ \noalign{\vskip0.2cm}
\overline{\NNc_P}\!\!\!&&&&\!\!\!\overline{\NNc_{P\gamma}}
\end{array}
\eeq where the projections correspond to `forgetting the flags'. This
seems to be what mathematicians call the {\it Hecke
correspondence}. \cite{Narasimhan-Ramanan}

\vspace{0.3cm} \underline{Examples on the Sphere}

We consider the case in which $H$ is simple. Recall that the set of
isomorphism classes of the holomorphic $\HC$-bundles over complex
projective line $\CP$ with flag structure at $z=0$ is indexed by the
lattice $\Pv$. The bundle with flag structure $\Ph_a$ indexed by $a\in
\Pv$ is described by the transition rule
$\sigma^{(a)}_0(z)=\sigma^{(a)}_{\infty}(z)z^{-a}$ where
$\sigma^{(a)}_0$ is an admissible trivialization over the
$z$-plane. The bundle $\Ph_a'$ obtained by the
$\gamma_j$-surgery\footnote{We denote by $\gamma_j$ the element of
$\Gmalcv$ represented by the loop
$\gamma_j(\theta)=e^{-i\mu_j\theta}n_{w_jw_0}$.} of $\Ph_a$ with
respect to $\sigma^{(a)}_0$ is then described by the transition
relation
${\sigma_0^{(a)}}'={\sigma_{\infty}^{(a)}}'z^{-a}z^{-\mu_j}n_{w_jw_0}$
of an admissible trivialization ${\sigma_0^{(a)}}'$ and a
trivialization ${\sigma_{\infty}^{(a)}}'$ over $\CP-\{0\}$. This shows
that the action of $\Gmalcv$ on the set of isomorphism classes is
represented by the action \beq (a,\gamma_j)\mapsto
(w_jw_0)^{-1}(a+\mu_j)\,, \eeq on the indexing set $\Pv$.

As we have seen in the preceding section, the moduli space
$\NNc_{P^{(j)},x}$ is one point. The unique element is represented by
the bundle $\Ph_j=\Ph_{(w_jw_0)^{-1}\mu_j}$ that is described by the
transition relation
$\sigma_0(z)=\sigma_{\infty}(z)z^{-\mu_j}n_{w_jw_0}$ of an admissible
section $\sigma_0$ on the $z$-plane and a section $\sigma_{\infty}$ on
the $w$-plane. Hence we see that the $\gamma_j$-surgery induces the
following map : \beq
\NNc_{P^{(i)},x}\stackrel{\gamma_{j,x}}{\longto}\NNc_{P^{(i')},x}\,,
\eeq where $\gamma_{i'}=\gamma_i\gamma_j$. Thus, the conjecture holds
on the sphere.

\vspace{0.3cm} \underline{Example with $\Sigma=$ Torus and $H=SO(3)$}

For $H=SO(3)$, we know that there is only one non-trivial element
$\gamma_1$ in $\Gmalcv\cong\Z_2$ which is represented by a path \beq
\gamma_1(\theta)=e^{-i\theta\nibun\sigma_3}n\,\qquad;\,\,\,n=\pmatrix{
0 & -1 \cr 1 & 0 \cr }, \eeq in $SU(2)$. We apply the
$\gamma_1$-surgery to the topologically trivial semi-stable bundles
with flag structure at $z=1$ that are listed in (\ref{listtriv}) and
we argue whether the conjecture holds true.

A $PSL(2,\C)$-bundle $\Ph^{(0)}$ in the list (\ref{listtriv}) is
described by the transition rule of the form
$\sigma(zq)=\sigma(z)h_q(z)$, and a flag is realized by a ray $\left
[\matrix{ x_1\cr x_2\cr }\right ]$ in $\C^2$. If we choose a matrix
$h_f\in SL(2,\C)$ such that $(h_f)^{\!1}_1=x_1$ and
$(h_f)^{\!2}_1=x_2$, then, $\sigma_0(z)=\sigma(z)h_f$ is an admissible
trivialization on a neighborhood $U_0$ of $z=1$. Hence, the
$\gamma_1$-surgery of $\Ph^{(0)}$ with respect to $\sigma_0$ leads to
a bundle $\Ph^{(1)}$ with an admissible trivialization $\sigma_0'$ on
$U_0$ and a trivialization $\sigma$ on $\C^*- q^{\Z}$ that are related
by \beqa \sigma_0'(z)&=&\sigma(z)h_f(z-1)^{-\nibun\sigma_3}n\,, \qquad
z\in U_0-\{z=1\}\,,\\ \sigma(zq)&=&\sigma(z)h_q(z)\,,\qquad z\notin
q^{\Z}\,.  \eeqa

If we could find $PSL(2,\C)$-valued functions $\tilde{\chi}$ on
$\C^*-q^{\Z}$ and $h'_q$ on $\C^*$ such that \beq
\left\{\begin{array}{l}
\tilde{\chi}(zq)=h_q(z)^{-1}\tilde{\chi}(z)h'_q(z)\\
\noalign{\vskip0.2cm}
\chi(z)=n^{-1}(z-1)^{\nibun\sigma_3}h_f^{-1}\tilde{\chi}(z)\quad\mbox{is
regular as $z\to 1$},
\end{array}\right.
\label{chitilde}
\eeq then, $\sigma'$ given by $\sigma'(z)=\sigma(z)\tilde{\chi}(z)$
determines a multivalued section of $\Ph^{(1)}$ satisfying
$\sigma'(zq)=\sigma'(z)h_q'(z)$. The flag at $z=1$ is given by
$f'=\sigma'(1)\chi(1)^{-1}\!B$.

The conservation $\Aut(\Ph^{(0)},f)\cong \Aut(\Ph^{(1)},f')$ of
symmetry groups under $\gamma_1$-surgery enables us to guess the form
of $h'_q(z)$ by looking at the lists (\ref{listflat}) and
(\ref{listnontriv}) of topologically non-trivial holomorphic bundles
with flag structure at $z=1$. After a calculation, we find the
solution listed below. Each arrow $\to$ signifies the
$\gamma_1$-surgery and $\tilde{\chi}$ indicates how the regular
multivalued section $\sigma'$ is related to the old singular
(multivalued) section $\sigma$ :

\vspace{0.2cm} \beq \hspace{-0.25cm}\begin{array}{rcl} \mbox{ If
$u\equiv\!\!\!\!\!\!/\,\,\,\, 0$}\hspace{0.5cm}&&\\
\noalign{\vskip0.1cm} \left(\Ph_u^{(0)},\left [\matrix{ 1\cr 1\cr
}\right ]\right)&\to&\left(\Ph_F^{(1)},\left [\matrix{ 1\cr y_u\cr
}\right ]\right)\qquad y_u=iq^{\frac{1}{4}}e^{2\pi i
u}\displaystyle{\frac{\vartheta(2\tau, 2u+\tau)}{\vartheta(2\tau,
2u)}}\,,\\ \noalign{\vskip0.3cm} \tilde{\chi}(z)&=&\pmatrix{
\tilde{R}_u(1)R_u(z) & ie^{-2\pi i
u}q^{-\frac{1}{4}}\tilde{R}_{-u}(1)R_{u-\frac{\tau}{2}}(z) \cr
-\tilde{R}_u(1)R_{-u}(z) & -ie^{-2\pi i
u}q^{-\frac{1}{4}}\tilde{R}_u(1)R_{-u-\frac{\tau}{2}}(z) \cr } \\
\noalign{\vskip0.25cm} &;&R_u(z)=\displaystyle{\frac{\vartheta(2\tau,
\zeta +2u+\tau)}{\left(\vartheta(\tau, \zeta
+\frac{\tau+1}{2})\right)^{\nibun}}}\qquad
\tilde{R}_u(z)=(z-1)^{\nibun}R_u(z)\,.
\end{array}
\label{heckeg1so3}
\eeq \beq \hspace{0.25cm}\begin{array}{rcl} \left(\Ph_{00}^{(0)},\left
[\matrix{ 0\cr 1\cr }\right ]\right)&\to&\left(\Ph_F^{(1)},\left
[\matrix{ 1\cr y_0\cr }\right ]\right)\\ \noalign{\vskip0.3cm}
\tilde{\chi}(z)=&&\hspace{-0.7cm}\pmatrix{ R_0(z)F(z) &\!\!
iq^{-\frac{1}{4}}R_{-\frac{\tau}{2}}(z)G(z) \cr R_0(z) &
\!\!iq^{-\frac{1}{4}}R_{-\frac{\tau}{2}}(z) \cr }\,\,
\begin{array}{l}
F(z)=2z\frac{\partial}{\partial
z}\log\vartheta(2\tau,\zeta\!+\!\tau)-1\\ \noalign{\vskip0.3cm}
G(z)=2z\frac{\partial}{\partial z}\log\vartheta(2\tau,\zeta).
\end{array}
\end{array}
\eeq \beq \hspace{-0.2cm}\begin{array}{rcl} \left(\Ph_u^{(0)},\left
[\matrix{ 1\cr 0\cr }\right ]\right)&\to&\left(\Ph_u^{(1)},\left
[\matrix{ 1\cr 0\cr }\right ]\right)\qquad
\tilde{\chi}(z)=n\left(\vartheta(\tau,\zeta+\frac{\tau+1}{2})\right)^{\nibun\sigma_3}.\\
\noalign{\vskip0.3cm} \left(\Ph_u^{(0)},\left [\matrix{ 0\cr 1\cr
}\right ]\right)&\to &\left(\Ph_{-u}^{(1)},\left [\matrix{ 1\cr 0\cr
}\right ]\right)\qquad
\tilde{\chi}(z)=\left(\vartheta(\tau,\zeta+\frac{\tau+1}{2})\right)^{\nibun\sigma_3}.\\
\noalign{\vskip0.4cm} \left(\Ph_{00}^{(0)},\left [\matrix{ 1\cr 0\cr
}\right ]\right)&\to&\left(\Ph_0^{(1)},\left [\matrix{ 1\cr 1\cr
}\right ]\right)\\ \noalign{\vskip0.1cm} \tilde{\chi}(z)&=&\pmatrix{
c(z)H(z) & -c(z)^{-1} \cr c(z) & 0 \cr }\,\,
\begin{array}{l}
c(z)=\left(\vartheta(\tau,\zeta+\frac{\tau+1}{2})\right)^{\nibun}\\
\noalign{\vskip0.2cm} H(z)=z\frac{\partial}{\partial
z}\log\vartheta(\tau,\zeta+\frac{\tau+1}{2}).
\end{array}
\end{array}
\eeq

\vspace{0.2cm}
\noindent The most important point to notice is that the compactified
moduli space $\overline{\NNc_{{\rm triv},x}}$ whose elements are
represented by $\left\{\left(\Ph_u^{(0)},\left [\matrix{ 1\cr 1\cr
}\right ]\right)\right\}_{\!u\equiv\!\!\!\!/\,\, 0}\!\!\!$ and
$\left(\Ph_{00}^{(0)},\left [\matrix{ 0\cr 1\cr }\right ]\right)$ is
mapped bijectively to the (compact) moduli space $\NNc_{{\rm
non-triv},x}$ whose elements are represented by the flag structures
$\left\{\,\left [\matrix{ 1\cr y\cr }\right ]\,\right\}_{\!y}$ on the
semi-stable bundle $\Ph_F^{(1)}$ : \beq \overline{\NNc_{{\rm
triv},x}}\stackrel{\gamma_{1,x}}{\longto}\NNc_{{\rm non-triv},x}\,.
\eeq In terms of the coordinates $u$ and $y$, this map is given by
$u\mapsto y_u$ where $y_u$ is given in (\ref{heckeg1so3}) and
satisfies $y_{-u}=y_{u}$, $y_{u+\frac{1}{2}}=-y_u$ and
$y_{u+\frac{\tau}{2}}=-y_u^{-1}$. Note that the orbifold points
$u\equiv \frac{1}{4},\frac{\tau}{4},\frac{\tau+1}{4}$ are mapped to
the orbifold points $y_{\frac{1}{4}}=0$, $y_{\frac{\tau}{4}}=i$ and
$y_{\frac{\tau+1}{4}}=1$. It should be noted also that the
compactification divisor of $\overline{\NNc_{{\rm triv},x}}$
represented by $\left(\Ph_{00}^{(0)},\left [\matrix{ 0\cr 1\cr }\right
]\right)$ is mapped to the smooth point $y_0$. If we decide to take
$u^2$ as the complex coordinate around that point, the above bijection
becomes an isomorphism and the conjecture holds also in this case.

It may be remarked that the points $\left\{\left(\Ph_u^{(0)},\left
[\matrix{ 1\cr 0\cr }\right ]\right)\right\}$ which do not lie in
$\overline{\NNc_{{\rm triv},x}}$ are mapped to the points
$\left\{\left(\Ph_u^{(1)},\left [\matrix{ 1\cr 0\cr }\right
]\right)\right\}$ that are not projected to the semi-stable point by
the `flag forgetful map' though the former are projected to the
semi-stable points $\left\{\Ph_u^{(0)}\right\}$.

\renewcommand{\theequation}{4.5.\arabic{equation}}\setcounter{equation}{0}
\vspace{0.4cm}
\begin{center}
{\sc 4.5 Field Identification}\label{4.5}
\end{center}
\hspace{1.5cm} We are now in a position to combine all the results
obtained in the previous arguments. We first construct an action of
the fundamental group $\pi_1(H)$ on the set of gauge invariant fields
and then, under some assumptions, observe the relation (\ref{FI})
which leads to the field identification.

\vspace{0.3cm} \underline{The Spectral Flow for the Total System and
  the Transformation of the Flag Partners}

The space of states for the total system is the tensor product of the
three spaces --- the space ${\cal H}^M$ of states for the matter field
theory $M$, the space $\Omega^0(L\HC/LH,{\cal L})$ for the $\HC/H$-WZW
model and the space ${\cal F}^{\rm gh}\ot\bar {\cal F}^{\rm gh}$ for
the adjoint ghost system. On each of these spaces there is a
representation of the central extension $\tilde{L\tilH}_{\!\C}$ of the
loop group $L\tilH_{\!\C}$, $\tilde{\gamma}\mapsto {\cal
J}(\tilde{\gamma})=J(\tilde{\gamma})\bar J(\tilde{\gamma})$, where
$\tilH$ is a certain covering group of $H$. Since the sum
$(kr_G^H+r_V^H)+(-\kh)+2\lieh^{\vee}$ of the levels vanishes, the
representation on the total space ${\cal H}^{\rm tot}={\cal
H}^M\ot\Omega^0({\cal L})\ot {\cal F}^{\rm gh}\ot\bar {\cal F}^{\rm
gh}$\label{page:Htot} descends to a representation of the loop group
$L\tilH_{\!\C}$ or of $LH_{\!\C}$.

We now recall the results on the spectral flow obtained in the
Chapters 1, 2 and 3 and consider the spectral flow on the total space
${\cal H}^{\tot}$. Let the loop $\gamma(\theta)=e^{-ia\theta}n$ in $H$
represent an element of the affine Weyl group $\Waffh$ of $LH$, where
$a\in\Pv$ and $n\in N_T$. We denote by $A_{\gamma}$\label{page:Agamma}
the basic gauge field $A_{n^{-1}a n}$ of the $H$-bundle $P_0$ over the
unit disc $D_0$ with a trivialization $s_0:D_0\to P_0$. Then, we can
find a horizontal frame $s$ defined on a neighborhood of the boundary
circle $S=\partial D_0$ such that
$s_0(\theta)=s(\theta)\gamma(\theta)$ on $S$. As we have seen in the
previous chapters, the screening by $A_{\gamma}$ of a local field $O$
inserted at $z=0$ corresponds to the spectral flow $\tilh_{\gamma}$ on
the space ${\cal H}^{\bf I}$ of states : \beq Z_{D_0}^{{\bf
I}(s)}(A_{\gamma}\,;\,O(0))=\tilh_{\gamma}Z_{D_0}^{{\bf
I}(s_0)}(0\,;\;O(0))\,, \eeq where ${\bf I}$ denotes one of the three
sectors --- matter $M$, $\HC/H$-WZW model or ghost system. If
$\sigma_0$ is a holomorphic trivialization of $(P_{0{\!\bf C}},
\bartial_{\!A_{\gamma}})$ such that $\sigma_0(0)=s_0(0)$, it is
related to the horizontal frame $s$ by a holomorphic trivialization
$\hgmm$, $\sigma_0(z)=s(z)\hgmm(z)$. This $\hgmm$ is a holomorphic
extension of the loop $\gamma$ up to multiplication by elements of
$T_{\!\C}$. Then, the spectral flow $\tilh_{\gamma}$ is determined by
this function $\hgmm$ up to a constant factor depending on the
behavior of $A_{\gamma}$ on the interior of $D_0$. If ${\bf I}$
denotes the total system however, since the level is zero, or since
the chiral anomally is absent, $\tilh_{\gamma}$ is uniquely determined
by the transition function $\hgmm$. In fact, it can be considered to
be the action ${\cal J}(\hgmm|_S)$ of the loop $\hgmm|_S\in
LH_{\!\C}$. Hence we may denote just by $\hgmm$\label{page2:hgmm} the
spectral flow $\tilh_{\gamma}$ on the total space ${\cal H}^{\tot}$.

Next, we see that a certain kind of spectral flow induces a
transformation of the subspace of ${\cal H}^{\tot}$ consisting of the
states corresponding to the flag partners of dressed gauge invariant
fields. For simplicity, we take $H$ to be simple. Let the loop
$\gamma(\theta)=e^{-i\mu_j\theta}n_{w_jw_0}$ represent an element of
the subgroup $\Gmalcv$ of the affine Weyl group $\Waffh$. The spectral
flow $\tilh_{\gamma}$ acting on each sector ${\cal H}^{\bf I}$
satisfies the following.

Since $\Gmalcv$ is the subgroup of $\Waffh$ consisting of elements
that preserve the alc\^ove $\alcv$, the spectral flow $\tilh_{\gamma}$
acting on ${\cal H}^M$ preserves the subspace
$\displaystyle{\bigoplus_{\Lmd,\lmd}}L_{\Lmd,\lmd}\ot
\overline{L_{\Lmd,\lmd}}$ consisting of highest weight vectors with
respect to the loop algebra $\tilde{L\lieh}_{\!\C}\oplus
\tilde{L\lieh}_{\!\C}$. Due to the same reason, $\tilh_{\gamma}$
acting on the ghost Fock space preserves the ray generated by
\label{page2:volFLH}$|\Omega\rangle=\displaystyle{\prod_{-\alpha<0}}c_0^{-\alpha}\bar
c_0^{-\alpha}|0\rangle_{\rm gh}$ which is characterized by
$b_n(v)|\Omega\rangle =0$ for $n\geq 1$, $v\in \h$ and
$b_0(v)|\Omega\rangle=0$ for $v\in {\rm Lie}(B)$. And finally, as we
have seen in Chapter 3, the state $|-\lmd-2\rho\rangle$ corresponding
to the field $\left| e^{\lmd+2\rho}(b(h))\right|^{\!2}$ is mapped by
$\tilh_{\gamma}$ to another state $|-\lmd'-2\rho\rangle$ with
$\lmd'=w_jw_0\lmd+(\kh-2\lieh^{\vee})\ttr\mu_j$.

The above three properties show that the spectral flow $\hgmm$ on
${\cal H}^{\tot}$ maps the state $\Phi_{\lmd}\ot|-\lmd-2\rho\rangle\ot
|\Omega\rangle$ corresponding to the flag partner $O_{\lmd}(s_0(0)B)$
to another state $\Phi'_{\lmd'}\ot |-\lmd'-2\rho\rangle\ot
|\Omega\rangle$ which also corresponds to another flag partner. We
should note that the ambiguity in determining the function $\hgmm$ is
just the ambiguity of constant multiplication $\hgmm\to \hgmm
c_{\varrho}^a$ by elements of $T_{\!\C}$. Since the state
corresponding to a flag partner has weight zero, the resulting state
is uniquely determined by the class $\gamma\in \Gmalcv$. The new flag
partner can now be denoted by $(\gamma O_{\lmd})(s_0(0)B)$.

This map $\gamma : O_{\lmd}\mapsto \gamma
O_{\lmd}$\label{page:gammaOlmd} describes the action of $\pi_1(H)\cong
\Gmalcv$ on the set of flag partners and hence on the set of gauge
invariant local fields.

\vspace{0.3cm} \underline{The Field Identification}

We have defined the action of the fundamental group $\pi_1(H)$ on the
set of gauge invariant local fields and on the set of isomorphism
classes of principal $H$-bundles. Now we can ask whether the relation
(\ref{FI}) holds true.

To start with, we fix the notations. Let $P$ be a principal $H$-bundle
over $\Sigma$. We choose a neighborhood $U_0$ of the insertion point
$x\in \Sigma$, a complex coordinate $z$ with $z(x)=0$, and a section
$s_0$ of $P|_{U_0}$. We assume that $U_0$ includes the unit disc
$D_0=\{ z\,;\,|z|\leq 1\}$. We denote by the same letter $\gamma$ the
element of $\pi_1(H)\cong \Gmalcv$ and a representative loop
$S=\partial D_0\to H$ where the coordinate $z$ determines the
parametrization of $S$. We realize the $H$-bundle $P\gamma$ by one
obtained from $P$ by cutting and gluing method as in the introductory
part of the section 4.4. That is, denoting the restriction $s_0|_S$ by
$s$, $P\gamma$ is obtained by attaching $D_0\times H$ to
$P-P|_{D_0^{\circ}}$ under the identification
$s_0'(\theta)=s(\theta)\gamma(\theta)$, where $s_0'$ is a
trivialization of $D_0\times H$.

We compare the left and right hand sides of (\ref{FI}) in their
integral expressions (\ref{newintexpr}). Let $\cal V$ be an open
coordinatized subset in $\NNc_{P,x}$ with a holomorphic family
$\{(A_v,f_v)\}_{v\in \VV}$ of representatives. We may take $A_v$ to be
flat on $U_0$ in such a way that $s_0$ is horizontal $d_{A_v}s_0=0$,
and also we may take the flag $f_v$ in such a way that $s_0$ is
admissible $f_v=s_0(x)B$. We choose a family of holomorphic
trivialisations $\sigma_{\infty}(v)$ on $\Sigma-D_0^{\circ}$ so that
the family of transition functions $h_{\infty0}(v)$ relating $s_0$ and
$\sigma_{\infty}(v)$ depends holomorphically on $v$,
$\displaystyle{\frac{\partial}{\partial \bar v^{\rA}}}h_{\infty
0}(v)=0$.

For each such family, we construct another family $\{(A'_v,f'_v)\}$ of
connections and flags at $x$ for the $H$-bundle $P\gamma$ : \beq
\begin{array}{rcl}
A'_v|_{D_0}&=&A_{\gamma}\,,\\
A'_v|_{\Sigma-D_0^{\circ}}&=&A_v|_{\Sigma-D_0^{\circ}}\,,
\end{array}
\quad\mbox{and}\,\,\,f'_v=s'_0(x)B.  \eeq The middle equation should
be understood under the identification
$P\gamma|_{\Sigma-D_0^{\circ}}=P|_{\Sigma-D_0^{\circ}}$. $A_{\gamma}$
in the first equation is the basic gauge field corresponding to
$\gamma$ with respect to the frame $s_0'$. Since the frame $s$ on $S$
is related to $s_0'$ by $s_0'=s\gamma$, it extends to a section
denoted also by $s$ on a neighborhood of $S$ which is horizontal with
repect to the connection $A_v'=A_{\gamma}$. Hence, we see that $A_v'$
determines a smooth connection of $P\gamma$.  If we choose a
holomorphic secton $\sigma_0'$ of
$(P\gamma_{\!\C},\bartial_{\!A'_v})|_{U_0}$ such that
$\sigma'_0(x)=s'_0(x)$, then, it is admissible and satisfies
$\sigma'_0(z)=s(z)\hgmm(z)$ where $\hgmm$ is a holomorphic extension
of $\gamma$ up to multiplication by an element of $T_{\!\C}$. This
shows that the holomorphic bundle $(P\gamma_{\!\C},\bartial_{\!A'_v})$
with flag structure $f'_v$ is obtained by the $\gamma$-surgery of
$(\bartial_{\!A_v},f_v)$. If we assume that the conjecture made in
section 4.4 holds true, $(\bartial_{\!A'_v},f'_v)$ represents a class
in $\overline{\NNc_{P\gamma,x}}$ and we may consider the coordinates
$v^1,\cdots,v^{\dNf}$ of $\VV$ as coordinates of the open subset
$\VV\gamma_x$ of $\overline{\NNc_{P\gamma,x}}$.

Note that the holomorphic section $\sigma_{\infty}(v)$ determines a
holomorphic section $\sigma_{\infty}'(v)$ of
$(P\gamma_{\!\C},\bartial_{\!A'_v})|_{\Sigma-D_0^{\circ}}$ by
$\sigma'_{\infty}(v)=\sigma_{\infty}(v)$ under the identification
$P\gamma|_{\Sigma-D_0^{\circ}}=P|_{\Sigma-D_0^{\circ}}$. Then, it is
related to $\sigma_0'$ by the transition relation \beq
\sigma'_0=\sigma_{\infty}'(v)h_{\infty 0}'(v)\qquad;\quad h_{\infty
0}'(v,z)=h_{\infty 0}(v,z)\hgmm(z)\,, \eeq where the transition
function $h_{\infty 0}'(v)$ also depends holomorphically on $v^{\rA}$.

The form $\Omega_{\Sigma,P,x}^M(\met\,;\,{\cal O}\,\gamma O)$ for the
integral expression of the right hand side of (\ref{FI}) is given by
(\ref{newform}) on $\VV$ where $O$ should be replaced by $\gamma O$
and $\nu_{\!\rA}(v)$ is given in (\ref{nuA}) with
$\sigma_0(v)=s_0$. We may as well assume here that the residual
gauge-fixing function $F_v(h)$ is independent of the behavior of
$h|_{D_0}$, that is, $F_v(h_1)=F_v(h_2)$ if $h_1=h_2$ on
$\Sigma-D_0$. Also we may assume that the contour integrals of the $b$
and $\bar b$-fields are placed in $U_0-D_0$ so that there are no field
insertion in $D_0$ other than $\gamma O(f_v)$.

Recalling the definition of the action of $\pi_1(H)$ on flag partners,
we have the following relation of the wave functions : \beq
Z_{D_0}^{\tot(s_0)}({A_v}|_{D_0};\,\gamma O(f_v))=Z_{D_0}^{\tot
(s)}(A_{\gamma}\,;\,O(f'_v))\,.
\label{preFI}
\eeq Introducing a family $\{F'_v\}$ of residual gauge-fixing
functions for the family of symmetry groups
$\Aut(\bartial_{\!A'_v},f'_v)$ by $F'_v(h')=F_v(h)$ if
$h'|_{\Sigma-D_0^{\circ}}=h|_{\Sigma-D_0^{\circ}}$, we see from the
above relation (\ref{preFI}) that the measure
$\Omega_{\Sigma,P,x}^M({\cal O}\,\gamma O)$ is expressed on $\VV$ as
\beqa \lefteqn{\Omega_{\Sigma,P,x}^M(\,\met\,;{\cal O}\,\gamma O)_v}
\\ &=&\prod_{\rA=1}^{\dNf}\dd
v^{\rA}\,Z_{\Sigma,P\gamma}^{\tot}\!\biggl(\,\met,A'_v\,\mbox{\Large
;} \,\delta\Bigl(F'_v(h)\Bigr)\!\prod_{i=1}^{\dSf}\!{F'}^i_{v,h} (c)
\bar{F}'^i_{v,h}(\bar c) \prod_{\rA=1}^{\dNf}\!\ooint_x
\!b\,\nu'_{\!\rA}(v)\ooint_x \!\bar b\, \bnu'_{\!\rA}(v)\,\tilde{\cal
O}\, O(f'_v)\,\biggr) .\nonumber \eeqa The section $\nu'_{\!\rA}(v)$
in this expression is given by \beq \nu'_{\!\rA}(v)=s\cdot h_{\infty
0}(v)^{-1}\frac{\partial}{\partial v^{\rA}}h_{\infty
0}(v)=\sigma'_0\cdot h_{\infty 0}'(v)^{-1}\frac{\partial}{\partial
v^{\rA}}h_{\infty 0}'(v)\,.  \eeq Hence we see that the form
$\Omega_{\Sigma,P,x}^M({\cal O}\,\gamma O)$ on $\VV$ is identical to
the form $\Omega_{\Sigma,P\gamma,x}^M({\cal O}\, O)$ on $\VV\gamma_x$
under the (conjectured) identification
$\displaystyle{\overline{\NNc_{P,x}}\stackrel{\gamma_x}{\longto}\overline{\NNc_{P\gamma,x}}}$. Repeating
the same thing on other open subsets, we see that the forms on
$\overline{\NNc_{P,x}}$ and on $\overline{\NNc_{P\gamma,x}}$ are
identical to each other under the same identification. Hence we have
the following relation leading to the field identification : \beq
\int_{\NNc_{P,x}}\Omega_{\Sigma,P,x}^M(\met\,;\,{\cal O}\,\gamma
O)=\int_{\NNc_{P\gamma,x}}\Omega_{\Sigma,P\gamma,x}^M(\met\,;\,{\cal
O}\, O)\,.  \eeq

\newpage
\renewcommand{\theequation}{5.0.\arabic{equation}}\setcounter{equation}{0}

{\large CHAPTER 5. SAMPLE CALCULATIONS}\label{ch.5}

\vspace{1cm} \hspace{1.5cm} In the final chapter, we calculate the
partition functions of the gauged WZW models on the torus for several
choices of target and gauge groups. Calculations for the topologically
trivial bundles are essentially done in the ref. \cite{GawKup}. Here,
we take good care of the overall normalization which has not
explicitly been done in that reference. The resulting partition
function is physically satisfactory in view of the field
identification if there is no pair of weights fixed by the spectral
flows (no field identification fixed point), and is problematical if
there are. Aiming at the resolution of the latter problem, a
topologically non-trivial bundle is shown to give possibly non-zero
contribution if there are identification fixed points. As the simplest
nontrivial example, we attempt to calculate the full torus partition
function of the level $(k_1,2)$ WZW model with the target group
$G=SU(2)\times SU(2)$ coupled to the $SO(3)$ gauge fields where the
gauge group $SO(3)$ acts on the target by $h:(g_1,g_2)\mapsto
(hg_1h^{-1},hg_2h^{-1})$. When the level $k_1$ is an even integer, the
topologically non-trivial bundle contributes as a constant term which
may be interpreted as a `half' of the Ramond ground state.

\renewcommand{\theequation}{5.1.\arabic{equation}}\setcounter{equation}{0}
\vspace{0.7cm}
\begin{center}
{\sc 5.1 Differential Equations for\\ Partition and Correlation
Functions of WZW Model on Torus}\label{5.1}
\end{center}
\hspace{1.5cm} The Sugawara construction of the energy momentum tensor
of the group $G$ WZW model leads with the aid of Ward identities to a
system of differential equations for the correlation functions with
respect to the modular parameters of the Riemann surfaces and the
holomorphic $G_{\!\C}/Z_G$-bundles. We derive some of them for the
partition and one point functions on the torus $\Sigma=\{\, (x,y)\in
\R^2\}/\Z^2$. We denote by $\metau$ the flat metric on $\Sigma$ given
by $\metau=d\z d\bz+d\bz d\z$ where $\z=x+\tau y$ is a complex
coordinate of the Riemann surface $\Sigma_{\tau}=\C/\Z+\tau \Z$.

\vspace{0.4cm} \underline{Ward Identities --- Topologically Trivial
Case}

We first take closer look at the Ward identities (\ref{onecurr}) and
(\ref{twocurr}) for $P$ a topologically trivial $G/Z_G$-bundle
$\Sigma\times G/Z_G$ over the torus where $G$ is a compact connected
simply connected Lie group and $Z_G$ denote its center. Since
semi-stable orbits are relevant in the consideration of gauge
theories, we take as the background gauge fields the following kind of
flat connections

\newpage \beq A_u=\frac{\pi}{\tau_2}ud\bz-\frac{\pi}{\tau_2}\bar u
d\z\,,
\label{flatu}
\eeq parametrized by $u$ belonging to the complexified Lie algebra
$\liet_{\!\C}$ of the maximal torus. We shall rewrite the identities
(\ref{onecurr}) and (\ref{twocurr}) in the forms that do not involve
the integration over the surface $\Sigma$. To avoid the doubling of
the description, we only look at the left movers, that is, we write
the Ward identities for insertions of the left current $J=J_{\z}d\z$.

The basic tool in the reformulation is the Green function for the
Cauchy-Riemann operator $\bartial_{\!A_u}$ acting on sections of the
adjoint bundle $\ad P_{\!\C}=\Sigma_{\tau}\times \g_{\!\C}$. For the
description, we introduce the multi-valued section $\sigma$ of $\PC$
given by $\sigma(z)=(\z,e^{\pitau u(\z-\bz)})$ where $z=e^{-2\pi i
\z}$. It is holomorphic with respect to the connection $A_u$ and
satisfies \beq \sigma(zq)=\sigma(z)g\qquad ; \,\,\, g=e^{-2\pi i u}\,.
\eeq Now, we put \beqa G_{\!w}(z)&=&\sgmad(w)\ot dz \sum_{n\in \Z}
\frac{(w-1)q^n \ad g^n}{(z-q^n)(z-q^n w)}\sgmad(z)^{-1}\\
\Gree(z)&=&\sgmad(z')\ot dz\sum_{n\in \Z}\frac{(q^{n-1}-q^n)\ad
g^n}{(z-q^n)(z-q^{n-1})}\sgmad(z)^{-1}
\label{page1:Gwz}
\eeqa where $\sgmad$ (resp. $\!\sgmad^{-1}$) is the multi-valued
holomorphic frame of the adjoint bundle $\ad \PC$ (resp. coadjoint
bundle $(\ad \PC)^*$) determined by the multi-valued section $\sigma$
of $\PC$. The maps $G_{\!w}:z\mapsto G_{\!w}(z)$ and $\Gree:z\mapsto
\Gree(z)$ are meromorphic single-valued sections of $(\ad \PC)_w\!\ot
\!K\!\ot\!  (\ad \PC)^*$ and of $(\ad \PC)_{z'}\!\ot\! K\!\ot\! (\ad
\PC)^*$ with the following singularities : \beqa
G_{\!w}(z)&\sim&\sgmad(w)\ot\frac{dz}{z-w}\sgmad(z)^{-1}\qquad
\mbox{as $z\to w$}\,,\\ &\sim&
\sgmad(w)\ot\frac{-dz}{z-1}\sgmad(z)^{-1}\qquad \mbox{as $z\to
1$}\,,\\ \Gree(z)&\sim& \sgmad(z')\ot \frac{dz}{z-1}(\ad
g-1)\sgmad(z)^{-1}\qquad \mbox{as $z\to 1$}\,.  \eeqa The Ward
identities (\ref{onecurr}) and (\ref{twocurr}) with the $s$-point
insertions $\OO=\prod_{l=1}^s O(x_l)$ are then equivalent to the
following equations \beq
Z(J\epsilon(z)\OO)=\sum_{l=1}^sZ(J_l(G_{\!x_l}\epsilon(z))\OO)-Z(\ooint_{C'}\!\!\!J\Gree\epsilon(z)\OO)\,,\label{Ward5.1}
\eeq \beqa Z(J\epsilon_1(z)J\epsilon_2(w)\OO)&=&k\Bigl\{
\trP\!\!\left(\epsilon_2(w)\partial_{\!A_u}^{(w)}\!G_{\!w}\epsilon_1(z)\right)-\ooint_{C'}\!\!\trP\!\!\left(\partial_{\!A_u}^{(z')}G_{\!z'}\epsilon_2(w)\Gree\epsilon_1(z)\right)\Bigr\}Z(\OO)\nonumber
\\
&+&\sum_{l=1}^sZ\Bigl(J_l(G_{\!x_l}[G_{\!w}\epsilon_1(z),\epsilon_2(w)]-\ooint_{C'}\!\!G_{\!x_l}[G_{\!z'}\epsilon_2(w),\Gree\epsilon_1(z)])\OO\,\Bigr)\nonumber
\\ &+&\sum_{l, l'=1}^s Z\Bigl(
J_l(G_{\!x_l}\epsilon_1(z))J_{l'}(G_{\!x_{l'}}\epsilon_2(w))\OO\,\Bigr)\nonumber
\\ &+&Z\Bigl(\ooint_{C'}\!J\Bigl\{
-\Gree[G_{\!w}\epsilon_1(z),\epsilon_2(w)]+\ooint_{C''}\!\!\Gree[G_{\!z''}\epsilon_2(w),\Green\epsilon_1(z)]\Bigr\}\OO\,)\nonumber
\\
&-&\sum_{l=1}^sZ\Bigl(\ooint_{C'}\!\!J\Gree\epsilon_1(z)J_l(G_{\!x_l}\epsilon_2(w))\OO+\ooint_{C'}\!\!J\Gree\epsilon_2(w)J_l(G_{\!x_l}\epsilon_1(z))\OO\,\Bigr)\nonumber
\\ &-&Z\Bigl(\ooint_{C'}J\Gree\epsilon_1(z)\ooint_{C''}J\Green
\epsilon_2(w)\OO\,\Bigr)\,,
\label{Ward5.2}
\eeqa where $\epsilon$, $\epsilon_1$ and $\epsilon_2$ are local
holomorphic sections of the adjoint bundle $(\ad \PC,
\bartial_{\!A_u})$ and $J_l(\epsilon(x_l))\OO$ denotes
$O(x_1)\cdot\cdot J(\epsilon)O(x_l)\cdot\cdot O(x_s)$ (read the
explanation below the eq. (\ref{twocurr}) for the definition of
$J(\epsilon)\OO$). The contour $C'$ or $C''$ in the above expressions
is given by $t\in [0,1]\mapsto C'(t)=re^{it}\in \C^*$ where we assume
the inequality $1<r<|q^{-1}|$. In the last term of the right hand side
of (\ref{Ward5.2}), $C''$ is slightly `smaller' than $C'$, that is,
$|z''|<|z'|$.

\vspace{0.3cm} \underline{The Differential Equations --- Topologically
Trivial Case}

Making use of the above Ward identities (\ref{Ward5.1}) and
(\ref{Ward5.2}), we derive from the Sugawara construction
(\ref{defTzzwzw}) (see also (\ref{normJJwzw})) the differential
equations in terms of the modular parameters $\tau$ and $u$ satisfied
by the partition and one point functions. This is essentially a review
of the work of D. Bernard \cite{B}.

We denote by $Z(O)$ the one point function $Z_{\Sigma,{\rm
triv.}}^{G,k}(\metau,A_u \,;\,O(x))$ under consideration. If we put
$O=1$, $Z(1)$ is the partition function. We first note the
identification of the derivatives $\frac{\partial}{\partial u^i}Z$,
$\frac{\partial^2}{\partial u^i\partial u^j}Z$ and
$\frac{\partial}{\partial \tau}Z$ with the contour integrals of the
current and the energy momentum tensor along the cycle $C$ given above
: \beqa Z(\oint_CJ\sigma\delta u O)&=&\delta_u Z(O)\,,\label{relJu}\\
Z(\oint_C J\sigma\delta_1 u \oint_{C'}J\sigma\delta_2 u
O)&=&\delta_{\!{}_1 u}\delta_{\!{}_2 u}Z(O)-\pitau k\,\tr(\delta_1
u\delta_2 u)Z(O)\,,\\ Z(\oint_C dz z T_{zz}
O)&=&\left(\frac{\partial}{\partial
\tau}-\phi^i\frac{\partial}{\partial u^i}\right)Z(O)\,,\label{relTtau}
\eeqa which are the consequence of the definition (\ref{defTJ}) of $J$
and $T$ with the aid of the Ward identities. $\phi^i$ in the equation
(\ref{relTtau}) is one of the `real coordinates' $\psi$ and $\phi$
belonging to $i\liet$ related to $u$ by $u=\psi -\tau \phi$. For each
root $\alpha$ of $G$, we also have from the Ward identities \beqa
Z(\ooint_CJ\sigma_{\alpha} O)&=&\frac{1}{1-e^{-2\pi i
\alpha(u)}}Z(J(\sigma_{\alpha})O)\,,\\ Z(\ooint_{C'
}\!\!J\sigma_{\alpha}\ooint_{C''}\!\!J\sigma_{-\alpha}
O)&=&\frac{1}{1-e^{-2\pi i\alpha(u)}}Z\Bigl(
\alpha(\phi)k\trP(\sigma_{\alpha}\sigma_{-\alpha})(x)O +\nipi
J([\sigma_{\alpha},\sigma_{-\alpha}])O \nonumber\\
&&\hspace{3cm}+\frac{1}{1-e^{2\pi i
\alpha(u)}}J(\sigma_{\alpha})J(\sigma_{-\alpha})O\,\Bigr)\,,\label{relJJalpha}
\eeqa where $\sigma_{\alpha}$ denotes the multivalued section
$\sigma_{\alpha}=\sgmad\cdot e_{\alpha}$ of $\ad \PC$ in which
$e_{\alpha}\in \g$ is a root vector.

With the aid of the Ward identities (\ref{Ward5.1}) and
(\ref{Ward5.2}) and of the relations (\ref{relJu}) $\sim$
(\ref{relJJalpha}), the Sugawara construction (\ref{defTzzwzw}) of the
energy momentum tensor leads to a differential equations for
$Z(O)$. To write it in a simple looking form, we introduce the
normalized function $\tilde{Z}(\tau, u |O)$ given by \beq
Z(\metau,A_u\,;\,O(x))=e^{\pinitau k\tr(u-\bar
u)^2}\frac{\tilde{Z}(\tau,u|O)}{|\Pi(\tau, u)|^2}\,, \eeq where
$\Pi(\tau,u)$ is the Weyl-Kac denominator defined by \beq \Pi(\tau,
u)=q^{\frac{\dim G}{24}}\prod_{\alpha\in \Delta_+}(e^{\pi
i\alpha(u)}-e^{-\pi i
\alpha(u)})\prod_{n=1}^{\infty}\Bigl\{(1-q^n)^l\!\prod_{\alpha\in
\Delta}(1-q^ne^{-2\pi i\alpha(u)})\,\Bigr\}\,.
\label{defdenom}
\eeq We also introduce the symbols $\delta_{\tau}=\nipi
\frac{\partial}{\partial \tau}$, $D_j=\nipi\frac{\partial}{\partial
u^j}$ and $t_{\alpha}=e^{-2\pi i\alpha(u)}$. Then, the differential
equation is written as \beqa
\lefteqn{\left\{\delta_{\tau}-\frac{\eta^{ij}}{2(k+\lieg^{\vee})}D_iD_j\right\}\tilde{Z}(\tau,u|O)}
\label{heateqn}\\ &&=\frac{-1}{2(k+\lieg^{\vee})}\sum_{\alpha\in
\Delta}\!\!\left(
\frac{t_{\alpha}}{(1-t_{\alpha})^2}+\sum_{n=1}^{\infty}\frac{nq^n}{1-q^n}(t_{\alpha}^n+t_{-\alpha}^n)\right)\!\tilde{Z}(\tau,u|J(\sigma_{\alpha})J(\sigma_{-\alpha})O)\,.\nonumber
\eeqa Note that the Weyl-Kac denominator $\Pi(\tau, u)$ satisfies the
similar differential equation \beq
\left\{\delta_{\tau}-\frac{\eta^{ij}}{2\lieg^{\vee}}D_iD_j\right\}\Pi(\tau,
u)=0.  \eeq

\vspace{0.3cm} \underline{The Differential Equations --- Topologically
Non-Trivial Case}

We next consider the partition function of the WZW model corresponding
to topologically non-trivial configurations. For simplicity and
concreteness, we restrict our attention to the $SU(2)$ WZW model
coupled to external $SO(3)$ gauge field of the non-trivial
$SO(3)$-bundle. Since the semi-stable orbit is relevant in the
consideration of gauge theories, we assume that the background gauge
field is the flat connection $A_F$ with the holonomies represented by
the matrices $\tilde{a}$ and $\tilde{b}$ given in
(\ref{holonomnontriv}).

We first rewrite the Ward identities (\ref{onecurr}) and
(\ref{twocurr}) in the forms that do not involve the integration over
the surface $\Sigma$. The basic tool is the Green function of the
Cauchy-Riemann operator $\bartial_{\!A_F}:\Omega^0(\Sigma,\ad \PC)\to
\Omega^{0,1}(\Sigma,\ad \PC)$. Using the multi-valued holomorphic
section $\sigma(z)=\sigma_F^{(1)}(z)$ which satisfies the relation
(\ref{transflatnontriv}), the Green function is expressed as
$G_{\!w}(z)=\sgmad(w)\!\ot\! dz g(w,z)\sgmad(z)^{-1}$\label{page2:Gwz}
where $g(w,z)$ is the ${\rm End}(\lieg)$-valued function having the
following representation matrix with respect to the base $\sigma_+$,
$\sigma_0$ and $\sigma_-$ : \beq g(w,z)=\pmatrix{
\sum\frac{q^n}{z-q^{2n}w} & 0 &
-\sum\frac{z^{-1}q^{n-\nibun}}{z-q^{2n-1}w} \cr 0 & f(w,z) & 0 \cr
-\sum\frac{wq^{n-\nibun}}{z-q^{2n-1}w} & 0 &
\sum\frac{wz^{-1}q^n}{z-q^{2n}w} \cr }\,,
\label{Greennontriv}
\eeq where the sums $\sum_n$ in the four entries are over all integers
and $f(w,z)$ is given by \beq
f(w,z)=\frac{1}{z-w}\frac{z+w}{2z}\prod_{n=1}^{\infty}\left\{\left(\frac{1+q^n\frac{w}{z}}{1+q^n}\right)\!\!\left(\frac{1+q^n\frac{z}{w}}{1+q^n}\right)\!\!\left(\frac{1-q^n\frac{w}{z}}{1-q^n}\right)^{\!\!\!-1}\!\!\!\left(\frac{1-q^n\frac{z}{w}}{1-q^n}\right)^{\!\!\!-1}\!\right\}\,.  \eeq

The Ward identities (\ref{onecurr}) and (\ref{twocurr}) are equivalent
to the following equations \beq
Z(J\epsilon(z)\OO)=\sum_{l=1}^sZ(J_l(G_{\!x_l}\epsilon(z))\OO)\,, \eeq
\beqa Z(J\epsilon_1(z) J\epsilon_2(w) \OO)&=&
k\trP\!\!\left(\partial^{(w)}_{\!A_F}G_{\!w}\epsilon_1(z)\epsilon_2(w)\right)\!Z(\OO)\nonumber\\
&&+\sum_{l=1}^sZ(J_l(G_{\!x_l}[G_{\!w}\epsilon_1(z),\epsilon_2(w)])\OO)\nonumber\\
&&+\sum_{l,l'=1}^sZ(J_l(G_{\!x_l}\epsilon_1(z))J_{l'}(G_{\!x_{l'}}\epsilon_2(w))\OO)\,,
\eeqa where $\epsilon$, $\epsilon_1$ and $\epsilon_2$ are local
holomorphic sections of $\ad \PC$.

With the aid of these identities, we have the following consequence of
the Sugawara construction (\ref{defTzzwzw}) of the energy momentum
tensor : \beq \frac{\partial}{\partial \tau}Z_{\Sigma,
P}(\metau,A_F)=0\,.
\label{diffeqnontriv}
\eeq That is, the partition function $Z_{\Sigma,P}(\metau, A_F)$ for
the non-trivial $SO(3)$ bundle $P$ is a constant (independent of the
modular parameter $\tau$).

\newpage
\renewcommand{\theequation}{5.2.\arabic{equation}}\setcounter{equation}{0}
\vspace{0.4cm}
\begin{center}
{\sc 5.2 Torus Partition Function for the Trivial Principal
Bundle}\label{5.2}
\end{center}
\hspace{1.5cm} We calculate the torus partition function of the gauged
WZW model with the compact connected simply connected target group $G$
and the the gauge group $H$ which is a connected subgroup of the
adjoint group $G/Z_G$ of $G$. In this section, we only consider
topologically trivial configurations. This is essentially a review of
the work of Gaw\c edzki and Kupiainen \cite{GawKup} which deals with
the case in which the gauge group $\bar H$ is a subgroup of the target
group $G$. However, if the center $Z_G$ has non-trivial intersection
with $\bar H$, the gauge transformation group for $H=\bar H/(\bar
H\cap Z_G)$ is different from that for $\bar H$ and there arises a
difference in the normalization of the partition function.

For simplicity, we assume that $H$ is also semi-simple. Since two
different groups $G$ and $H$ are considered at the same time, we shall
put indices `$G$' and `$H$' on suitable places --- $\Pv_G$, $\trH$,
$\Delta_+(H)$, etc. We choose the maximal tori $T_G$ and $T_H$ of $G$
and $H$ in such a way that $T_H$ is a subgroup of $T_G/Z_G$. In this
section, we denote by $\Pv_H$ the kernel of the exponential map
$i\lieth\to T_H$ given by $v\mapsto e^{2\pi i v}$ though it may be
possible that $\Pv_H$ is not the dual lattice of the root lattice
$\QQ_H$.\label{page1:Q} For each level $k$ of the $G$-WZW model which
is a multiple of integers $(k_1,\cdots,k_M)$ ($k_j$ is the level for
the $j$-th simple factor $G_j$ of $G$), we denote by
$\tilk=(\tilk_1,\cdots, \tilk_N)$ the corresponding level for $H$
defined by $\tilk\trH(v_1v_2)=k\trG(\imath v_1\imath v_2)$ for
$v_1,v_2\in \h$ where $\imath$ is the embedding map of $\h$ to $\g$
(see eq. (\ref{reltrace})).\footnote{The symbol $k\trG$ should be
understood as the normalized trace such that
$k\trG(XY)=k_j\tr_{\!{}_{G_j}}(XY)$ if $X,Y\in {\rm Lie}(G_j)$ and
$k\trG(XY)=0$ if $X$ and $Y$ belong to different simple factors.}

\vspace{0.3cm} \underline{The Partition Function of the WZW Model with
  Target $G$}\label{5.2.1}

Let us denote simply by $Z^{G,k}(\tau, u)$ the partition function
$Z_{\Sigma,{\rm triv}}^{G,k}(\metau,A_u)$ of the level $k\in \N$
target $G$ WZW model for the trivial $G/Z_G$-bundle on the torus where
$\metau$ is the metric introduced in the preceding section and $A_u$
is the flat gauge field (\ref{flatu}) parametrized by
$u\in\lietg_{\!\C}$. It satisfies the following conditions : \beqa
&&\mbox{(gauge invariance)}\quad Z^{G,k}(\tau, wu+n+\tau
m)=Z^{G,k}(\tau, u),\quad w\in W_G,\,\,\, n,m\in \Pv_G,\\
\noalign{\vskip0.2cm} &&\mbox{(modular invariance)}\quad
Z^{G,k}\!\!\left(\frac{-b+a\tau}{d-c\tau},\frac{u}{d-c\tau}\right)\!=Z^{G,k}(\tau,u),\,\,\,\pmatrix{
a & b \cr c & d \cr }\!\in SL(2,\Z),\hspace{1cm}\\
\noalign{\vskip0.1cm} &&\mbox{(heat equation)}\quad
\left\{\delta_{\tau}
-\frac{\eta^{ij}}{2(k+\lieg^{\vee})}D_iD_j\right\}\tilde{Z}^{G,k}(\tau,u)=0,\\
\noalign{\vskip0.3cm} &&\mbox{(reality)}\qquad
Z^{G,k}(\tau,u)^*=Z^{G,k}(\tau,u)\,.  \eeqa where
$\tilde{Z}^{G,k}(\tau,u)$ is given by $e^{-\pinitau k\trG(u-\bar
u)^2}Z^{G,k}(\tau,u)|\Pi^G(\tau,u)|^2$. The first is the consequence
of the invariance under the gauge transformation by $h=n_w^{-1}e^{2\pi
i (mx-ny)}$. The second is the consequence of the diffeomorphism
invariance. The third comes from the Sugawara construction as we have
seen in the preceding section. The final is required because of the
identity $I(A,g)^*=I(A,g^{-1})$ satisfied by the classical action and
the Haar property of the measure for the functional integration.

The solution of the heat equation together with the requirement of the
invariance under $(\Qv_G+\tau \Qv_G)\semidir W_G$ is given by \cite{B}
\beq Z^{G,k}(\tau,u)=e^{\pinitau k \trG(u-\bar
u)^2}\!\!\!\!\sum_{\Lmd,\bar \Lmd\in \PPpk(G)}N_{\bar
\Lmd,\Lmd}\Chi_{(\Lmd,k)}^G(\tau,u)\overline{\Chi_{(\bar
\Lmd,k)}^G(\tau,u)}\,,
\label{toruspartWZW}
\eeq where $\Chi_{(\Lmd,k)}^G$\label{page:character} is the character
of the irreducible highest representation $L_{(\Lmd,k)}$ of the loop
group $\tilde{LG}$ : \beq
\Chi_{(\Lmd,k)}^G(\tau,u)=\tr^{}_{L_{(\Lmd,k)}}(q^{L_0}e^{2\pi i
J_0(u)})\,.
\label{defaffch}
\eeq In the above expression, $L_0$ corresponds to the zero mode (with
respect to $z$) of the energy momentum tensor derived by the variation
of the metric $\metau$.\footnote{ In the conventional notation, this
one should be denoted by $L_0-\frac{c_{G,k}}{24}$ so that the highest
weight vector has the $L_0$-eigenvalue
$\sDelta_{\Lmd}=\frac{(\Lmd,\Lmd+2\rho)}{2(k+\lieg^{\vee})}$.} Since
we have $\Chi_{(\Lmd,k)}^G(\tau,u+n)=e^{2\pi i
\Lmd(n)}\Chi_{(\Lmd,k)}^G(\tau,u)$ for $n\in \Pv_G$, $N_{\bar
\Lmd,\Lmd}=0$ unless $\bar \Lmd-\Lmd\in \QQ_G$. Recall that for each
$\gamma=e^{-i\mu \theta}w\in \Gmalcv^G$, the spectral flow
$\tilh_{\gamma}$ transforms the space
$L_{(\Lmd,k)}\!\ot\!\overline{L_{(\Lmd,k)}}$ to the space
$L_{(\gamma\Lmd,k)}\!\ot\!\overline{L_{(\gamma\Lmd,k)}}$ where
$\gamma\Lmd\in \PPpk(G)$ is given by $\gamma\Lmd=w\Lmd+k\ttr\mu$. With
the aid of the transformation rule (\ref{spctrlwzwLn}) and
(\ref{spctrlwzwJnt}), this shows the following relation : \beq
\Chi_{(\gamma\Lmd,k)}^G(\tau, u)=q^{\frac{k}{2}\trG\mu^2}e^{2\pi i
k\trG(\mu u)}\Chi_{(\Lmd,k)}^G(\tau, w^{-1}u+\tau w^{-1}\mu)\,.
\label{characterspec}
\eeq Then, the gauge invariance requires $N_{\gamma\bar
\Lmd,\gamma\Lmd}=N_{\bar \Lmd,\Lmd}$. Finally, the modular invariance
requires $\sum_{\bar \Lmd',\Lmd'}\overline{S_{\bar \Lmd',\bar
\Lmd}^G}N_{\bar \Lmd',\Lmd'}S_{\Lmd',\Lmd}^G=N_{\bar \Lmd,\Lmd}$ where
$\Bigl(S_{\Lmd',\Lmd}^G\Bigr)$ is the matrix representing the
transformation of characters under $(\tau,u)\mapsto
(-\frac{1}{\tau},\frac{u}{\tau})$. It is given by \cite{Kac} \beq
S_{\Lmd',\Lmd}^G=i^{|\Delta_+(G)|}\left|\PP_G/(k+\lieg^{\vee})\ttr\Qv_G\right|^{-\nibun}\sum_{w\in
W_G}(-1)^{l(w)}e^{-\frac{2\pi
i}{k+\lieg^{\vee}}\left(\Lmd'+\rho_{\!{}_G},w(\Lmd+\rho_{\!{}_G})\right)}.
\label{modularS}
\eeq Among several solutions, it seems natural to take the following
as the solution in our case of simply connected target group $G$ :
\beq N_{\bar \Lmd,\Lmd}=\delta_{\bar \Lmd,\Lmd}\,.
\label{mass1conn}
\eeq

\vspace{0.3cm} \underline{The Partition Function of the $\HC/H$-WZW
Model}\label{5.2.2}

As we shall see, to calculate the partition function of the gauged WZW
model, we need to know the following partition function of the level
$-\kh=-\tilk-2\lieh^{\vee}$ WZW model with the target $\HC/H$ for
generic value of $u\in \lieth_{\!\C}$ : \beq
Z^{-\tilk-2\coxh}\!(\tau,u)=Z_{\Sigma,{\rm
triv.}}^{\Hc/H,-\tilk-2\coxh}\!\!\Bigl(\,\metau,
A_u\,;\,\frac{\delta^{(l)}(\varphi(x_0))}{\Vol T_H}\Bigr)\,.
\label{page:varphi}
\eeq where $x_0$ is any base point and $l$ is the rank of
$H$. $\varphi$ in the above expression is a field valued in $i\lieth$
induced by the Iwasawa decomposition $h=n_+e^{\frac{\varphi}{2}}\un$
of the field $h$ where $n_+$ is valued in $N_0^+(H)$, $\varphi$ is
valued in $i\lieth$ and $\un$ is valued in $H$.

The integration over the fields $n_+$ with the weight $e^{\kh
I^{\tilH}(A_u, hh^*)}$ where $hh^*$ is expressed as
$n_+e^{\varphi}n_+^*$ is known \cite{GawKup} to serve the following
factor \beq e^{-\frac{i}{4\pi}(\tilk+\coxh)\int_{\Sigma}\trH \partial
\varphi \bartial \varphi}\prod_{\alpha\in
\Delta_+(H)}\det\Bigl(\bartial_{\!\alpha(u)}^{\dag}\bartial_{\!\alpha(u)}\Bigr)^{\!-1}\,,
\eeq where $\bartial_{\!\alpha(u)}$ is the Cauchy-Riemann operator
$d\bz \Bigl( \frac{\partial}{\partial\bz}+\pitau \alpha(u)\Bigr) :
\Omega^0(\Sigma_{\tau})\to \Omega^{0,1}(\Sigma_{\tau})$. The
determinants in the above expression can be defined with the use of
the zeta function regularization.\footnote{ For a certain Laplace
operator $D^{\dag}\!D$ on a compact space, the corresponding zeta
function $\zeta_D$ is defined by
$\zeta_D(s)=\sum_{\lmd>0}m_{\lmd}\lmd^{-s}$ where the sum runs over
the (discrete) set of positive eigenvalues of $D^{\dag}\!D$ and
$m_{\lmd}$ denotes the dimension of ${\rm Ker}(D^{\dag}\!D-\lmd)$. The
determinant of $D^{\dag}\!D$ is defined by
$\det'(D^{\dag}\!D)=e^{-\zeta_D'(0)}$.\label{regdet}} Now we are left
with the following path-integral : \beqa
Z^{-\tilk-2\coxh}\!(\tau,u)&=&\prod_{\alpha>0}\!\det\Bigl(\bartial_{\!\alpha(u)}^{\dag}\bartial_{\!\alpha(u)}\Bigr)^{\!-1}\!\!\int{\cal
D}\varphi
e^{-\frac{i}{4\pi}(\tilk+\coxh)\int\trH\partial\varphi\bartial
\varphi}\frac{\delta^{(l)}(\varphi(x_0))}{\Vol T_H}\\
\noalign{\vskip0.2cm}
&=&\frac{(2(\tilk+\coxh))^{\frac{l}{2}}}{(2\pi)^l\Vol
T_H}{\tau_2}^{\frac{l}{2}}\left({\det}'\bartial^{\dag}\bartial\right)^{\!\!-\frac{l}{2}}\!\prod_{\alpha>0}\!\det\Bigl(\bartial_{\!\alpha(u)}^{\dag}\bartial_{\!\alpha(u)}\Bigr)^{\!-1}\,,
\eeqa where we have used the fact that
$\det'(e^{t}D^{\dag}\!D)=e^{\zeta_D(0)t}\det'(D^{\dag}\!D)$ for a
Laplace operator $D^{\dag}\!D$ and that $\zeta_D(0)=-1$ for
$D=\bartial$ and $\zeta_D(0)=0$ for $D=\bartial_{\alpha(u)}$ with
$\alpha(u)\equiv\!\!\!\!\!\!/\,\, 0$ (see \cite{Bost}). The
determinants in the expression above have been calculated by Ray and
Singer \cite{Ray-Singer}. Using their result, we have \beq
Z^{-\tilk-2\coxh}\!(\tau,u)=\frac{(2(\tilk+\coxh))^{\frac{l}{2}}}{(4\pi)^l\Vol
T_H}{\tau_2}^{\!-\frac{l}{2}}\,e^{-\pinitau \coxh \trH(u-\bar
u)^2}|\Pi^H(\tau,u)|^{-2}\,.
\label{toruspartHCWZW}
\eeq It is easy to see that
$\tilde{Z}^{-\tilk-2\coxh}\!(\tau,u)=e^{-\pinitau
(-\tilk-2\coxh)\trH(u-\bar
u)^2}Z^{-\tilk-2\coxh}\!(\tau,u)|\Pi^H(\tau,u)|^2$ satisfies the heat
equation \beq
\left\{\delta_{\tau}-\frac{\eta^{ij}}{2(-\tilk-2\coxh+\coxh)}D_iD_j\right\}\!\tilde{Z}^{-\tilk-2\coxh}\!(\tau,u)=0\,.  \eeq
This is not surprising because, to derive the equation
(\ref{heateqn}), we have used only the Ward identities and the
Sugawara construction which we expect to hold also for the $\HC/H$-WZW
model (see (\ref{WardJHC}) and (\ref{SugHC})). We can also see that
$Z^{-\tilk-2\coxh}\!(\tau,u)$ is invariant under the gauge
transformations $(\Pv_H+\tau \Pv_H)\semidir W_H$.

\vspace{0.3cm} \underline{The Partition Function of the Gauged WZW
Model}

Now let us calculate the partition function of the gauged WZW model
with the target group $G$ and the gauge group $H\subset G/Z_G$. We use
the method developed in Chapter 3. The symmetry group
$S_u=\Aut\bartial_{\!A_u}$ is the cylinder $(T_H)_{\!\C}$ for any $u$
which satisfies $\alpha(u)\not\equiv 0$ for any root $\alpha$ of
$H$. Since $T_H\subset (T_H)_{\!\C}$ is compact, half of the symmetry
$S_u$ serves the non-divergent factor $\Vol T_H$. Hence, as the
residual gauge-fixing term $\delta^{(2l)}\left(F_u(h)\right)$, we may
take the following \beq \frac{1}{\Vol
T_H}\delta^{(l)}(\varphi(x_0))\,, \eeq which we have considered in
advance. Consequently, we have the following expression for the
partition function $Z_{{\rm triv}}(\tau)=Z_{\Sigma,{\rm
triv}}(\metau)$ : \beq Z_{{\rm
triv}}(\tau)=\int_{\NNc_H}\!\prod_{j=1}^l \dd u^j \,Z^{G,k}(\tau,
\imath u)\,Z^{-\tilk-2\coxh}\!(\tau,u) \,Z^{\rm gh}(\tau,u)\,,
\label{torusparttotal}
\eeq where the contribution of the adjoint ghost system is given by
\beqa Z^{\rm gh}(\tau,u)&=&Z^{\rm gh}\Bigl(\,\metau,
A_u\,;\,\prod_{i=1}^lc^i(x_0)\bar c^i(x_0)
\prod_{j=1}^l\frac{i}{2\pi}\!\int_{\Sigma}\!b\frac{\partial
A_u''}{\partial u^j}\,\frac{i}{2\pi}\!\int_{\Sigma}\!\bar
b\frac{\partial A_u'}{\partial \bar u^j}\,\Bigr)\\
\noalign{\vskip0.2cm}
&=&\left(\pitau\right)^{\!2l}\!{\det}'_{\Omega^0(\ad P_{\!\bf
c}\!)}\!\Bigl(\,\bartial_{\!A_u}^{\dag}\bartial_{\!A_u}\Bigr)=(2\pi)^{\!2l}e^{\pinitau
2\coxh \trH(u-\bar u)^2}|\Pi^H(\tau,u)|^4\,.\label{toruspartgh} \eeqa

According to the branching rule (\ref{branchingrule}) of the
representation $L_{(\Lmd,k)}$ of $\tilde{L\lieg}_{\!\C}$ with respect
to the subalgebra $\tilde{L\lieh}_{\!\C}$, we have the following
expansion of the character $\Chi_{(\Lmd,k)}^G(\tau,\imath u)$ for the
integrable weight $\Lmd\in \PPpk(G)$ in terms of the characters of the
irreducible representations of $\tilde{L\lieh}_{\!\C}$ : \beq
\Chi_{(\Lmd,k)}^G(\tau,\imath
u)=\!\!\!\!\sum_{\,\,\,\,\,\,\,\lmd\in\PPptilk(H)}b_{\Lmd,
\lmd}(\tau)\,\Chi_{(\lmd,\tilk)}^H(\tau,u)\,.
\label{branch}
\eeq $b_{\Lmd,\lmd}$ in the above expression is called the {\it
branching function}. If $L_n^G$ denote the operators on $L_{(\Lmd,k)}$
corresponding to the mode expansion of the energy momentum tensor,
they are expressed essentially as bilinear forms of the generators of
$J(\tilde{L\lieg}_{\!\C})$ (see eq. (\ref{virgenwzw})). If we apply
such algebraic construction to the sub-generators
$J(\tilde{L\lieh}_{\!\C})$ and denote the result by $L_n^H$, the
differences $L_n^G-L_n^H$ denoted by $L_n^{G:H }$ generate a Virasoro
algebra \cite{GKO} which acts on the subspace $L_{\Lmd,\lmd}$. The
branching function is then expressed as the character \beq
b_{\Lmd,\lmd}(\tau)=\tr^{}_{L_{\Lmd,\lmd}}\!\!\left(\,q^{L_0^{G:H}}\right)\,.
\eeq
The generators $L_n^{G:H}$ are invariant under the spectral flow
$\tilh_{\gamma}$ representing an element $\gamma$ of the affine Weyl
group $\Waff(H)$ of $LH$ (see eq. (\ref{spctrlwzwLn})). If $\gamma$ is
an element of the isotropy subgroup $\Gmalcv(H)$ at the alc\^ove
$\alcv_H$, the spectral flow $\tilh_{\gamma}$ maps the subspace
$L_{\Lmd,\lmd}\ot \overline{L_{\Lmd,\lmd}}$ to another subspace
$L_{\gamma_G\Lmd,\gamma
\lmd}\ot\overline{L_{\gamma_G\Lmd,\gamma\lmd}}$ where $\gamma_G\Lmd$
and $\gamma\lmd$ are given in (\ref{newweight1}). Here, $\gamma_G$
denotes an element of $\Gmalcv(G)$ that is equivalent to $\gamma$
modulo $\Waff(G)$ where $\gamma$ is considered as an element of
$\Waff(G/Z_G)$. The commutativity of $L_0^{G:H}$ and
$\tilde{h_{\gamma}}$ shows that \beq
b_{\gamma_G\Lmd,\gamma\lmd}(\tau)=b_{\Lmd,\lmd}(\tau)\,.
\label{branchspec}
\eeq

Combining the equations (\ref{toruspartWZW}), (\ref{mass1conn}),
(\ref{branch}), (\ref{toruspartHCWZW}) and (\ref{toruspartgh}), we can
rewrite the expression (\ref{torusparttotal}) as \beqa Z_{\rm
triv}(\tau)&=&\left(\frac{2\pi^2(\tilk+\coxh)}{\tau_2}\right)^{\!\!\frac{l}{2}}\frac{1}{\Vol
T_H}\int_{\NNc_H}\!\prod_{j=1}^l\dd u^j\,e^{\pinitau
(\tilk+\coxh)\trH(u-\bar u)^2}\label{inttoruspart}\\
&&\times\,\,\,\sum_{\stackrel{\scriptstyle \Lmd\in \PPpk(G)}{\lmd,\bar
\lmd\in \PPptilk(H)}}b_{\Lmd,\lmd}(\tau)\overline{b_{\Lmd,\bar
\lmd}(\tau)}\Chi_{(\lmd,\tilk)}^H(\tau,u)\overline{\Chi_{(\bar
\lmd,\tilk)}^H(\tau,u)}|\Pi^H(\tau,u)|^2.\nonumber \eeqa The equation
(\ref{branchspec}) together with the identity (\ref{characterspec})
applied to the characters $\Chi_{(\lmd,\tilk)}^H$ shows that the
integrand of (\ref{inttoruspart}) is indeed a well-defined measure on
the moduli space $\NN_H=\lieth_{\!\C}\lslash (\Pv_H+\tau
\Pv_H)\semidir W_H$ of semi-stable topologically trivial
$\HC$-bundles. Since the moduli space is the quotient of
$\NN_{\tilH}=\lieth_{\!\C}\lslash (\Qv_H+\tau \Qv_H)\semidir W_H$ by
$\left(\Pv_H/\Qv_H\right)^{\!2}$, the integral in the expression
(\ref{inttoruspart}) may be replaced by \beq
\frac{1}{|\Pv_H/\Qv_H|^2}\int_{\NN_{\tilH}}\prod_{j=1}^l\dd u^j\,
\cdots \eeq where $\tilH$ is the simply connected covering group of
$H$. If we use now the orthogonality relation of the characters which
is proved in Appendix 3, we have the following expression \beqa Z_{\rm
triv}(\tau)&=&\frac{(2\pi)^l\Vol(i\lieth/\Qv_H)}{|\Pv_H/\Qv_H|^2 \Vol
T_H}\sum_{\stackrel{\scriptstyle \Lmd\in \PPpk(G)}{\lmd\in
\PPptilk(H)}}\left|\,b_{\Lmd,\lmd}(\tau)\right|^2\\
&=&\frac{1}{|\pi_1(H)|}\sum_{\stackrel{\scriptstyle \Lmd\in
\PPpk(G)}{\lmd\in \PPptilk(H)}}\left|\,b_{\Lmd,\lmd}(\tau)\right|^2\,,
\eeqa where, in proceeding to the right hand side, we have used the
relations $\Pv_H/\Qv_H\cong \pi_1(H)$ and $\Vol
T_H=(2\pi)^l\Vol(i\lieth/\Pv_H)$.

In view of the equation (\ref{branchspec}), $Z_{\rm triv}(\tau)$ can
be seen to be written as \beq Z_{\rm
triv}(\tau)={\sum_{\Lmd,\lmd}}^{\!\sim}\frac{1}{|{\cal
S}_{\Lmd.\lmd}|}\left|\,b_{\Lmd,\lmd}(\tau)\right|^2\,, \eeq where the
sum $\sum_{\Lmd,\lmd}^{\sim}$ is over the quotient
$\Bigl(\PPpk(G)\!\times\!\PPptilk(H)\Bigr)\lslash \Gmalcv(H)$ and
${\cal S}_{\Lmd,\lmd}$\label{page:symfac} is the isotropy subgroup of
$\Gmalcv(H)$ at $(\Lmd,\lmd)$.

In particular, if the action of $\Gmalcv(H)$ on the set
$\PPpk(G)\!\times\!\PPptilk(H)$ is free, the partition function for
the trivial $H$-bundle is given by \beq Z_{\rm triv}(\tau)=\tr_{{\cal
H}/\!\sim}^{}\!\!\left(\,q^{L_0^{G:H}}\bar q^{L_0^{G:H}}\right)\,,
\label{partfree}
\eeq where ${\cal H}/\!\sim$ is the space of states corresponding to
the gauge invariant fields in which only one state is contained among
the states related by the spectral flows $\Gmalcv(H)$. As we shall see
in the following section, in such free case, there is no contribution
from the topologically non-trivial $H$-bundles and hence, the above
$Z_{\rm triv}(\tau)$ is the full partition function.

If the action of $\Gmalcv(H)$ is not free, the coefficients of some
terms in the $q\bar q$ expansion of $Z_{\rm triv}(\tau)$ may be
fractional and the physical interpretation such as (\ref{partfree}) is
impossible. In the following section, by choosing simplest examples,
we argue that such problem may be resolved by taking into accout the
contribution of topologically non-trivial $H$-bundles. This seems to
provide a geometric method for the `fixed point resolution' in coset
conformal field theories \cite{LVW}, \cite{Schell-Yank}.

\renewcommand{\theequation}{5.3.\arabic{equation}}\setcounter{equation}{0}
\vspace{0.4cm}
\begin{center}
{\sc 5.3 Torus Partition Functions for Non-Trivial Principal
Bundles}\label{5.3}
\end{center}
\hspace{1.5cm} In this section, we consider the torus partition
functions of the gauged WZW models for topologically non-trivial
configurations. First, using the field identification and the fusion
rule, we give a criterion for the non-vanishing of such partition
functions. A simplest non-trivial example is seen to be provided by
the level $(k_1,k_2)$ $SU(2)\times SU(2)$ WZW model ($k_1,k_2\in 2\N$)
coupled to $SO(3)$ gauge fields. The calculation of the partition
function for the non-trivial $SO(3)$ bundle is attempted in three
different methods --- (i) With the use of the differential equation
for the partition functions of the constituent field theories for the
non-trivial $SO(3)$-bundle, (ii) With the use of the differential
equations for the one point functions for the trivial $SO(3)$-bundle,
(iii) Direct evaluation of the $\HC/H$-sector integration of the one
point function for the trivial $SO(3)$-bundle. The latter two methods
use the field identification.

\vspace{0.3cm} A Criterion for the Non-Vanishing of the Partition
Function

\underline{\hspace{6cm}------ Field Identification Fixed Points}

Let $G$ be a compact connected and simply connected Lie group and let
$H$ be a closed subgroup of the adjoint group. We give a criterion for
the non-vanishing of the torus partition function of the group $G$ WZW
model coupled to the gauge field of the non-trivial $H$-bundles. For
simplicity, we only consider the case in which $H$ is semi-simple,
though the result also applies to general compact Lie groups including
such groups that contain $U(1)$-factors.

By the field identification (\ref{FI}), for each $\gamma\in \pi_1(H)$,
the partition function for the $H$-bundle $P_{\gamma}=P_{\rm
triv}\gamma$ coincides with the one point function for the trivial
$H$-bundle $P_{\rm triv}$ with the insertion of $\gamma(1)$ at any
point in $\Sigma$. Since the identity field $1$ corresponds to the
state belonging to the subspace $L_{0,0}\ot \overline{L_{0,0}}$
labeled by the pair of weights $(0,0)\in
\PPpk(G)\!\times\!\PPptilk(H)$, the gauge invariant field $\gamma(1)$
corresponds to the state $\Phi_{\gamma(1)}$ belonging to the subspace
$L_{\gamma_G0,\gamma 0}\ot \overline{L_{\gamma_G0,\gamma 0}}$ labeled
by the pair $(\gamma_G 0,\gamma 0)\in
\PPpk(G)\!\times\!\PPptilk(H)$. Let $V_{\gamma 0}$ denote the finite
dimensional irreducible $\tilH$-module with the highest weight $\gamma
0$.

We denote by $(O^G_{\gamma(1)})$ the ${\rm End}(V_{\gamma 0})$-valued
field of the group $G$ WZW model corresponding to the vacuum
representation of the space of the $J(\tilde{L\lieh}_{\!\C})\times\bar
J(\tilde{L\lieh}_{\!\C})$-descendants of $\Phi_{\gamma(1)}$. Then, for
the non-vanishing of the partition function for the $H$-bundle
$P_{\gamma}$, the ${\rm End}(V_{\gamma 0})$-valued one point function
\beq Z_{\Sigma,{\rm triv}}^{G,k}(\,\metau, A_H\,;\,(O^G_{\gamma(1)}))
\label{onepointinv}
\eeq of the level $k$ group $G$ WZW model must be non-vanishing for
generic value of the $H$-gauge field $A_H$. Since the entries
$(O^G_{\gamma(1)})^{\bar m}_{\, m}$ are the $G$-current descendants of
the ${\rm End}(V_{\gamma_G 0})$-valued field $(O_{\gamma_G 0})$
corresponding to the vacuum subspace of $L_{(\gamma_G 0,k)}\ot
\overline{L_{(\gamma_G 0,k)}}$, the one point function \beq
Z_{\Sigma,{\rm triv}}^{G,k}(\,\metau,A_G\,;\,(O_{\gamma_G 0}))
\label{onepointG}
\eeq must also be non-vanishing for generic value of the $G$-gauge
field $A_G$.

Let us introduce the normalized one point function
$\Bigl(Z_{\gamma(1)}^G\Bigr)$ by \beq Z_{\Sigma, {\rm
triv}}^{G,k}(\,\metau,A\,;\,(O^G_{\gamma(1)}))=e^{\frac{i\tilk}{2\pi}\int_{\Sigma}\trH
A'A''}\left( Z_{\gamma(1)}^G(A',A'')\right)\,, \eeq where $A'=A_zdz$
and $A''=A_{\bar z}d\bar z$ are the $(1,0)$ and the $(0,1)$ components
of the $H$-gauge field $A$. Recall that $h^{-1}A'' h+h^{-1}\bartial h$
and $h^*A'h^{*-1}+h^*\partial h^{*-1}$ are denoted by ${A^h}''$ and
${A^h}'$ respectively. The important thing to notice is that the
normalized one point function satisfies \beq
\left(Z_{\gamma(1)}^G({A^{h_R}}',{A^{h_L}}'')\right)=e^{\tilk
I(A'',h_L)+\tilk I(A',h_R)}\Rho\!\left(h^*_R(x)\right)\!\left(
Z_{\gamma(1)}^G(A',A'')\right)\!\Rho\!\left(h_L(x)\right)\,, \eeq
where $x$ is the insertion point and the representation of $\tilH$ on
the space $V_{\gamma 0}$ is denoted by $\Rho :\tilH\to GL(V_{\gamma
0})$. The following space of $V_{\gamma 0}^*$ valued holomorphic
functions of $A''$ \beq \left\{ \Psi :\Omega^{0,1}(\Sigma,
\lieh_{\!\C})\to V_{\gamma 0}^*\,;\,\Psi({A^h}'')=e^{\tilk
I(A'',h)}\Rho(h(x))^t \Psi(A'')\,\right\}\,, \eeq is called the
conformal block \cite{GawQuad}, \cite{Witten2}, \cite{EMSS} and is
known to be of dimension $\sum_{\lmd\in \PPptilk(H)}N_{\gamma 0
\lmd}^{\lmd}$ where $N_{\lmd_1 \lmd_2}^{\lmd_3}$ for $\lmd_1$,
$\lmd_2$, $\lmd_3\in \PPptilk(H)$ is the fusion coefficient given by
\cite{Verlinde} \beq N_{\lmd_1\lmd_2}^{\lmd_3}=\sum_{\lmd\in
\PPptilk(H)}\frac{S_{\lmd_1,\lmd}S_{\lmd_2,\lmd}S_{\lmd_3,\lmd}^*}{S_{0,\lmd}}\,,
\label{fusioncoeff}
\eeq in which $S_{\lmd',\lmd}=S^H_{\lmd',\lmd}$ is the matrix
representing the transfomation under $(\tau,u)\mapsto
(\frac{-1}{\tau},\frac{u}{\tau})$ of the characters of integrable
representations of $\tilde{L\lieh}_{\!\C}$ at level $\tilk$. Hence,
for the non-vanishing of the one point function given in
(\ref{onepointinv}), the weight $\gamma 0$ must satisfy \beq
\sum_{\,\,\,\lmd\in \PPptilk(H)}\!N_{\gamma 0 \,\lmd}^{\lmd}\ne 0\,.
\label{nonvanishH}
\eeq By exactly the same argument, for the non-vanishing of
(\ref{onepointG}), $\gamma_G 0$ must satisfy \beq \sum_{\Lmd\in
\PPpk(G)}\!N_{\gamma_{\!{}_G}\!0\, \Lmd}^{\Lmd}\ne 0\,,
\label{nonvanishG}
\eeq where $N_{\Lmd_1\Lmd_2}^{\Lmd_3}$ is the fusion coefficient for
the group $G$ WZW model at level $k$.

We shall prove that the {\it conditions (\ref{nonvanishH}),
(\ref{nonvanishG}) for the non-vanishing of the partition function
$Z_{\Sigma,P\gamma}(\metau)$ are equivalent to the existence of the
fixed points of the action of $\gamma$ on the set $\PPpk(G)\times
\PPptilk(H)$ given by $\gamma : (\Lmd,\lmd)\mapsto (\gamma_G\Lmd,
\gamma\lmd)$}. Note that $\hat{\rho}_H=(0,\rho_H, \coxh)$ is invariant
under the action of $\Gmalcv(H)$, which follows from the statement
that an affine root $\hat{\alpha}$\label{page:hatalpha} of $H$ is
positive if and only if $(\hat{\rho}_H,\hat{\alpha})>0$ with respect
to the scalar product (\ref{scalarproduct}). Using this fact, we can
see from the expression (\ref{modularS}) that
$S_{\gamma\lmd',\lmd}=(-1)^{l(w)}e^{-2\pi
i(\lmd+\rho_{\!{}_H})(\mu)}S_{\lmd',\lmd}$ if $\gamma$ is represented
by a loop $\gamma(\theta)=e^{-i\theta\mu}n_w$.\footnote{This was
pointed out by D. Gepner \cite{Gep}.}  Putting such identity into the
expression (\ref{fusioncoeff}), we obtain the relation \beq
N_{\gamma_1\!\lmd_1\gamma_2\lmd_2}^{\gamma_1\!\gamma_2\lmd_3}=N_{\lmd_1
\lmd_2}^{\lmd_3}\,.  \eeq Hence, the dimension of the conformal block
is given by \beq \sum_{\lmd\in
\PPptilk(H)}\!\!N_{0\,\lmd}^{\gamma^{-1}\!\!\lmd}=\sharp \!\left\{
\lmd\in \PPptilk(H)\,;\,\gamma\lmd=\lmd\,\right\}\,, \eeq where we
have used the obvious identity
$N_{0\,\lmd}^{\lmd'}=\delta_{\lmd}^{\lmd'}$. Doing the same thing for
$\gamma_G 0\in \PPpk(G)$, we see that the above assertion holds.

\vspace{0.3cm} \underline{\underline{Example ------ $SU(2)\times
    SU(2)$ mod. $SO(3)$ Gauged WZW Model}}

\vspace{0.2cm} As the simplest non-trivial example with field
identification fixed points, we consider the following system with the
gauge group $SO(3)$. The gauged WZW model with the target
$G=SU(2)\times SU(2)$ on which the gauge group $H=SO(3)$ acts by
$h:(g_1,g_2)\mapsto (hg_1h^{-1},h g_2 h^{-1})$. If the level $k$ of
the $SU(2)\times SU(2)$ WZW model is given by $k=(k_1,k_2)$, the
corresponding number $\tilk$ for $H$ is $\tilk=k_1+k_2$. The action of
the group $\Gmalcv(H)=\Z_2$ of spectral flows on the set
$\PP_{\!+}^{(k_1)}\times\PP_{\!+}^{(k_2)}\times\PP_{\!+}^{(\tilk)}$ is
generated by $(j_1,j_2,j_3)\mapsto
(\frac{k_1}{2}-j_1,\frac{k_2}{2}-j_2, \frac{\tilk}{2}-j_3)$. Hence, if
both $k_1$ and $k_2$ are even integers, there is a unique fixed point
$(\frac{k_1}{4},\frac{k_2}{4},\frac{\tilk}{4})$. In this section, we
shall attempt to calculate the torus partition function of the model
with even $k_1$, $k_2$ for the topologically non-trivial
$SO(3)$-bundle $P$ : \beq Z_{\rm
non-triv}(\tau)=Z_{\Sigma,P}(\metau)\,.  \eeq We do it in three
different ways.

\vspace{0.3cm} \underline{(i) Using the Differential Equations}

Since there is only one semi-stable bundle $\Ph_F^{(1)}$ whose
automorphism group is the finite group $\Z_2\times \Z_2$ (see
(\ref{symflatnontriv})) of order $4$, we have the following expression
for the partition function \beq Z_{\rm
non-triv}(\tau)=\prod_{i=1}^2Z_{\Sigma,P}^{SU(2),k_i}(\,\metau,A_F)\,Z_{\Sigma,P}^{\Hc/H,-\tilk-4}\Bigl(\,\metau,A_F\,;\frac{1}{4}\Bigr)\,Z_{\Sigma,P}^{\rm
gh}(\,\metau, A_F)\,.  \eeq

For the partition function of the $SU(2)$ WZW model, the differential
equation (\ref{diffeqnontriv}) says that \beq Z^{SU(2),k_i}_{\Sigma,
P}(\,\metau,A_F)=\mbox{a constant},\qquad \mbox{for $i=1,2$.}  \eeq

Since the proof uses only the Ward identities and the Sugawara
construction of the energy momentum tensor which we expect to hold
also for the $PSL(2,\C)/PU(2)$ WZW model, we also have \beq
Z^{\Hc/H,-\tilk-4}_{\Sigma,P}(\,\metau,A_F)=\mbox{a constant}.  \eeq

In the course of the derivation of such differential equations, we
have actually obtained and used the following property of the Green
function $G_w(z)$ given in (\ref{Greennontriv}) : \beq \lim_{w\to
z}\left\{\trP\!\!\left(\partial^{(w)}_{\!A_F}G_w\sigma_a(z)\eta^{ab}\sigma_b(w)\right)-\frac{3(dz)^2}{(z-w)^2}\right\}=\frac{(dz)^2}{2z^2}\,.  \eeq
Since we have $Z^{\rm gh}(\,\metau,A_F\,;\,c(w) b(z))=G_w(z)Z^{\rm
gh}(\,\metau,A_F)$, we see from the above equation that \beq Z^{\rm
gh}_{\Sigma,P}\Bigl(\,\metau,
A_F\,;\,:\!b_z^{\sigma}(z)(\partial_z+A^{\sigma}_z)c^{\sigma}(z)\!:\,\Bigr)=\frac{1}{2z^2}Z^{\rm
gh}_{\Sigma,P}(\,\metau,A_F)\,.  \eeq Then, putting this together with
$\lmd=1$, $V=\spl(2,\C)$ and $\tr((A^{\sigma}_z)^2)=\frac{1}{8z^2}$
into the expression (\ref{emmulti}) of the energy momentum tensor, we
have \beq Z^{\rm \!gh}_{\Sigma,P}(\,\metau,A_F\,;\,T_{zz})=0\,.  \eeq
Hence, we see that $Z_{\rm non-triv}(\tau)$ is independent of $\tau$,
that is, it is a constant.

\vspace{0.3cm} \underline{(ii) Using the Differential Equations via
the Field Identification}

The second and the third methods use the field identification. The
dressed gauge invariant field for $\gamma_1(1)$ is given by \beq
\frac{1}{\tilk +1}\tr_{\frac{\tilk}{2}}\!\left(
(hh^*)^{-1}_{\frac{\tilk}{2}}\Bigl[
(g_1)_{\frac{k_1}{2}}\ot(g_2)_{\frac{k_2}{2}}\Bigr]_{\frac{\tilk}{2}}\right)\,,
\eeq where $\tr_j$ is the trace in the spin $j$ representation $V_j$
of $SL(2,\C)$, $(g)_j$ denotes the representation of $g$ on $V_j$ and
$[A_{j_1}\ot A_{j_2}]_{j_3}$ denotes the endomorphism of the spin
$j_3$ component with respect to the decomposition $V_{j_1}\ot
V_{j_2}=\oplus_j V_j$ into irreducible representations. Using these
expressions, the relation (\ref{FI}) shows that \beq Z_{\rm
non-triv}(\tau)=\int_{\NN_H}\dd \up
\prod_{i=1}^2Z_{\frac{k_i}{2}}^{k_i}(\tau,
\up)\,\,Z_{\frac{\tilk}{2}}^{-\tilk-4}(\tau,\up)\,Z^{\rm gh}_{\up\bar
\up}(\tau,\up)\,, \eeq \beqa \noalign{\vskip-0.25cm}
\mbox{where}\qquad\qquad\quad
Z^{k_i}_{\frac{k_i}{2}}(\tau,\up)&=&Z_{\Sigma,{\rm
triv}}^{SU(2),k_i}\!\Bigl(\,\metau,A_u\,;\,\Bigl((g)_{\frac{k_i}{2}}\Bigr)^{\!0}_{0}\Bigr)\,,\\
\noalign{\vskip0.1cm}
Z^{-\tilk-4}_{\frac{\tilk}{2}}(\tau,\up)&=&Z_{\Sigma,{\rm
triv}}^{\Hc/H,
-\tilk-4}\Bigl(\,\metau,A_u\,;\frac{\delta(\varphi(x_0))}{\Vol
T_H}\Bigl((hh^*)^{-1}_{\frac{\tilk}{2}}\Bigr)^{\!0}_{0}\Bigr)\frac{1}{\tilk
+1}\,,\hspace{1.5cm}\\ Z^{\rm gh}_{\up\bar
\up}(\tau,\up)&=&\frac{\pi^2}{2\tau_2^2}{\det}'_{\Omega^0_{\ad
P_{\!\bf
c}}}\!\!\Bigl(\,\bartial_{\!A_u}^{\dag}\bartial_{\!A_u}\Bigr)\,,\\
\noalign{\vskip0.2cm} \mbox{and}\qquad\quad\quad\,\, A_u''=\pitau
ud\bz&;&\quad u=\pmatrix{ \frac{\up}{2} & 0 \cr 0 & -\frac{\up}{2} \cr
}\,.
\label{page:upsilon}
\eeqa Note that the weight zero components of the matrices
$(g_i)_{\frac{k_i}{2}}$ and $(hh^*)^{-1}_{\frac{\tilk}{2}}$ are
selected out due to the symmetry $T_{\!\C}$ of the background gauge
fields $A_u$.

The ghost contribution $Z^{\rm gh}_{\up\bar \up}$ is already
calculated (see (\ref{toruspartgh})) : \beq Z_{\up\bar \up}^{\rm
gh}(\tau,\up)=2\pi^2(q\bar q)^{4\phi^2}|\Pi(\tau,\up)|^4\,, \eeq where
we introduce the real coordinates $\phi$ and $\psi$ by
$\frac{\up}{2}=\psi-\tau \phi$ and we denote the Weyl-Kac denominator
$\Pi(\tau,u)$ by $\Pi(\tau,\up)$.

If we introduce the normalized one point functions
$\tilde{Z}^{k_i}_{\frac{k_i}{2}}$ and
$\tilde{Z}^{-\tilk-4}_{\frac{\tilk}{2}}$ by \beqa
Z_{\frac{k_i}{2}}^{k_i}(\tau,\up)&=&(q\bar
q)^{k_i\phi^2}\frac{\tilde{Z}_{k_i/2}^{k_i}(\tau,\up)}{|\Pi(\tau,\up)|^2}\,,\\
Z_{\frac{\tilk}{2}}^{-\tilk-4}(\tau,\up)&=&(q\bar
q)^{-(\tilk+4)\phi^2}\frac{\tilde{Z}^{-\tilk-4}_{\tilk/2}(\tau,\up)}{|\Pi(\tau,\up)|^2}\,,
\eeqa we have the following differential equations (see
(\ref{heateqn}) : \beqa
&&\left\{\delta_{\tau}-\frac{1}{k_i+2}D^2-\frac{k_i}{4}D(\Pi^{-1}D\Pi)\right\}\!\tilde{Z}_{\frac{k_i}{2}}^{k_i}(\tau,\up)=0\,,\label{heatki}\\
&&\left\{\delta_{\tau}+\frac{1}{\tilk+2}D^2+\frac{\tilk}{4}D(\Pi^{-1}D\Pi)\right\}\!\tilde{Z}_{\frac{\tilk}{2}}^{-\tilk-4}(\tau,\up)=0\,,
\eeqa where $\delta_{\tau}=\frac{1}{2\pi i}\frac{\partial}{\partial
\tau}$ and $D=\nipi\frac{\partial}{\partial \up}$. Since there is only
one fixed point $(\frac{k_1}{4},\frac{k_2}{4},\frac{\tilk}{4})$, the
dimension of the conformal block is one which implies that
$\tilde{Z}^{k_i}_{k_i/2}(\tau, \up)$ is a product of holomorphic and
anti-holomorphic functions of $\tau$ and $\up$. The solution
satisfying the equation (\ref{heatki}) is given by
$\tilde{Z}^{k_i}_{k_i/2}(\tau,\up)=\Bigl|\,\Pi(\tau,\up)^{\frac{k_i}{2}+1}\Bigr|^2$. Hence,
we have \beq Z^{k_i}_{\frac{k_i}{2}}(\tau,\up)=C_{k_i}(q\bar
q)^{k_i\phi^2}\Bigl|\,\Pi(\tau,\up)^{\frac{k_i}{2}}\Bigr|^{2}\,, \eeq
where $C_{k_i}$ is a numerical constant. Therefore, we see that \beq
Z_{\rm non-triv}(\tau)=C\int_{\NN_H}\dd \up
\,\Bigl|\,\Pi(\tau,\up)^{\frac{\tilk+2}{2}}\Bigr|^{2}\tilde{Z}^{-\tilk-4}_{\frac{\tilk}{2}}(\tau,\up)\,,
\label{intint}
\eeq where the constant $C$ is given by $2\pi^2C_{k_1}C_{k_2}$. Noting
that the real coordinates $\psi$ and $\phi$ of the moduli space
$\NN_H$ of flat connections are independent of the complex structure
$\tau$ of the surface, we see that the differentiation of the
expression (\ref{intint}) by $\tau$ gives \beq \delta_{\tau}Z_{\rm
non-triv}(\tau)=C\int_{\NN_H}\dd \up \left(-\frac{1}{4\pi\tau_2}-2\phi
D+\delta_{\tau}\right)\Bigl|\,\Pi^{\frac{\tilk+2}{2}}\Bigr|^{\!2}\tilde{Z}^{-\tilk-4}_{\frac{\tilk}{2}}\,.  \eeq
It is a straightforward calculation to check that the integrand is a
total derivative \beq \left(-\frac{1}{4\pi\tau_2}-2\phi
D+\delta_{\tau}\right)\Bigl|\,\Pi^{\frac{\tilk+2}{2}}\Bigr|^{\!2}\tilde{Z}^{-\tilk-4}_{\frac{\tilk}{2}}=\frac{1}{\tilk+2}DF(\tau,\up)\,,
\eeq where $F(\tau,\up)$ is a single-valued function on $\NN_H$ given
by \beq
F(\tau,\up)=C'\left\{DZ^{\tilk}_{\frac{\tilk}{2}}Z^{-\tilk-4}_{\frac{\tilk}{2}}-Z^{\tilk}_{\frac{\tilk}{2}}DZ^{-\tilk-4}_{\frac{\tilk}{2}}\right\}\!Z^{\rm
gh}_{\up\bar \up}(\tau,\up)\,, \eeq where $C'$ is a numerical
constant. Hence we again see that $Z_{\rm non-triv}(\tau)$ is a
constant.

\vspace{0.3cm} \underline{(iii) Direct Evaluation of the Path-Integral
  via the Field Identification}

Instead of using the differential equations, we may calculate the
path-integral $Z^{-\tilk-4}_{\tilk/2}(\tau,\up)$ directly as in the
case of the calculation of the partition function for the
topologically trivial bundle.

We choose the following parametrization of the field $hh^*$ : \beq
hh^*=\pmatrix{ e^{\varphi}(1+|w|^2) & w \cr \bar w & e^{-\varphi} \cr
}\,.  \eeq Then, the field
$\Bigl((hh^*)^{-1}_{\frac{\tilk}{2}}\Bigr)^0_{\,0}$ is expressed as
\beq
\Bigl((hh^*)^{-1}_{\frac{\tilk}{2}}\Bigr)^0_{\,0}=\sum_{l=1}^{\frac{\tilk}{2}}\left
(\matrix{ \tilk\cr \frac{\tilk}{2}\cr }\right
)^{\!2}(1+|w|^2)^{\frac{\tilk}{2}-l}|w|^{2l}=\left (\matrix{ \tilk\cr
\frac{\tilk}{2}\cr }\right )|w|^{\tilk}+\cdots\,, \eeq where $+\cdots$
denotes a power series of $|w|^2$ of order lower than
$\frac{\tilk}{2}$. Using the same parameters, the classical action is
expressed as \beq
(-\tilk-4)I(A_u,hh^*)=\frac{i}{2\pi}(\tilk+4)\int_{\Sigma}\left\{\partial
\varphi\bartial\varphi+\partial_{\!\up,\varphi}\bar
w\bartial_{\!\up,\varphi} w\right\}\,, \eeq where
$\bartial_{\!\up,\varphi}$ is the Cauchy-Riemann operator
$\bartial_{\!\up,\varphi}=\bartial +\pitau \up d\bz +\bartial\varphi$
acting on complex functions on $\Sigma_{\tau}$.

Then, the path-integral is given by \beqa
Z^{-\tilk-4}_{\frac{\tilk}{2}}(\tau,\up)&=&\frac{{\tilk\choose
\tilk/2}}{\tilk+1}\int {\cal D}\varphi
\frac{\delta(\varphi(x_0))}{\Vol T_H}{\cal D}^2\!w\,
e^{-\frac{i(\tilk+4)}{2\pi}\int_{\Sigma}(\partial\varphi\bartial\varphi
+\partial_{\!\up,\varphi}\bar
w\bartial_{\!\up,\varphi}w)}\Bigl\{|w(x)|^{\tilk}\!+\cdots\Bigr\}\hspace{0.8cm}\nonumber\\
&=&\frac{{\tilk\choose
\tilk/2}}{\tilk+1}\det\Bigl(\bartial_{\!\up}^{\dag}\bartial_{\!\up}\Bigr)^{-1}\!\!\!\int
{\cal D}\varphi \frac{\delta(\varphi(x_0))}{\Vol
T_H}e^{-\frac{i(\tilk+2)}{2\pi}\int_{\Sigma}\partial\varphi\bartial\varphi
}\Bigl\{G_{\up,\varphi}(x,x)^{\frac{\tilk}{2}}+\cdots
\Bigr\}\,,\nonumber \eeqa where $G_{\!\up,\varphi}(x_1,x_2)$ is the
Green function for the operator $\bartial_{\!\up,\varphi}$ such that
\beq
G_{\!\up,\varphi}(x,x)=e^{-2\phi(x)}\int_{\Sigma}e^{2\phi(\z)}\frac{\dd
\z }{2\pi (\tilk+4)}\Bigl|\,G_{\!\up}(\z,\z(x))\,\Bigr|^2\,, \eeq in
which $G_{\!\up}(\z,\xi)$ is the Green function for the operator
$\bartial_{\!\up}=\bartial_{\!\up,0}$ that has the following
expression in terms of the function $E(\z)=\vartheta(\tau,
\zeta+\frac{\tau+1}{2})$\label{fcnE} : \beq
G_{\!\up}(\z,\xi)=e^{\pitau \up (\bz-\bar
\xi-\z+\xi)}\frac{E(\z-\xi-\up)}{E(\z-\xi)}\frac{E'(0)}{E(-\up)}\,.
\eeq

Hence, we have \beqa
Z^{-\tilk-4}_{\frac{\tilk}{2}}(\tau,\up)&=&\frac{{\tilk\choose
\tilk/2}}{\tilk+1}\det\Bigl(\bartial_{\!\up}^{\dag}\bartial_{\!\up}\Bigr)^{-1}\!\!\!\int_{\Sigma^{\frac{\tilk}{2}}}\prod_{j=1}^{\frac{\tilk}{2}}\frac{\dd
\z_j}{2\pi(\tilk+4)}\Bigl|\,G_{\!\up}(\z_j,\z(x))\,\Bigr|^2
\nonumber\\ &&\times\,\int {\cal D}\varphi
\frac{\delta(\varphi(x_0))}{\Vol
T_H}e^{-\frac{i(\tilk+2)}{2\pi}\int_{\Sigma}\partial\varphi\bartial\varphi
-\tilk\varphi(x)+\sum_{j=1}^{\frac{\tilk}{2}}2\varphi(\z_j)}+\cdots
\eeqa To perform this Gaussian integral, we have to take care of the
divergence coming from the singularity $\sim \log|\z-\xi|^2$ as $\z\to
\xi$ of the Green function for the scalar laplacian given by \beq
G(\z,\xi)=\log|\z-\xi|^2+\pinitau \Bigl(\z-\xi-\bz+ \bar \xi
+\frac{\tau-\bar \tau}{2}\Bigr)^2+c\,, \eeq where $c$ is a
constant. We define the path-integral by first applying the point
splitting regularization $G(\z,\z)\to G(\z+\Delta\z,\z)$ together with
the renormalization by multiplying the factor \beq
e^{\frac{\tilk}{4}\log|\Delta\z|^2}\,, \eeq and then by taking the
limit $\Delta\z\to 0$. This enables us to neglect the lower order
terms $+\cdots$ and we have the following expression \beqa
Z^{-\tilk-4}_{\frac{\tilk}{2}}(\tau,\up)&=&\frac{(2(\tilk+2))^{\nibun}{\tilk\choose
\tilk/2}}{2\pi(\tilk+1)\Vol
T_H}\tau_2^{\nibun}\Bigl({\det}'\bartial^{\dag}\bartial\Bigr)^{-\nibun}\!\!\det\Bigl(\bartial_{\!\up}^{\dag}\bartial_{\!\up}\Bigr)^{-1}\label{exprint}\\
&\times&\int_{\Sigma^{\frac{\tilk}{2}}}\prod_{j=1}^{\frac{\tilk}{2}}\frac{\dd
\z_j
}{2\pi(\tilk+4)}\Bigl|\,G_{\!\up}(\z_j,\z(x))\,\Bigr|^2\,e^{-\frac{\tilk}{4}G_{\rm
reg}-\frac{2}{\tilk+2}\sum_{i<j}G(\z_i,\z_j)+\frac{\tilk}{\tilk+2}\sum_{j=1}^{\frac{\tilk}{2}}G(\z_j,\z(x))}\,,\nonumber
\eeqa where $G_{\rm reg}$ is the regularized value of the Green
function at the coincident point : \beq G_{\rm reg}=\lim_{\z\to
\xi}\Bigl(G(\z,\xi)-\log|\z-\xi|^2\Bigr)=\log\Bigl|\,2\pi
\eta(\tau)^3\Bigr|^2+c\,, \eeq in which $\eta(\tau)$ is the Dedekind
$\eta$ function. It is easy to see that the expression (\ref{exprint})
is independent of the choice of the constant $c$.

We now recall the Jacobi's triple product identity \beq
E(\z)=(1-e^{-2\pi i\z})\prod_{n=1}^{\infty}(1-q^n)(1-q^ne^{2\pi
i\z})(1-q^ne^{-2\pi i\z})=e^{-\pi i\z}q^{-\frac{1}{8}}\Pi(\tau,\z)\,,
\eeq which shows that $E'(0)=q^{-\frac{1}{8}}2\pi i \eta(\tau)^3$. We
also recall the identity \beq
\tau_2^{\nibun}\Bigl({\det}'\bartial^{\dag}\bartial\Bigr)^{-\nibun}\!\!\det\Bigl(\bartial_{\!\up}^{\dag}\bartial_{\!\up}\Bigr)^{-1}=\nibun
\tau_2^{-\nibun}e^{-\pinitau (\up-\bar \up)^2}|\Pi(\tau,\up)|^{-2}\,.
\eeq Then, denoting $\up$ by $-\z_0$ and putting $\z(x)=\z_0$, we have
\beqa
Z^{-\tilk-4}_{\frac{\tilk}{2}}(\tau,\up)&=&\frac{(2(\tilk+2))^{\nibun}{\tilk\choose
\tilk/2}}{4\pi(\tilk+1)\Vol
T_H}\tau_2^{-\nibun}\left|\frac{\eta(\tau)^3}{\tilk+4}\right|^{\!\frac{\tilk}{2}}\!p(\tau,\up)^{\!-\frac{\tilk+2}{2}}\!\int_{\Sigma^{\frac{\tilk}{2}}}\frac{\displaystyle{\prod_{j=1}^{\frac{\tilk}{2}}\dd
\z_j p(\tau,\z_j)}}{\displaystyle{\prod_{0\leq i<j\leq
\frac{\tilk}{2}}p(\tau,\z_i-\z_j)^{\frac{2}{\tilk+2}}}}\nonumber\\
&;&p(\tau,\z)=e^{\pinitau (\z-\bz)^2}|\Pi(\tau,\z)|^2\,.  \eeqa

The moduli space $\NN_H$ is the $\up=-\z_0$-plane with the
identification $\z_0\equiv \pm \z_0+n+\tau m$ where $n,m\in \Z$. Since
we have $Z^{k_i}_{k_i/2}(\tau \up)=C_{k_i}p(\tau,\up)^{\frac{k_i}{2}}$
and $Z^{\rm gh}_{\up\bar \up}(\tau,\up)=2\pi^2p(\tau,\up)^2$, we
obtain the following expression for the partition function \beq Z_{\rm
non-triv}(\tau)=\frac{(\tilk+2)^{\nibun}{\tilk\choose
\tilk/2}}{4(\tilk
+1)}C_{k_1}C_{k_2}\tau_2^{-\nibun}\left|\frac{\eta(\tau)^3}{\tilk+4}\right|^{\!\frac{\tilk}{2}}\int_{\Sigma^{\frac{\tilk}{2}+1}}\frac{\displaystyle{\prod_{j=0}^{\frac{\tilk}{2}}\dd
\z_j p(\tau,\z_j)}}{\displaystyle{\prod_{0\leq i<j\leq
\frac{\tilk}{2}}p(\tau,\z_i-\z_j)^{\frac{2}{\tilk+2}}}}.  \eeq The
result of numerical calculations imply that this is independent of the
modular parameter $\tau$ though we do not know the proof yet.

\vspace{0.3cm} \underline{The Full Partition Function}

Let us consider the full partition function of the model with even
levels $k_1$, $k_2$. It is given as the sum \beq Z(\tau)=Z_{\rm
triv}(\tau)+Z_{\rm non-triv}(\tau)\,, \eeq of contributions from the
trivial and the non-trivial $SO(3)$-bundles.

As we have seen in the preceding section, the partition function for
the topologically trivial configuration is given by \beq Z_{\rm
triv}(\tau)=\frac{1}{2}\sum_{\stackrel{\scriptstyle j_1\in
\PP_{\!+}^{(k_1)}\!\!,\,j_2\in\PP_{\!+}^{(k_2)}}{j_3\in\PP_{\!+}^{(\tilk)}\!\!\!,\,j_3-j_1-j_2\in
\Z}}\Bigl|\,b_{(j_1,j_2),j_3}(\tau)\Bigr|^2\,, \eeq where we consider
$\PPpk$ as the set $\{0,\nibun,1,\cdots, \frac{k}{2}\}$ of spins
integrable at level $k$. Due to the invariance under the replacement
$((j_1,j_2), j_3)\to
((\frac{k_1}{2}-j_1,\frac{k_2}{2}-j_2),\frac{\tilk}{2}-j_3)$, the sum
can be rewritten as \beq \sum_{(j_1,j_2,j_3)\in
F^{\circ}}\Bigl|\,b_{(j_1,j_2),j_3}(\tau)\Bigr|^2+\nibun\Bigl|\,b_{(\frac{k_1}{4},\frac{k_2}{4}),\frac{\tilk}{4}}(\tau)\Bigr|^2.  \eeq
where $F^{\circ}$ is a fundamental domain of the action of $\Z_2$ on
$\PP_{\!+}^{(k_1)}\times
\PP_{\!+}^{(k_2)}\times\PP_{\!+}^{(\tilk)}-\{(\frac{k_1}{4},\frac{k_2}{4},\frac{\tilk}{4})\}$. As
we have noted, the term
$\nibun\Bigl|\,b_{(\frac{k_1}{4},\frac{k_2}{4}),\frac{\tilk}{4}}(\tau)\Bigr|^2$
is physically unsatisfactory since it may occur that the coefficients
of some terms in the $q\bar q$ expansion are not integers.

It may be possible that the addition of $Z_{\rm non-triv}(\tau)$ (a
constant) makes the full partition function $Z(\tau)$ physically
acceptable. In the following, we shall see that this is the case for
the model with $k_2=2$.

The models with $k_2=2$ are known as the unitary $N=1$ supersymmetric
minimal models. The existence of the super current $G_0$ satisfying
$G_0^2=L_0$ in the Ramond sector implies that the positive eigenspaces
of $L_0$ are degenerate with even multiplicities.\footnote{As before,
this operator $L_0$ should be expressed as $L_0-\frac{c}{24}$ in the
conventional notations where $c$ is the central charge of the theory.}
Indeed, the branching function
$b_{(\frac{k_1}{4},\nibun),\frac{\tilk}{4}}(\tau)$ is given by
\cite{GKO} \beqa
b_{(\frac{k_1}{4},\nibun),\frac{\tilk}{4}}(\tau)&=&\frac{\eta(2\tau)}{\eta(\tau)^2}\sum_{n\in\Z}\Bigl(q^{4Nn^2}-q^{4N(n+\nibun)^2}\Bigr)\\
\noalign{\vskip0.2cm}
&=&\frac{1+2\sum_{n=1}^{\infty}(-1)^nq^{Nn^2}}{1+2\sum_{n=1}^{\infty}(-1)^nq^{n^2}}\,,
\eeqa where $N$ is an integer defined by
$N=\frac{\tilk}{2}(\frac{\tilk}{2}+1)/2$. Hence, if we could show that
$Z_{\rm non-triv}(\tau)=\nibun$, the coefficients of all terms in the
$q\bar q$ expansion of the full partition function would be integers,
and the model may be interpreted as a supersymmetric model with unique
Ramond ground state.

 In the literature \cite{Cappelli}, such a constant term is
interpreted as the contribution of the periodic boundary condition of
the corresponding statistical model. The above full partition function
is the (A,A) modular invariant in the list in that reference.

\newpage {\large CONCLUDING REMARKS}\label{conclude}

\vspace{1cm} \hspace{1.5cm} We summarize what is done in this
paper. We realized with certain geometrical meaning the spectral flows
in such field theories that appear in quantization of the gauged WZW
models or of models of free fermions coupled to gauge fields. We have
expressed a correlation function of the model as an integral over the
moduli space of holomorphic principal bundles with flag structure at
one point. The geometrical interpretation of the spectral flow is
conjectured to induce an identification of the moduli spaces
corresponding to different topological types. The conjecture is
checked on the sphere with general simple structure group and on the
torus with the structure group $PSL(2,\C)$. Under the assumption that
the conjecture holds true, we have obtained the relation between
correlators that lead to the field identification. As an application,
the condition of the non-vanishing of the torus partition function for
topologically non-trivial configurations is determined in terms of the
spectral flow.

\vspace{0.2cm} It is clear that we need general proof of the
conjecture : The map induced by the spectral flow transforms a moduli
space relevant in the path integral to another relevant moduli
space. For the proof, it seems that we need a new characterization of
the holomorphic principal bundles with flag structure that represent
points in the `moduli space relevant in the path integral'. It may be
possible that a certain parabolic stability \cite{Mehta-Seshadri} will
play the role.

\vspace{0.2cm} In recent years, `mirror symmetry' has attracted
interests of many physicists and mathematicians. It was first
suggested\cite{LVW} by the existence of an automorphism of $N=2$ super
conformal algebra which is also called the spectral flow. It implies a
natural map (called the mirror map) between moduli space of K\"ahler
structures of one complex manifold and the moduli space of complex
structures of another smooth manifold. Though it seems very
speculative, the present work may shed some light on the construction
of such mirror map, or conversely the study of mirror symmetry may
give a deeper understanding of the field identification phenomena.

\newpage {\large ACKNOWLEDGEMENTS}\label{acknowledge}

\vspace{1cm} I would like to thank Prof. Yoichi Kazama for continuous
encouragement. I thank Kenji Mohri for discussion in the early stage
of the work. I feel grateful for sharing similar point of view with
Toshio Nakatsu and Yuji Sugahara. I thank Prof. Hiraku Nakajima for
explaining me on the Hecke correspondence. I would like to express my
thank to all the members of Institute of Physics, Komaba, for thier
kind support. Finally, I have to and I wish to appreciate the
continuous influence of Prof. Graeme Segal and Prof. Krzysztof Gaw\c
edzki.

\newpage
\renewcommand{\theequation}{A.1.\arabic{equation}}\setcounter{equation}{0}

{\large {\bf Appendix 1. Lifting the Adjoint Action}}\label{a.1}

Let $G$ be a compact simple simply connected Lie group with center
$Z_{G}$ and $H$ be the quotient group $H=G/Z_{G}$. In this appendix,
we prove that the adjoint acion of $LH$ on $LG_{\!\C}$ lifts to an
action on the semigroup $\LWZ^k$ introduced in Chapter 2.

We start with the requirement that the weight
$e^{-kI_{\Sigma_{0}}(A,g)}$ for a surface $\Sigma_{0}$ with boundary
$\partial \Sigma_{0}\cong S^1$ is gauge covariant. (See (\ref{cov}) in
Chapter 2 and the remark following that formula.)  Assuming that there
exists such a transformation $\ad\gamma^{-1}:\LWZ^k\to \LWZ^k$ for
$\gamma\in LH$, we can show that it is an automorphism of $\LWZ^k$ :
\beqa
e^{-kI_{\Sigma_{0}}(A^{h},h^{-1}\!g_{1}h)}e^{-kI_{\Sigma_{0}}(A^{h},h^{-1}\!g_{2}h)}&=&e^{-kI_{\Sigma_{0}}(A^h,h^{-1}\!g_{1}g_{2}h)+k\Gamma_{\Sigma_{0}}(A^h,h^{-1}\!g_{1}h,h^{-1}\!g_{2}h)}\nonumber\\
&=&\gamma^{-1}e^{-kI_{\Sigma_{0}}(A,g_{1}g_{2})}\gamma \,
e^{k\Gamma_{\Sigma_{0}}(A,g_{1},g_{2})}\nonumber \\
&=&\gamma^{-1}\left(\,
e^{-kI_{\Sigma_{0}}(A,g_{1})}e^{-kI_{\Sigma_{0}}(A,g_{2})}\,\right)\gamma\,,
\eeqa where we have used the reation
$e^{-kI_{\Sigma_{0}}(A,g_{1})}e^{-kI_{\Sigma_{0}}(A,g_{2})}=e^{-kI_{\Sigma_{0}}(A,g_{1}g_{2})+k\Gamma_{\Sigma_{0}}(A,g_{1},g_{2})}$.

 The requirement of gauge covariance is equivalent to the following :
\beq
\gamma^{-1}\{(\check{g},1)\}\gamma=\{(\check{h}^{-1}\!\check{g}\check{h},e^{-kI_{\hat{\Sigma}_{0}}\left(0\star
A^{h},
(\check{h}^{-1}\check{g}\check{h})\star(h^{-1}\!gh)\right)+kI_{\hat{\Sigma}_{0}}(0\star
A,\check{g}\star g)})\}\,\,,
\label{app1}
\eeq where $\hat{\Sigma}_{0}$ is a closed Riemann surface obtained by
capping $\Sigma_{0}$ by the disc $D_{\infty}$. Maps $g:\Sigma_0\to
G_{\!\C}$ and $\check{g}:D_{\!\infty}\to G_{\!\C}$ are related by
$\check{g}|_S=g|_S$ and the values are different from $1$ only on a
neighborhood $U$ of $S$ in $\hat{\Sigma}_0$. Maps $h:\Sigma_0\cap U\to
H$ and $\check{h}:D_{\!\infty}\cap U\to H$ are related by
$\check{h}|_S=h|_S=\gamma\in LH$. By a calculation we see that
$I_{\bar \Sigma_{0}}\!\left(0\star A^{h}\!,
(\check{h}^{-1}\check{g}\check{h})\star(h^{-1}gh)\right)\!-\!I_{\bar
\Sigma_{0}}(0\star A,\check{g}\star g)=c(\check{h},\check{g})$ is
independent of $(h_0,A,g_0)$ :\label{page2:cocycle} \beq
c(\check{h},\check{g})\!=\!K_{\!D_{\!\infty}}(\check{h}^{-1}\!\check{g}\check{h})\!-\!K_{\!D_{\!\infty}}(\check{g})\!-\!\frac{i}{4\pi}\!\int_{D_{\!\infty}}\!\!\!\!
\tr\!\left\{
(d\check{g}\check{g}^{-1}\!\!+\check{g}^{-1}\!d\check{g})\check{h}d\check{h}^{\!-1}\!\!+\check{h}d\check{h}^{\!-1}\!\check{g}\check{h}d\check{h}^{\!-1}\!\check{g}^{-1}\right\}\!,
\eeq where
$K_{D_{\!\infty}}(\check{g})=\frac{i}{4\pi}\int_{D_{\!\infty}}\tr(\partial
\check{g}^{-1}\bartial \check{g})$. This shows that {\it the adjoint
action of the loop group $LH$ on $LG_{\!\bf C}$ lifts to an
automorphic action on the semigroup $\LWZ^k$ by the rule} \beq
\gamma^{-1}\{(\check{g},
u)\}\gamma=\{(\check{h}^{-1}\check{g}\check{h},
ue^{-kc(\check{h},\check{g})})\}.  \eeq

{\it Remark}. As is shown in \cite{P-S}, the lift is unique. In fact,
the above transformation law can be derived also by transporting the
transformation law given in that reference through the isomorphism of
$\tilde{LG}_{\!\bf C}$ and $\bigl(\LWZ\bigr)^{\times}$ given by
$$[(\gamma,p,c)]\mapsto \{(g(p),c
e^{\int_{p}\beta\,-K_{\!D_{\!\infty}}\!(g(p))})\}\,\,,$$ where
$\gamma\in LG_{\!\bf C}$, $p$ is a path in $LG_{\!\bf C}$ from $1$ to
$\gamma$, $g(p):D_{\!\infty}\to G_{\!\bf C}$ is given by
$g(p)(re^{i\theta})=p(r^{-1})(e^{-i\theta})$ and $\beta$ is a one form
on $LG_{\!\bf C}$ given by $$\beta_{\gamma}(\delta \gamma
)=\frac{i}{4\pi}\int_{S^1}\tr(\gamma^{-1}\!d\gamma
\,\gamma^{-1}\!\delta \gamma )\,\,.$$

\newpage
\renewcommand{\theequation}{A.2.\arabic{equation}}\setcounter{equation}{0}

{\large {\bf Appendix 2. Root Systems and Weyl Groups}}\label{a.2}

We describe here the basic facts on root systems and Weyl
groups.\cite{Bourbaki} The notation here is the same as in the
Appendix 1. We choose and fix the maximal tori $T_{G}\subset G$,
$T_{H}=T_{G}/Z_{G}\subset H$. We identify their Lie algebras
$\liet=\lietg=\lieth$ and denote by $\V=i\liet$ the purely imaginary
part of its complexification. We introduce four lattices in $\V^*$ and
in $\V$ : \beq
\begin{array}{ccc}
\V^*\supset \PP\hspace{-0.05cm}&\cdots\cdots&\hspace{-1.05cm}\Qv \\
\cup \hspace{-1.08cm}& &\hspace{-1.1cm}\!\!\cap\\
\QQ\hspace{-1.08cm}&\cdots\cdots&\hspace{-0.12cm}\Pv\subset \V
\end{array}
\eeq where $A\cdots\cdots B$ means that $A$ is the dual lattice of
$B$. The lattices $\Qv$ and $\Pv$ are determined in such a way that
the exponential mappings induce isomorphisms $\liet/2\pi
i\Qv\stackrel{\simeq}{\to}T_{G}$ and $\liet/2\pi
i\Pv\stackrel{\simeq}{\to}T_{H}$. $\PP$ is the {\it weight lattice} of
$T_G$. $\QQ$ is called the {\it root lattice} since it is generated by
the set $\Delta$ of $roots$ of $\g$, that is, the set of weights of
adjoint representation of $G$ or of $H$ on $\g$. The center $Z_{G}$ of
$G$ is isomorphic to $\Pv/\Qv$.

\vspace{0.2cm}
\noindent{\it (2.1)}\underline{ Finite Weyl
  group.}\label{page:Weylgrp} The {\it Weyl group} $W$ of $(G,T_{G})$
  is defined by $W=N_{T_{G}}/T_{G}$ where $N_{T_{G}}$ is the
  normalizer of $T_{G}$ in $G$. $W$ acts adjointly on $T_{G}$ and
  hence has faithful representations on $\V^*$ and $\V$ leaving
  invariant the four lattices and $\Delta$. A $W$-invariant inner
  product $(\,\,,\,\,)$ on $\V^*$ is normalized as $(\alpha,
  \alpha)=2$ for a long root $\alpha$ and defines a linear isomorphism
  $\iota:\V\to \V^*$. For each root $\alpha \in \Delta$, we introduce
  a hyperplane $\hp_{\alpha}=\{ x\in V; \alpha(x)=0\}$ and we denote
  by $s_{\alpha}$ the orthogonal reflections with respect to
  $\hp_{\alpha}$. Since $W$ preserves $\Delta$, the family
  $\cup_{\alpha \in \Delta}\hp_{\alpha}\subset \V$ of hyperplanes is
  invariant by $W$. A $chambre$ of $\Delta$ is, by definition, a
  connected component of $\V-\cup_{\alpha\in
  \Delta}\hp_{\alpha}$. Then, the following is known
\begin{th}
(1) $W$ acts simply transitively on the set of chambres.\\ (2) If
\(\hp_{1}, \cdots ,\hp_{l}\) are walls of a chambre $\Ch$, for each
$i$ there exist a unique root $\alpha_{i}$ such that
$\hp_{\alpha_{i}}=\hp_{i}$ and that $\alpha_{i}$ takes positive values
on $\Ch$.\\ (3) The set ${\rm B}(\Ch)=\{ \alpha_{1},\cdots
,\alpha_{l}\}$ forms a base of $\V^*$.  \\ (4) The set $S(\Ch)=\{
s_{\alpha_{1}},\cdots , s_{\alpha_{l}}\}$ generate $W$.\\ (5) Any root
$\alpha\in \Delta$ is expressed as
$\alpha=\sum_{i=1}^{l}n_{i}\alpha_{i}$ where $n_{i}$ are all
non-negative integers or all non-positive integers.
\label{page:theoremWeyl}
\end{th}
We choose and fix a chambre $\Ch$.  {\it (3)} and {\it (5)} shows that
${\rm B}(\Ch)=\{ \alpha_{1},\cdots, \alpha_{l}\}$ forms a base of
$\QQ$ and that the subset $\{ \mu_1,\cdots, \mu_l\}$ of $\V$ such that
$\alpha_{i}(\mu_j)=\delta_{i,j}$ forms a base of $\Pv$ dual to ${\rm
B}(\Ch)$. {\it (5)} of the theorem shows that $\Delta$ is decomposed
as a disjoint union of the set $\Delta_{+}$ of {\it positive roots}
and the set $\Delta_{-}=-\Delta_{+}$ of {\it negative roots} where a
root is {\rm positive} if it takes positive values on $\Ch$. We see
from {\it (1)} that there exists a unique element $w_{0}\in W$ such
that $w_{0}\Delta_{+}=\Delta_{-}$. It is the longest element with
respect to the length $l$ on $W$ determined by $S(\Ch)$ where
$l(w)$\label{page:length} is the minimun length $n$ of such sequence
$(s_{i_{1}},\cdots, s_{i_{n}})$ of $S(\Ch)$ that $w=s_{i_{1}}\cdots
s_{i_{n}}$.

  The highest weight of the adjoint representation is called $highest$
$root$ and is denoted by $\tilde{\alpha}$.\label{page2:highestroot} It
can be shown that, for any $\alpha \in \Delta$, $\tilde{\alpha}
-\alpha$ is a linear combination of elements $\alpha_{1}, \cdots,
\alpha_{l}$ of ${\rm B}(\Ch)$ with non-negative coefficients and in
particular, $\tilde{\alpha}$ is expressed as
$\tilde{\alpha}=\sum_{i=1}^{l}n_{i}\alpha_{i}$ for $n_{i}\geq 1$. We
define a subset ${\cal J}$\label{a.2calJ} of $\{ 1, \cdots , l\}$ by
$j\in {\cal J}\Leftrightarrow n_{j}=1$. Then, we see the following
\begin{pn}
 $a\in \Pv$ satisfies $|\alpha(a)|=1$ or $0$ for any $\alpha\in
\Delta$ if and only if $a=0$ or $a=w\mu_j$ for $w\in W$ and $j\in
{\cal J}$. Moreover, any $\Qv$ orbit in $\Pv$ contains one and only
one element of $\{0\}\cup\{ \,\mu_j\,;j\in {\cal J}\}$.
\label{page:prop}
\end{pn}
The latter part can be understood after we introduce the group
$\Gamma_{\alcv}$.

\vspace{0.3cm}

\noindent {\it (2.2)}\underline{ Infinite Weyl groups.} The {\it
  affine Weyl group} $\Waff$ of a loop group $LG$ is defined by
  $\Waff=N_{U(1)\times \tilde{T}_{G}}/(U(1)\!\times \!\tilde{T}_{G})$
  where $N_{U(1)\times \tilde{T}_{G}}$ is the normalizer of
  $U(1)\!\times\! \tilde{T}_{G}$ in $U(1)\semidir \tilde{LG}$ where
  the group $U(1)$ of rigid rotations acts on $\tilde{LG}$ by
  $(u\gamma)(\hat{z})=\gamma(u^{-1}\!\hat{z})$.\label{page:a2.Waff} We
  have the isomorphism $\Waff\cong{\rm Hom}(U(1),T_{G})\semidir W\cong
  \Qv\semidir W$. $\Waff$ can naturally be considered as a normal
  subgroup of the affine Weyl group $\Waffh\cong{\rm
  Hom}(U(1),\tilde{T}_{H})\semidir W\cong \Pv\semidir W$ of $LH$. The
  adjoint action of $\Waffh$ on the Lie algebra $i
  \hat{\V}=i{\R}_{rot}\!\oplus\! i\V\!\oplus \!i{\R}_K$ of the torus
  $U(1)\!\times \!\tilde{T}_{H}$ and the action on its dual
  $\hat{\V}^*={\R}\!\oplus\! \V^*\!\oplus\!{\R}$ is given by \beqa
  e^{-ia\theta}w: (\, x\, , t\,,y\, )\in \hat{\V}
  \!\!&\longmapsto&\!\!\! \Bigl(\, x\,, wt-xa\,,y-\tr(awt)+\frac{x}{2}
  \tr(a^2\!) \Bigr)\in \hat{\V},\label{affact}\\ e^{-ia\theta}w : (\,
  n\,, \lmd \,,k\,)\in \hat{\V}^* \!\!&\longmapsto& \!\!\!\Bigl(\,
  n+w\lmd(a)+\frac{k}{2}\tr(a^2\!)\,,w\lmd+k\ttr a \,, k\, \Bigr)\in
  \hat{\V}^*\,.
\label{affact*}
\eeqa This action preserves the following non-degenerate symmetric
bilinear form on $\hat{\V}^*$ : \beq
\Bigl((\,n_1\,,\lmd_1\,,k_1\,),(\,n_2\,,\lmd_2\,,k_2\,)\Bigr)=(\,\lmd_1\,,\lmd_2\,)-n_1k_2-k_1n_2\,.\label{scalarproduct}
\eeq

An {\it affine root} of $LG$ is, by definition, a weight of the
adjoint action of $U(1)\times \tilde{T}_{H}$ on the Lie algebra
$L\lieg_{\C}$. The set $\Daff\subset \hat{\V}^*$\label{a.2:Daff} of
non-zero affine roots is invariant under the action of $\Waffh$ and is
given by $\Daff= {\Z}_{\neq 0}\!\!\times \!\{0\}\!\times\!\{0\}\cup
{\Z}\!\times\! \Delta\!\times\!\{0\} $.

The hyperplane $\V_{x}=\{x\}\!\times\! \V$ in $\Vaff=\hat{\V}/{\R}_K$
is $\Waffh$-invariant for each $x\in {\R}_{rot}$. Identifying $\V$
with $\V_{-1}$ we see that $\Waffh$ forms an affine transformation
group of $\V$ (see (\ref{affact})). Since the family
$\{\hat{\hp}_{\hat{\alpha}}\}_{\hat{\alpha}\in
\Daff}$\label{page2:hatalpha} of hyperplanes in $\Vaff$ is
$\Waffh$-invariant where $\hat{\hp}_{\hat{\alpha}}=\{ \, {\bf
v}\!\in\! \Vaff\, ;\, \hat{\alpha}({\bf v})=0\,\}$, the family
$\{\hp_{\hat{\alpha}}\}_{\hat{\alpha}\in {\Z}\!\times \!\Delta\!\times
\!\{0\}}$ of hyperplanes in $\V$ is invariant under the affine action
of $\Waffh$, where $\hp_{(n,\alpha,0)}=\hat{\hp}_{(n,\alpha,0)}\cap
\V_{-1}\cong\{\, t\!\in\! \V ; \alpha(t)=n\}$. We denote by
$s_{\hat{\alpha}}$ the orthogonal reflection with respect to
$\hp_{\hat{\alpha}}\neq \emptyset$. An {\it alc\^ove} is, by
definition, a connected component of $\V-\cup_{\hat{\alpha}\in
\Daff}\hp_{\hat{\alpha}}$. Then, we have the

\begin{th}
(1) $\Waff$ acts simply transitively on the set of alc\^oves.\\ (2) If
$\hp_{0},\hp_{1},\cdots, \hp_{l}$ are walls of an alc\^ove $\alcv$,
for each $i$ there exists a unique affine root $\hat{\alpha}_{i}$ such
that $\hp_{\hat{\alpha}_{i}}=\hp_{i}$ and that $\hat{\alpha}_{i}$
takes positive values on $\alcv$.\\ (3) The set ${\rm B}(\alcv)=\{
\hat{\alpha}_{0}, \hat{\alpha}_{1}, \cdots, \hat{\alpha}_{l}\}$ forms
a base of ${\R}\oplus\V^*\oplus\{0\}$.\\ (4) The set $S(\alcv)=\{
s_{\hat{\alpha}_{0}},s_{\hat{\alpha}_{1}},\cdots,
s_{\hat{\alpha}_{l}}\}$ generates $\Waff$.\\ (5) Any affine root
$\hat{\alpha}\in \Daff$ is expressed as
$\hat{\alpha}=\sum_{i=0}^{l}n_{i}\hat{\alpha}_{i}$ where $n_{i}$ are
all non-negative integers or all non-positive integers.
\label{theoremaffWeyl}
\end{th}
A chambre $\Ch$ determines an alc\^ove $\alcv=\{ \, t\in \Ch\, ; \,
\tilde{\alpha}(t)<1\, \}$ which gives ${\rm B}(\alcv)=\{
(-1,\alpha_{0},0), (0,\alpha_{1},0), \cdots , (0,\alpha_{l},0)\}$
where $\alpha_{0}=-\tilde{\alpha}$. {\it (5)} of the theorem shows
that the set $\Daff$ is decomposed as a disjoint union of the set
$\Delta_{{\rm aff} +}$ of {\it positive affine roots} and the set
$\Delta_{{\rm aff} -}=-\Delta_{{\rm aff} +}$ of {\it negative affine
roots}\label{a.2:Daffpm} where an affine root is positive if it takes
positive values on the alc\^ove $\alcv$. The root vectors
$e_{\hat{\alpha}}$ for $\hat{\alpha}\in \Delta_{{\rm aff} +}$ generate
a Lie subalgebra of $L\lieg_{\C}$ corresponding to the subgroup $N^+$
of $LG_{\!\C}$.

$\Waffh$ also acts on the set of alc\^oves and we denote by
$\Gamma_{\alcv}\subset \Waffh$ the isotropy group at $\alcv$. Then we
see that $\Waffh$ decomposes into semi-direct product of $\Waff$ and
$\Gamma_{\alcv}$ : \beq \Waffh \cong \Waff \semidir
\Gamma_{\alcv}\,\,.
\label{Waff'decompo}
\eeq The subgroup $\Gamma_{\alcv}$ preserves the decomposition
$\Daff=\Delta_{{\rm aff}+}\cup \Delta_{{\rm aff}-}$ which shows with
the aid of {\it (5)} of the theorem that $\Gamma_{\alcv}$ permutes the
elements $\hat{\alpha}_{0},\cdots , \hat{\alpha}_{l}$ of ${\rm
B}(\alcv)$. Looking at the transformation rule (\ref{affact*}), we see
that the homogeneous part of $\Gamma_{\alcv}$ permutes the distinct
elements $\alpha_{0},\alpha_{1}, \cdots , \alpha_{l}$ of
$\Delta$. Since the relative disposition of these $l+1$ roots is used
to construct the extended Dynkin diagram, $\Gamma_{\alcv}$ can be
identified with a subgroup of the group of Dynkin diagram
automorphisms.

We describe the group $\Gamma_{\alcv}$ more explicitly. For each $i\in
{\cal J}\subset\{1,\cdots,l\}$, we take a subgroup $G_{i}$ of $G$
sharing a maximal torus with $G$ whose roots constitute the set
$\Delta_{i}=\{\alpha\in\Delta ; \alpha(\mu_{i})=0\}$. We think of its
Weyl group $W_{i}$ as a subgroup of $W$ and we take the element
$w_{i}\in W_{i}$ of maximal length with respect to the length
determined by $S_{i}=\{ s_{\alpha} \in W_{i} ; \alpha \in {\rm
B}_{i}\}$ where ${\rm B}_{i}={\rm B}(\Ch)-\{\alpha_{i}\}$. Then a
direct calculation shows the
\begin{pn}
 The group $\Gamma_{\alcv}$ is given by $\Gamma_{\alcv}=\{\, 1\,
\}\cup \{ \,\, e^{-i\mu_{j}\theta}w_{j}w_{0}\in \Waffh \, ; \,\, j\in
{\cal J}\,\, \}$\,.
\label{a.2:propo}
\end{pn}
Note that the embedding $\Pv \hookrightarrow \Waffh$ induces the
isomorphism $\Pv/\Qv\cong \Waffh /\Waff \cong \Gamma_{\alcv}$. Now we
see that the latter part of the proposition 2 holds true.

\newpage
\renewcommand{\theequation}{A.3.\arabic{equation}}\setcounter{equation}{0}

{\large {\bf Appendix 3. Orthogonality of Characters}}\label{a.3}

We give a proof of the orthogonality formula between characters of
integrable representations of the loop group which is essentially due
to \cite{GawKup}. We use notations introduced in Appendix 2.

Let $\Chi_{(\Lmd,k)}(\tau,u)$ be the character of the representation
$L_{(\Lmd,k)}$ of the group $\tilde{LG}_{\!\C}$ defined in
(\ref{defaffch}) and let $\NN_G$ be the moduli space of flat
connections of the (trivial) $G$-bundle on the torus which is realized
by $\V_{\!\C}\lslash (\Qv+\tau\Qv)\semidir W$. We shall prove the
equation \beq \int_{\NN_G}\prod_{j=1}^l\dd u^j\, e^{\pinitau
(k+\coxg)\tr(u-\bar
u)^2}\Chi_{(\Lmd_1,k)}(\tau,u)\overline{\Chi_{(\Lmd_2,k)}(\tau,u)}|\Pi(\tau,u)|^2=\frac{\Vol(\V/\Qv)}{\left(\frac{\tau_2}{2}(k+\coxg)\right)^{\!\frac{l}{2}}}\delta_{\Lmd_1,\Lmd_2}\,,
\label{orthogonality}
\eeq where $u^1,\cdots, u^l$ are the coefficients of
$u=\sum_{j=1}^lu^je_j$ with resect to the orthonormal base $\tr(e_i
e_j)=\delta_{i,j}$ and the volume of the torus $\V/\Qv$ is determined
with respect to the metric `$\tr$'.

To prove it, we make essential use of the Weyl-Kac character formula
\beq \Chi_{(\Lmd,k)}(\tau,u)=\sum_{w\in
W}(-1)^{l(w)}\Theta_{w(\Lmd+\rho),k+\coxg}(u,\tau)\lslash\!\Pi(\tau,u)\,,
\label{WeylKac}
\eeq where $\Pi(\tau,u)$ is the Weyl-Kac denominator defined in
(\ref{defdenom}) and $\Theta_{\Lmd,k}$ is the classical Theta function
of degree $k$ with characteristic $\Lmd$ given by \beq
\Theta_{\Lmd,k}(\tau,u)=\sum_{n\in
k\ttr\Qv+\Lmd}\!\!q^{\frac{1}{2k}(n,n)}e^{2\pi i n(u)}\,.  \eeq

As the fundamental domain of the group $(\Qv+\tau \Qv)\semidir W$ in
$\V_{\!\C}$, we take the following : \beq W\alcv\times (-\tau)\alcv\,,
\eeq where $\alcv$ is an arbitrarily chosen alc\^ove. We introduce the
real coordinates $\psi^j$ and $\phi^j$ related to $u$ by
$u^j=\psi^j-\tau\phi^j$. Since $W\alcv$ is the fundamental domain of
$\Qv$, we have \beq \int_{W\alcv}\prod_{j=1}^ld\psi^j\,e^{2\pi
i(m_1(u)-m_2(\bar u))}=\delta_{m_1,m_2}e^{4\pi
\tau_2m_1(\phi)}\,\Vol(W\alcv)\,, \eeq for $m_1,m_2\in \PP$. With the
aid of this relation, the integral ${\cal I}$ in the left hand side of
(\ref{orthogonality}) is expressed as \beqa {\cal
I}&=&\sum_{w_1,w_2\in W}{\sum_{\stackrel{\scriptstyle n_i\in
(k+\coxg)\ttr\Qv}{+w_i(\Lmd_i+\rho)}}}^{\!\!\!\!\!\!\!\!\!\!\!\!\!\!\!(i=1,2)}(-1)^{l(w_1)+l(w_2)}\!\!\int_{\alcv}\prod_{j=1}^l2d\phi^j\,e^{-2\pi
\tau_2(k+\coxg)\tr\phi^2}q^{\frac{(n_1,n_1)}{2(k+\coxg)}}\bar
q^{\frac{(n_2,n_2)}{2(k+\coxg)}}\nonumber\\ \noalign{\vskip-0.4cm}
&&\hspace{8cm}\times \delta_{n_1,n_2}e^{4\pi \tau_2
n_1(\phi)}\,\Vol(W\alcv)\,.  \eeqa Note that $\delta_{n_1,n_2}\ne 0$
only if $w_1(\Lmd_1+\rho)-w_2(\Lmd_2+\rho)\in (k+\coxg)\ttr\Qv$. Since
$\Lmd_i+\rho$ lies in $(k+\coxg)\ttr\Qv$ for $i=1,2$ and since $\alcv$
is a fundamental domain of $\Waff=\Qv\semidir W$, we must have
$w_1=w_2$ and $\Lmd_1=\Lmd_2$. Then, the integral is given by \beqa
{\cal
I}&=&\Vol(\V/\Qv)\,\delta_{\Lmd_1,\Lmd_2}\!\!\!\!\!\!\!\sum_{\stackrel{\scriptstyle
w\in
W}{n\in(k+\coxg)\ttr\Qv+w(\Lmd+\rho)}}\!\!\!\!\!\!\!\int_{\alcv}\prod_{j=1}^l
2d\phi^j\, e^{-2\pi \tau_2(k+\coxg)\tr(\phi-n)^2}\\
\noalign{\vskip0.2cm}
&=&\Vol(\V/\Qv)\,\delta_{\Lmd_1,\Lmd_2}\int_{\V}\prod_{j=1}^l
2d\phi^j\, e^{-2\pi \tau_2(k+\coxg)\tr(\phi+\cdots)^2}\,.  \eeqa This
Gaussian integral is easily seen to coincide with the right hand side
of (\ref{orthogonality}).

\newpage
\label{reference}

\twocolumn

{\large {\bf Index of Notation}}\label{indexofnotation}

\vspace{0.3cm}

{\small $A'$, $A''$\hfill \pageref{trivwzw}

$A_a$\hfill \pageref{basic}

$A_{\gamma}$\hfill \pageref{page:Agamma}

$A_u$\hfill \pageref{flatconnSO(3)triv}, \pageref{page:holfam},
\pageref{flatu}

$\A_P$\hfill \pageref{ch.3}

$\A^{\circ}_P$\hfill \pageref{page:localparam}

$\A_{ss}$, $\A_{s}$\hfill \pageref{page:Ass}, \pageref{page:As}

$\ad P$, $\ad \PC$, $\ad P^*$\hfill \pageref{page:adP1},\pageref{ch.2}

$\ad_GP$, $\ad_{G_{\!\bf c}}P$\hfill \pageref{ch.2}

$\alpha$\hfill \pageref{page1:roots}, \pageref{page:Weylgrp}

$\tilde{\alpha}$\hfill \pageref{page1:highestroot},
\pageref{page2:highestroot}

$\hat{\alpha}$\hfill \pageref{page1:BCaff}, \pageref{page:hatalpha},
\pageref{page2:hatalpha}

${\rm B}(\Ch)$, ${\rm B}(\alcv)$\hfill
\pageref{page1:BCaff},\pageref{page:theoremWeyl},
\pageref{theoremaffWeyl}

$B=B_0^+$\hfill \pageref{page:B0}, \pageref{page:B}

$B^+$\hfill \pageref{page:B+}

$b$ (ghost)\hfill \pageref{pullback2}

$b(h)$, $b_f(h)$ (`Borel-part' of $h$)\hfill \pageref{ch3partner},
\pageref{flagmeasuregh}

$b_{\Lmd,\lmd}(\tau)$\hfill \pageref{branch}

$\Ch$\hfill \pageref{page:B+}, \pageref{page:theoremWeyl}

$\alcv$\hfill \pageref{page1:Gmalcv}, \pageref{theoremaffWeyl}

$c$ (ghost)\hfill \pageref{pullback2}

$c(\check{h},\check{g})$\hfill \pageref{page1:cocycle},
\pageref{page2:cocycle}

$c_{\lmd}$\hfill \pageref{page:clmd}

$c_{\lmd,V}$\hfill \pageref{page:clmdv}

$c_1(L)$\hfill \pageref{chern}

$\tilde{\chi}(z)$\hfill \pageref{chitilde}

$\Chi_{(\Lmd,k)}(\tau,u)$\hfill \pageref{page:character},
\pageref{WeylKac}

$D_0$\hfill \pageref{basic}

$d_a$, $d_{\mu}$\hfill \pageref{page:da}, \pageref{page:dmu}

$Det_{H_+}$, $Det_{H_+}^*$\hfill \pageref{page:Det}

${\det}'D^{\dag}\!D$\hfill \pageref{pullback}, \pageref{regdet}

$\DiffS$\hfill \pageref{page:DiffS}

$\DiffoS$\hfill \pageref{actiondiffoslc}

$\DiffoSC$\hfill \pageref{page:DiffoSC}

$\sDelta$, $\sDelta_{\Lmd}$ (conformal dimension)\hfill
\pageref{page1:confdim}, \pageref{cwzw}, \pageref{page2:confdim}

${}^a\!\sDelta$\hfill \pageref{zxcv}, \pageref{xcvb}

$\Delta$ (set of roots)\hfill \pageref{page1:roots}, \pageref{a.2}

$\Delta_+$, $\Delta_-$\hfill \pageref{page1:Delta+},
\pageref{page:theoremWeyl}

$\Delta_{{\rm aff}}$\hfill \pageref{page1:Gmalcv}, \pageref{a.2:Daff}

$\Delta_{{\rm aff}+}$, $\Delta_{{\rm aff}-}$\hfill
\pageref{page1:Gmalcv}, \pageref{a.2:Daffpm}

$\delta_{\tau}$\hfill \pageref{defdenom}

$E$\hfill \pageref{vctbdle}

$E(\z)$\hfill \pageref{fcnE}

$e_{\tinI}$, $e^{\tinI}$\hfill \pageref{page:eI}

$e_{\alpha}$\hfill \pageref{page1:roots}

$e_{{\rm a}}$\hfill \pageref{normJJwzw}

$e^{-kI_{\Sigma_0}(A_0,g)}$,
$e^{-kI_{\Sigma_{\infty}}(A_{\infty},g)}$\hfill \pageref{weightWZWD0},
\pageref{weightWZWDinfty}

$\epsilon_i(u)$\hfill \pageref{page:epsilonu}

$\eta^{{\rm a b}}$\hfill \pageref{normJJwzw}

${\cal F}_{H_+}$\hfill \pageref{def:spaceofstates}

$Fl(H)$\hfill \pageref{4.1}

$Fl(P_x)$\hfill \pageref{page:FLPx}

$F_u$, $F_{u,h}$\hfill \pageref{page:Fu}

$F_{v}$, $F_{v,h}$\hfill \pageref{page:Fv}

$\tilde{f}_t$, $f_t^{\flat}$, $f_t^{\sharp}$\hfill
\pageref{page1:ftilde}, \pageref{page2:ftilde}

$\Phi_{\Lambda}$\hfill \pageref{hws}

$\phi$, $\phi^i$ (modular parameter)\hfill
\pageref{flatconnSO(3)triv}, \pageref{relTtau}

$\varphi$\hfill \pageref{page:varphi}

$G$, $G_{\!\C}$\hfill \pageref{ch.2}

$\lieg^{\vee}$\hfill \pageref{ch.2}

$\G_P$, $\GPC$\hfill \pageref{ch.3}

$\G_{P,G}$\hfill \pageref{2.1}

$G_w(z)$, $G_{\bullet}^{z'}\!(z)$\hfill \pageref{page1:Gwz},
\pageref{page2:Gwz}

$\gamma O$\hfill \pageref{ch.4}, \pageref{page:gammaOlmd}

$\gamma O_{\lmd}(f)$\hfill \pageref{page:gammaOlmd}

$\met$\hfill \pageref{ch.1}

$\met_0$\hfill \pageref{ch1.FS}

$\metau$\hfill \pageref{5.1}

$Gr_{\HH_+}$\hfill \pageref{page:Gr}

$GL_{res}$, $G\tilde{L\:\:}_{\!\!\!\!res}$\hfill
\pageref{actiondiffoslc}, \pageref{page:Det}

$\Gmalcv$\hfill \pageref{page1:Gmalcv}, \pageref{Waff'decompo}

$\Gamma_{\Sigma}(g,h)$, $\Gamma_{\Sigma}(A,g,h)$\hfill
\pageref{page:GammaSigma}

$\tilde{\gamma}^*$\hfill \pageref{page:gamma*}

$\gamma_x$\hfill \pageref{page:gammax}

$\gamma\Lmd$, $\gamma{\cal E}$\hfill \pageref{newweight1},
\pageref{newweight2}

$H$ ($=G/Z_G$)\hfill \pageref{ch.2}, \pageref{a.1}, \pageref{a.2}

$H$ ($\subset G/Z_G$)\hfill \pageref{ch.3}

$\HH_{+}$, $\HH_{-}$\hfill \pageref{page:H+}

$\HH^{(\lmd,\pm)}$\hfill \pageref{page:Hlmd-}

$\HH^{(\lmd,V^*)}$\hfill \pageref{page:HlmdV}

${\cal H}^{(G,k)}$\hfill \pageref{page:spstWZW}

$\check{{\cal H}}^{(G,k)}$\hfill \pageref{page2:spstWZW}

${\cal H}^M$\hfill \pageref{invfields}

${\cal H}^{\rm tot}$\hfill \pageref{page:Htot}

$h_a$\hfill \pageref{page:ha(z)}

$h_{\gamma}$\hfill \pageref{page:hgmm}, \pageref{page2:hgmm}

$\tilh$\hfill \pageref{ch1.tilh}, \pageref{ch2.tilh},
\pageref{ch3.tilh}

$h_{\smet}$\hfill \pageref{2.1}

$I_{\Sigma,L}$\hfill \pageref{ch.1}

$I_E$\hfill \pageref{page:clmdv}

$I_{\Sigma}(A,g)$\hfill \pageref{trivwzw}

$I_{\Sigma,P}(A,g)$\hfill \pageref{ch.2}, \pageref{nontrivwzw}

${\cal J}$, ${\cal J}_0$\hfill \pageref{page1:calJ0},
\pageref{page2:calJ0}, \pageref{a.2calJ}

$\Jac\Sigma$\hfill \pageref{page:JacSigma}

$J$ (current)\hfill \pageref{1.1}

$J(\epsilon){\cal O}$, $\bar J(\epsilon){\cal O}$\hfill
\pageref{eqn:Jepsi}

$J^{\sigma}(X)$\hfill \pageref{holJ}

$J^{\sigma}_n(X)$\hfill \pageref{page:Jsgmn}

$J(\tilde{\gamma})$, $\bar J(\tilde{\gamma})$\hfill
\pageref{wzwdef:reprJ}

${\cal J}(\tilde{\gamma})$\hfill \pageref{HCwzwdef:reprJ}

$:(J,J):$\hfill \pageref{defnormWZW}

$k$\hfill \pageref{2.1}

$\kh$\hfill \pageref{neglev}

$K$\hfill \pageref{ch.1}

$L$\hfill \pageref{ch.1}

$\pounds_{h}$ (left translation on $Fl(H)$)\hfill \pageref{page:relR}

$L_v{\cal O}$, $\bar L_v{\cal O}$\hfill \pageref{eqn:Lv}

$L\C^*$\hfill \pageref{actiondiffoslc}

$\tilde{L\C}^*$\hfill \pageref{page:DiffoSC}

$LG$, $LG_{\!\C}$\hfill \pageref{page:LGC}

$\tilde{LG}_{\!\C}$\hfill\pageref{page:GammaSigma}

$LU(1)$\hfill \pageref{page:DiffS}

$\LWZ^k$, $\LWZ^{*k}$\hfill \pageref{weightWZWD0},
\pageref{weightWZWDinfty}

$\LWZ^{-1}\to L\R$\hfill \pageref{page:LWZ-}

${\cal L}\to L\HC/LH$\hfill \pageref{page:calL}

$L_{-\lmd}$\hfill \pageref{def:L-lmd}

$L_{(\Lambda,k)}$, $\overline{L_{(\Lmd,k)}}$\hfill
\pageref{page:LLmdk}

$L_{\Lmd,\lmd}$\hfill \pageref{branchingrule}

$l(w)$\hfill \pageref{page:length}

$\Lambda^*$\hfill \pageref{page:Lmd*}

$\lmd$ (spin)\hfill \pageref{ch.1}

$\lmd$ (weight)\hfill \pageref{page:weightlmd}

$M$\hfill \pageref{ch.3}

$\mu_j$\hfill \pageref{ch3.muj}, \pageref{page2:muj},
\pageref{page:prop}

$\mu$ (type)\hfill \pageref{page:type}

$\mu(E)$ (slope)\hfill \pageref{page:slope}

$N_{T_G}$\hfill \pageref{page:NTG}

$N_0^+$\hfill \pageref{page:B0}

$N^+$\hfill \pageref{page:B0}

$N_{\lmd_1\,\lmd_2}^{\,\lmd_3}$\hfill \pageref{fusioncoeff}

$\NN_P$\hfill \pageref{page:As}

$\NNc_P$\hfill \pageref{page:NNcP}

$\NNc_{P,x}$\hfill \pageref{defdomain}

$\NNc_{P^{(j)},x}$\hfill \pageref{page:NNcPjx}

$\NNc_{{\rm triv},x}$, $\NNc_{{\rm non-triv},x}$

\hspace{0.3cm}(for $H=SO(3)$ and $\Sigma=$ torus)\hfill
\pageref{trivnontriv}

$\lnu_a(u)$\hfill \pageref{page:nuu}

$\nu_{\alpha}(f)$\hfill \pageref{page:nuf}

$\nu_{\rA}(v)$\hfill \pageref{nuA}

$(O_{\lmd})^{\!\lmd}_{\lmd}$\hfill \pageref{branchingrule}

$O_{\lmd}(f)$\hfill \pageref{flagmeasuregh}

$|O_{\Lambda}\rangle$\hfill \pageref{aaaa}

${\cal O}$\hfill \pageref{defcorr}, \pageref{wzwdef:fieldinsertion}

$\tilde{\cal O}$\hfill \pageref{page:Otilde}

$h{\cal O}$\hfill \pageref{page:clmd}, \pageref{intPW}

${\cal O}_{\Sigma}$\hfill \pageref{page:calOSigma}

${\cal O}^{\times}_{\Sigma}$\hfill \pageref{page:calOSigma}

${\cal O}(x)^a$\hfill \pageref{page:calOx}

$\Omega$\hfill \pageref{page:volFLH}

$|\Omega\rangle$\hfill \pageref{page2:volFLH}

$\Omega_{\lmd}(hh^*)$\hfill \pageref{flagmeasure1}

$\Omega_{\Sigma,P}^M(\met\,;{\cal O})$\hfill \pageref{ch3.intmeasure}

$\Omega_{\Sigma,P,x}^M(\met\,;{\cal O}\,O\,)$\hfill \pageref{newform}

$\omega_z$\hfill \pageref{page:ch1.omegaz}

$\PP^{\vee}$\hfill \pageref{page:ha(z)}, \pageref{a.2}

$\Pv_j$\hfill \pageref{page:Pvj}

$\PP$\hfill \pageref{a.2}

$\PPpk$\hfill \pageref{page:Pk+}

$\Pic\Sigma$\hfill \pageref{page:calOSigma}

$P$, $P_{\!\C}$\hfill \pageref{page1:P}, \pageref{page2:P}

$P^{(j)}$\hfill \pageref{page1:calJ0}, \pageref{page2:calJ0}

${\cal P}_{[a]}$\hfill \pageref{adiag}, \pageref{page2:calJ0}

${\cal P}_a$\hfill \pageref{page:Pha}

${\cal P}^{(1)}_F$\hfill \pageref{transflatnontriv}

${\cal P}^{(0)}_u$, ${\cal P}_{00}^{(0)}$\hfill
\pageref{ch4.transtrivu}

${\cal P}^{(1)}_u$\hfill \pageref{transnonflnontrivu}

$\Pi(\tau, u)$\hfill \pageref{defdenom}, \pageref{a.3}

$\psi$, $\psi^i$ (modular parameter)\hfill
\pageref{flatconnSO(3)triv}, \pageref{relTtau}

$\psi^m$ (section of $L_{-\lmd}$)\hfill \pageref{ch4:holsec}

$\psi(z)$ \hfill \pageref{page:psi(z)}

$\psi_n^{(+)}$, $\psi_n^{(-)}$\hfill \pageref{eqn:psi-}

$\psi_n^{\tinI(+)}$, $\psi_{n,\tinI}^{(-)}$\hfill
\pageref{eqn1.4:psipm}

$\psi_S$\hfill \pageref{ch1:monom}

$:\psi^{\sigma}_-(z)\psi^{\sigma}_+(w)\!:$\hfill \pageref{eqn:ord}

$:\psi_m^{(-)}\psi_n^{(+)}\!:$\hfill \pageref{ch1:normalord}

$:\psi_-\psi_+\!:$, $:\psi_-\partial_{\!A}\psi_+\!:$,
$:\partial_{\!A}\psi_-\psi_+\!:$\hfill \pageref{defnorm}

$\QQ^{\vee}$\hfill \pageref{page1:Qv}, \pageref{a.2}

$\QQ$\hfill \pageref{page1:Q}, \pageref{a.2}

$q$ ($=e^{2\pi i \tau}$)\hfill \pageref{page:agen.q}

$\sq$ (charge or weight) \hfill \pageref{page1:confdim},
\pageref{page2:sq}

${}^a\!\sq$\hfill \pageref{zxcv}, \pageref{xcvb}

$\rho$\hfill \pageref{page2:confdim}, \pageref{page:halfroots}

$S$\hfill \pageref{page1:S}, \pageref{page2:S}

$S_{\Lmd,\Lmd'}$\hfill \pageref{modularS}

${\cal S}$, ${\cal S}_d$, ${\rm S}$\hfill \pageref{ch1:monom}

${\cal S}_{\Lmd,\lmd}$\hfill \pageref{page:symfac}

$\Sigma$\hfill \pageref{ch.1}

$\Sigma_{\tau}$\hfill \pageref{ex.ch3}

$\sgmV$, $\sigma_{\!\ad}$\hfill \pageref{page:sigmav}

$\sigma^{(1)}_F$\hfill \pageref{transflatnontriv}

$\sigma^{(0)}_u$, $\sigma^{(0)}_{00}$\hfill \pageref{ch4.transtrivu}

$\sigma^{(1)}_u$\hfill \pageref{transnonflnontrivu}

$\displaystyle{\ooint}$\hfill \pageref{flagmeasuregh}

$T$ (energy-momentum tensor)\hfill \pageref{1.1}

$T$ (maximal torus),$T_H$\hfill \pageref{basic}, \pageref{a.2}

$T_G$\hfill \pageref{page:NTG}, \pageref{a.2}

$\tau$ (geodesic)\hfill \pageref{defnorm}

$\tau$ (modular parameter), $\tau_2$\hfill \pageref{ex.ch3}

$\tr$\hfill \pageref{trivwzw}

$\trP$\hfill \pageref{ch.2}

$\ttr\mu$\hfill \pageref{newweight1}

$\vartheta(\tau, \z)$\hfill \pageref{thetafcn}

$\Theta_{\Lmd, k}(\tau,u)$\hfill \pageref{a.3}

$u$\hfill \pageref{flatconnSO(3)triv}, \pageref{page:holfam},
\pageref{flatu}

$\up$\hfill \pageref{page:upsilon}

$\V$, $\V^*$\hfill \pageref{page:B+}, \pageref{a.2}

$\hat{\V}$, $\hat{\V}^*$\hfill \pageref{page:weightlmd}, \pageref{a.2}

$v$\hfill \pageref{page:vvv}

${\rm W}^{(\lmd,\pm)}_{\Sigma_0}$\hfill \pageref{page:Hlmd-}

$W$\hfill \pageref{page:B+}, \pageref{a.2}

$\Waff$\hfill \pageref{page1:Gmalcv}, \pageref{page:a2.Waff}

$\Waffh$\hfill \pageref{page1:Gmalcv}, \pageref{page:a2.Waff}

$w_0$\hfill \pageref{page:Lmd*}, \pageref{page:theoremWeyl}

$w_jw_0$\hfill \pageref{page1:wjw0}, \pageref{page:NNcPjx},
\pageref{a.2:propo}

$w_{\bullet}$ \hfill \pageref{ch1:monom}

$y_u$\hfill \pageref{heckeg1so3}

$Z_G$\hfill \pageref{ch.2}

$Z_{\Sigma, P}(\met\,;{\cal O})$\hfill \pageref{ch.3}

$Z_{\Sigma,L}(\met,A\,;{\cal O})$\hfill \pageref{defcorr}

$Z_{\Sigma,P}(\met,A\,;{\cal O})$\hfill \pageref{2.1}

$Z^{G,k}(\tau, u)$\hfill \pageref{5.2.1}

$Z^{-\tilk-2\lieh^{\vee}}(\tau,u)$\hfill \pageref{5.2.2}

$Z_{\Sigma_0}(A_0;{\cal O}_0)$,
$Z_{\Sigma_{\infty}}\!(A_{\infty};{\cal O}_{\infty})$\hfill
\pageref{wf}, \pageref{2.3}

$Z_{\Sigma,P}^{\rm gh}(\met,A\,;{\cal O})$\hfill \pageref{pullback2}

$Z^{\rm gh}(\tau, u)$\hfill \pageref{toruspartgh}

$Z_{\Sigma,P}^{\rm tot}(\met,A\,;{\cal O})$\hfill \pageref{4.1.fl},
\pageref{page:Ztot}

$\z$\hfill \pageref{ex.ch3}

$(\,\,,\,\,)$ on $\Omega^{p,q}(\Sigma,\ad \PC)$\hfill \pageref{3.2}

$\langle\,\,,\,\,\rangle$ pairing of $\Omega^{1,0}_{(\ad \PC)^*}$ and
$\Omega^{0,1}_{\ad \PC}$\hfill \pageref{ch3:pair}

$(\,\,,\,\,)_{Fl(H)}$ on $H^0(Fl(H),L_{-\lmd})$\hfill
\pageref{page:volFLH}

$(\,\,,\,\,)$ on $\hat{\V}^*$\hfill \pageref{scalarproduct}

\end{document}